SISSA - International School for Advanced Studies

# Evolution of low mass stars: lithium problem and $\alpha$-enhanced tracks and isochrones

Dissertation Submitted for the Degree of
"Doctor Philosophiæ"

**Supervisors**
Prof. Alessandro Bressan
Dr. Paolo Molaro
Dr. Léo Girardi

**Candidate**
Xiaoting FU

*We are the dust of stars.*

# Contents

# Abstract

PARSEC (PAdova-TRieste Stellar Evolution Code) is the updated version of the stellar evolution code used in Padova. It provides stellar tracks and isochrones for stellar mass from 0.1 $M_{\odot}$ to 350 $M_{\odot}$ and metallicity from Z=0.00001 to 0.06. The evolutionary phases are from pre-main sequence to the thermally pulsing asymptotic giant branch.

The core of my Ph.D. project is building a new PARSEC database of $\alpha$ enhanced stellar evolutionary tracks and isochrones for *Gaia*. Precise studies on the Galactic bulge, globular cluster, Galactic halo and Galactic thick disk require stellar models with $\alpha$ enhancement and various helium contents. It is also important for extra-Galactic studies to have an $\alpha$ enhanced population synthesis. For this purpose we complement existing PARSEC models, which are based on the solar partition of heavy elements, with $\alpha$-enhanced partitions. We collect detailed measurements on the metal mixture and helium abundance for the two populations of 47Tuc (NGC 104) from literature, and calculate stellar tracks and isochrones with these chemical compositions that are *alpha*-enhanced. By fitting precise color-magnitude diagram with *HST ACS/WFC* data from low main sequence till horizontal branch, we calibrate some free parameters that are important for the evolution of low mass stars like the mixing at the bottom of the convective envelope. This new calibration significantly improves the prediction of the RGB bump brightness. We also check that the He evolutionary lifetime is correctly predicted by the current version of PARSEC . As a further result of this calibration process, we derive an age of 12.00±0.2 Gyr, distance modulus $(m\text{-}M)_0=13.22^{+0.02}_{-0.01}$, reddening $E(V\text{-}I)=0.035^{-0.008}_{+0.005}$, and red giant branch mass loss around 0.172 $M_{\odot}$ ~ 0.177 $M_{\odot}$ for 47Tuc. We apply the new calibration and $\alpha$-enhanced mixtures of the two 47Tuc populations ( [$\alpha$/Fe] ~0.4 and 0.2) to other metallicities. The new models reproduce the RGB bump observations much better than previous models, solving a long-lasting discrepancy concerning its predicted luminosity. This new PARSEC database, with the newly updated $\alpha$-enhanced stellar evolutionary tracks and isochrones, will also be part of the new product for *Gaia*. Besides the $\alpha$ enhanced metal mixture in 47Tuc, we also calculate evolutionary tracks based on $\alpha$ enhanced metal mixtures derived from ATLAS9

APOGEE atmosphere model. The full set of isochrones with chemical compositions suitable for globular clusters and Galactic bulge stars, will be soon made available online after the full calculation and calibration are performed.

PARSEC is able to predict the evolution of stars for any chemical pattern of interest. Lithium is one of the most intriguing and complicated elements. The lithium abundance derived in metal-poor main sequence stars is about three times lower than the value of primordial Li predicted by the standard Big Bang nucleosynthesis when the baryon density is taken from the CMB or the deuterium measurements. This disagreement is generally referred as the "cosmological lithium problem". We here reconsider the stellar Li evolution from the pre-main sequence to the end of the main sequence phase by introducing the effects of convective overshooting and residual mass accretion. We show that $^7$Li could be significantly depleted by convective overshooting in the pre-main sequence phase and then partially restored in the stellar atmosphere by a tail of matter accretion which follows the Li depletion phase and that could be regulated by extreme ultraviolet (EUV) photo-evaporation. By considering the conventional nuclear burning and microscopic diffusion along the main sequence, we can reproduce the Spite plateau for stars with initial mass $m_0 = 0.62 - 0.80\ M_{\odot}$, and the Li declining branch for lower mass dwarfs, e.g, $m_0 = 0.57 - 0.60\ M_{\odot}$, with a wide range of metallicities (Z=0.00001 to Z=0.0005), starting from an initial Li abundance $A(\mathrm{Li}) = 2.72$. This environmental Li evolution model also offers the possibility to interpret the decrease of Li abundance in extremely metal-poor stars, the Li disparities in spectroscopic binaries and the low Li abundance in planet hosting stars.

# Acronyms

Here, we provide a brief list of acronyms.

| | |
|---|---|
| ACS | Advanced Camera for Surveys; |
| AGB | asymptotic giant branch; |
| BBN | Big Bang Nucleosynthesis; |
| CMB | cosmic microwave background; |
| CMD | color-magnitude diagram; |
| dSph | dwarf spheroidal; |
| EMPS | extremely metal-poor stars; |
| EOS | equation of state; |
| EOV | envelope overshooting; |
| EUV | extremely Ultra Violet; |
| GC | globular cluster; |
| GCRs | Galactic cosmic rays; |
| HB | horizontal branch; |
| HRD | Hertzsprung-Russell diagram; |
| HST | Hubble Space Telescope; |
| ISM | interstellar medium; |
| IMF | initial mass function; |
| MLT | mixing length; |
| MS | main sequence; |
| PARSEC | PAdova-TRieste Stellar Evolution Code; |
| POP II | population II; |
| POP I | population I; |
| PMS | pre-main sequence; |
| RGB | red giant branch; |
| RGBB | red giant branch bump; |
| SFH | star formation history; |
| TO | turn-off; |
| TP-AGB | thermally-pulsing asymptotic giant branch; |
| YSO | young stellar object; |
| ZAMS | zero-age main sequence; |
| ZAHB | zero-age horizontal branch. |

# Chapter 1

# Introduction

Stars are the archaeology book of the cluster or the galaxy they reside in. Understanding their composition, evolution and nucleosynthesis is the key to probing the evolution of the host galaxies and clusters, the history of their formation, as well as the nature of the stars themselves. In the introduction chapter, I will briefly review the historic progress in understanding stars (Chap. 1.1), summarize modern efforts on numerical stellar evolution, including basic calculations and stellar codes currently being adopted (Chap. 1.2), and then introduce how my Ph.D work goes a bit further in these fields (Chap. 1.3 and 1.4).

## 1.1 Historic progress in understanding stars

In 1609, for the first time in the human history, at his home in Padova, Galileo Galilei observed stars in the Milky Way with a telescope. For the next 300 years astronomers routinely observed the Galactic stars, recorded their brightness and location, and classified them into different categories. However, the underlying physics of stars was still poorly understood until the modern physics shed light on it.

In the early 19th century when astronomers started to measure the distances of stars with parallaxes, they found it impossible to investigate the composition and characteristics of stars with the foreseeable technology because, a part for the Sun, even the nearest one is too far away to reach. In 1835 the French philosopher Auguste Comte wrote down his prediction on stellar study:

*"We understand the possibility of determining their shapes, their distances, their sizes and their movements; whereas we would never know how to study by any means their chemical composition, or their mineralogical structure, and, even more so, the nature of any organized beings that might live on their surface."* (Hearnshaw, 2010).

Today this quotation is kind of popular in the community because though we can not physically reach the stars still, astronomers have proved that we do can study their structure and their elemental abundances, from both observational efforts and theoretical calculations.

The first work connecting the "pure observational astronomy" and stellar astrophysics is the so-called Hertzsprung-Russell diagram (HRD; i.e., Rosenberg, 1910; Russell, 1914). Danish astronomer Dane Ejnar Hertzsprung (1873–1967) and American astronomer Henry Norris Russell (1877–1957) independently found that, when their luminosity and their spectral type are plotted together in a figure, the stars are broadly divided into two sequences: dwarf and giant stars. This powerful diagram, which is still used today to study stars and stellar clusters, was a hint indicating that the sequences shown in the diagram track the evolution of stars (DeVorkin, 1984).

The HRD inspired an English astronomer, Sir Arthur Stanley Eddington (1882-1944), when Russell visited London and presented his diagram at a meeting of the Royal Astronomical Society in 1913 (Eisberg, 2002). At the time, Eddington was the chief assistant of the Royal Greenwich Observatory. In 1926 Eddington offered the first stellar model able to explain the physics behind the HRD, in his book *"The Internal Constitution of the Stars"* (Eddington, 1926). This model, almost entirely based on the theory of radiative transfer, described the relationship between the stellar mass and luminosity, and successfully reproduced the empirical H-R diagram derived from observational data. Also in this model, based on the formulae of the equilibrium structure of a self-gravitating gaseous sphere, Eddington first proposed that the star is supported not only by gas pressure against gravitational collapse but also by radiation from a source of energy in the center of the star. However, at that time, it was not yet clear what exactly could power the luminosity of the stars.

One of the proton-proton nuclear reactions (pp chain), $p + p \rightarrow {}^{2}_{1}D + e^{+} + \nu_e$ , was proposed as the possible energy source of the Sun by George Gamow (1904–1968) and Carl Friedrich von Weizsäcker (1912–2007). After attending a meeting with Gamow in 1938, Hans Bethe (1906–2005), a German American nuclear physicist, improved the pp chain (as in the schematic diagram of Fig. 1.1 ) and announced the carbon-nitrogen-oxygen cycle (CNO cycle, as in the schematic diagram of Fig. 1.2), in 1939. The energy released in these two sets of nuclear reactions showed agreement with the observed stellar luminosites. Bethe's work was a breakthrough in the understanding of the energy source of the stars, and Bethe was awarded the Nobel Prize in Physics in 1967.

The discovery of the pp chain and the CNO cycle solved the doubts of Eddington on the energy source of stars. But it led to another question: Helium could be synthesized in the stellar interior, but where do the other elements in the stars come from? George Gamow used to believe that all elements were synthesized during the Big Bang, and that stars' composition should reflect the distribution that resulted as the Universe cooled (this belief allowed him to predict the existence of the microwave background radiation, though). Sir Fred Hoyle (1915–2001), a British astronomer, disliked the big bang theory so much (he named the theory "big bang" in order to taunt over it) and argued that stars should be considered the generators of all heavy elements instead. With his original idea, in 1957 the famous $B^2FH$ paper (Burbidge et al., 1957) refreshed people's awareness of the elements origin, and set the cornerstone of stellar nucleosynthesis. In this paper, Margaret Burbidge, Geoffrey Burbidge, William Fowler, and Fred Hoyle published an extensive survey of nuclear reactions that could contribute to element building in stars. They were able to predict not only the observed isotopic abundances which can be identified with the absorption lines in optical spectroscopic observations, but also the existence of the p-process (nucleosynthesis process which are responsible for the proton-rich isotopes), r-process (rapid neutron capture process), and s-process (slow-neutron-capture-process), to account for many of the elements heavier than iron. In contrast to Gamow's static chemical composition, $B^2FH$ predicts that the chemical composition evolve from that in the early Universe. They argue that as a star dies, it enriches the interstellar medium (ISM) with "heavy elements" it produces, from which newer stars are formed. This is consistent with our modern picture of galactic chemical evolution.

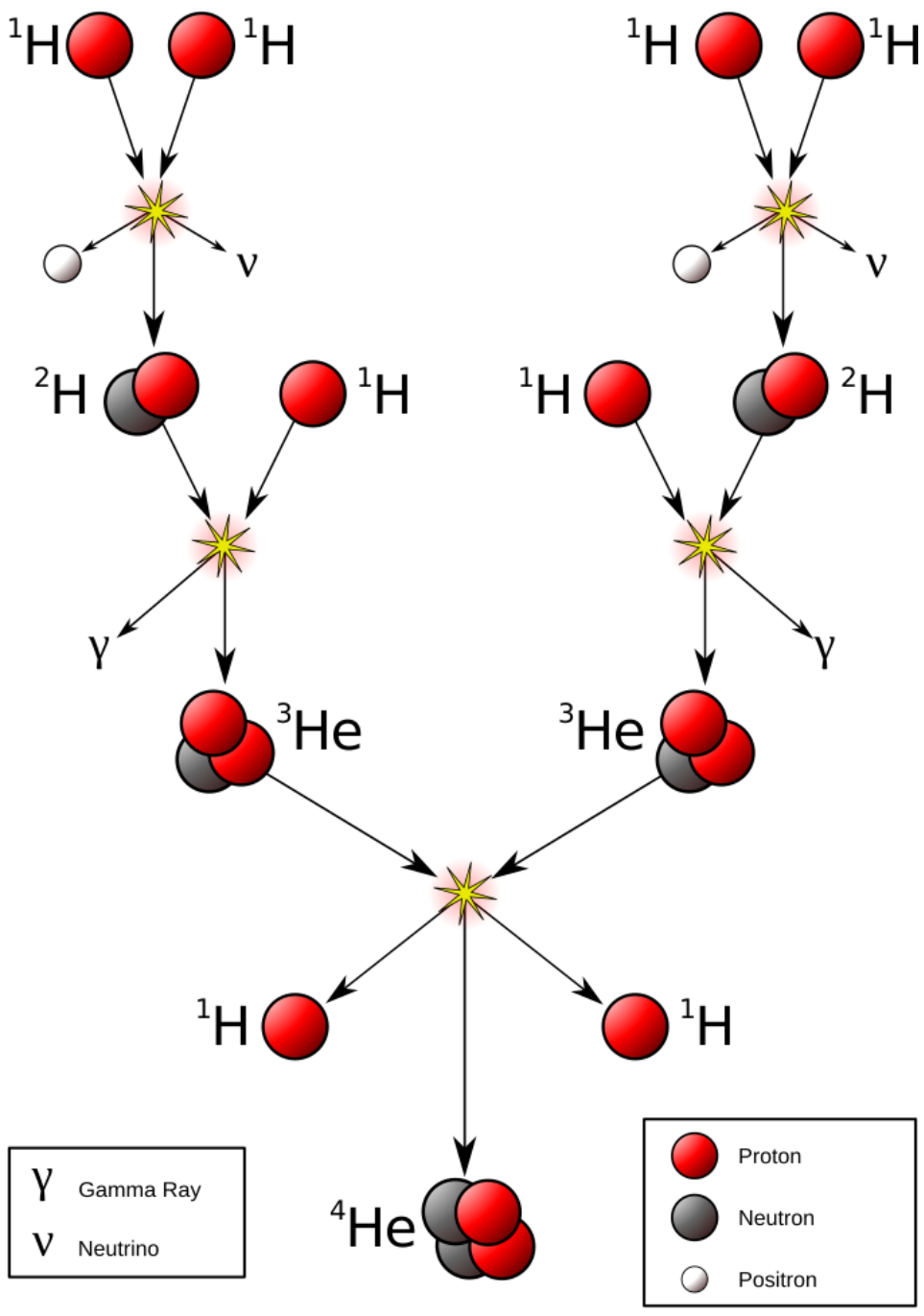

Figure 1.1: Schematic of PP chain. The figure is made by Borb, under CC BY-SA 3.0. https://commons.wikimedia.org/w/index.php?curid=680469

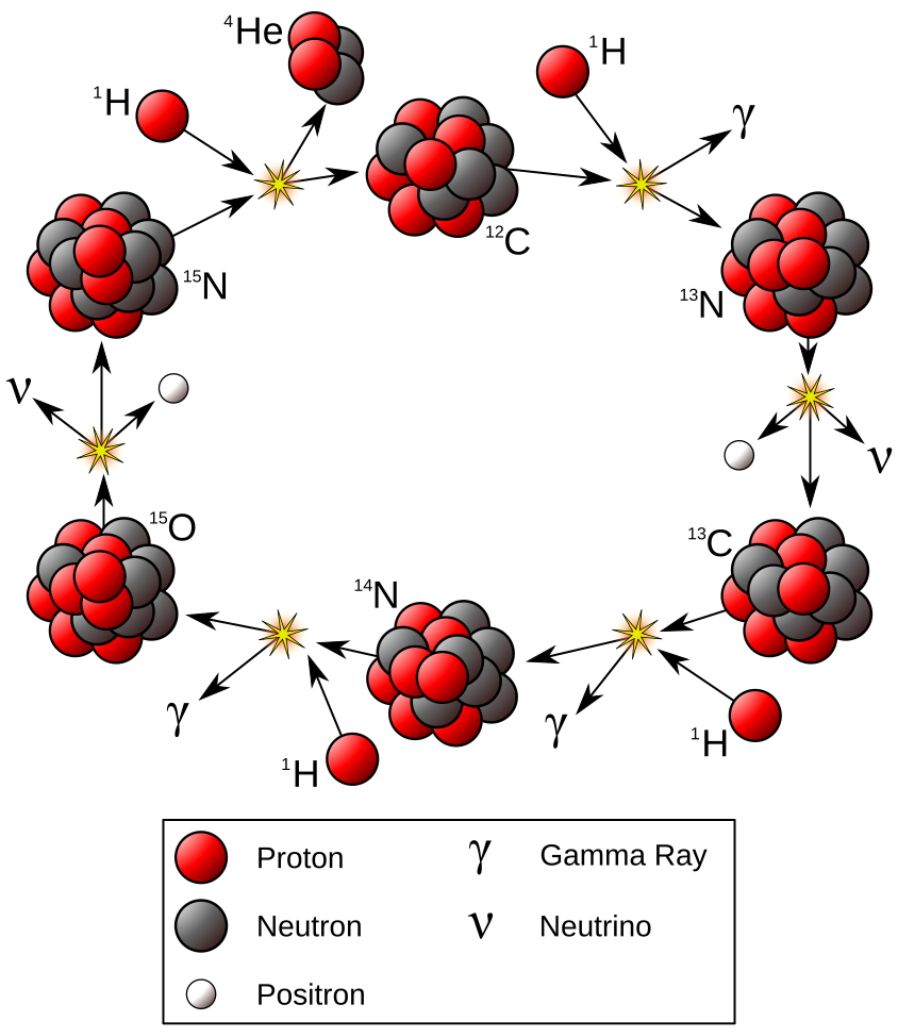

Figure 1.2: Schematic of CNO cycle. By Borb, CC BY-SA 3.0, https://commons.wikimedia.org/w/index.php?curid=691758

Today many evidence show that the big bang nucleosynthesis (BBN) produced helium-4 ($^4$He), deuterium (D), helium-3 ($^3$He), and a very small amount of lithium-7 ($^7$Li), which is an isotope of lithium (Fields, Molaro & Sarkar, 2014). All elements heavier than lithium are synthesized in stars, either during their placid evolution or during the phase of supernova explosion.

## 1.2 Modern stellar evolution models

Modern stellar models are developed on the basis of the aforementioned efforts and started from the modeling of the Sun. The first evolutionary model for the Sun was published in 1957 (Schwarzschild, Howard & Härm, 1957), beginning from the zero-age main sequence. The mixing length theory, as one of the foundation of standard stellar models, was first applied to solar model in 1958 in a paper written in German (Böhm-Vitense, 1958). The MLT parameter and the helium abundance are adjusted to produce a model with observed solar radius and luminosity, then adopted in the models of the other stars (Demarque & Larson, 1964). The MLT involves stellar parameters that can not be observed directly and can only be calibrated by observations of the stellar radius, luminosity and effective temperature. The semi-empirical MLT developed in Böhm-Vitense (1958) is still the most popular MLT in present-day. Here is the basic idea of MLT: A unit of fluid elements travels over a mean free path and dissolves into the surrounding medium, sharing its energy content with the surrounding matter. This mean free path $l_{\mathrm{MLT}}$ is called the mixing length, which is customarily parameterized in units of the pressure scale as $l_{MLT} = \alpha_{MLT} \times H_P$. With $\alpha_{\mathrm{MLT}}$, the convective energy transport can be analytically formulated and used to solve the equations of stellar structure.

Sandage (1962) was the first author to use the isochrone method in studying a star cluster. The isochrone is ”similar” to stellar evolutionary track in HRD, but instead of displaying the evolution of star of a given mass, it illustrates effective temperature and luminosity of stars with different masses at the same age. The Isochrone method is very useful to understand the formation history and evolution of star clusters (Demarque & Larson, 1964; Sandage & Eggen, 1969).

The precursor of the modern stellar evolution codes was developed by Rudolf Kippenhahn, Andreas Weigert, and Emmi Hofmeister (Kippenhahn, Andreas; & Hofmeister, 1967). They built numerical FORTRAN programs to solve the equations of stellar structure. This was the first step to calculate stellar evolution models, which could be extended to derive all other physical properties of stars and generate observable quantities to be tested in observations. In the following I will briefly summarise the formulations used to calculate stellar structure and evolution, then introduce our stellar evolution code `PARSEC` (PAdova and TRieste Stellar Evolution Code) and other codes.

### 1.2.1 Equations of stellar structure and evolution

Solving the equations of a stellar structure is usually based on the following assumptions:
The star is assumed to be spherical and effects of magnetic fields, tidal forces, and rotation are generally neglected (though more recent models include these effects). All quantities thus depend only on their distance from the center of the star. The interior matter is assumed to be in thermal (negligible variation of temperature), mechanical (quasi-static change of radius) and chemical (negligible variation of chemical composition) equilibrium. Of course this cannot hold for the whole star but only on any selected interior volume element and, for this reason, the star is said to be in Local Thermodynamic Equilibrium (LTE). In LTE all thermodynamic quantities depend only on state functions (like temperature and pressure) in the given interior element.

The equations of stellar structure in LTE are:

$$\text{Mass conservation:} \quad \frac{\partial m}{\partial r} = 4\pi r^2 \rho, \tag{1.1}$$

$$\text{Hydrostatic equilibrium:} \quad \frac{\partial P}{\partial r} = -\rho\frac{Gm}{r^2}, \tag{1.2}$$

$$\text{Thermal conservation:} \quad \frac{\partial L}{\partial r} = 4\pi r^2 \rho q, \tag{1.3}$$

$$\text{Radiative energy transfer:} \quad \frac{\partial T}{\partial r} = -\frac{3}{4ac}\frac{\kappa\rho}{T^3}\frac{L}{4\pi r^2}, \tag{1.4}$$

$$\text{Convective energy transfer:} \quad \frac{\partial T}{\partial r} = \frac{C_V}{C_P}\frac{T}{P}\frac{\partial P}{\partial r}, \tag{1.5}$$

$$\text{Nuclear reaction:} \quad \frac{\partial X_i}{\partial t} = \frac{A_i}{\rho}(\Sigma r_{ji} - \Sigma r_{ik}), \quad i = 1, ..., I. \tag{1.6}$$

The symbols are: stellar mass $m$, radial coordinate $r$, density $\rho$, pressure $P$, Newton's gravitational constant G ($G = 6.6738 \times 10^{-11}\,\mathrm{m^3\,kg^{-1}\,s^{-2}}$), luminosity $L$, nuclear energy generation rate $q$, opacity $\kappa$, radiation constant $a$ ($a = 7.5657 \times 10^{-16}\,\mathrm{J\,m^{-3}\,K^{-4}}$), light velocity $c$ ($c = 2.9979 \times 10^8\,\mathrm{m\,s^{-1}}$), specific heat at constant pressure $C_\mathrm{P}$, specific heat per unit volume $C_\mathrm{V}$, mass fraction $X_\mathrm{i}$ of any element $i$ considered, the nuclear reaction rate $r_{ji}$ between elements $i$ and $j$, and the corresponding atomic mass $A_i$ of the element $i$. Equation 1.4 and 1.5 are temperature gradients under energy transfer of radiation and convection, respectively.

Among all the above quantities, the radial coordinate $r$ and the evolutionary time $t$ are independent variables because the spherical symmetry assumption makes the structure determination a one-dimensional problem, all functions depending only upon the distance from the center ($r$), and the evolutionary time $t$.

However in practice the equations are solved with $m$ (Lagrange coordinate) as the independent variable rather than with $r$ (Euler coordinate). This is because during the evolution the total mass of the star remains almost constant except for the red giant branch (RGB) and asymptotic giant branch (AGB) phases and for the massive stars. Whilst the stellar radius can change by hundreds or even thousands of times, compared to the radius on the main sequence. So in practice the above equations of stellar structure are re-written as:

$$\frac{\partial r}{\partial m} = \frac{1}{4\pi r^2 \rho}, \tag{1.7}$$

$$\frac{\partial P}{\partial m} = -\frac{Gm}{4\pi r^4}, \tag{1.8}$$

$$\frac{\partial L}{\partial m} = q, \tag{1.9}$$

$$\frac{\partial T}{\partial m} = -\frac{3}{4ac}\frac{\kappa}{T^3}\frac{L}{16\pi^2 r^4}. \tag{1.10}$$

$$\frac{\partial T}{\partial m} = \frac{C_V}{C_P}\frac{T}{P}\frac{\partial P}{\partial m}. \tag{1.11}$$

where $m$ varies between 0 and the total mass of the star. The formula of nuclear reaction rate is the same as in Equation 1.6. Each point in the spherical structure can be labeled by the value $m$ of the mass interior to that point and any property inside the star can be expressed as a function of $m$ and $t$. One advantage of using $m$ as the independent variable is the specification of chemical compositions. Imaging a star without nucleosynthesis, when it expands or contracts, the composition parameter $X_{\rm i}(m)$ remains constant because the coordinate $m$ moves with the gas sphere, whereas the function $X_{\rm i}(r)$ changes simply by virtue of describing a different mass element.

In addition, $T$, $P$, and $\rho$ are related to one another through the equation of state (EOS). The EOS, opacities and nuclear reaction rates are generally calculated separately and are used in the numerical alculations via interpolation tables or fitting functions.

In order to solve the above equations one requires suitable boundary conditions, that are described below.

### 1.2.2 Boundary conditions

One needs to handle a two boundary condition problems in the calculation of stellar structure, in the stellar center and at the surface.

In the stellar center, two boundary conditions are set with:

$$m = 0 : \ r = 0 \tag{1.12}$$

$$m = 0 : \ L = 0 \tag{1.13}$$

To be distinguished from the local variables, hereafter we use $\mathscr{M}$ and $\mathscr{L}$ for the surface value of mass $M$ and luminosity $L$. For the surface boundary condition, the simplest choices are :

$$m = \mathscr{M} : \ P = 0 \tag{1.14}$$

$$m = \mathscr{M} : \ t = 0 \tag{1.15}$$

The choices of equations 1.14 and 1.15 are commonly referred as the "zero boundary conditions". A surface effective temperature, $T_{eff}$, can be defined from the surface luminosity $\mathscr{L}$ and the radius R at $\mathscr{M}$ by

$$\sigma T_{eff}^4 = \frac{\mathscr{L}}{4\pi R^2} \tag{1.16}$$

where $\sigma$ is the StefanBoltzmann constant. It is possible to show that, as long as radiation is the main energy transfer mechanism, say, for stars on the upper main sequence, the zero boundary condition is adequate for the stars and provide a value of the surface effective temperature without significant error.

For stars with lower surface temperature the zero boundary condition is no longer adequate. If hydrogen is largely neutral at the surface, the opacity is very large under the surface. So the temperature gradient must accordingly be determined at the surface with convection. However, the adiabatic approximation to the convective temperature gradient (Equation 1.11) is not suitable for the low-density surfaces of cool stars (e.g., RGB, AGB). Instead, the temperature gradient must be considered super-adiabatic and radiation may carry a sufficient energy flux. To set the boundary conditions in this case, one should set a "photospheric" boundary condition which is estimated from the theory of stellar atmospheres. The photosphere is defined where the Rosseland mean optical depth $\tau_{Ross}$ in a given atmosphere model equals 2/3:

$$\tau_{Ross} \equiv \int_R^\infty \kappa\rho dr = \frac{2}{3} \tag{1.17}$$

and we have the following boundary conditions:

$$T = T(\tau_{Ross} = 2/3) \tag{1.18}$$

$$P = P(\tau_{Ross} = 2/3) \tag{1.19}$$

In general the external boundary conditions are regarded as "tentative boundary conditions", yielding a trial value of the effective surface temperature $T_{\rm eff}$ (from the Planck law) and a trial value of the surface gravity $g$ (from the law of gravity). These tentative boundary conditions are used to integrate inward through the hydrogen ionization zone, up to where one encounters the internal solution obtained by the integration of the full system of the interior stellar structure (equation 1.7, 1.8, 1.9, 1.10, and 1.11). This internal system is relaxed to the true solution, by an iterative method that accounts for the match with the external solutions, finally providing the whole structure from the center to the surface.

### 1.2.3 Opacity

The opacity, $\kappa$, is a physical quantity that describes the reduction ($dI$) of the radiation intensity $I$ by the matter along the propagation path length $dr$:

$$dI = -\kappa I\rho dr. \tag{1.20}$$

In the stellar matter, the main sources of opacity $\kappa$ are the following:

- **Electron scattering:** The opacity induced by electron scattering is independent of the frequency. It represents the minimum value of the opacity in stars at temperature $\gtrsim 10^4$ K (Ezer & Cameron, 1963). If the stellar matter is fully ionized then the opacity is

$$\kappa_\nu = \frac{8\pi}{3}\frac{r_e^2}{\mu_e m_u} = 0.20(1+X)\mathrm{cm}^2\mathrm{g}^{-1}, \tag{1.21}$$

where $r_e$ is the classical electron radius and $m_u$ is the atomic mass unit (= 1amu = $1.66053\times10^{-24}$g). $X$ is the mass fraction of Hydrogen. The second equality is assumed for elements heavier than helium with molecular weight-to-charge number-ratio $(Z_i\,\mu_i/Z_i) \approx 2$. Because of the frequency independence of the electron scattering opacity, the Rosseland mean (defined below) is the same as in equation (1.21).

- **Free-free absorption:** When a free electron passes by an ion, the system formed by the electron and the ion can absorb (or emit) radiation. The opacity resulted from this process is

$$\kappa_\nu \sim Z^2\rho T^{-1/2}\nu^{-3}, \tag{1.22}$$

and the corresponding Rosseland mean is

$$\kappa_{\rm ff} \propto \rho T^{-7/2}. \tag{1.23}$$

It is known as the Kramers opacity.

- **Bound-free absorption:** A bound system composed of electrons and a nucleus can absorb the radiation and releases an electron (or a successive release of electrons). The bound-free absorption opacity also takes the above Kramers opacity form, but with different coefficients and different dependence on the chemical composition as:

$$\kappa_{bf} \approx \kappa_{ff}\frac{10^3 Z}{X+Y}. \tag{1.24}$$

Compared to the bound-free absorption, the free-free absorption is more important at low metallicity.

- **Bound-bound absorption:** A bound system can absorb the radiation and an electron is excited to a higher bound state. The bound-bound absorption can become a major contributor to the opacity at $T < 10^6$ K.

- $H^-$: The $H^-$ ions are formed through the reaction $H + e^- \rightleftharpoons H^-$. The dissociation energy is 0.75 eV, which corresponds to 1.65 $\mu m$. The condition for the formation of $H^-$ is that the temperature should be 3000 K $\lesssim$Teff $\lesssim$ 8,000 K. The upper limit of the temperature range ensures a significant amount of neutral Hydrogen, while the lower limit prevents the amount of free electrons, contributed by the metals and neutral Hydrogen, from being depleted by molecules ($H_2$, $H_2O$, etc.). The approximation form of $H^-$ opacity is $\kappa_{H^-} \propto Z\rho^{1/2}T^9$. $H^-$ is the major opacity source in the Solar atmosphere. $H^-$ can be treated in the same way as normal atoms or ions, to take into account its contribution to the bound-free and free-free opacities.

- **Molecules:** In the outer atmosphere of stars with effective temperatures Teff $\leq$ 4000 K, molecules dominate the opacity. Among them, the most important ones are $H_2$, CN, CO, $H_2O$, and TiO. The related transitions are vibrational and rotational transitions, as well as photon dissociation. An approximate form for the opacity of molecules is $\kappa_{\rm mol} \propto T^{-30}$. To appreciate the complexity of the molecular opacities, the reader can refer to the figure 4 of Marigo & Aringer (2009).

- **Dust:** The dust opacity contribution can be important in cool giants or supergiants. Suppose a dust particle is of size $A_d$ and of density $\rho_d$, and a fraction $X_d$ of the stellar material mass is locked in dust, then the dust opacity is:

$$\kappa_d = \left\{ \begin{array}{ll} \frac{3X_d}{4A_d\rho_d}, & \lambda << A_d \\ \left(\frac{\lambda}{A_d}\right)^{-\beta}, & \lambda \geq A_d \end{array} \right\}, \tag{1.25}$$

with $\beta = 4$ for simple Rayleigh scattering with dust particles in smooth spherical shape of constant size $A_d$, and $\beta \approx 1 - 2$ for the compound dust with a size and shape distribution (Owocki, 2013; Li, 2005).

- **Rosseland mean opacity:** As we have seen, only the opacity by electron scattering is frequency independent, while the others may change abruptly with the frequency. It is quite useful to define some frequency averaged opacity to evaluate the global effect to the radiation field by the stellar matter. One of the mostly used is the Rosseland opacity, which is defined as

$$\frac{1}{\kappa_{\rm Ross}} = \frac{\int_0^\infty \frac{1}{\kappa_\nu} u_\nu d\nu}{\int_0^\infty u_\nu d\nu}, \tag{1.26}$$

where $u_\nu$ is the radiation energy density. By nature of this harmonic average, the Rosseland mean is weighted towards the frequency ranges of maximum energy flux throughput. For example, the stellar photosphere is defined at a radius where $\kappa_{\rm Ross} = 2/3$. As $e^{-2/3} \approx 0.5$, it means half of the flux can escape from this radius freely. Another common average/representative opacity is the opacity at the wavelength of 5000 Å, but this is not useful for stellar evolution calculation.

### 1.2.4 PARSEC (PAdova and TRieste Stellar Evolution Code)

PARSEC (PAdova-TRieste Stellar Evolution Code) is the result of a thorough revision and update of the stellar evolution code used in Padova (Bressan et al., 1993; Bertelli et al., 1994; Fagotto et al., 1994; Girardi et al., 1996; Girardi et al., 2000; Marigo et al., 2001; Girardi et al., 2002; Marigo & Girardi, 2007; Marigo et al., 2008; Bertelli et al., 2009).

The series of Padova stellar database is one of the most popular evolutionary tracks and isochrones used by the community. From the early models containing several chemical compositions (Bressan et al., 1993; Bertelli et al., 1994; Fagotto et al., 1994; Girardi et al., 2000) to the large grids of isochrones online (starting from Girardi et al., 2002), the code has been complemented constantly with up-to-date input physics. Marigo et al. (2001) extended the evolutionary models to initial zero-metallicities, covering stellar mass from 0.7 $M_\odot$ to 100 $M_\odot$ . Detailed thermally-pulsing asymptotic giant branch (TP-AGB) is included by Marigo & Girardi (2007); Marigo et al. (2008). Low-temperature gas opacities Æ SOPUS code was introduced to the model in 2009 (Marigo & Aringer, 2009). Bertelli et al. (2009) allow varies helium content from 0.23 to 0.46 for low mass stars from ZAMS to the TP-AGB. In 2010 Girardi et al. (2010) made corrections for the low mass, low metallicity AGB tracks. And in 2012, with all the update in EOS, opacity, and nucleosynthesis, Bressan et al. (2012) presented the new version in this series, PARSEC .

PARSEC is widely used in the astronomical community. It provides a population synthesis tool to study resolved and unresolved star clusters and galaxies (e.g. Bruzual & Charlot, 2003), and offers reliable model for many other field of studies, such as to derive black hole mass when observing gravitational wave (e.g. Spera, Mapelli & Bressan, 2015; Belczynski et al., 2016), to study the chemical evolution in galaxies (e.g. Ryde et al., 2015; Vincenzo et al., 2016), to get host star parameters for exoplanet (Santos et al., 2013; Maldonado et al., 2015, etc.), to explore the mysterious "cosmological lithium problem" (Fu et al., 2015), to derive the main parameters of star clusters (for instance, Donati et al., 2014; Borissova et al., 2014; San Roman et al., 2015) and Galactic structure (e.g. Küpper et al., 2015; Li et al., 2016; Balbinot et al., 2016; Ramya et al., 2016), to study dust formation (e.g. Nanni et al., 2013; Nanni et al., 2014), to constrain dust extinction (e.g. Schlafly et al., 2014; Schultheis et al., 2015; Bovy et al., 2016), and to understand the stars themselves (e.g. Kalari et al., 2014; Smiljanic et al., 2016; Gullikson, Kraus & Dodson-Robinson, 2016; Reddy & Lambert, 2016; Casey et al., 2016), etc.

There are now four versions of PARSEC available online [1]: The very first version PARSEC v1.0 provides isochrones for 0.0005≤Z≤0.07 (-1.5≤[M/H]≤+0.6) with the mass range 0.1 $M_\odot$ ≤M<12 $M_\odot$ from pre-main sequence to the thermally pulsing asymptotic giant branch (TP-AGB). In PARSEC v1.1 we expanded the metallicity range down to Z=0.0001 ([M/H]=-2.2). PARSEC v1.2S including big improvement both on the very low mass stars and massive stars: Calibrated with the mass-radius relation of dwarf stars, Chen et al. (2014) improves the surface boundary conditions for stars with mass $M \lesssim 0.5$ $M_\odot$ ; Tang et al. (2014) introduce mass loss for massive star M≥14 $M_\odot$ ; Chen et al. (2015) improve the mass-loss rate when the luminosity approaches the Eddington luminosity and supplement the model with new bolometric corrections till M=350 $M_\odot$ . In a later version ( PARSEC v1.2S + COLIBRI PR16) we add improved tracks of TP-AGB (Marigo et al., 2013; Rosenfield et al., 2016).

A set of evolutionary tracks and isochrones with $\alpha$ enhancement will be added to the website soon, Chap.1.3 gives the introduction while the detail of the calculations are presented in Part I.

### 1.2.5 Other stellar models

There are many other groups who are working on the stellar evolution. Here below we briefly summary some of the most popular ones.

Generally there are not significant differences in the basic equations used for stellar structure, but different group use different numerical methods, different MLT calibrations, different prescriptions for internal mixing and different composition patterns. For example, Fig. 1.3 displays the relationships between the initial helium mass fraction (Y) and the initial metallicity Z, adopted by different stellar evolution groups.

#### Yonsei-Yale

The Yonsei-Yale ($Y^2$)[2] standard models (Yi et al., 2001; Kim et al., 2002; Yi, Kim & Demarque, 2003; Demarque et al., 2004) contain solar mixure and [$\alpha$/Fe] =0.3 isochrones to the tip of the giant branch for metallicities 0.00001≤ $Z$ ≤0.08 and age 0.1–20 Gyr. The primordial helium fraction is 0.23 and helium contents are calculated according to $\Delta Y/\Delta Z = 2.0$. Helium diffusion and convective core overshoot have also been taken into consideration. The new models on low mass stars (Spada et al., 2013) (0.1–1.25 $M_\odot$ ) cover metallicities [Fe/H]= 0.3 to −1.5 but with no $\alpha$ enhancement. The Solar abundance is from Grevesse & Sauval (1998).

[1]CMD 2.8 input form: http://stev.oapd.inaf.it/cgi-bin/cmd_2.8
[2]http://www.astro.yale.edu/demarque/yyiso.html.

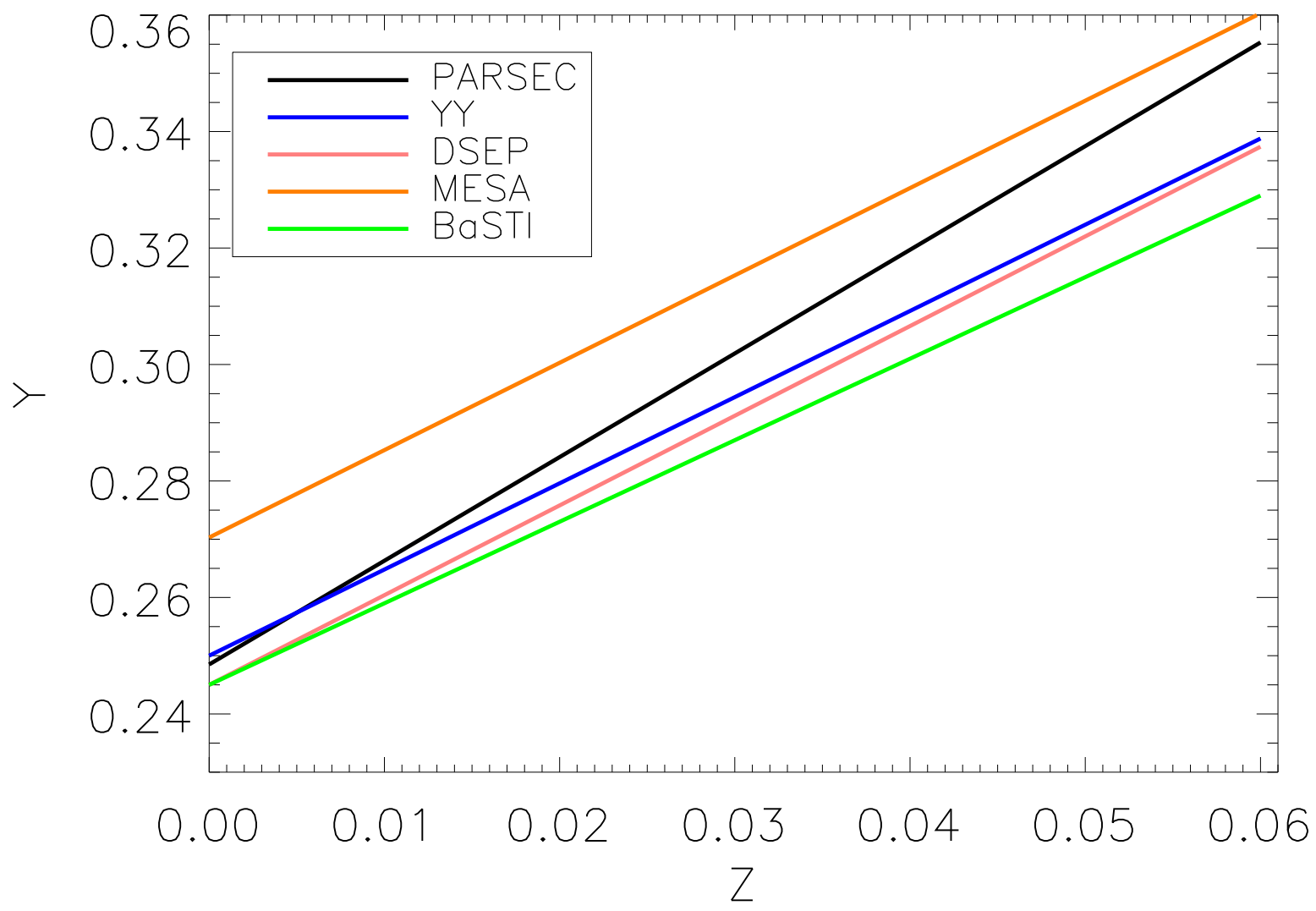

Figure 1.3: Comparison of the helium enrichment law in different models.

The primordial helium fraction is 0.25, and helium contents are calculated according to $\Delta Y/\Delta Z = 1.48$. The mixing length is 1.743. Overshooting is ignored. The OPAL opacities are used at high temperature and Ferguson et al. (2005) opacities are used at low temperature. OPAL EOS is used. Phoenix BT-Settl models (Rogers, Swenson & Iglesias, 1996) are used as the boundary conditions.

**DSEP**

The DSEP (The Dartmouth Stellar Evolution Program) [3] (Dotter, 2007; Dotter et al., 2007, 2008) is derived from the Yale stellar evolution code (Guenther et al., 1992). It provides stellar evolutionary tracks in the mass range of 0.1–4 $M_\odot$, and isochrones of age 0.25–15 Gyr. The metallicities are from $[Fe/H] = +0.5$ to $-2.5$, and $\alpha$ enhancement from $[\alpha/\mathrm{Fe}] = 0.8$ to $-0.2$. The Helium abundance are scaled with the relation $Y = 0.245 + 1.54Z$. They also provide $Y = 0.33$ and 0.4 models. They use the Solar abundance from Grevesse & Sauval (1998). Their definition of $\alpha$ enhancement is keeping [Fe/H] unchanged but adding the $[\alpha/\mathrm{Fe}]$ value to each of the $\alpha$-elements (O, Ne, Mg, Si, S, Ca, and Ti). The mixing length used is $\alpha_{MLT} = 1.938$ (Solar-calibrated). They parameterize the core overshooting as a function of stellar composition and the overshooting grows with the size of con-vective core. They also use the EOS from FreeEOS. The opacities from OPAL are used for the high temperature, while low temperature opacity tables are calculated by themselves. They employ the Phoenix and Castelli & Kurucz (2003) atmosphere models as the surface boundary conditions.

[3] http://stellar.dartmouth.edu/models/index.html.

### Geneva

The Geneva stellar evolution model grids[4] provide three different sets of evolutionary tracks. A standard version with rotation for stellar mass from 0.8 $M_\odot$ to 120 $M_\odot$ and Z = 0.014, 0.006, 0.002 (Ekström et al., 2012; Georgy et al., 2013a; Yusof et al., 2013), a version with rotation for B-type star (1.7–15 $M_\odot$ , at Z = 0.014, 0.006, and 0.002, Georgy et al., 2013b), and a version without rotation for low mass stars, from pre-main sequence to carbon burning (0.5–3.5 $M_\odot$ , Z = 0.006, 0.01, 0.014, 0.02, 0.03, Mowlavi et al., 2012). The solar metallicity in Geneva models is Z=0.014, the mixing length is $\alpha_{MLT}$ = 1.6467, and the Helium abundance is scaled with the relation $Y = 0.248 + 1.2857Z$.

### FRANEC

The FRANEC (Frascati Raphson Newton Evolutionary Code, Chieffi & Straniero, 1989; Degl'Innocenti et al., 2008) is an evolutionaty stellar code developed in Frascati, Italy in 1970s and updated by many works (Dominguez et al., 1999; Domínguez, Straniero & Isern, 1999; Cariulo, Degl'Innocenti & Castellani, 2004). Databases of stellar models and isochrones partially available online. [5] It offers models from pre-main sequence to the cooling sequence of white dwarfs, covering a mass range from 0.1 to 25 $M_\odot$ for several metallicities and helium abundances (Degl'Innocenti et al., 2008). The last update (Chieffi & Limongi, 2013) provides Solar metallicity (Grevesse & Sauval, 1998) models of mass from 13 to 120 $M_\odot$ . It contains both rotating and non-rotating models. The mixing length adopted is $\alpha_{MLT}$ =2.3 and an overshooting of $0.2H_P$ is included. The opacities are from OPAL and LAOL. The equation of state is from Straniero, Chieffi & Limongi (1997).

### BasTi

BaSTI (A Bag of Stellar Tracks and Isochrones) [6] (Pietrinferni et al., 2013) is a modification of the FRANEC code. It contains models in the mass range of 0.5–10 $M_\odot$ and metallicities Z=0.04 to 0.0001. The solar metallicity is from Grevesse &

[4] http://obswww.unige.ch/Recherche/evol/Geneva-grids-of-stellar-evolution.
[5] http://astro.df.unipi.it/SAA/PEL/Z0.html
[6] http://basti.oa-teramo.inaf.it/

Noels (1993) and the $\alpha$ enhancement [$\alpha$/Fe] =0.4 follows that from Salaris & Weiss (1998). The opacities are from OPAL for $T > 10,000$K, whereas those from Alexander & Ferguson (1994) are used for lower temperatures. Opacities of Alexander & Ferguson (1994) include the contributions from molecular and dust grain. The mixing-length parameter from the Solar-calibration (to reproduce the Solar radius, luminosity, metallicity at an age of 4.57 Gyr) is 1.913 (there are also models computed with 1.25). The overshooting is switched off. Helium contents follows the enrichment law: $Y = 0.245 + 1.4Z$.

**MESA**

MESA (Modules for Experiments in Stellar Astrophysics) [7] (Paxton et al., 2011) is the first open source stellar structure and evolution code. It provides the community the flexibility to modify or improve the code. Users can download the code and modify the detailed parameters and setups. It has been extended for a wide application: from planets to massive stars, binaries, rotations, oscillations, pulsations, and explosions. Their product MIST (MESA Isochrones and Stellar Tracks Choi et al., 2016) assumes a linear enrichment law to the protosolar helium abundance $Y = 0.2703 + 1.5Z$, and a mixing length $\alpha_{MLT}$=1.82.

## 1.3 Solar-scaled mixture and $\alpha$ enhancement

Our Sun, as the nearest and most understood star, offers a unique benchmark and the best test field to stellar evolution models. Many stellar models are developed with the solar-scaled metal mixture, i.e. the initial partition of heavy elements keeps always the same relative number density as that in the Sun. Though being the most widely used mixture, the solar-scaled metal mixture is not universally applicable for all types of stars. In fact, one of the most important group of elements, the so called $\alpha$-elements group, tracing the nucleosynthesis products of stellar evolution and other stellar properties, is not always observed in solar proportions. This group is important because it is the result of consecutive fusion processes involving helium, $\alpha$-captures, and it is known to occur in massive stars that are the main contributors to the metal enrichment. The so called "alpha process elements" are thus the most representative elements of the enrichment produced by massive stars. Many studies have confirmed the existence of an "enhancement" of $\alpha$-elements in the Milky Way halo (e.g. Zhao & Magain, 1990; Nissen et al., 1994; McWilliam et al., 1995). Venn et al. (2004) compiled a large sample of the $\alpha$-to-iron ratio ([Mg/Fe], [Ca/Fe], and [Ti/Fe]) measurements of stars in the Galactic halo and on the disk, which show greater values than that of the sun ( [$\alpha$/Fe]$_\odot$=0), with an increasing trend at decreasing [Fe/H]. Kirby et al. (2011) obtained data of eight dwarf spheroidal (dSph) Milky Way satellite galaxies and found that the $\alpha$-abundance trend is different compared to the Galactic halo stars, possibly indicating different star formation paths. Enhanced $\alpha$-elements abundance is also confirmed in globular clusters (e.g. Carney, 1996; Sneden, 2004; Pritzl, Venn & Irwin, 2005), in the Galactic Bulge (Gonzalez et al., 2011; Johnson et al., 2014, for instance), and in the Galactic thick disk (e.g. Fulbright, 2002; Reddy, Lambert & Prieto, 2006; Ruchti et al., 2010).

[7] http://mesa.sourceforge.net/.

It is likely that the origin of this $\alpha$-abundance enhancement comes from the time scale difference between the metal enrichment coming form Type II and Type Ia supernovae. The core collapse (mostly type II) supernovae (SNe), evolve from massive stars and mainly produce $\alpha$-elements (O, Ne, Mg, Si, S, Ar, Ca, and Ti). On the contrary, Type Ia SNe originate from binary evolution after at least one white dwarf star has been formed. They mainly synthesize iron-peak elements (V, Cr, Mn, Fe, Co and Ni) in a thermonuclear incineration of the compact star. Since massive stars evolve much faster (a few, to few tens of Myrs) than binary white dwarfs (from $>$ 40 Myrs to a few Gyrs Maoz et al., 2011), they recycle the interstellar medium earlier than Type Ia SNe. The alpha-to-iron ratio [$\alpha$/Fe] in ISM is thus initially higher than that of the Sun (that formed only 4.5 Gyr ago). It then decreases as the star cluster/galaxy evolves and the Type Ia SNe begin to pollute the ISM with Fe-peak elements. Thus the evolution profile of [$\alpha$/Fe] records the star formation history and leaves the imprint in stars. An alternative explanation could be that the Initial Mass Function of the $\alpha$-enhanced stellar populations was much richer in massive stars than the one from which our Sun was born. However there is no clear evidence in support of this alternative possibility.

In order to model star clusters and galaxies more precisely, the previous Padova isochrone database offered a few sets of $\alpha$-enhanced models, for four relatively high metallicities (Salasnich et al., 2000). Now, with the thorough revision and update input physics, we introduce $\alpha$-enhanced metal mixtures in `PARSEC` . These models are particularly suited for old and metal poor stellar populations and constitute the first part of my work, Part. I. In this part I will present and discuss the new `PARSEC` database of the $\alpha$-enhanced stellar evolutionary tracks and isochrones.

## 1.4 Lithium in stars, origin and problems

In parallel with the work on `PARSEC` $\alpha$-enhanced stellar models I have analyzed another interesting problem related to the evolution of low metallicity stars, which is related to their Lithium content.

Stars write their history with elemental abundances. Different elements may be produced by star of different mass and in different environmental conditions and, by measuring the stellar chemical abundances, one can trace the formation history and investigate the physical mechanisms that take place in and around the stars.

Lithium, fragile and scarce, sensitive and primitive, is one of the most complicated elements in stellar physics. Its abundance in stars, both on the main sequence phase and the giant branch phase, has plagued our current understanding of cosmology, stellar evolution, and metal sources of the interstellar medium (ISM). Here I first briefly summarise the basic characteristics of lithium, then introduce the problem and the puzzle it brings to the community.

Compared with the hydrogen fusion, lithium is much easier to be destroyed by proton capture $^{7}\mathrm{Li}\,(\mathrm{p}\,,\gamma)\,^{4}\mathrm{He} + ^{4}\mathrm{He}$. The nuclear reaction rate is very high even at a temperature of a few million Kelvin. Here is how the above nuclear reaction rate is defined:

$$rate = n_{Li}\, n_p\, \frac{8\sqrt{2}}{9\sqrt{3}} \frac{S(E_0)}{b\sqrt{m}} (\frac{3E_0}{k_B T})^2 exp(-\frac{3E_0}{k_B T}) \qquad (1.27)$$

where $n_{Li}$ and $n_p$ are the number densities of $^{7}$Li and proton particles that evolve in the reaction, $E_0$ corresponds to the energy of the nuclei dominating the reaction, $S(E_0)$ is the cross-section factor that essentially describes the energy dependence of the reaction once the nuclei have penetrated the potential barrier, $m$ is the reduced mass, $T$ is the environment temperature, and $b$, $k_B$ are constants. The cross-section $S(E_0)$ of lithium burning reaction is very high and the reduced mass $m$ is relatively low so that the reaction rate can be important even at low temperature (a few million Kelvin).

Because of its sensitivity to the temperature, the lithium abundance is a good tracer of the stellar structure: lithium on the surface of the star will be destroyed if the surface convective zone reaches the hot stellar interiors.

Unlike other metal elements, which are mostly produced rather than being destroyed in stars, the destruction rates of Li exceed its creation rates in most stars. Today Li is among the least abundant elements lighter than Zinc. Since Li abundance is very low, it is not easy to measure its abundance in stars. Only a $2p$–$2s$ resonance doublet at 670.8 nm and a $3p$–$2p$ triplet at 610.4 nm can eventually be detected in optical spectra, though still with a low transition probability (Wiese & Fuhr, 2009). Fig. 1.4 shows the Grotrian diagram for Li I adopted from Carlsson et al. (1994).

## 1.4.1 Cosmological lithium problem

$^{7}$Li, together with $^{4}$He, $^{3}$He, and D, are the four isotopes synthesized in the primordial nucleosynthesis of the Big Bang. Their primordial abundances depend

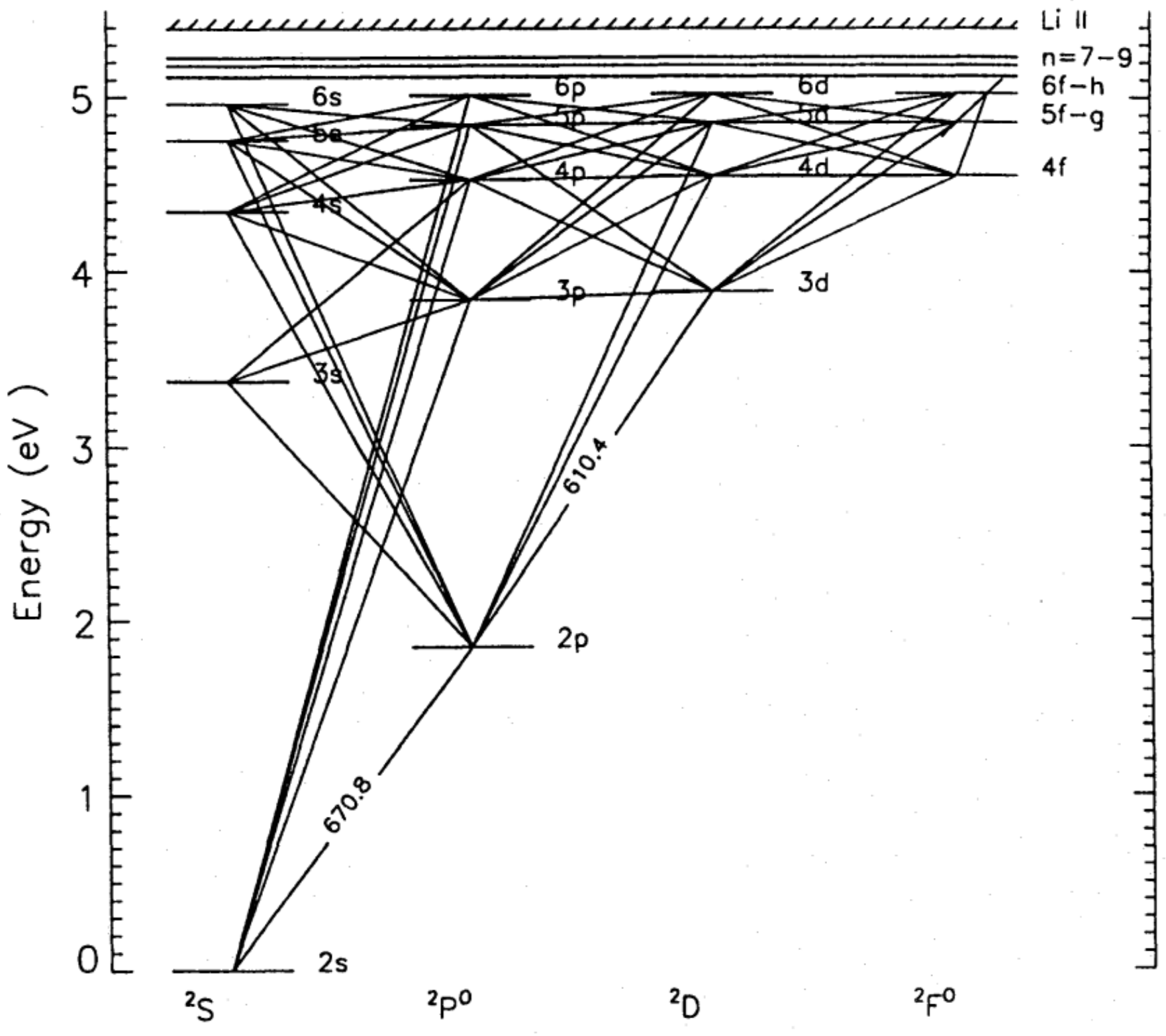

Figure 1.4: Figure 3 from Carlsson et al. (1994) with its original caption: `Grotrian diagram for Li I. For clarity,all permitted downward transitions from n=7-9 are omitted here but they are included in the model atom, which contains 21 levels, 70 lines and 20 bound-free transitions in total. The 670.8 nm line corresponds to the` $2p$–$2s$ `resonance doublet, the 610.4 nm line to the` $3p$–$2p$ `triplet, the 323.3 nm pumping line to` $3p$–$2s$. `The` $2s$ `bound-free edge is at 229.9nm, the` $2p$ `edge at 349.8 nm.`

mainly on the baryon-to-photon ratio (or put it another way, baryon density) with only a minor sensitivity to the universal speed-up expansion rate, namely the number of neutrino families (Fields, 2011; Fields, Molaro & Sarkar, 2014). The universal baryon density can be obtained either from the acoustic oscillations of the cosmic microwave background (CMB) observation or, independently, from the primordial deuterium abundance measured in un-evolved clouds of distant quasar spectra (Adams, 1976). Figure 1.5, adopted from Fields, Molaro & Sarkar (2014), shows how the primordial elemental abundances depend on the baryon density.

Observations based on the Wilkinson Microwave Anisotropy Probe (WMAP; Komatsu et al., 2011) predict a primordial $^7$Li abundance $A(\mathrm{Li}) = 2.72$[8] (Coc et al., 2012). From the baryon density measured by the Planck mission (Planck Collaboration et al., 2014) Coc, Uzan & Vangioni (2014) calculated the primordial value of $^7$Li/H to be $4.56 \sim 5.34 \times 10^{-10}$ (A(Li)$\approx$ 2.66 − 2.73).

Population II (POP II) main sequence (MS) stars show a constant $^7$Li abundance (Spite & Spite, 1982), which was interpreted as an evidence that these stars carry the primordial $^7$Li abundance because of their low metallicity. In the past three decades, observations of metal-poor main sequence stars both in the Milky Way halo (Spite & Spite, 1982; Sbordone et al., 2010), and in the globular clusters (Lind et al., 2009; Monaco et al., 2010) have confirmed that the $^7$Li abundance remains $A(\mathrm{Li}) \approx 2.26$ (Molaro, 2008). Figure 1.6, adopted from Molaro et al. (2012) (their Fig. 1), shows that for a wide range of effective temperatures, the lithium abundance lies on a plateau. This abundance, which defines the so-called Spite plateau, is three times lower than the primordial value predicted from the big bang nucleosynthesis (BBN). This discrepancy is the long-standing "lithium problem".

There are several lines of study which have been pursued to provide possible solutions to the problem: *i)* nuclear physics solutions which alter the reaction flow into and out of atomic mass-7 (Coc et al., 2012); *ii)* new particle physics where massive decaying particles could destroy $^7$Li (Olive et al., 2012; Kajino et al., 2012); *iii)* Chemical separation by magnetic field in the early structure formation that reduces the abundance ratio of Li/H (Kusakabe & Kawasaki, 2015); *iv)* $^7$Li depletion during main sequence evolution. It has been argued that certain physical processes, which may occur as the stars evolve on the main sequence, could cause the observed lithium depletion. Among these processes we recall gravitational settling (e.g. Salaris & Weiss (2001); Richard, Michaud & Richer (2005) and Korn et al. (2006)) or possible coupling between internal gravity waves and rotation-induced mixing (Charbonnel & Primas, 2005).

Considering the stellar Li evolution from the pre-main sequence phase to the end of the main sequence phase, in Chapter 7 I will discuss a new environmental solution to this long-standing problem: lithium was first almost completely destroyed and then re-accumulated by residual disk mass accretion. Specifically, $^{7}$Li can be significantly depleted by the convective overshoot in the pre-main sequence phase, and then partially be restored in the stellar atmosphere by accretion of the residual disk. This accretion could be regulated by extreme ultra-violet photo-evaporation. When the stars evolve to the ages we observe, this model can perfectly re-produce the observed Li abundance. This environmental Li evolution model not only provides a solution to the lithium problem, but also offers the possibility to interpret the decrease of Li abundance in extremely metal-poor stars, the Li disparities in spectroscopic binaries and the low Li abundance in planet hosting stars.

[8] $A(\mathrm{Li}) = 12 + \log[n(\mathrm{Li})/n(\mathrm{H})]$ where $n$ is number density of atoms and 12 is the solar hydrogen abundance.

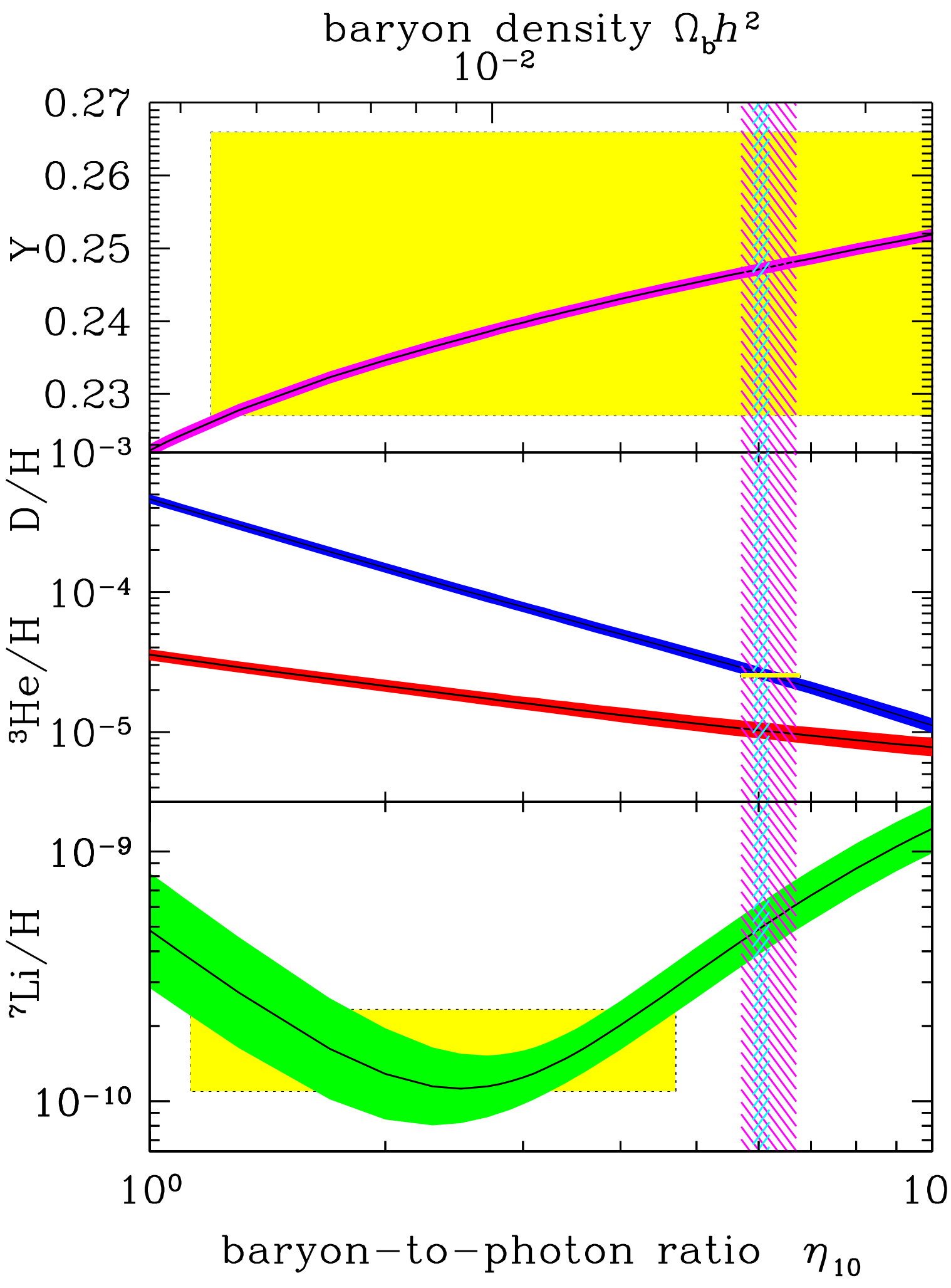

Figure 1.5: Figure 1.1 from Fields, Molaro & Sarkar (2014) with the original caption: `The abundances of` $^4$`He, D,` $^3$`He, and` $^7$`Li as predicted by the standard model of Big-Bang nucleosynthesis. The bands show the 95% confidence level range. Boxes indicate the observed light element abundances. The narrow vertical band indicates the CMB measure of the cosmic baryon density, while the wider band indicates the BBN concordance range (both at 95% confidence level).`

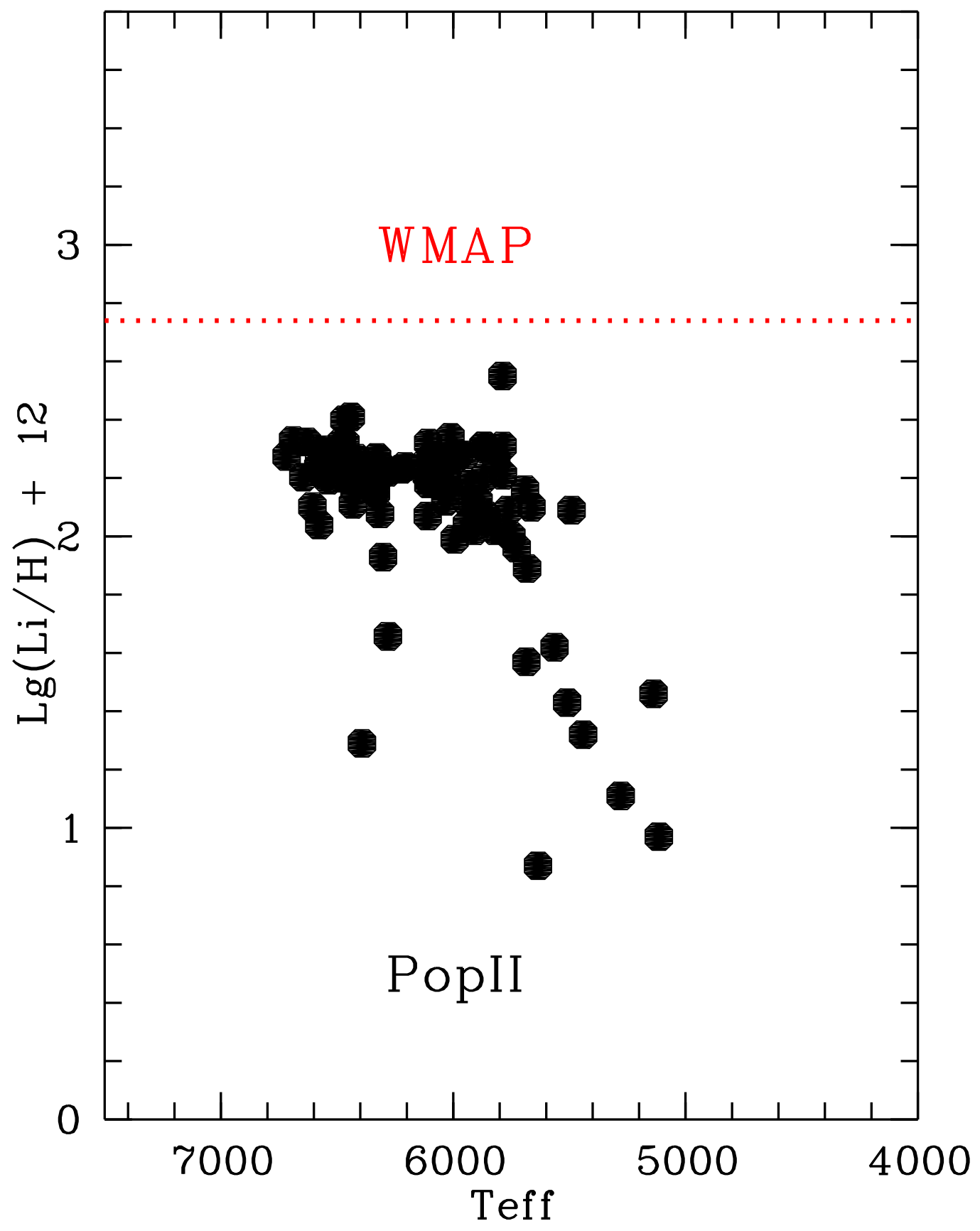

Figure 1.6: Figure 1 from Molaro et al. (2012) with the caption: `Li observations in Pop-II stars. The red line marks the WMAP Li prediction.`

# Part I:

## PARSEC tracks and isochrones with $\alpha$ enhancement

'Stars have a life cycle much like animals. They get born, they grow, they go through a definite internal development, and finally they die, to give back the material of which they are made so that new stars may live.'

-Hans Bethe

Studies on Galactic bulge, thick disk, halo, and globular clusters require stellar models with $\alpha$-enhancement, because observations show that stars have $\alpha$-to-Iron number ratios different from what observed in the Sun. In particular stars in the above galactic sub-components are $\alpha$-enhanced with $[\alpha/\mathrm{Fe}] > 0$. The $[\alpha/\mathrm{Fe}]$ ratio not only affects important observables like spectral lines (and also continuum) but also, mainly because of the opacity, it affects the main parameters of a star, like luminosity, effective temperature and life-times. As already elaborated in introduction massive stars, both during their evolution and in the final explosion, are responsible of the $\alpha$-elements production while Type Ia supernovae, mainly synthesize iron-peak elements. The shorter life time of massive stars, with respect to that of binary evolution of intermediate and low mass stars, leads to an early release of $\alpha$-elements in the ISM, from which the next generation of stars forms. As the star clusters or the galaxies age, Type Ia supernovae begin to inject more and more iron into the ISM causing $[\alpha/\mathrm{Fe}]$ to decrease. Of course the time evolution of the $[\alpha/\mathrm{Fe}]$ ratio depend also from the initial mass function (IMF) and the star formation history. $\alpha$-enhanced (or even depressed) models are thus an essential step to interpret observations of stars of the different galactic components and in different galaxies besides the Milky Way. In my work I have extended the `PARSEC` models from the standard solar-scaled composition to $\alpha$ enhanced mixtures, aiming at offering more accurate stellar tools to investigate these stars, to trace back their formation history, and possibly to calibrate the IMF in different environments.

We first check and calibrate the new `PARSEC` $\alpha$-enhanced stellar evolutionary tracks and isochrones using the well-studied globular cluster 47 Tucanae (NGC 104). Then we apply the calibrated parameters to obtain models of other metallicities. In particular features like the red giant branch bump and the horizontal branch morphology of 47 Tucanae are discussed in detail because they may provide important information on internal mixing, mass-loss and He abundance. Chapter 2 describes the input physics. Chapter 3 introduce the calculation with 47Tuc in details, including the isochrone fitting and luminosity function, envelope overshooting calibration with red giant branch bump, and mass loss in red giant branch from horizontal branch morphology. Chapter 4 compares the new `PARSEC` models with other stellar models and show its improvement on RGB bump prediction. In chapter 5 we introduce $\alpha$-enhanced models based on APOGEE ATLAS9 metal mixture and in Chapter 6 we show test on various MLT.

# Chapter 2

# Input physics

## 2.1 Nuclear reaction rates

In the latest versions of PARSEC (Fu et al., 2016), we update the nuclear reaction rates from JINA REACLIB database (Cyburt et al., 2010) with their April 6, 2015 new recommendations. In addition, more reactions, 52 instead of 47 as described in Bressan et al. (2012) for the previous versions of PARSEC , are taken into account. They are all listed in Table 2.1 together with the reference from which the reaction is taken. In the updated reaction network more isotope abundances are considered, in total $N_{el} = 29$: $^{1}$H, D, $^{3}$He, $^{4}$He,$^{7}$Li, $^{8}$Be, $^{8}$B, $^{12}$C, $^{13}$C, $^{14}$N, $^{15}$N, $^{16}$N, $^{17}$N, $^{17}$O, $^{18}$O, $^{18}$F, $^{19}$F, $^{20}$Ne, $^{21}$Ne, $^{22}$Ne, $^{23}$Na, $^{24}$Mg, $^{25}$Mg, $^{26}$Mg, $^{26}$Al$^{m}$, $^{26}$Al$^{g}$, $^{27}$Al, $^{27}$Si, and $^{28}$Si.

Table 2.1: Nuclear reaction rates adopted in this work and the reference from which we take their reaction energy $Q$.

| **reactions** | **$Q$ reference** |
|---|---|
| $p\,(p\,,\beta^{+}\,\nu)\,D$ | Betts, Fortune & Middleton (1975) |
| $p\,(D\,,\gamma)\,^{3}He$ | Descouvemont et al. (2004) |
| $^{3}He\,(^{3}He\,,\gamma)\,2\,p\,+^{4}\,He$ | Angulo et al. (1999) |
| $^{4}He\,(^{3}He\,,\gamma)\,^{7}Be$ | Cyburt & Davids (2008) |
| $^{7}Be\,(e^{-}\,,\gamma)\,^{7}Li$ | Cyburt et al. (2010) |
| $^{7}Li\,(p\,,\gamma)\,^{4}He\,+^{4}He$ | Descouvemont et al. (2004) |
| $^{7}Be\,(p\,,\gamma)\,^{8}B$ | Angulo et al. (1999) |
| $^{12}C\,(p\,,\gamma)\,^{13}N$ | Li et al. (2010) |
| $^{13}C\,(p\,,\gamma)\,^{14}N$ | Angulo et al. (1999) |
| $^{14}N\,(p\,,\gamma)\,^{15}O$ | Imbriani et al. (2005) |
| | continue to next page |

**Table 2.1 – continued from previous page**

| reactions | $Q$ reference |
|---|---|
| $^{15}$N (p , $\gamma$) $^{4}$He + $^{12}$C | Angulo et al. (1999) |
| $^{15}$N (p , $\gamma$) $^{16}$O | Iliadis et al. (2010) |
| $^{16}$O (p , $\gamma$) $^{17}$F | Iliadis et al. (2008) |
| $^{17}$O (p , $\gamma$) $^{4}$He + $^{14}$N | Iliadis et al. (2010) |
| $^{17}$O (p , $\gamma$) $^{18}$F | Iliadis et al. (2010) |
| $^{18}$O (p , $\gamma$) $^{4}$He + $^{15}$N | Iliadis et al. (2010) |
| $^{18}$O (p , $\gamma$) $^{19}$F | Iliadis et al. (2010) |
| $^{19}$F (p , $\gamma$) $^{4}$He + $^{16}$O | Angulo et al. (1999) |
| $^{19}$F (p , $\gamma$) $^{20}$Ne | Angulo et al. (1999) |
| $^{4}$He (2 $^{4}$He , $\gamma$) $^{12}$C | Fynbo et al. (2005) |
| $^{12}$C ($^{4}$He , $\gamma$) $^{16}$O | Cyburt (2012) |
| $^{14}$N ($^{4}$He , $\gamma$) $^{18}$F | Iliadis et al. (2010) |
| $^{15}$N ($^{4}$He , $\gamma$) $^{19}$F | Iliadis et al. (2010) |
| $^{16}$O ($^{4}$He , $\gamma$) $^{20}$Ne | Constantini & LUNA Collaboration (2010) |
| $^{18}$O ($^{4}$He , $\gamma$) $^{22}$Ne | Iliadis et al. (2010) |
| $^{20}$Ne ($^{4}$He , $\gamma$) $^{24}$Mg | Iliadis et al. (2010) |
| $^{22}$Ne ($^{4}$He , $\gamma$) $^{26}$Mg | Iliadis et al. (2010) |
| $^{24}$Mg ($^{4}$He , $\gamma$) $^{28}$Si | Strandberg et al. (2008) |
| $^{13}$C ($^{4}$He , n) $^{16}$O | Heil et al. (2008) |
| $^{17}$O ($^{4}$He , n) $^{20}$Ne | Angulo et al. (1999) |
| $^{18}$O ($^{4}$He , n) $^{21}$Ne | Angulo et al. (1999) |
| $^{21}$Ne ($^{4}$He , n) $^{24}$Mg | Angulo et al. (1999) |
| $^{22}$Ne ($^{4}$He , n) $^{25}$Mg | Iliadis et al. (2010) |
| $^{25}$Mg ($^{4}$He , n) $^{28}$Si | Angulo et al. (1999) |
| $^{20}$Ne (p , $\gamma$) $^{21}$Na | Iliadis et al. (2010) |
| $^{21}$Ne (p , $\gamma$) $^{22}$Na | Iliadis et al. (2010) |
| $^{22}$Ne (p , $\gamma$) $^{23}$Na | Iliadis et al. (2010) |
| $^{23}$Na (p , $\gamma$) $^{4}$He + $^{20}$Ne | Iliadis et al. (2010) |
| $^{23}$Na (p , $\gamma$) $^{24}$Mg | Iliadis et al. (2010) |
| $^{24}$Mg (p , $\gamma$) $^{25}$Al | Iliadis et al. (2010) |
| $^{25}$Mg (p , $\gamma$) $^{26}$Al$^{g}$ | Iliadis et al. (2010) |
| $^{25}$Mg (p , $\gamma$) $^{26}$Al$^{m}$ | Iliadis et al. (2010) |
| $^{26}$Mg (p , $\gamma$) $^{27}$Al | Iliadis et al. (2010) |
| $^{26}$Al$^{g}$ (p , $\gamma$) $^{27}$Si | Iliadis et al. (2010) |
| $^{27}$Al (p , $\gamma$) $^{4}$He + $^{24}$Mg | Iliadis et al. (2010) |
| $^{27}$Al (p , $\gamma$) $^{28}$Si | Iliadis et al. (2010) |
| $^{26}$Al (p , $\gamma$) $^{27}$Si | Iliadis et al. (2010) |
| $^{26}$Al (n , p) $^{26}$Mg | Tuli (2011) |

continue to next page

**Table 2.1 – continued from previous page**

| reactions | $Q$ reference |
|---|---|
| $^{12}$C ($^{12}$C , n) $^{23}$Mg | Caughlan & Fowler (1988) |
| $^{12}$C ($^{12}$C , p) $^{23}$Na | Caughlan & Fowler (1988) |
| $^{12}$C ($^{12}$C , $^{4}$He) $^{20}$Ne | Caughlan & Fowler (1988) |
| $^{20}$Ne ($\gamma$ , $^{4}$He) $^{16}$O | Constantini & LUNA Collaboration (2010) |

## 2.2 Equations of state

For the equation of state (EOS), as described in Bressan et al. (2012), we use the FreeEOS code developed by A.W. Irwin. The code is available under the GPL licence.[1] The FreeEOS package is fully implemented in our code for different approximations and levels of accuracy.

The EOS calculation accounts for contributions of several elements, namely: H, He, C, N, O, Ne, Na, Mg, Al, Si, P, S, Cl, Ar, Ca, Ti, Cr, Mn, Fe, and Ni. For any specified distribution of heavy elements $\{X_i/Z\}$, several values of the metallicity $Z$ are considered. For each value of $Z$ we pre-compute tables containing all thermodynamic quantities of interest (e.g. mass density, mean molecular weight, entropy, specific heats and their derivatives, etc.) over suitably wide ranges of temperature and pressure. In the same way as for the opacity, given the total metallicity $Z$ and the distribution of heavy elements $\{X_i/Z\}$, we construct two sets of tables, to which we simply refer to as "H-rich" and "H-free". A "H-rich" set contains $N_X = 10$ tables each characterised by different H abundances, and a "H-free" set consisting of 31 tables, which are designed to describe He-burning and He-exhausted regions. In practice we consider 10 values of the Helium abundance, from $Y = 0$ to $Y = 1 - Z$. For each $Y$ we compute three tables with C and O abundances determined by ratios: $R_C = X_C/(X_C + X_O) = 0.0, 0.5, 1.0$.

Multi-dimensional interpolations (in the variables $Z$, $X$ or $Y$ and $R_C$) are carried out with the same scheme adopted for the opacities: Interpolation over "H-rich" tables is performed in four dimensions, i.e. using $R$ ($R = \rho/T_6^3$; $T_6 = T/10^6$), $T$, $X$, and $Z$ as the independent variables. While the interpolation in $R$ and $T$ is bilinear, we adopt a parabolic scheme for both $X$ and $Z$ interpolation. Interpolation over "H-free" tables is performed in five dimensions, i.e. involving $R$, $T$, $Y$, $R_C = X_C/(X_C + X_O)$, and $Z$. Interpolation is bilinear in $R$ and $T$, linear in $R_C$, while we use as before a parabolic scheme for the interpolation in $Z$. All interesting derivatives are pre-computed and included in the EOS tables.

[1] http://freeeos.sourceforge.net/

Our procedure is to minimize the effects of interpolation by computing a set of EOS tables exactly with the partition of the new set of tracks, at varying global metallicity. This set is then inserted into the EOS database for interpolation when the global metallicity $Z$ of the star changes during the evolution.

## 2.3 Solar model

In order to calibrate the solar model in our code, we obtain a set of solar data obtained from the literature which are summarized in Table 2.2. We generate a large grid of 1 $M_\odot$ tracks with varying initial composition of the Sun from pre-main sequence (PMS) to 4.8 Gyr, and varying the initial composition of the Sun, $Z_{\rm initial}$ and $Y_{\rm initial}$, the mixing length parameter $\alpha_{\rm MLT}$ and the extent of the adiabatic overshoot at the base of the convective envelope $\Lambda_{\rm e}$, in order to compare with the present day surface solar parameters in Table 2.2. The mixing length parameter, $\alpha_{\rm MLT}$, that will be used to compute all the stellar evolutionary sets, is also obtained by this process. The calibration has been obtained exactly with the same set-up used for the calculations of the other tracks, i.e. with tabulated EOS and opacities and using microscopic diffusion. Finally we have adopted the parameters of the best fit obtained with the tabulated EOS, and we changed only the solar age, within the allowed range, in order to match as much as possible the solar data. From the initial values of the metallicity and Helium abundance of the Sun, $Y_{\rm initial}, Z_{\rm initial}$, and adopting for the primordial He abundance $Y_{\rm p} = 0.2485$ (Komatsu et al., 2011), we obtain also the Helium-to-metals enrichment ratio, $\Delta Y/\Delta Z = 1.78$. The parameters of our best model are listed in the lower part of Table 2.3.

In PARSEC we use the elemental abundances of our Sun based on Grevesse & Sauval (1998) and revision of Caffau et al. (2011b). In table 2.4 I list the solar composition mixture used in PARSEC . $N_i/N_Z$ represents the number density over the total number of metal elements, and $Z_i/Z_{tot}$ is the mass fraction for each element.

Table 2.2: Data used to calibrate the solar model. $L_\odot$, $R_\odot$, and $T_{\text{eff},\odot}$ represent for the solar luminosity, radius, and effective temperature, respectively. $Z_\odot$ and $Y_\odot$ are the solar metallicity and helium content today. $(Z/X)_\odot$ is the solar metallicity-to-hydrogen mass ratio in the sun. $R_{\text{ADI}}/R_\odot$, $\rho_{\text{ADI}}$, and $C_{\text{S,ADI}}$ represent the adiabatic radius, density, and sound speed.

| Solar data | Value | error | reference |
|---|---|---|---|
| $L_\odot$ ($10^{33}$erg s$^{-1}$) | 3.846 | 0.005 | Guenther et al. (1992) |
| $R_\odot$ ($10^{10}$ cm) | 6.9598 | 0.001 | Guenther et al. (1992) |
| $T_{\text{eff},\odot}$ (K) | 5778 | 8 | from $L_\odot$ & $R_\odot$ |
| $Z_\odot$ | 0.01524 | 0.0015 | Caffau et al. (2011b) |
| $Y_\odot$ | 0.2485 | 0.0035 | Basu & Antia (2004) |
| $(Z/X)_\odot$ | 0.0207 | 0.0015 | from $Z_\odot$ & $Y_\odot$ |
| $R_{\text{ADI}}/R_\odot$ | 0.713 | 0.001 | Basu & Antia (1997) |
| $\rho_{\text{ADI}}$ | 0.1921 | 0.0001 | Basu et al. (2009) |
| $C_{\text{S,ADI}}/10^7$cm/s | 2.2356 | 0.0001 | Basu et al. (2009) |

Table 2.3: Parameters of our best solar model. The distribution of heavy elements is from Caffau et al. (2011b). The stellar age here includes the pre-main sequence phase. $Z_{\text{initial}}$ and $Y_{\text{initial}}$ are the initial metallicity and helium content of the sun at the time when it was born. $\alpha_{\text{MLT}}$ is the parameter of mixing length. $\Lambda_e$ is the envelope overshoot parameter. All other parameters list here represent for the same physical quantities as itemized in table 2.2

| Model | tabulated EOS |
|---|---|
| $L$ ($10^{33}$erg s$^{-1}$) | 3.848 |
| $R$ ($10^{10}$ cm) | 6.9584 |
| $T_{\text{eff}}$ (K) | 5779 |
| $Z_\odot$ | 0.01597 |
| $Y_\odot$ | 0.24787 |
| $(Z/X)_\odot$ | 0.02169 |
| $R_{\text{ADI}}/R_\odot$ | 0.7125 |
| $\rho_{\text{ADI}}$ | 0.1887 |
| $C_{\text{S,ADI}}/10^7$cm/s | 2.2359 |
| Age(Gyr) | 4.593 |
| $Z_{\text{initial}}$ | 0.01774 |
| $Y_{\text{initial}}$ | 0.28 |
| $\alpha_{\text{MLT}}$ | 1.74 |
| $\Lambda_e$ | 0.05 |

Table 2.4: Solar composition mixture used in PARSEC .

| $Z_i$ | **Element** | $N_i/N_Z$ | $Z_i/Z_{tot}$ |
|---|---|---|---|
| 3 | Li | 8.906547E-9 | 3.564854E-9 |
| 4 | Be | 2.087418E-8 | 1.084685E-8 |
| 5 | B | 2.948556E-7 | 1.838182E-7 |
| 6 | C | 0.2627903 | 0.1819942 |
| 7 | N | 0.06021564 | 0.04863079 |
| 8 | O | 0.4781997 | 0.4411401 |
| 9 | F | 3.017236E-5 | 3.305146E-5 |
| 10 | Ne | 0.08701805 | 0.1012064 |
| 11 | Na | 0.001776681 | 0.002355094 |
| 12 | Mg | 0.03160162 | 0.04429457 |
| 13 | Al | 0.002453067 | 0.00381628 |
| 14 | Si | 0.02948556 | 0.04774824 |
| 15 | P | 2.397228E-4 | 4.28122E-4 |
| 16 | S | 0.01201737 | 0.02221761 |
| 17 | Cl | 2.627903E-4 | 5.372569E-4 |
| 18 | Ar | 0.002087418 | 0.004808005 |
| 19 | K | 1.070802E-4 | 2.414152E-4 |
| 20 | Ca | 0.001904186 | 0.004400113 |
| 21 | Sc | 1.229729E-6 | 3.187575E-6 |
| 22 | Ti | 8.701805E-5 | 2.402248E-4 |
| 23 | V | 8.310159E-6 | 2.440877E-5 |
| 24 | Cr | 3.887849E-4 | 0.00116562 |
| 25 | Mn | 2.040372E-4 | 6.463187E-4 |
| 26 | Fe | 0.02751752 | 0.08860858 |
| 27 | Co | 6.913681E-5 | 2.349273E-4 |
| 28 | Ni | 0.001477778 | 0.005002078 |
| 29 | Cu | 1.34806E-5 | 4.939354E-5 |
| 30 | Zn | 3.309096E-5 | 1.247568E-4 |
| 31 | Ga | 6.3039E-7 | 2.534041E-6 |
| 32 | Ge | 2.13604E-6 | 8.945245E-6 |
| 33 | As | 1.94854E-7 | 8.41745E-7 |
| 34 | Se | 2.13604E-6 | 9.728528E-6 |
| 35 | Br | 3.54576E-7 | 1.633638E-6 |
| 36 | Kr | 1.697498E-6 | 8.202048E-6 |
| 37 | Rb | 3.309096E-7 | 1.630713E-6 |
| 38 | Sr | 7.755492E-7 | 3.917953E-6 |
| | | | continue to next page |

**Table 2.4 – continued from previous page**

| $Z_i$ | **Element** | $N_i/N_Z$ | $Z_i/Z_{tot}$ |
|---|---|---|---|
| 39 | Y | 1.444473E-7 | 7.40464E-7 |
| 40 | Zr | 3.309096E-7 | 1.74053E-6 |
| 41 | Nb | 2.185795E-8 | 1.170898E-7 |
| 42 | Mo | 6.913681E-8 | 3.822478E-7 |
| 43 | Tc | 0.0 | 0.0 |
| 44 | Ru | 5.750549E-8 | 3.351053E-7 |
| 45 | Rh | 1.095997E-8 | 6.502976E-8 |
| 46 | Pd | 4.071077E-8 | 2.498277E-7 |
| 47 | Ag | 7.239513E-9 | 4.502651E-8 |
| 48 | Cd | 4.894511E-8 | 3.172688E-7 |
| 49 | In | 3.79935E-8 | 2.51527E-7 |
| 50 | Sn | 8.310159E-8 | 5.689158E-7 |
| 51 | Sb | 8.310159E-9 | 5.834109E-8 |
| 52 | Te | 1.444473E-7 | 1.062965E-6 |
| 53 | I | 2.689734E-8 | 1.968115E-7 |
| 54 | Xe | 1.229729E-7 | 9.309188E-7 |
| 55 | Cs | 1.121268E-8 | 8.592435E-8 |
| 56 | Ba | 1.121268E-7 | 8.878322E-7 |
| 57 | La | 1.229729E-8 | 9.849019E-8 |
| 58 | Ce | 3.160162E-8 | 2.553036E-7 |
| 59 | Pr | 4.26196E-9 | 3.462647E-8 |
| 60 | Nd | 2.627903E-8 | 2.185559E-7 |
| 61 | Pm | 0.0 | 0.0 |
| 62 | Sm | 8.505686E-9 | 7.374216E-8 |
| 63 | Eu | 2.751752E-9 | 2.411098E-8 |
| 64 | Gd | 1.095997E-8 | 9.937569E-8 |
| 65 | Tb | 6.600994E-10 | 6.048766E-9 |
| 66 | Dy | 1.147385E-8 | 1.075046E-7 |
| 67 | Ho | 1.5122E-9 | 1.43805E-8 |
| 68 | Er | 7.073093E-9 | 6.821333E-8 |
| 69 | Tm | 8.310159E-10 | 8.094522E-9 |
| 70 | Yb | 9.99331E-9 | 9.970383E-8 |
| 71 | Lu | 9.543537E-10 | 9.627844E-9 |
| 72 | Hf | 6.160406E-9 | 6.339998E-8 |
| 73 | Ta | 6.160406E-10 | 6.427278E-9 |
| 74 | W | 1.070802E-8 | 1.135056E-7 |
| 75 | Re | 1.583468E-9 | 1.700139E-8 |
| 76 | Os | 1.904186E-8 | 2.088668E-7 |

continue to next page

**Table 2.4 – continued from previous page**

| $Z_i$ | Element | $N_i/N_Z$ | $Z_i/Z_{tot}$ |
|---|---|---|---|
| 77 | Ir | 1.860841E-8 | 2.062354E-7 |
| 78 | Pt | 5.243356E-8 | 5.897767E-7 |
| 79 | Au | 8.505686E-9 | 9.659752E-8 |
| 80 | Hg | 1.121268E-8 | 1.296874E-7 |
| 81 | Tl | 6.600994E-9 | 7.778933E-8 |
| 82 | Pb | 7.406437E-8 | 8.849169E-7 |
| 83 | Bi | 4.26196E-9 | 5.135457E-8 |
| 84 | Po | 0.0 | 0.0 |
| 85 | At | 0.0 | 0.0 |
| 86 | Rn | 0.0 | 0.0 |
| 87 | Fr | 0.0 | 0.0 |
| 88 | Ra | 0.0 | 0.0 |
| 89 | Ac | 0.0 | 0.0 |
| 90 | Th | 9.99331E-10 | 1.337004E-8 |
| 91 | Pa | 0.0 | 0.0 |
| 92 | U | 2.627903E-10 | 3.606644E-9 |
| 93 | Np | 0.0 | 0.0 |
| 94 | Pu | 0.0 | 0.0 |
| 95 | Am | 0.0 | 0.0 |
| 96 | Cm | 0.0 | 0.0 |
| 97 | Bk | 0.0 | 0.0 |
| 98 | Cf | 0.0 | 0.0 |
| 99 | Es | 0.0 | 0.0 |

## 2.4 $\alpha$ enhancement

Three sets of $\alpha$-enhanced metal mixtures are considered in this work. Two are constructed to match the metal mixture of the two populations of 47Tuc ( [$\alpha$/Fe] ~0.4 and [$\alpha$/Fe] ~0.2), and the third set is derived from the one adopted by the APOGEE ATLAS9 atmosphere models with [$\alpha$/Fe] =0.4. Chap.3.1 will give the detailed description and references of 47Tuc metal mixtures while Chap. 5 will introduce the [$\alpha$/Fe] =0.4 mixture based on APOGEE ATLAS9 (Mészáros et al., 2012).

It is worth noting here that when we change the heavy element number fractions ($N_i/N_Z$) to obtain a new metal partition in `PARSEC` , their fractional abundance by mass ($Z_i/Z_{tot}$) is re-normalized in such a way that the global metallicity, $Z$, is kept constant. Hence, compared to the solar partition at the same total metallicity, a model with enhanced $\alpha$-elements shows a depression of Fe and the related elements, because the total metallicity remains unchanged by construction. By doing this, one can distinguish the evolutionary effects caused by the sole $\alpha$-enhanced mixture from those caused by a variation of the total metallicity. A comparison between the $\alpha$-enhanced tracks and the solar-scaled ones is displayed in Fig. 2.1. The HRD in the left panel of this figure shows that, with the same total metallicity $Z$ and Helium content $Y$, $\alpha$-enhanced stars are slightly hotter than the solar-scaled ones both on the main sequence and on the red giant branch. This is because in the $\alpha$-enhanced stars a depressed iron mixture leads to a relatively smaller opacity, making the star slightly hotter. Moreover, the higher temperature of the $\alpha$-enhanced stars also leads to a faster evolution, as illustrated in the right panel of Fig. 2.1.

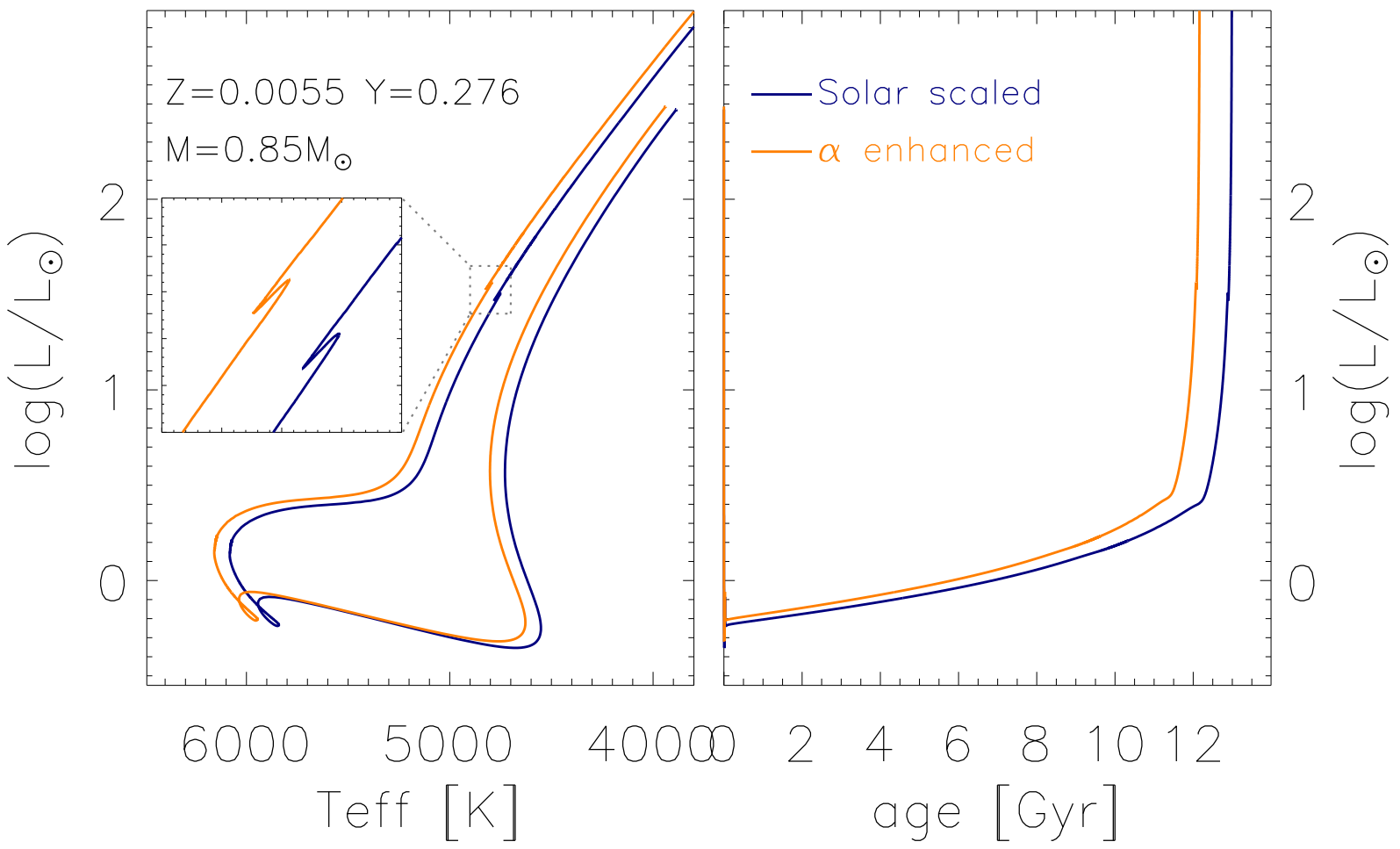

Figure 2.1: Evolutionary tracks with solar-scaled chemical composition (blue line) and with $\alpha$ enhanced mixture (orange line). Both of the tracks are for M=0.85 $M_\odot$ star and the total metallicity and Helium content are the same (Z=0.0055, Y=0.276). The left panel is HRD with sub figure zoom-in around the RGB bump region. The right panel shows how the luminosity of the star evolve with time.

## 2.5 Helium content

Various initial Helium abundance values, for a given metallicity, are allowed in the new version of `PARSEC`. In the previous versions the initial Helium mass fraction of the stars was obtained from the Helium to metals enrichment law:

$$Y = Y_p + \frac{\Delta Y}{\Delta Z} Z \ = \ 0.2485 + 1.78 * Z \tag{2.1}$$

where $Y_p$ is the primordial Helium abundance (Komatsu et al., 2011), and $\Delta Y/\Delta Z$ is the Helium-to-metal enrichment ratio. Because of differences in the adopted primordial and solar calibration He and metallicity values by different authors, the above two parameters are slightly different in different stellar evolution codes, as already shown in Fig. 1.3. The latest YY isochrone (Spada et al., 2013) adopts the relation $Y = 0.25 + 1.48Z$; DSEP (Dotter, 2007; Dotter et al., 2007, 2008) uses $Y = 0.245 + 1.54Z$; MESA (Choi et al., 2016) gives $Y = 0.2703 + 1.5Z$, and BaSTI (Pietrinferni et al., 2006) adopts $Y = 0.245 + 1.4Z$. However, observations reveal that the Helium content does not always follow a single relation. Differences in Helium abundance have been widely confirmed in globular clusters between stellar populations with very similar metallicity. The evidence include the direct He I measurement on blue horizontal branch star (Villanova, Piotto & Gratton, 2009; Mucciarelli et al., 2014; Marino et al., 2014; Gratton et al., 2015, for instance), on giant stars (Dupree, Strader & Smith, 2011; Pasquini et al., 2011), and the splitting of sequences in CMD both of GC in Milky Way (e.g. Bedin et al., 2004; Villanova et al., 2007; Piotto et al., 2007; Milone et al., 2008; Di Criscienzo et al., 2010) and in Magellanic Cloud clusters (Milone et al., 2015a, 2016). Bragaglia et al. (2010) found that the brightness of the RGB bump, which should increase with He abundance, is fainter in first generation than second generation in 14 globular clusters. Indeed, He variation is considered one of the key parameters (and problems) to understand multiple populations in GCs (see the review by Gratton, Carretta & Bragaglia, 2012, and the references therein). In the new version of `PARSEC`, we allow different Helium contents at any given metallicity Z.

# Chapter 3

# Calibration with 47Tuc

Globular Clusters (GCs) have been traditionally considered as the paradigm of a single stellar population, a coeval and chemically homogeneous population of stars covering a broad range of evolutionary phases, from the low-mass main sequence, to the horizontal branch (HB) and white dwarf sequences. For this reason they were considered the ideal laboratory to observationally study the evolution of low mass stars and to check and calibrate the stellar evolution theory. This image has been challenged during the last two decades by photometric and spectroscopic evidence of the presence of multiple populations in most, if not all, globular clusters (for instance NGC 6397 (Gratton et al., 2001; Milone et al., 2012), NGC 6752 (Gratton et al., 2001; Milone et al., 2010), NGC 1851 (Carretta et al., 2014), NGC 2808 (D'Antona et al., 2005; Carretta et al., 2006; Piotto et al., 2007; Milone et al., 2015b), NGC 6388 (Carretta et al., 2007), NGC 6139 (Bragaglia et al., 2015), M22 (Marino et al., 2011), etc.). Nevertheless, Globular Clusters remain one of the basic workbenches for the stellar model builders, besides their importance for dynamical studies and, given the discovery of multiple populations, also for the early chemo-dynamical evolution of stellar systems.

47Tuc, a relatively metal-rich Galactic Globular Cluster, also shows evidence of the presence of at least two different populations: *i)* bimodality in the distribution of CN-weak and CN-strong targets, not only in red giant stars (Briley, 1997; Norris & Freeman, 1979; Harbeck, Smith & Grebel, 2003) but also in MS members (Cannon et al., 1998); *ii)* luminosity dispersion in the sub-giant branch, low-main sequence and HB (Anderson et al., 2009; Di Criscienzo et al., 2010; Nataf et al., 2011) indicating a dispersion in He abundance; *iii)* anti-correlation of Na-O in RGB and HB stars (Carretta et al., 2009b, 2013; Gratton et al., 2013) and also in MS-TO ones (D'Orazi et al., 2010; Dobrovolskas et al., 2014).

The presence of at least two different populations with different chemical composition seems irrefutable (even if their origin is still under debate). Particularly convincing is the photometric study by Milone et al. (2012, and references therein), which concludes, in good agreement with other works (Carretta et al., 2009b, 2013), that for each evolutionary phase, from MS to HB, the stellar content of 47Tuc belongs to two different populations, "first generation" and "second generation" ones (thereafter, FG and SG respectively). The FG population represents $\sim$ 30% of the stars, and it is more uniformly spatially distributed than the SG population, which is more concentrated in the central regions of star clusters.

Choosing 47Tuc as a reference to calibrate PARSEC stellar models, requires therefore computation of stellar models with metal mixtures corresponding to the two identified populations. In the next section we describe the sources to derive the two different metal mixtures that will be used for the opacity and EOS tables in the stellar model computations, and in the follow-up isochrone fitting.

## 3.1 Metal mixtures

Chemical element abundances are given in the literature as the absolute values, A(X),[1] or as [X/Fe],[2] the abundance with respect to the iron content, and referred to the same quantity in the Sun. Since the solar metal mixture has changed lately and since there is still a hot debate about the chemical composition of the Sun, it is important to translate all the available data to absolute abundances, taking into account the solar mixture considered in each source. We follow that procedure to derive the metal mixtures for the first and second generation in 47Tuc.

Estimating the abundance of C and N (which together with O are the main contributors to Z) is quite challenging. It is often done from CN measurements, and assuming a known abundance of C (i.e. Carretta et al., 2013, for 47Tuc RGB stars). The abundances of some elements may change during stellar evolution because of standard (convection) and non-standard (i.e. rotational mixing) transport processes. Therefore, CNO abundances will be compiled from available measurements for MS/TO stars. Other elements which are usually included in the $\alpha$ elements are not expected to be affected by mixing processes during stellar evolution, and so, we will use the values measured in hundreds of stars, mainly in the red giant phase.

Carbon and nitrogen abundances relative to iron ([C/Fe], and [N/Fe]) are adopted from Cannon et al. (1998) for MS/TO 47Tuc stars. These values are also supported by the photometric study presented in Milone et al. (2012).

The separation between the two populations based on photometric colours done by Milone et al. (2012) agree with the separation based on Na-O anticorrelation by Carretta et al. (2009b) and Gratton et al. (2013). We decide hence to use the same criteria to classify the star as FG or SG member.

[1] $A(X) = \log(N_X/N_H) + 12$, with $N_X$ is the abundance in number for the element X.

[2] $[X/Fe] = \log(N_X/N_{Fe}) - \log(N_X/N_{Fe})_\odot$

O and Na abundances for stars close to the MS have been measured by D'Orazi et al. (2010) and Dobrovolskas et al. (2014). Both papers find the anti-correlation O-Na, and hence the sign of stellar multi-populations. The difference of Na and O abundances between the two populations is of the same order in both papers ( 0.25 dex), however, the absolute values in each work are significantly different. Dobrovolskas et al. (2014) derives [O/Fe] and [Na/Fe] for each population including recent NLTE corrections and 3D atmosphere modelling. That procedure is similar to that adopted in (Caffau et al., 2011b) for the solar mixture currently adopted in `PARSEC` . We decide hence to take the average O and Na abundances from Dobrovolskas et al. (2014) for the two populations in 47Tuc: $[O/Fe]_{FG} = 0.42dex$, $[Na/Fe]_{FG} = -0.12dex$, $[O/Fe]_{SG} = 0.17dex$ and $[Na/Fe]_{SG} = 0.1dex$.

Mg, Si, Ca, and Ti abundances are presented in different spectroscopic follow-up for RGB and HB stars in 47Tuc (Carretta et al., 2009b, 2013; Gratton et al., 2013; Cordero et al., 2014; Thygesen et al., 2014). There is no clear abundance difference of these elements between the two populations, and they are enhanced with respect to the Sun by 0.2–0.32 dex. Values of Al abundance taken from different literature sources show a large scatter. Carretta et al. (2013) (for RGB stars) provides values systematically larger than other works for the same target. They also find different values of Al abundance for each population: [Al/Fe]=0.45 and [Al/Fe]=0.64. Thygesen et al. (2014) warned about the important effects of NLTE corrections, both decreasing the scatter and shifting down the Al abundance estimate. We decide hence to adopt Al abundance from measurements by Thygesen et al. (2014) (13 RGBs) and Cordero et al. (2014) (150 RGBs).

Because no measurement on Ne and S abundance is available in the literature, we adopt [Ne/Fe]=0.4 and [S/Fe]=0.4 as estimated values.

Concerning He mass fraction, the scatter in luminosity seen in some evolutionary phases has been attributed to different amounts of He in the stellar plasma (see references above). The analyses presented in Milone et al. (2012) suggests that the best fitting of the colour difference between the two populations is obtained with a combination of different C, N and O abundances, plus a small increase of He in the SG (0.015-0.02). These results agree with those presented in Di Criscienzo et al. (2010) ($\Delta$ He=0.02–0.03), and rule out the possibility of explaining the 47Tuc CMD only with the variation of He abundance.

The resulting metallicities, assuming a [Fe/H]=-0.76 (Carretta et al., 2009a), are $Z_{FG} = 0.0056$ for the first generation and $Z_{SG} = 0.0055$ for the second generation, respectively. (Following Milone et al., 2012), the assumed He abundances are Y=0.256 and Y=0.276 for FG and SG respectively. Table 3.1 lists the general metal mixture information for the two stellar populations, including Z, Y, [M/H], [Fe/H], and $[\alpha/Fe]$ . The referred solar abundance is derived from Caffau et al. (2011b), as described in Bressan et al. (2012). Figure 3.1 illustrates the number density difference among FG, SG, and the solar mixture from Li to Co. For the de-tails of metal mixtures, table 3.2 and table 3.3 provide the abundances in number fraction and mass fraction, respectively for each chemical element.

Table 3.1: Adopted metal partitions of the two stellar populations (FG and SG) of 47Tuc. See the text for the references and table 3.2 and table 3.3 for more details.

| | FG | SG |
|---|---|---|
| Z | 0.0056 | 0.0055 |
| Y | 0.256 | 0.276 |
| [M/H] | -0.4308 | -0.4143 |
| [Fe/H] | -0.7621 | -0.7620 |
| [$\alpha$/Fe] | 0.4071 | 0.1926 |

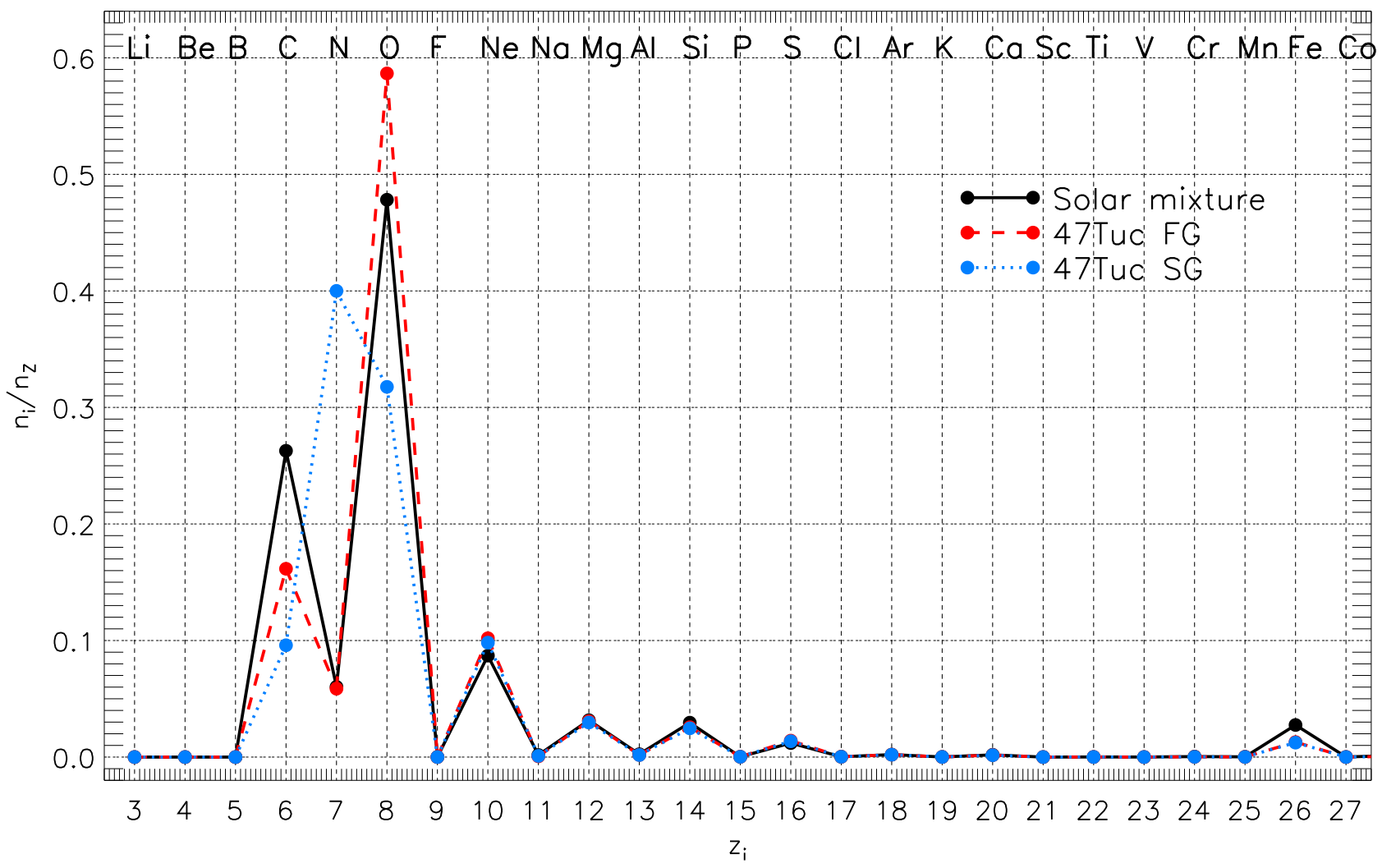

Figure 3.1: Comparison of PARSEC solar mixture (black solid line), 47Tuc FG (red dashed line) and 47Tuc SG (blue dotted line). X axis is the atomic number of each element, the name of the element is displayed in the top of the figure. Y axis shows the number density $n_i/n_Z$ (the number fraction of particular element over the total metal elements). Since the number density of elements with atom number larger than Fe is very small, here we only illustrate the comparison from Li to Co. More details for all metal elements are given in Table 3.2 and Table 3.3.

Table 3.2: Composition of the first stellar generation of 47Tuc. Z=0.0056, Y=0.256, [$\alpha$/Fe] ~0.4

| $Z_i$ | **Element** | $N_i/N_Z$ | $Z_i/Z_{tot}$ |
|---|---|---|---|
| 3 | Li | 3.299375E-07 | 5.366212E-07 |
| 4 | Be | 9.735398E-09 | 5.120059E-09 |
| 5 | B | 1.375161E-07 | 8.676808E-08 |
| 6 | C | 1.615674E-01 | 1.132476E-01 |
| 7 | N | 5.867508E-02 | 4.796040E-02 |
| 8 | O | 5.866157E-01 | 5.477066E-01 |
| 9 | F | 2.979305E-05 | 3.303113E-05 |
| 10 | Ne | 1.019421E-01 | 1.199995E-01 |
| 11 | Na | 6.285701E-04 | 8.432952E-04 |
| 12 | Mg | 3.079312E-02 | 4.368395E-02 |
| 13 | Al | 1.813233E-03 | 2.855037E-03 |
| 14 | Si | 2.560670E-02 | 4.196902E-02 |
| 15 | P | 2.367091E-04 | 4.278586E-04 |
| 16 | S | 1.407841E-02 | 2.634322E-02 |
| 17 | Cl | 2.594866E-04 | 5.369264E-04 |
| 18 | Ar | 2.061175E-03 | 4.805047E-03 |
| 19 | K | 1.057341E-04 | 2.412667E-04 |
| 20 | Ca | 1.653688E-03 | 3.867544E-03 |
| 21 | Sc | 1.214269E-06 | 3.185614E-06 |
| 22 | Ti | 6.432113E-05 | 1.797171E-04 |
| 23 | V | 3.875732E-06 | 1.152172E-05 |
| 24 | Cr | 1.813233E-04 | 5.502100E-04 |
| 25 | Mn | 9.515984E-05 | 3.050831E-04 |
| 26 | Fe | 1.283375E-02 | 4.182609E-02 |
| 27 | Co | 3.224436E-05 | 1.108932E-04 |
| 28 | Ni | 6.892134E-04 | 2.361141E-03 |
| 29 | Cu | 6.287148E-06 | 2.331533E-05 |
| 30 | Zn | 1.543312E-05 | 5.888919E-05 |
| 31 | Ga | 6.224650E-07 | 2.532482E-06 |
| 32 | Ge | 2.109186E-06 | 8.939742E-06 |
| 33 | As | 1.924044E-07 | 8.412272E-07 |
| 34 | Se | 2.109186E-06 | 9.722543E-06 |
| 35 | Br | 3.501184E-07 | 1.632633E-06 |
| 36 | Kr | 1.676158E-06 | 8.197002E-06 |
| 37 | Rb | 3.267495E-07 | 1.629709E-06 |

Continued on next page

**Table 3.2 – continued from previous page**

| $Z_i$ | **Element** | $N_i/N_Z$ | $Z_i/Z_{tot}$ |
|---|---|---|---|
| 38 | Sr | 7.657993E-07 | 3.915543E-06 |
| 39 | Y | 1.426313E-07 | 7.400084E-07 |
| 40 | Zr | 3.267495E-07 | 1.739459E-06 |
| 41 | Nb | 2.158316E-08 | 1.170178E-07 |
| 42 | Mo | 6.826765E-08 | 3.820127E-07 |
| 44 | Ru | 5.678256E-08 | 3.348992E-07 |
| 45 | Rh | 1.082219E-08 | 6.498975E-08 |
| 46 | Pd | 4.019897E-08 | 2.496740E-07 |
| 47 | Ag | 7.148501E-09 | 4.499881E-08 |
| 48 | Cd | 4.832980E-08 | 3.170736E-07 |
| 49 | In | 3.751587E-08 | 2.513722E-07 |
| 50 | Sn | 8.205687E-08 | 5.685658E-07 |
| 51 | Sb | 8.205687E-09 | 5.830520E-08 |
| 52 | Te | 1.426313E-07 | 1.062311E-06 |
| 53 | I | 2.655920E-08 | 1.966904E-07 |
| 54 | Xe | 1.214269E-07 | 9.303461E-07 |
| 55 | Cs | 1.107172E-08 | 8.587149E-08 |
| 56 | Ba | 1.107172E-07 | 8.872860E-07 |
| 57 | La | 1.214269E-08 | 9.842960E-08 |
| 58 | Ce | 3.120434E-08 | 2.551465E-07 |
| 59 | Pr | 4.208380E-09 | 3.460517E-08 |
| 60 | Nd | 2.594866E-08 | 2.184214E-07 |
| 62 | Sm | 8.398756E-09 | 7.369679E-08 |
| 63 | Eu | 2.717158E-09 | 2.409614E-08 |
| 64 | Gd | 1.082219E-08 | 9.931455E-08 |
| 65 | Tb | 6.518009E-10 | 6.045045E-09 |
| 66 | Dy | 1.132961E-08 | 1.074385E-07 |
| 67 | Ho | 1.493190E-09 | 1.437166E-08 |
| 68 | Er | 6.984173E-09 | 6.817136E-08 |
| 69 | Tm | 8.205687E-10 | 8.089542E-09 |
| 70 | Yb | 9.867678E-09 | 9.964249E-08 |
| 71 | Lu | 9.423559E-10 | 9.621921E-09 |
| 72 | Hf | 6.082960E-09 | 6.336097E-08 |
| 73 | Ta | 6.082960E-10 | 6.423324E-09 |
| 74 | W | 1.057341E-08 | 1.134357E-07 |
| 75 | Re | 1.563561E-09 | 1.699093E-08 |
| 76 | Os | 1.880247E-08 | 2.087383E-07 |
| 77 | Ir | 1.837448E-08 | 2.061085E-07 |

Continued on next page

**Table 3.2 – continued from previous page**

| $Z_i$ | **Element** | $N_i/N_Z$ | $Z_i/Z_{tot}$ |
|---|---|---|---|
| 78 | Pt | 5.177439E-08 | 5.894139E-07 |
| 79 | Au | 8.398756E-09 | 9.653809E-08 |
| 80 | Hg | 1.107172E-08 | 1.296076E-07 |
| 81 | Tl | 6.518009E-09 | 7.774147E-08 |
| 82 | Pb | 7.313327E-08 | 8.843725E-07 |
| 83 | Bi | 4.208380E-09 | 5.132297E-08 |
| 90 | Th | 9.867678E-10 | 1.336182E-08 |
| 92 | U | 2.594866E-10 | 3.604425E-09 |
| 93 | Np | 0.000000D+00 | 0.000000D+00 |
| 94 | Pu | 0.000000D+00 | 0.000000D+00 |
| 95 | Am | 0.000000D+00 | 0.000000D+00 |
| 96 | Cm | 0.000000D+00 | 0.000000D+00 |
| 97 | Bk | 0.000000D+00 | 0.000000D+00 |
| 98 | Cf | 0.000000D+00 | 0.000000D+00 |
| 99 | Es | 0.000000D+00 | 0.000000D+00 |

Table 3.3: Composition of the first stellar generation of 47Tuc. Z=0.0055, Y=0.276, [$\alpha$/Fe] ~0.2.

| $Z_i$ | **Element** | $N_i/N_Z$ | $Z_i/Z_{tot}$ |
|---|---|---|---|
| 3 | Li | 3.205999E-07 | 5.366185E-07 |
| 4 | Be | 9.374764E-09 | 5.074106E-09 |
| 5 | B | 1.324221E-07 | 8.598933E-08 |
| 6 | C | 9.593130E-02 | 6.920121E-02 |
| 7 | N | 4.000003E-01 | 3.364863E-01 |
| 8 | O | 3.176584E-01 | 3.052338E-01 |
| 9 | F | 2.894909E-05 | 3.303096E-05 |
| 10 | Ne | 9.816583E-02 | 1.189225E-01 |
| 11 | Na | 1.004524E-03 | 1.386961E-03 |
| 12 | Mg | 2.965243E-02 | 4.329188E-02 |
| 13 | Al | 1.746065E-03 | 2.829413E-03 |
| 14 | Si | 2.465814E-02 | 4.159235E-02 |
| 15 | P | 2.300037E-04 | 4.278565E-04 |
| 16 | S | 1.355690E-02 | 2.610678E-02 |
| 17 | Cl | 2.521360E-04 | 5.369237E-04 |
| 18 | Ar | 2.002788E-03 | 4.805023E-03 |
| 19 | K | 1.027389E-04 | 2.412655E-04 |
| 20 | Ca | 1.592430E-03 | 3.832832E-03 |
| 21 | Sc | 1.179872E-06 | 3.185598E-06 |
| 22 | Ti | 6.193845E-05 | 1.781041E-04 |
| 23 | V | 3.732161E-06 | 1.141831E-05 |
| 24 | Cr | 1.746065E-04 | 5.452718E-04 |
| 25 | Mn | 9.163478E-05 | 3.023450E-04 |
| 26 | Fe | 1.235835E-02 | 4.145070E-02 |
| 27 | Co | 3.104991E-05 | 1.098979E-04 |
| 28 | Ni | 6.636825E-04 | 2.339950E-03 |
| 29 | Cu | 6.054250E-06 | 2.310608E-05 |
| 30 | Zn | 1.486142E-05 | 5.836065E-05 |
| 31 | Ga | 6.048322E-07 | 2.532470E-06 |
| 32 | Ge | 2.049439E-06 | 8.939698E-06 |
| 33 | As | 1.869541E-07 | 8.412230E-07 |
| 34 | Se | 2.049439E-06 | 9.722495E-06 |
| 35 | Br | 3.402005E-07 | 1.632625E-06 |
| 36 | Kr | 1.628677E-06 | 8.196962E-06 |
| 37 | Rb | 3.174935E-07 | 1.629701E-06 |

Continued on next page

**Table 3.3 – continued from previous page**

| $Z_i$ | **Element** | $N_i/N_Z$ | $Z_i/Z_{tot}$ |
|---|---|---|---|
| 38 | Sr | 7.441062E-07 | 3.915523E-06 |
| 39 | Y | 1.385910E-07 | 7.400048E-07 |
| 40 | Zr | 3.174935E-07 | 1.739450E-06 |
| 41 | Nb | 2.097176E-08 | 1.170172E-07 |
| 42 | Mo | 6.633380E-08 | 3.820108E-07 |
| 44 | Ru | 5.517405E-08 | 3.348975E-07 |
| 45 | Rh | 1.051562E-08 | 6.498943E-08 |
| 46 | Pd | 3.906024E-08 | 2.496727E-07 |
| 47 | Ag | 6.946002E-09 | 4.499859E-08 |
| 48 | Cd | 4.696074E-08 | 3.170720E-07 |
| 49 | In | 3.645314E-08 | 2.513710E-07 |
| 50 | Sn | 7.973241E-08 | 5.685630E-07 |
| 51 | Sb | 7.973241E-09 | 5.830491E-08 |
| 52 | Te | 1.385910E-07 | 1.062306E-06 |
| 53 | I | 2.580684E-08 | 1.966894E-07 |
| 54 | Xe | 1.179872E-07 | 9.303415E-07 |
| 55 | Cs | 1.075808E-08 | 8.587107E-08 |
| 56 | Ba | 1.075808E-07 | 8.872816E-07 |
| 57 | La | 1.179872E-08 | 9.842911E-08 |
| 58 | Ce | 3.032040E-08 | 2.551452E-07 |
| 59 | Pr | 4.089167E-09 | 3.460500E-08 |
| 60 | Nd | 2.521360E-08 | 2.184203E-07 |
| 62 | Sm | 8.160841E-09 | 7.369643E-08 |
| 63 | Eu | 2.640188E-09 | 2.409602E-08 |
| 64 | Gd | 1.051562E-08 | 9.931406E-08 |
| 65 | Tb | 6.333371E-10 | 6.045015E-09 |
| 66 | Dy | 1.100867E-08 | 1.074380E-07 |
| 67 | Ho | 1.450891E-09 | 1.437159E-08 |
| 68 | Er | 6.786329E-09 | 6.817102E-08 |
| 69 | Tm | 7.973241E-10 | 8.089502E-09 |
| 70 | Yb | 9.588152E-09 | 9.964200E-08 |
| 71 | Lu | 9.156614E-10 | 9.621873E-09 |
| 72 | Hf | 5.910645E-09 | 6.336066E-08 |
| 73 | Ta | 5.910645E-10 | 6.423292E-09 |
| 74 | W | 1.027389E-08 | 1.134352E-07 |
| 75 | Re | 1.519270E-09 | 1.699084E-08 |
| 76 | Os | 1.826985E-08 | 2.087372E-07 |
| 77 | Ir | 1.785397E-08 | 2.061075E-07 |

Continued on next page

**Table 3.3 – continued from previous page**

| $Z_i$ | **Element** | $N_i/N_Z$ | $Z_i/Z_{tot}$ |
|---|---|---|---|
| 78 | Pt | 5.030775E-08 | 5.894110E-07 |
| 79 | Au | 8.160841E-09 | 9.653762E-08 |
| 80 | Hg | 1.075808E-08 | 1.296070E-07 |
| 81 | Tl | 6.333371E-09 | 7.774109E-08 |
| 82 | Pb | 7.106159E-08 | 8.843681E-07 |
| 83 | Bi | 4.089167E-09 | 5.132272E-08 |
| 90 | Th | 9.588152E-10 | 1.336175E-08 |
| 92 | U | 2.521360E-10 | 3.604407E-09 |
| 93 | Np | 0.000000D+00 | 0.000000D+00 |
| 94 | Pu | 0.000000D+00 | 0.000000D+00 |
| 95 | Am | 0.000000D+00 | 0.000000D+00 |
| 96 | Cm | 0.000000D+00 | 0.000000D+00 |
| 97 | Bk | 0.000000D+00 | 0.000000D+00 |
| 98 | Cf | 0.000000D+00 | 0.000000D+00 |
| 99 | Es | 0.000000D+00 | 0.000000D+00 |

In order to be consistent with the $\alpha$-element measurements in the large stellar spectroscopic surveys, we adopt the same $\alpha$-elements definition to calculate the total $\alpha$ enrichment [$\alpha$/Fe] , as APOGEE does. In APOGEE the six $\alpha$-elements are: O, Mg, Si, S, Ca, and Ti. Thus for the FG stars of 47Tuc, [Z=0.0056, Y=0.256], [$\alpha$/Fe] = $0.4071dex$ and, for the SG stars, [Z=0.0055, Y=0.276], [$\alpha$/Fe] = $0.1926dex$. These two values, approximately ~ 0.4 and 0.2 respectively, are the typical [$\alpha$/Fe] values observed in alpha-enriched stars. Finally we note that we will adopt the metal partitions of these two alpha-enriched generations to calculate stellar evolutionary tracks and isochrones also for other metallicities.

## 3.2 Isochrones fitting and Luminosity function

With the detailed metal mixture and helium abundance of 47Tuc, we calculate new sets of evolutionary tracks and isochrones and transform them into the observational color-magnitude diagram (CMD) in order to fit the data. This process is used to calibrate other stellar parameters in the model as described below.

### 3.2.1 Low main sequence to turn-off

Kalirai et al. (2012) provides deep images of 47Tuc taken with the Advanced Camera for Surveys (ACS) on *Hubble Space Telescope* (HST). The corresponding colour-magnitude diagrams cover the whole main sequence of this cluster, till the faintest stars. Figure 3.2 shows our isochrone fitting of their photometric data, i.e. V band (F606W) and I band (F814W). In order to display the relative density of stars on CMD the data are plotted with the Hess diagram (binsize 0.025 mag in color and 0.1 mag in I magnitude). By assuming a standard extinction law (Cardelli, Clayton & Mathis, 1989), we derive, from the isochrone fitting, an age of 12.00±0.2 Gyr, a distance modulus of $(m\text{-}M)_0$ $13.22^{+0.02}_{-0.01}$, and a reddening of E(V-I)=$0.035^{-0.008}_{+0.005}$. Isochrones with these three sets of parameters (middle value and the two extreme values) are plotted in Figure 3.2, though for the entire main sequence they are almost superimposed. The main difference of these three sets is in the RGB phase which we will discuss in Chap. 3.2.3. The relations between the three parameters are illustrated in Figure 3.3.

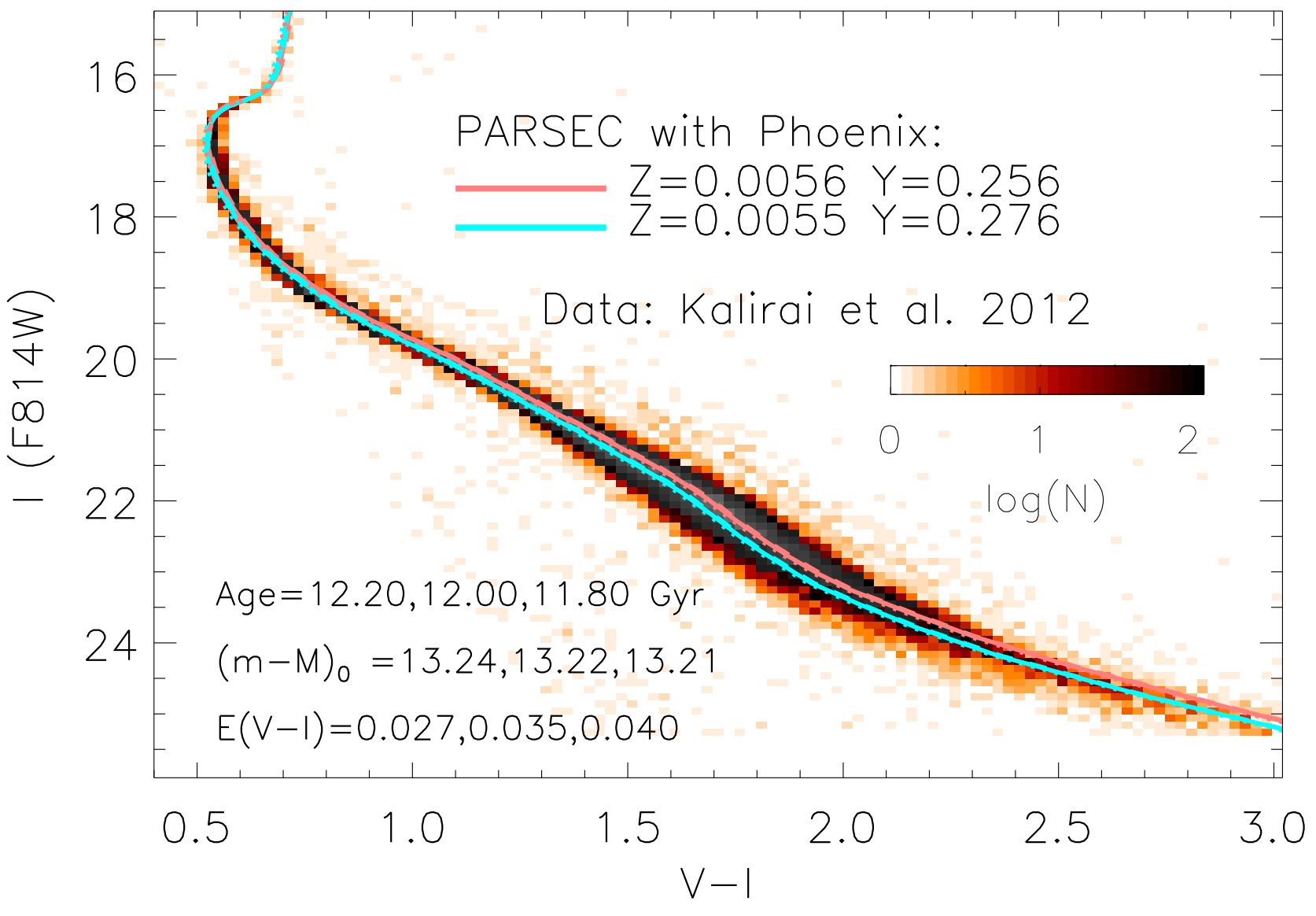

Figure 3.2: Isochrone fitting with Hess diagram of 47Tuc data for the low main sequence(Kalirai et al., 2012). The bin size of Hess diagram is 0.025 mag in color and 0.1 mag in I(F814W) magnitude. Three sets with different fitting parameters (age, $\eta$, $(m\text{-}M)_0$, and E(V-I)) listed in the legend are plotted with solid line, dotted line, and dashed line, respectively. The corresponding isochrones are almost superimposed in the figure and the do not show visible difference.

We note that our best fit parameters are different from the HST proper motion results (Watkins et al., 2015), which gives 4.15 kpc ( $(m\text{-}M)_0 \sim 13.09$), and the eclipsing binary distance measurement in the cluster (Thompson et al., 2010) $(m\text{-}M)_0 = 13.35$. Gaia will release the parallaxes and proper motions including stars in 47Tuc in its DR2 at the end of 2017, and will help to solve the distance problem. However we will show in the following section that our results offer the very best global fitting, from the very low main sequence till the red giant and horizontal branches.

### 3.2.2 RGB bump and envelope overshooting calibration

Some GC features in CMD are very sensitive to stellar model parameters which are, otherwise, hardly constrained from observations directly. This is the case of the efficiency of mixing below the convective envelope (envelope overshooting), that is known to affect the luminosity of the red giant branch bump (RGBB). In this section we will use the 47Tuc data to calibrate the envelope overshooting to be used in low mass stars by PARSEC .

The RGB bump is one of the most intriguing features in the CMD. When a star evolves to the "first dredge-up" in the red giant phase, its surface convective zone deepens while the burning hydrogen shell moves outwards. Figure 3.4 depicts the structure evolution during the first dredge-up for a 0.85 $M_\odot$ star with [Z=0.0055, Y=0.276]. When the hydrogen burning shell encounters the chemical composition discontinuity left by the previous penetration of the convective zone, the sudden increase of H affects the efficiency of the burning shell and the star becomes temporarily fainter. Soon after a new equilibrium is reached, the luminosity of the star raises again. Since the evolutionary track crosses the same luminosity three times in a short time, there is an excess of star counts in a small range of magnitudes, making a "bump" in the star number distribution (luminosity function) along the red giant branch. This is because the number of stars in the post main sequence phases is proportional to the evolutionary time of the stars in these phases. The longer the crossing time of the chemical composition discontinuity by the burning shell, the more the stars accumulate in that region of the RGB.

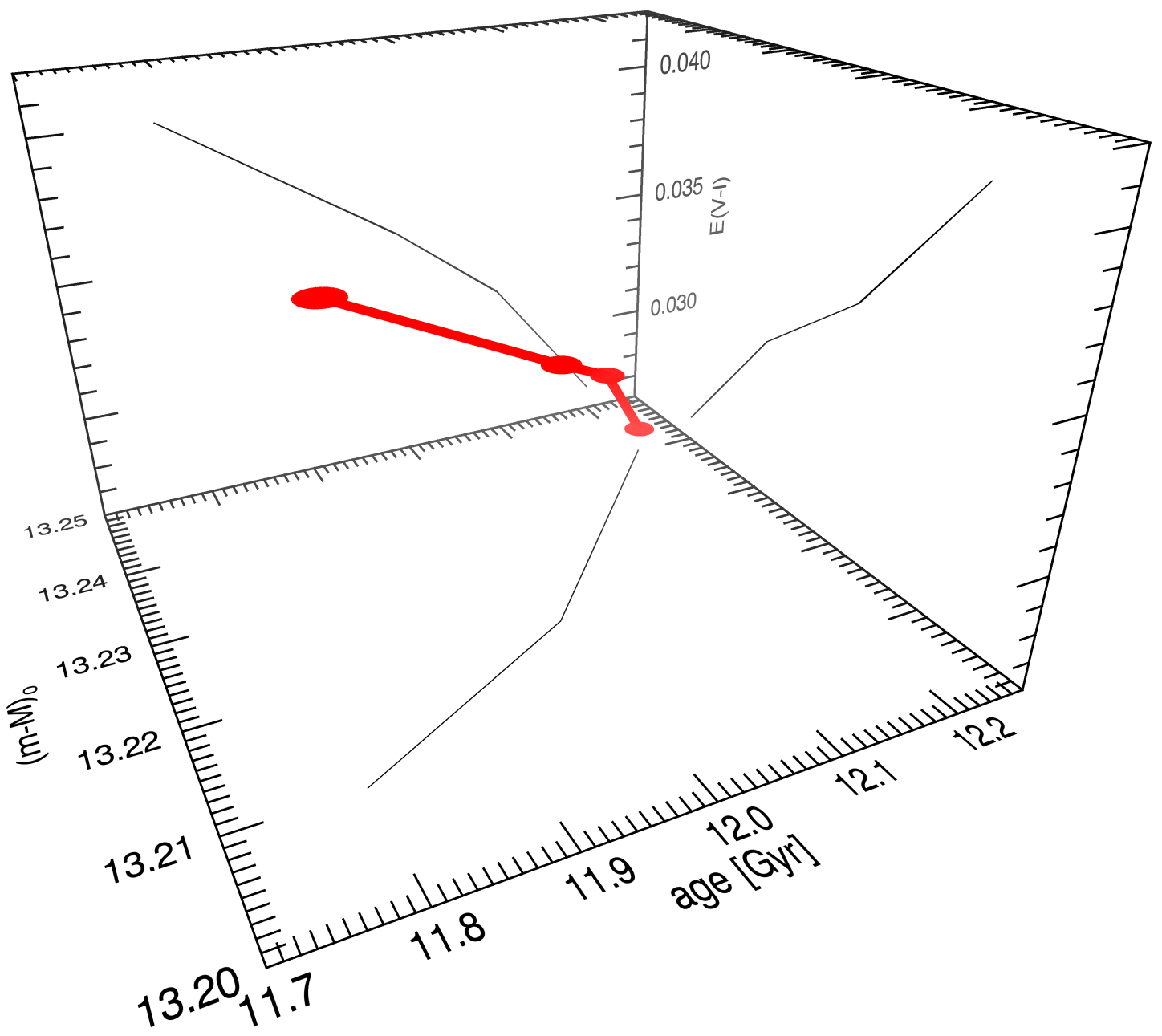

Figure 3.3: Relations between the three fitting parameters, age, distance modulus $(m\text{-}M)_0$, and reddening E(V-I). The relations are derived from our best fits to 47Tuc: age=12.00±0.2 Gyr, $(m\text{-}M)_0=13.22^{+0.02}_{-0.01}$, and reddening $E(V\text{-}I)=0.035^{-0.008}_{+0.005}$.

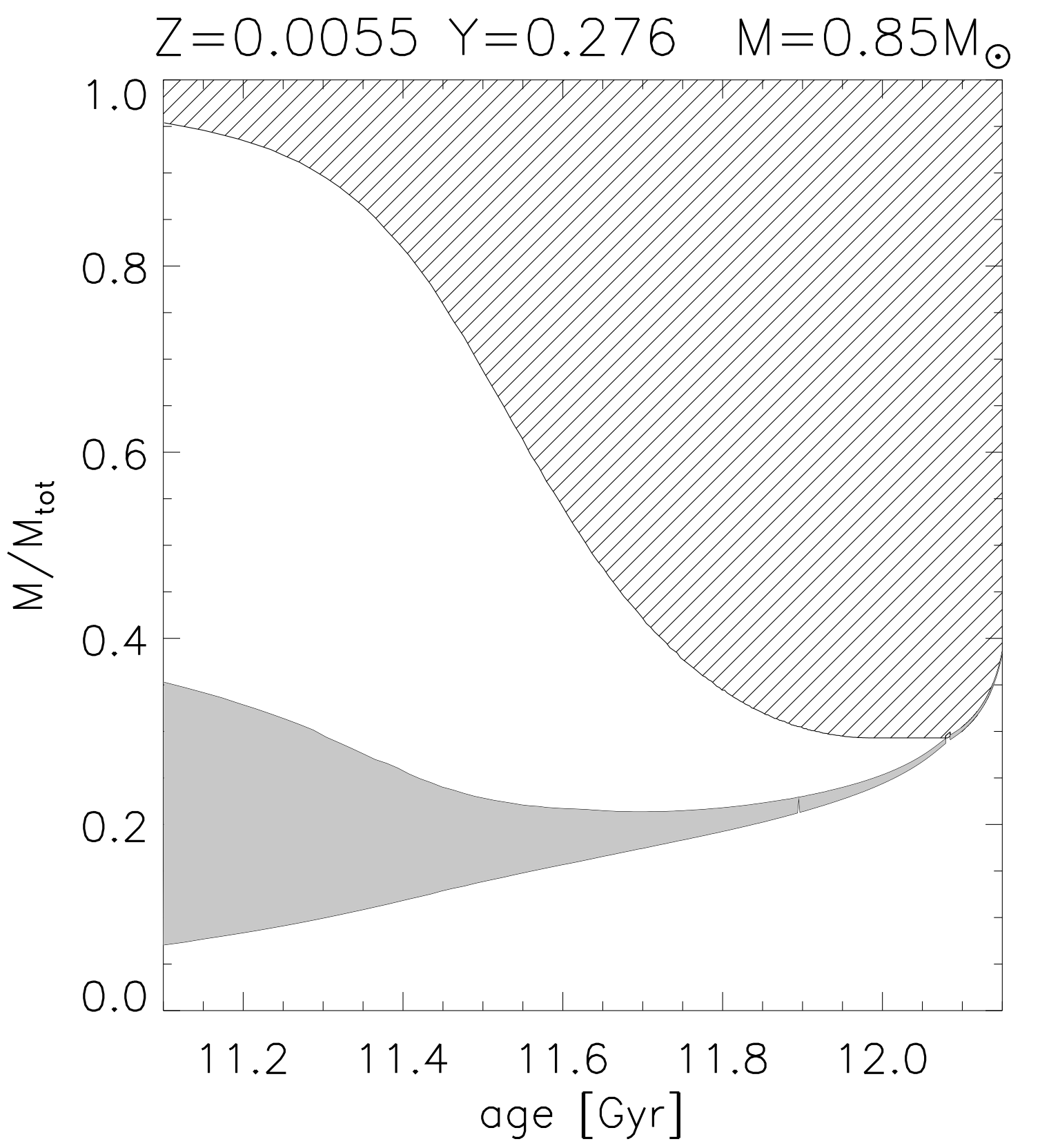

Figure 3.4: Kippenhahn diagrams around the first dredge-up time. Y axis is mass coordinate, X axis is the age of the star. Region filled with oblique lines represents the surface convective zone, and layer filled with grey is the hydrogen burning shell.

The properties of RGBB, including the brightness and the extent, are important to study the stellar structure and to investigate the nature of GCs. 47Tuc was the first GC where the existence of the RGBB was confirmed (King, Da Costa & Demarque, 1985). Since then, many works, both theoretical and observational (for instance, Alongi et al., 1991; Cassisi & Salaris, 1997; Zoccali et al., 1999; Bono et al., 2001; Cassisi, Salaris & Bono, 2002; Bjork & Chaboyer, 2006; Salaris et al., 2006; Cecco et al., 2010; Bragaglia et al., 2010; Cassisi et al., 2011; Nataf et al., 2013), have studied the features of RGBB.

The intrinsic brightness and extent of the RGBB are sensitive to:

- Total metallicity and metal partition.

  Nataf et al. (2013) propose an empirical function of RGBB extent to metallicity: the more metal-poor the globular cluster is, the smaller is the extent of the RGBB. From the theoretical point of view, stars with lower total metallicity are brighter compared to the higher metallicity stars, causing their hydrogen burning shell to move outwards faster. Since they are also hotter, the surface convective envelope is thinner and the chemical composition discontinuity is smaller and less deep. As a consequence, their RGBB is very brief and covers a small range of magnitudes at higher luminosity. This is why RGBB in metal-poor globular clusters is very difficult to be well sampled.

  The metal partition also affects the features of RGBB, even with the total metallicity remains the same. As already shown in Fig. 2.1 and in Sec. 2, a stellar track with $\alpha$-enhancement is hotter than the solar scaled one because of different opacity, leading to a brighter RGBB.

- Helium content

  A larger helium content renders the star hotter and brighter so that it has a shorter main sequence lifetime (Fagotto et al., 1994). Bragaglia et al. (2010) studied the RGBB of 14 globular clusters and found that the more He-rich second generation shows brighter RGBB than the first generation. Similar to the mechanism in metal-poor stars, hot He-rich stars have less deep convective envelopes and their high luminosity makes the hydrogen burning shells to move faster across the discontinuity. Hence the RGBB of He-rich stars are brighter and less extended Salaris et al. (2006)

  Figure 3.5 and Table 3.4 present an example of the evolution of a 0.85 $M_\odot$ star during the RGBB phase. Different metallicity and helium content have been considered. Comparing the track with Z=0.0055, Y=0.296 (green line) to that with Z=0.0055, Y=0.296 (blue line), and that with Z=0.0056, Y=0.276 (orange line) with the one with Z=0.0056, Y=0.256 (red line), in the left panel of Figure 3.5, one can see that, with the same stellar mass and total metallicity Z, a He-rich star is brighter and evolves faster than a star with lower helium abundance. The luminosity range between the two filled dots for each stellar track corresponds to the extent of the RGBB. The right panel of this figure displays their behavior in the HRD, and we see that He-rich stars are hotter. Table 3.4 lists in detail the mean luminosity $log(\bar{L}/L_{\odot})_{RGBB}$, evolutionary time $\Delta$ $t_{RGBB}$, and the luminosity extent $\Delta$ $\log(L)_{RGBB}$ of the RGBB.

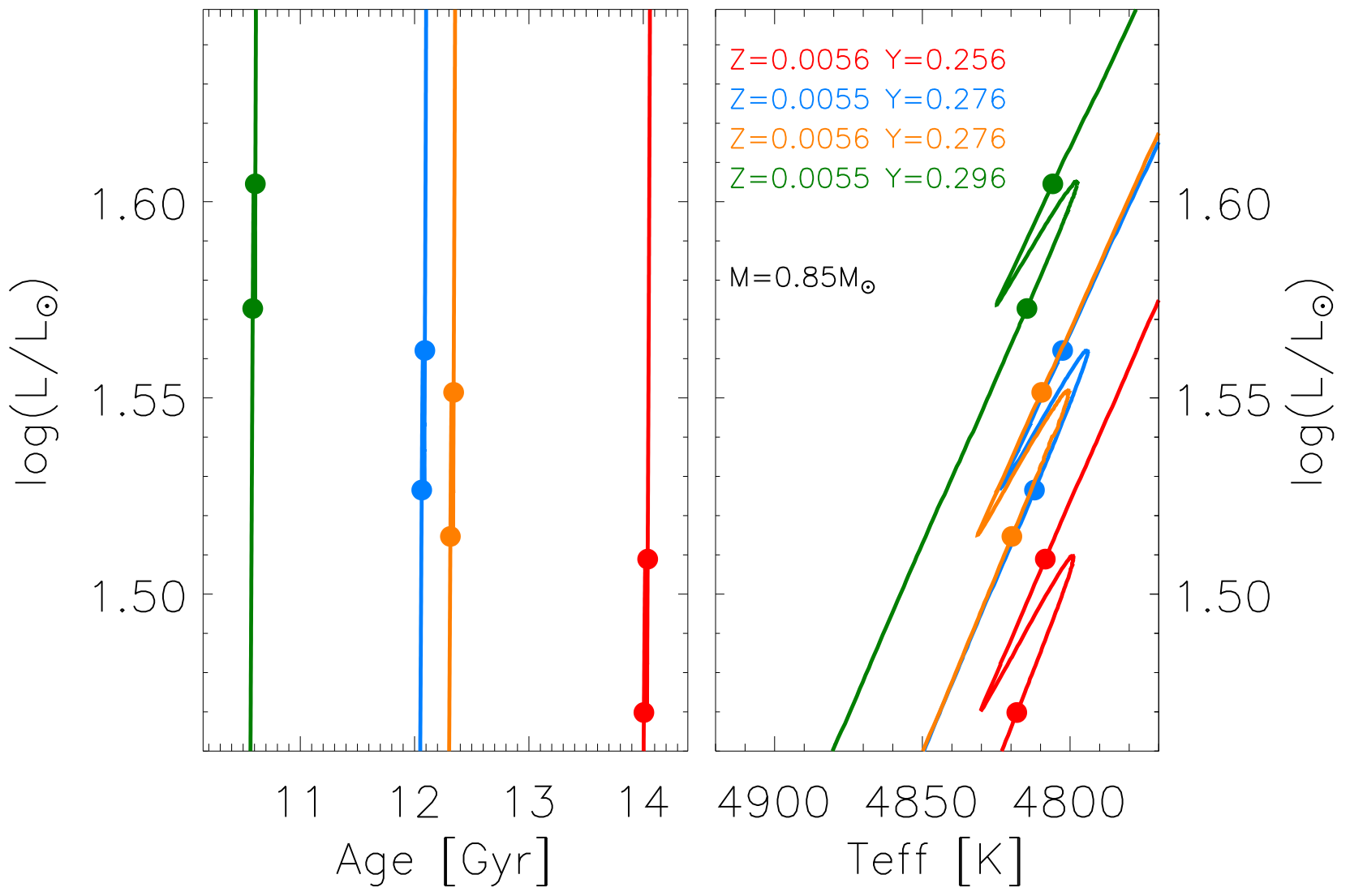

Figure 3.5: RGBB evolution and HRD for different metallicities and helium contents. The red, blue, orange, and green lines represent stars of 0.85 $M_{\odot}$ with [Z=0.0056, Y=0.256], [Z=0.0055, Y=0.276], [Z=0.0056, Y=0.276], and [Z=0.0055, Y=0.296], respectively. Tracks with Z=0.0056 are calculated with [$\alpha$/Fe] ~0.4 and with Z=0.0055 are calculated with [$\alpha$/Fe] ~0.2. The filled dots mark the maximum and minimum luminosity of RGBB for each track. Table 3.4 lists the evolutionary parameters for the RGBB in more detail.

- Age.

  Stars with younger age are hotter, with their thinner convective envelope, their RGBB are brighter. GC with multiple populations born in different ages will show a more extent RGBB compared with the single age one with the same metal mixture and He content.

Table 3.4: RGBB parameters of a 0.85 $M_\odot$ star for different metallicities and helium contents. Mean luminosity $log(\bar{L}/L_\odot)_{RGBB}$ , luminosity extent $\Delta$ log(L)$_{RGBB}$, RGBB beginning time $t_{0,RGBB}$, and last time $\Delta$ $t_{RGBB}$ are listed. The RGBB morphology and HRD is shown in Fig.3.5

| Z | Y | $log(\bar{L}/L_\odot)_{RGBB}$ | $\Delta$ log(L)$_{RGBB}$ | $t_{0,RGBB}$ (Gyr) | $\Delta$ $t_{RGBB}$ (Myr) |
|---|---|---|---|---|---|
| 0.0056 | 0.256 | 1.4894 | 0.03909 | 14.006 | 33.024 |
| 0.0056 | 0.276 | 1.5331 | 0.03672 | 12.313 | 28.220 |
| 0.0055 | 0.276 | 1.5443 | 0.03557 | 12.062 | 27.159 |
| 0.0055 | 0.296 | 1.5887 | 0.03177 | 10.584 | 22.637 |

- Mixing efficiency.

  The mixing efficiency of the star, both mixing length and envelope overshooting (EOV), affects the brightness and evolutionary time of RGBB. The more efficient the mixing is, the earlier the hydrogen-burning shell meets the discontinuity left by the surface convective zone and the fainter the RGBB is. For the mixing length, we adopt the solar-calibrated value $\alpha_{MLT}$=1.74 in PARSEC as described in (Bressan et al., 2012). The EOV is calibrated with the new stellar tracks against the observations of the RGBB of 47Tuc.

Overshooting is the non-local mixing that may occur at the borders of any convectively unstable region (i.e., Bressan et al. (2015) and references therein). The extent of the overshooting at the base of the convective envelope is called envelope overshooting (with parameter $\Lambda_e$). At the base of the convective envelope of the Sun, models with an overshooting region of $\Lambda_e \approx 0.3 \sim 0.5\ H_p$ (where $H_p$ is the pressure scale height) provide a better agreement with the helioseismology data (Christensen-Dalsgaard et al., 2011). There are also other observations that can be better explained with overshooting, for instance: the blue loops of intermediate and massive stars (Alongi et al., 1991; Tang et al., 2014), the carbon stars luminosity functions in the Magellanic Clouds, that require a more efficient third dredge-up in AGB stars (Herwig, 2000; Marigo & Girardi, 2007), the surface abundance of light elements (Fu et al., 2015), and asteroseismic signatures in HD52265 (Lebreton & Goupil, 2012).

In Figure 3.6 we compare the RGBB evolution of models computed with different EOV values, $\Lambda_e$, at the same stellar mass and composition. Like in Figure 3.5, every pair of filled dots marks the brightness extent of the RGBB. The figure shows that a larger envelope overshooting not only makes the RGBB fainter, but also of longer duration, leading to more populated RGBBs.

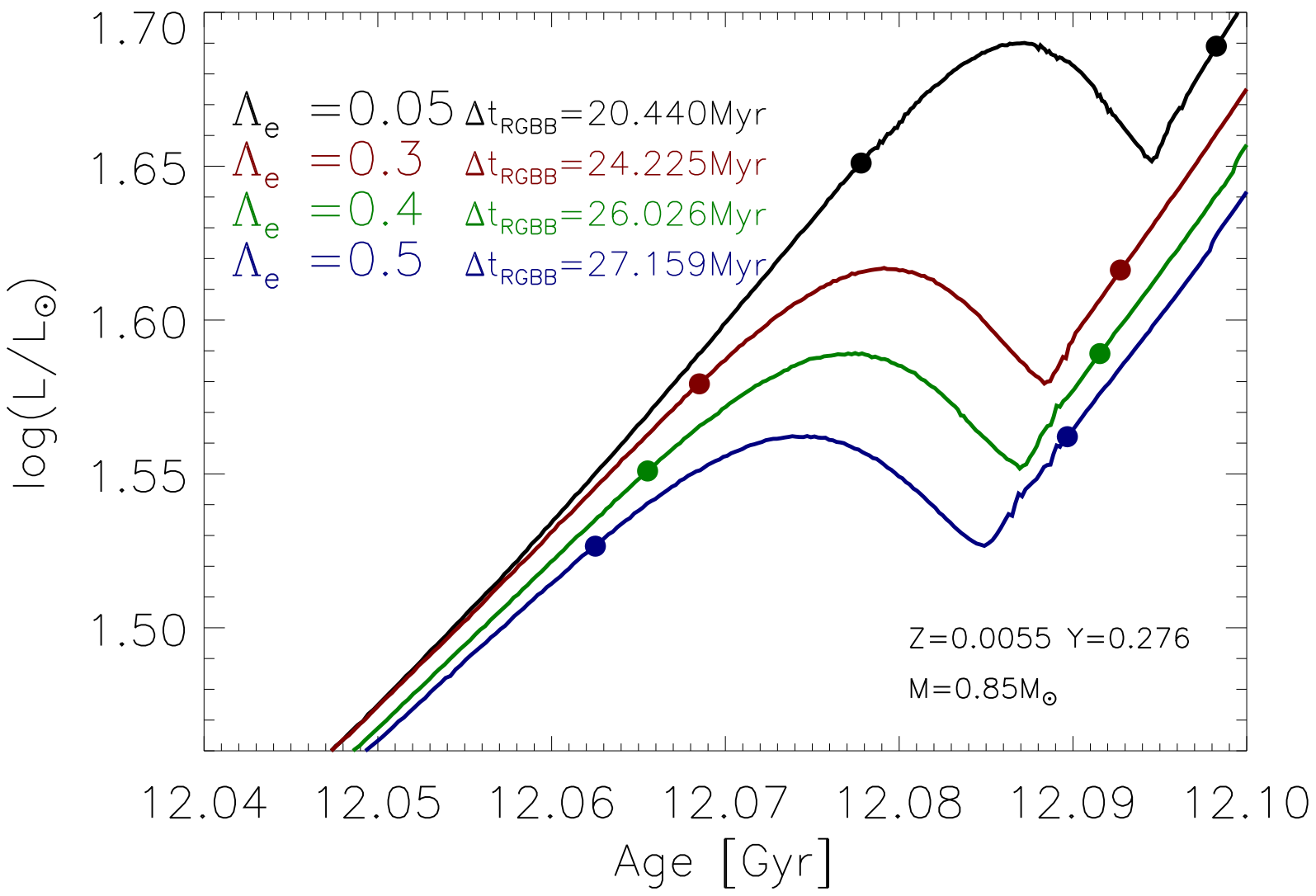

Figure 3.6: The RGBB luminosity as a function of stellar age for a 0.85 $M_{\odot}$ star but with different EOV. The black, red, green, and blue line from top to bottom represent tracks with $\Lambda_e$ = 0.05, 0.3, 0.4, and 0.5. The filled dots mark the minimum and maximum luminosity of RGBB for each track, and $\Delta t_{RGBB}$ is the evolution time from the minimum luminosity to the maximum one.

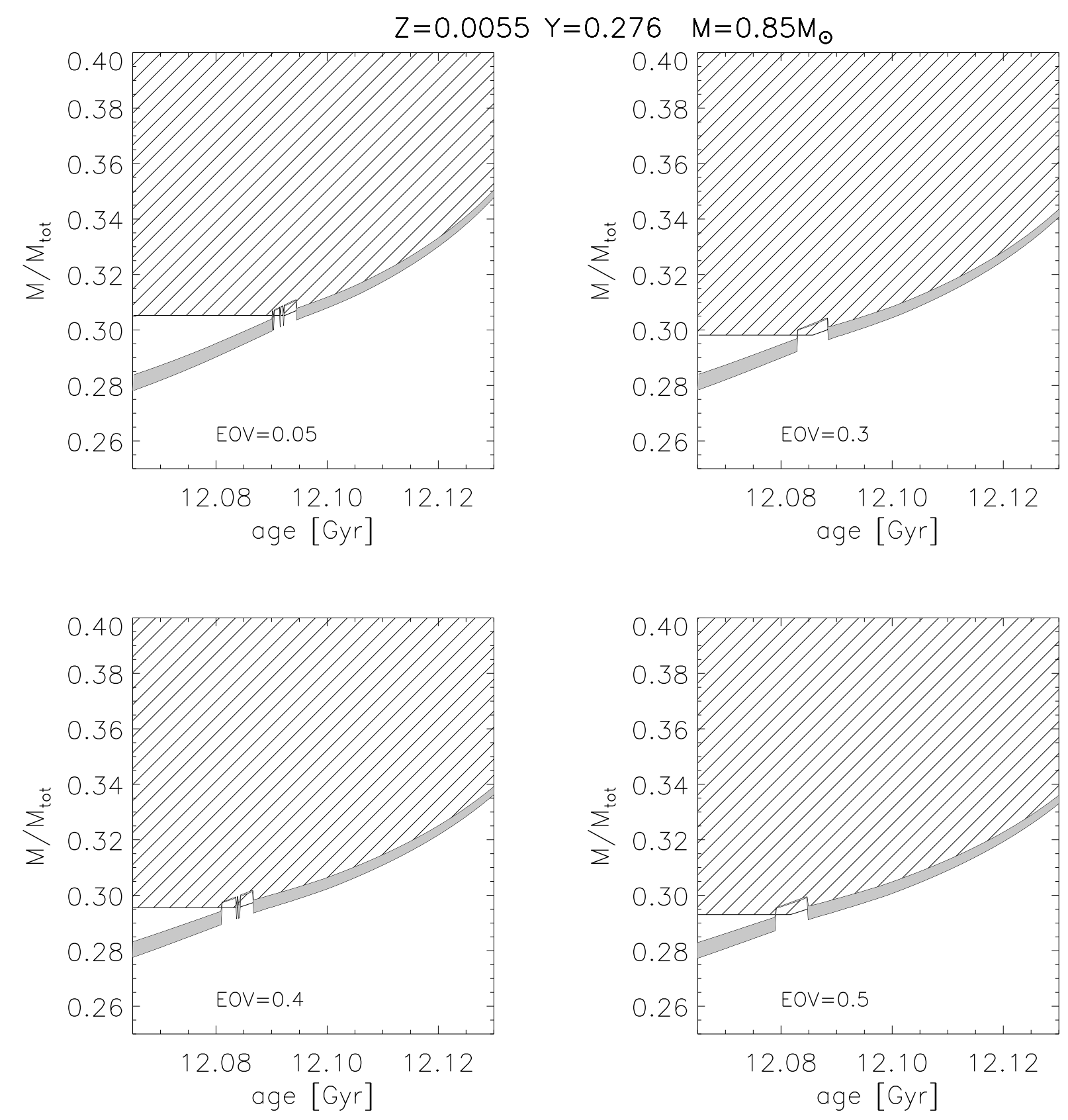

Figure 3.7: Kippenhahn diagrams around the RGBB for 0.85 $M_\odot$ stars with [Z=0.0055, Y=0.276] for different envelope overshooting value. Y axis is mass coodindate, X axis is the age of the star. Region filled with oblique lines represents the surface convective zone, and layer filled with grey is the hydrogen burning shell.

Figure 3.7 shows the evolution of the structure of a 0.85 $M_{\odot}$ [Z=0.0055, Y=0.276] star with differet EOV, during the RGBB. A larger EOV value leads to a deeper surface convective zone, and the hydrogen burning shell encounters the chemical discontinuity earlier.

The Luminosity Function (LF) is a useful tool to compare the observed morphology of RGBB with that predicted by the theory. Taking into account that the 47Tuc population contribution is 30% from the FG and 70% from the SG as suggested by Milone et al. (2012) and Carretta et al. (2009b), we simulated the LF of 47Tuc with our isochrones with different EOV values. The comparison between the observed and predicted LFs is shown in Figure 3.8. For the observed LF we have used data from the HST/ACS survey of globular clusters (Sarajedini et al., 2007). Both observations and models are sampled in bins of 0.05 magnitudes. The three sub-figures represent three different sets of fitting parameters: the two extreme values and the middle one of age=12.00±0.2 Gyr, $(m\text{-}M)_0=13.22^{+0.02}_{-0.01}$, $E(V\text{-}I)=0.035^{-0.008}_{+0.005}$. In each sub-figure, the model LF (orange histogram) are calculated with envelope overshooting $\Lambda_e$=0.3 in the upper panel and with $\Lambda_e$=0.5 in the lower panel. It is evident that the LF computed adopting the small envelope overshooting value $\Lambda_e$=0.3, shows RGBB too bright compared to data (black histogram filled with oblique lines), in all sets of best fitting parameters. We find that the agreement between observations and models is reached when one adopts a value of $\Lambda_e = 0.5H_p$ below the convective border, for all the above sets of parameters. This provides a robust calibration of the envelope overshooting parameter. This envelope overshooting calibration will be applied to all other stellar evolution calculations of low mass stars.

### 3.2.3 Red Giant Branch

Unlike Kalirai et al. (2012) who focus on the faint main sequence part as shown in Figure 3.2, another dataset of 47Tuc, HST/ACS survey of globular clusters (Sarajedini et al., 2007), is devoted to the Horizontal Branch (Anderson et al., 2008) with the same instrument. In Figure 3.9, Figure 3.10, and Figure 3.11 we show the entire fit of ACS data of 47Tuc, from main sequence up to the red giant branch and HB. The three figures represent three sets of fitting parameters, covering the range of the best fits we derived: age=12.00±0.2 Gyr, distance modulus $(m\text{-}M)_0=13.22^{+0.02}_{-0.01}$, and reddening $E(V\text{-}I)=0.035^{-0.008}_{+0.005}$. These parameters are the same as those we used to fit the lower main sequence data in Figure 3.2. In all these three figures, the Hess diagram is used for the global fitting (the left panel) with bin size 0.025 mag in color and 0.1 mag in I(F814W) magnitude. The HB region and the turn-off region are zoomed with scatter plots in the two right panels. Thanks to the detailed composition derived from the already quoted observations, and the new computed models, we are able to perform a global fit of the CMD of

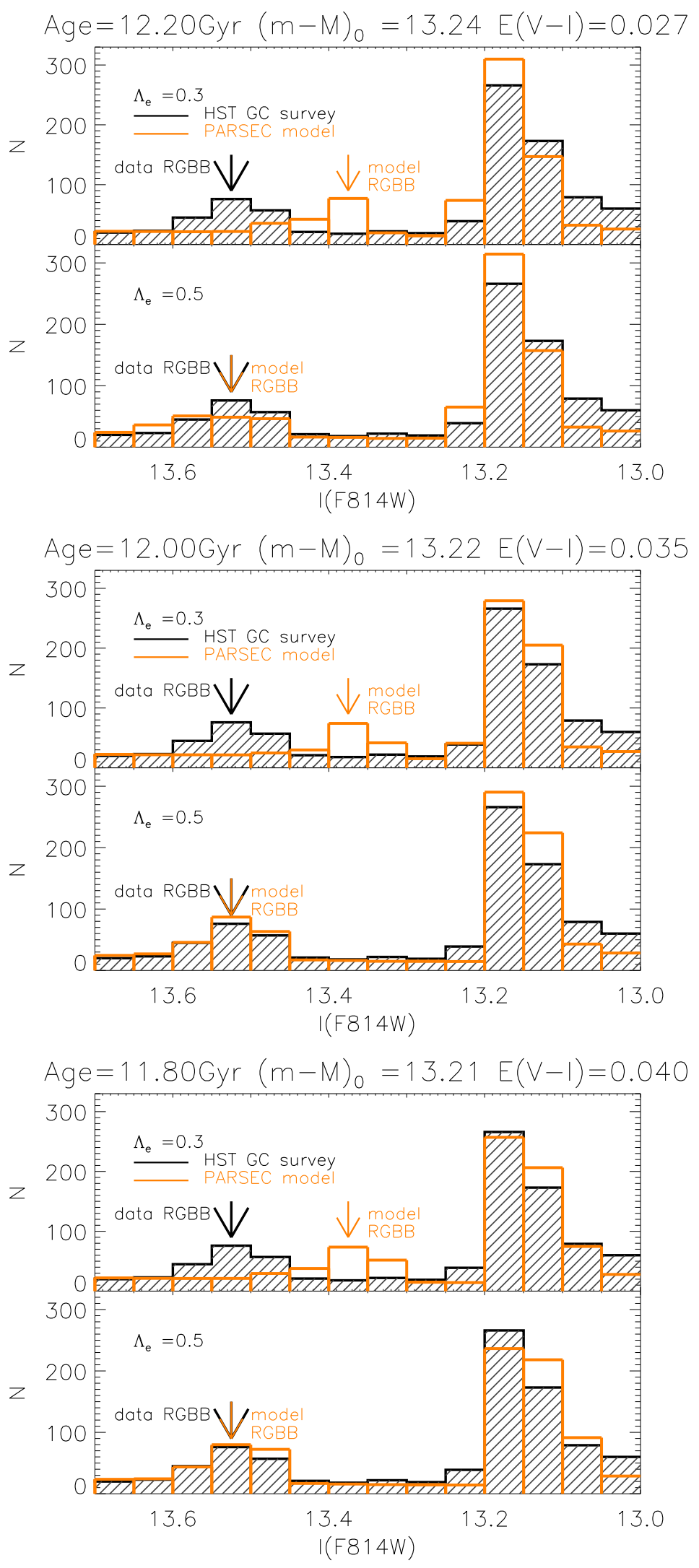

Figure 3.8: Comparison between LF of 47Tuc data (Sarajedini et al., 2007) and the new PARSEC isochrone with different EOV. Three different sets of fitting parameters (age, (m-M)$_0$, and E(V-I)) are plotted as the same as in Fig.3.2. The black histogram filled with oblique lines is the data LF, whilst orange histogram is LF derived from new PARSEC isochrones with 30% contribution from the FG of 47Tuc and 70% from the SG. The upper panel isochrones of each subfigure are calculated with EOV value $\Lambda_e$=0.3, and the lower panel are the ones with $\Lambda_e$=0.5. Orange arrow and black arrow mark the location of RGBB in model and in data, respectively. The bin size of the LF is 0.05 mag.

47Tuc covering almost every evolutionary phase over a range of about 13 Magnitudes. This must be compared with other fits that can be found in literature and that usually are restricted to only selected evolutionary phases (Kim et al., 2002; Salaris et al., 2007; VandenBerg et al., 2013, 2014; Chen et al., 2014; McDonald & Zijlstra, 2015, etc.)

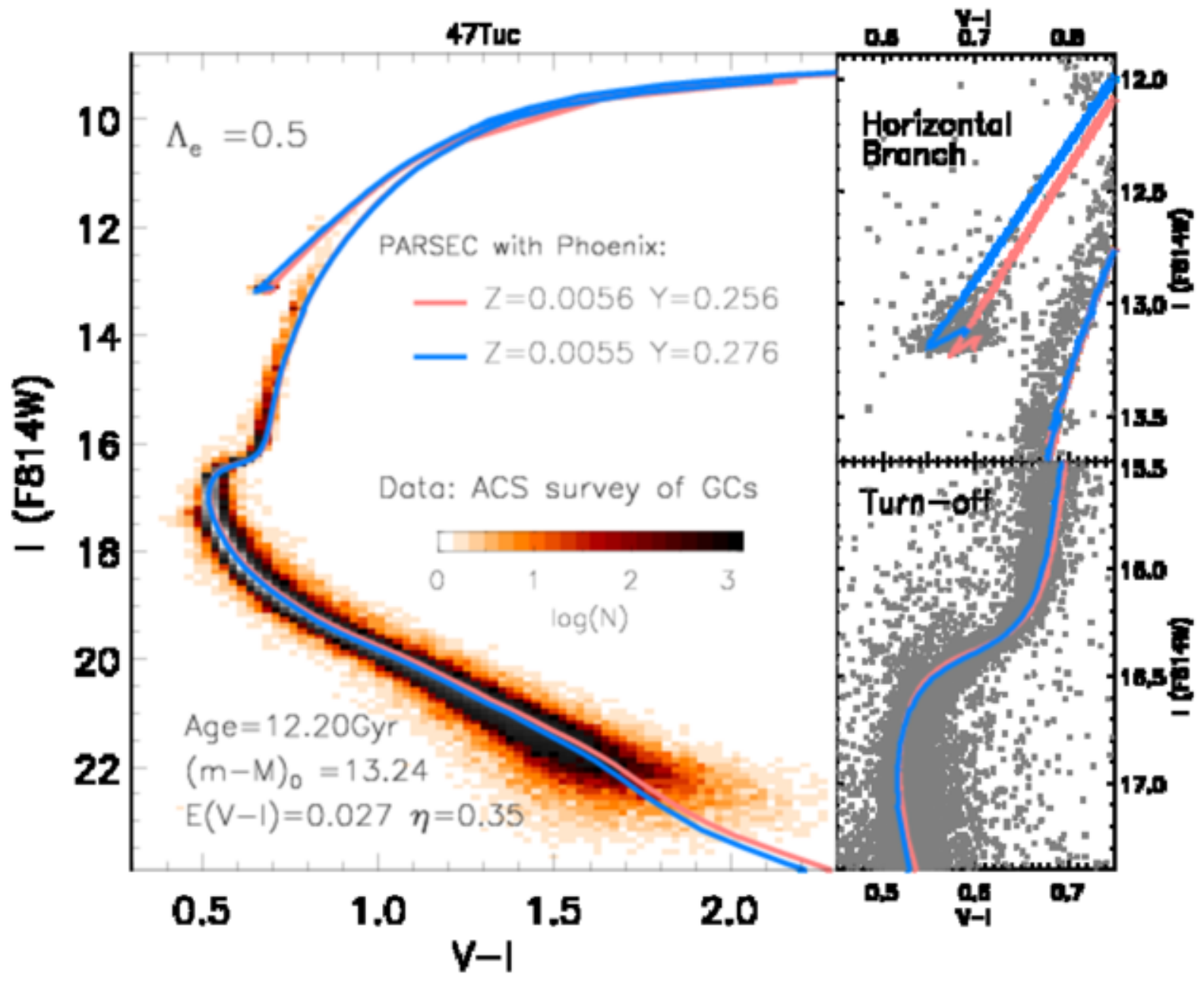

Figure 3.9: Isochrone fitting with Hess diagram (the left panel) of 47Tuc data (Sarajedini et al., 2007) for all the main evolutionary phases, and with scatter plots highlighting the horizontal branch region (the upper right panel) and the turn-off region (the lower right panel). The red line and blue line represent the isochrones of the first and second generation, respectively. The fitting parameters are: age=12.2 Gyr, $(m\text{-}M)_0$=13.24, E(V-I)=0.027. [4]

[4]The figure is in low resolution because of the size limit of arXive. SISSA library offers the full thesis with high-resolution figures: http://urania.sissa.it/xmlui/handle/1963/35251

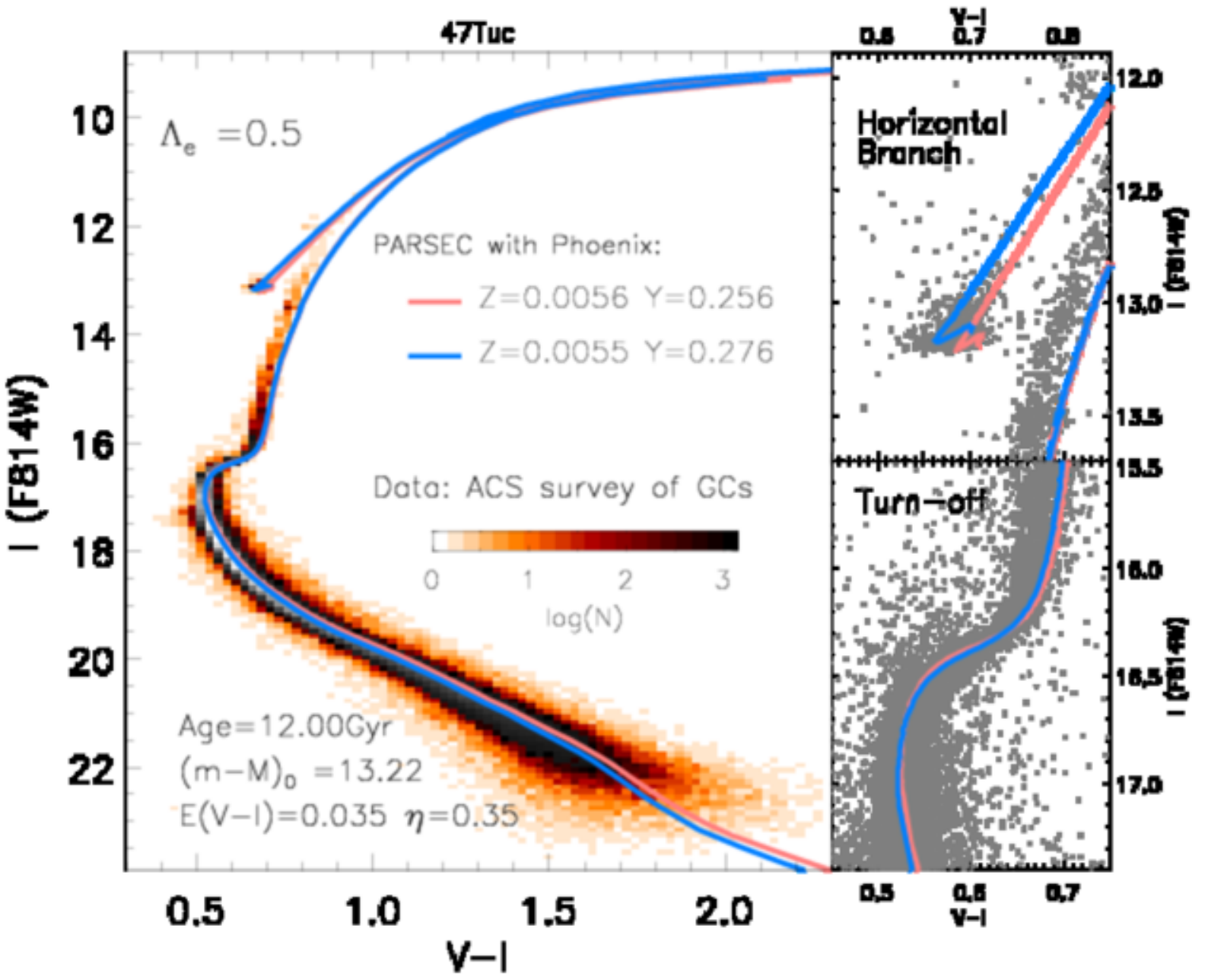

Figure 3.10: Isochrone fitting with Hess diagram (the left panel) of 47Tuc data (Sarajedini et al., 2007) for all the evolutionary phases, and with scatter plots highlighting the horizontal branch region (the upper right panel) and the turn-off region (the lower right panel). The red line and blue line represent isochrones of the first and second generation, respectively, as the legend shows. The fitting parameters are: age=12.00 Gyr, $(m\text{-}M)_0$=13.22, E(V-I)=0.035. [6]

[6]The figure is in low resolution because of the size limit of arXive. SISSA library offers the full thesis with high-resolution figures: http://urania.sissa.it/xmlui/handle/1963/35251

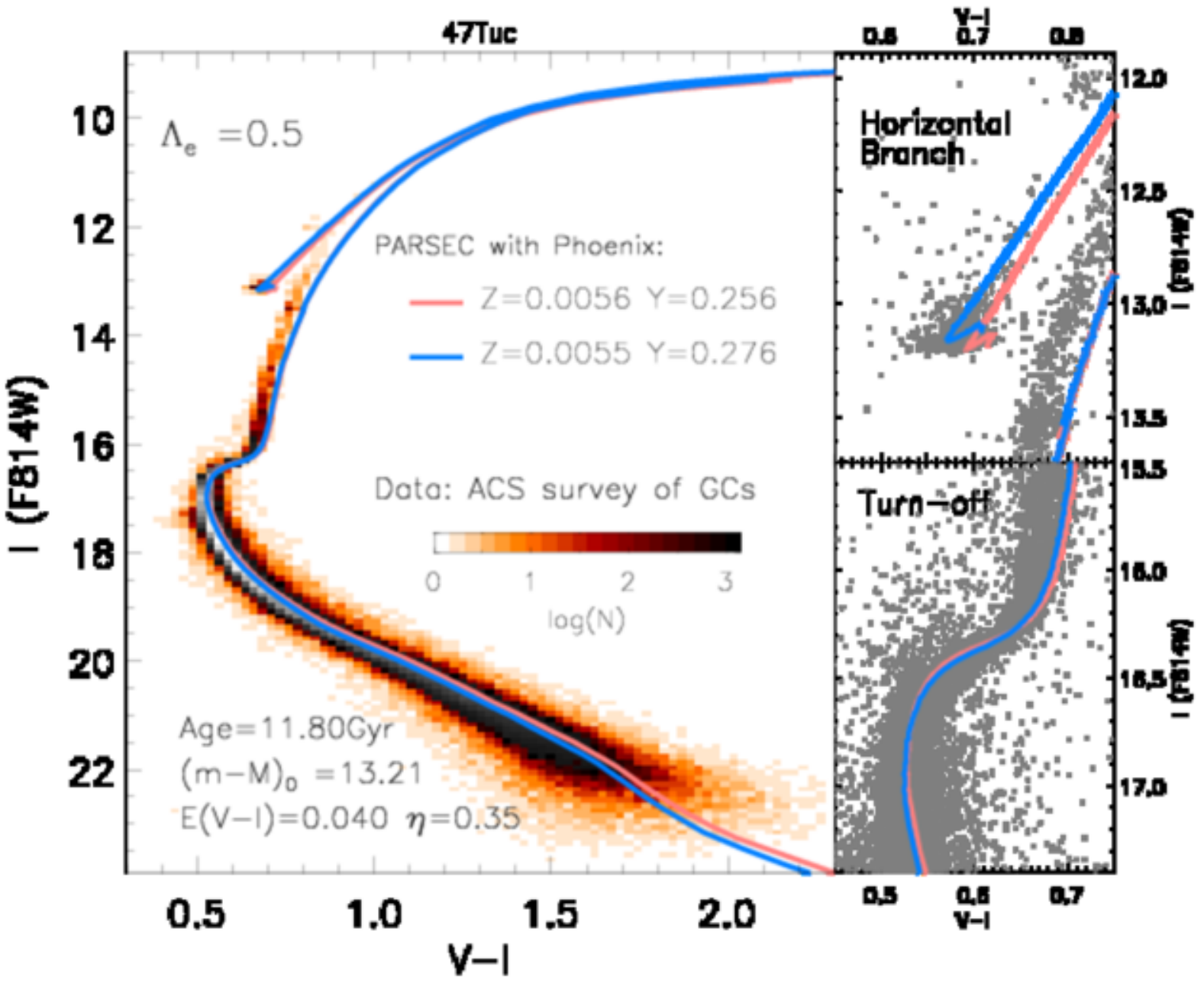

Figure 3.11: Isochrone fitting with Hess diagram (the left panel) of 47Tuc data (Sarajedini et al., 2007) for all the evolutionary phases, and with scatter plots highlighting the horizontal branch region (the upper right panel) and the turn-off region (the lower right panel). The red line and blue line represent isochrones of the first and second generation, respectively, as the legend shows. The fitting parameters are: age=11.80 Gyr, $(m\text{-}M)_0$=13.21, E(V-I)=0.040. [8]

The only phase that cannot be fitted in a very satisfactory way is the RGB phase. As the upper right panels of these three figures show, in each case the isochrones corresponding to both stellar generations run on the red side of the data in the RGB phase. Part of the discrepancy could be explained by the bolometric correction used. Here we are using bolometric correction from `PHOENIX` atmosphere models as described in Chen et al. (2015) for `PARSEC v1.2S`, where only the total metallicity is considered in the transformation of log(L) .vs. log(Teff) into I .v.s V-I. As the metallicities of the two 47Tuc populations (Z=0.0056 and Z=0.0055) show only a marginal difference, we adopt for the two populations the same bolometric corrections. Thus Figure 3.9, Figure 3.10, and Figure 3.11 reflect basically the difference of the two populations in the theoretical log(L) .vs. log(Teff) HR diagram. This "RGB-too-red" problem also exists in Dotter et al.

[8] The figure is in low resolution because of the size limit of arXive. SISSA library offers the full thesis with high-resolution figures: http://urania.sissa.it/xmlui/handle/1963/35251

(2007), when they fit the same set of data using DSEP models (see their Figure 12), as they apply bolometric correction from PHOENIX as well.

To minimize this discrepancy, we used ATLAS12 code (Kurucz, 2005) to compute new atmosphere models with the detailed chemical compositions of the two 47Tuc populations. We adopt these ATLAS12 models for the new fits to 47Tuc, but only for models with Teff hotter than 4000 K ((V-I) ~ 1.3). For lower Teff we still use PHOENIX because ATLAS12 models may be not reliable at cooler temperatures (Chen et al., 2014). Here we show the fit obtained with ATLAS12+PHOENIX bolometric correction in Figure 3.12 only for the case age=12.00 Gyr, $(m\text{-}M)_0$=13.22, E(V-I)=0.035 as an example. We see that with the same fitting parameters as in Figure 3.10, the prediction of the RGB colors is improved by applying new ATLAS12 bolometric correction. The two stellar generation are split on RGB phase in Figure 3.12. We see that the SG (Z=0.0055), which is the main contributor as suggested by Milone et al. (2012); Carretta et al. (2009b), is consistent with the denser region of the RGB data. In other evolutionary phases the new ATLAS12 bolometric corrections do not bring noticeable changes.

Since ATLAS12 only slightly affects the color of the RGB base, and the remainder of the chapter deals with the LF of the bump and of the HB, in the following discussion, we will continue to use the standard atmosphere models of PARSEC v1.2S. PARSEC isochrones with ATLAS12 atmosphere models will be discussed in detail in another following work (Chen et al. in prep.)

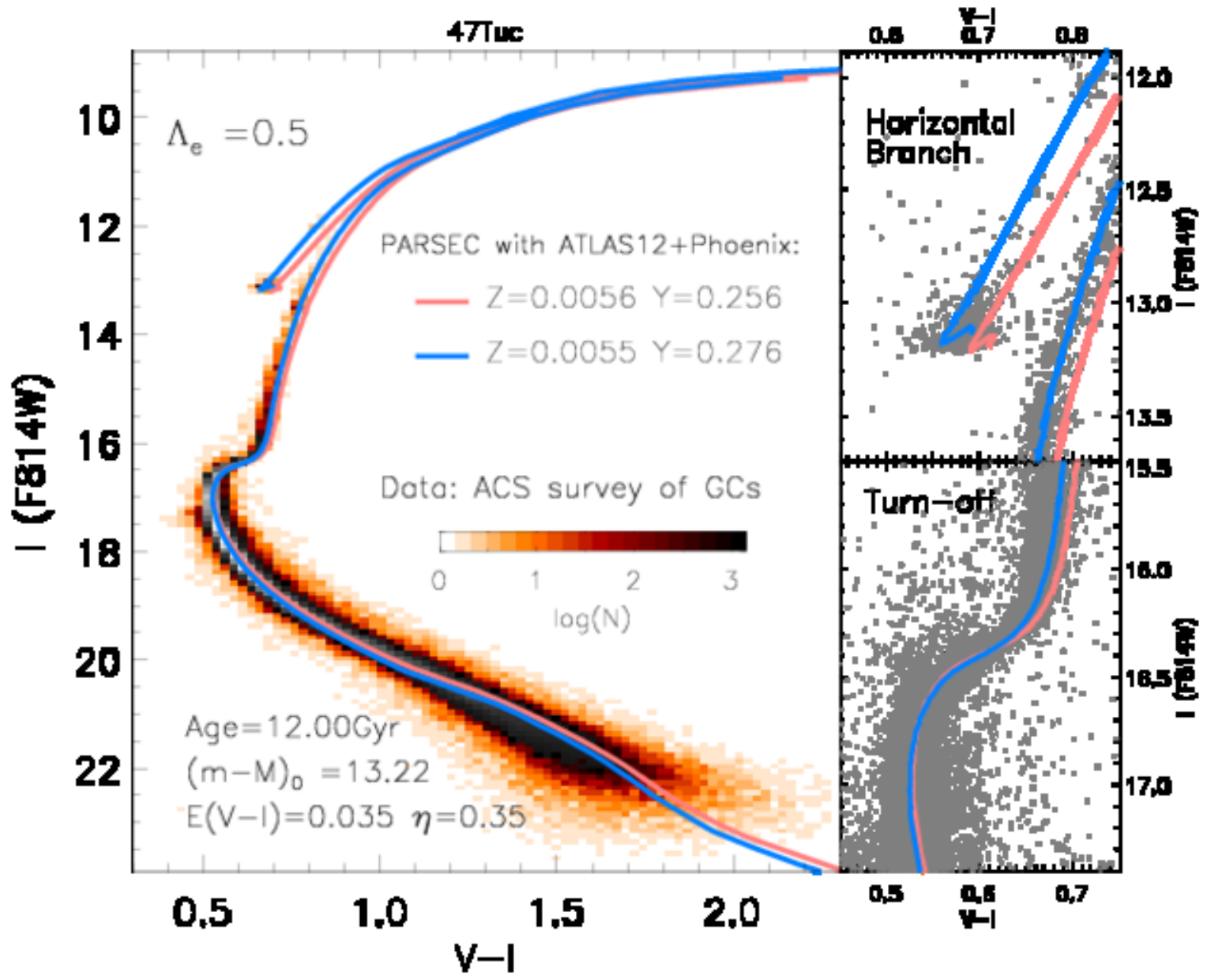

Figure 3.12: The same isochrone fitting with Hess diagram and scatter plots of 47Tuc data (Sarajedini et al., 2007) as in Figure 3.10, but with atmosphere models from ATLAS12 for Teff hotter than 4000 K. [10]

Mass loss by stellar winds during the RGB phase has been considered for low mass stars, using the empirical formula by Reimers (1975) multiplied by an efficiency factor $\eta$. In Figure 3.13 and Figure 3.14 we show the mass lost by the stars at the tip of the RGB phase, in unit of $M_\odot$, with different efficiency factor $\eta$ and ages for the 47Tuc FG and SG respectively. $\Delta M$ in the figure is the difference between the initial mass and current mass of the tip RGB star: $\Delta M = M_{initial} - M_{current}$. The mass lost is an increasing function of the age (smaller initial mass at the RGB tip). It is very difficult to derive observationally the mass lost in RGB stars directly since an accurate mass is not easy to derive and the RGB tip is hard to identify. However, the RGB mass loss charaterises the HB morphology, and this will be discussed in next chapter.

[10]The figure is in low resolution because of the size limit of arXive. SISSA library offers the full thesis with high-resolution figures: http://urania.sissa.it/xmlui/handle/1963/35251

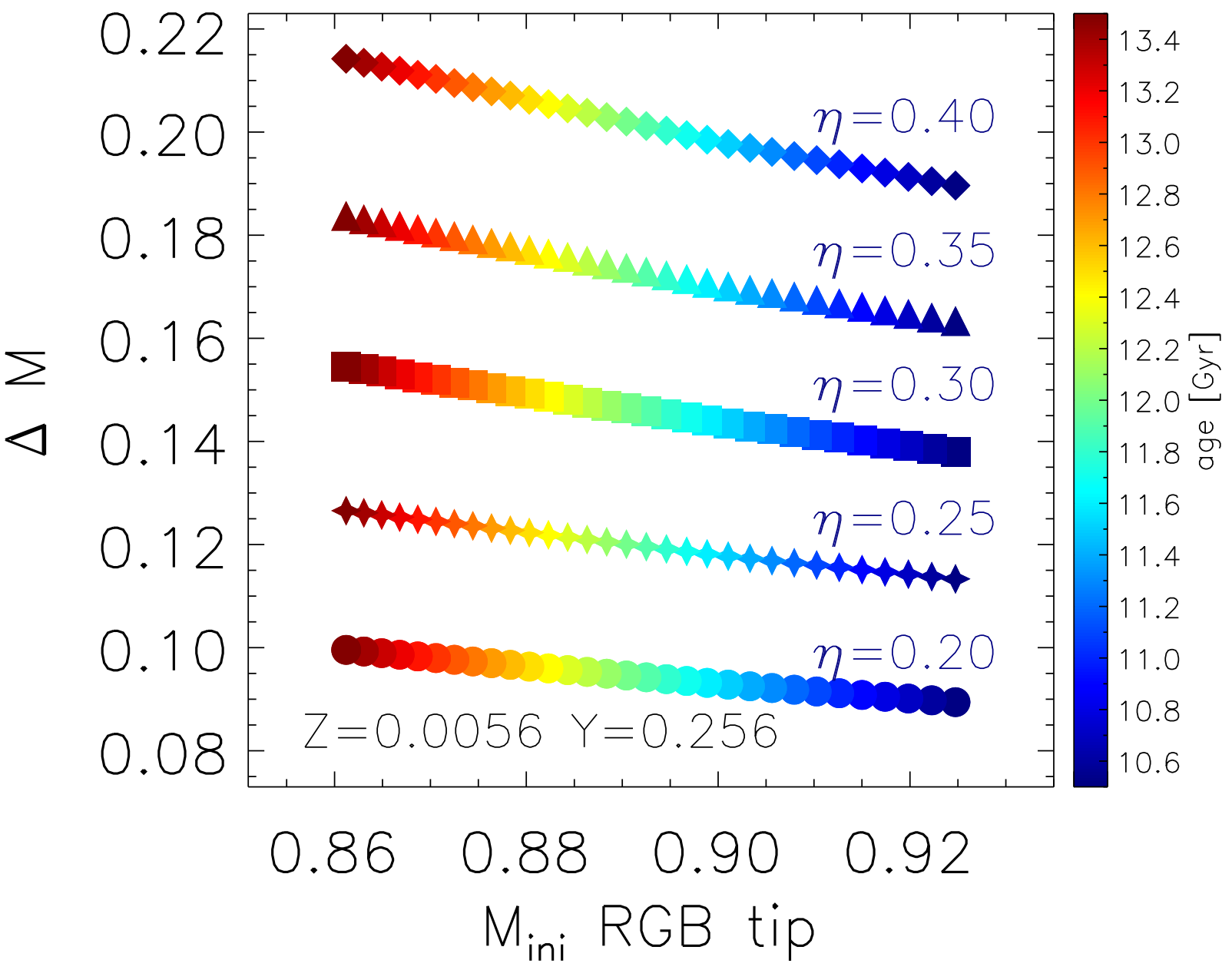

Figure 3.13: RGB mass lost in unit of $M_\odot$ for FG of 47Tuc (Z=0.0056, Y=0.256). The X axis is the initial mass of the tip RGB star, and the Y axis shows the mass lost in this star during RGB phase. Five different efficiency factor $\eta$ are illustrated, from top to bottom $\eta$=0.40 (filled diamond), $\eta$=0.35 (filled triangle), $\eta$=0.30 (filled square), $\eta$=0.25 (filled star), and $\eta$=0.20 (filled dots). The color code displays the age, as shown in the color bar.

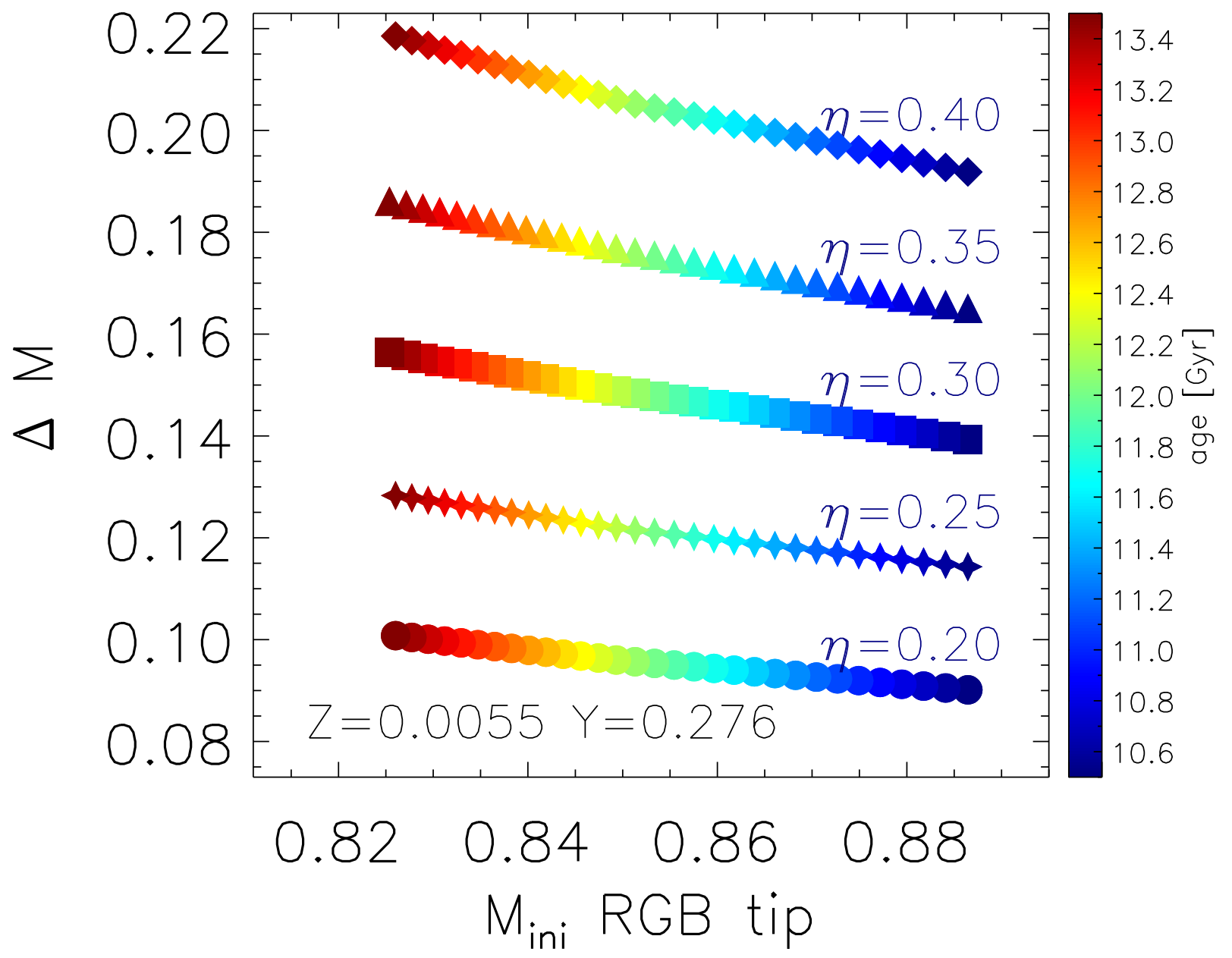

Figure 3.14: RGB mass lost in unit of $M_{\odot}$ for SG of 47Tuc (Z=0.0055, Y=0.276). The X axis is the initial mass of the tip RGB star, and the Y axis shows the mass lost in this star during RGB phase. Five different efficiency factor $\eta$ are illustrated, from top to bottom $\eta$=0.40 (filled diamond), $\eta$=0.35 (filled triangle), $\eta$=0.30 (filled square), $\eta$=0.25 (filled star), and $\eta$=0.20 (filled dots). The color code displays the age, as shown in the color bar.

### 3.2.4 Horizontal Branch morphology

The morphology of the Horizontal Branch in globular clusters is widely studied since the "second parameter problem" (that is, the colour of the HB is determined not only by metallicity, van den Bergh, 1967; Sandage & Wildey, 1967) was introduced. Aside from metallicity as the "first parameter", age , He content, mass-loss, and cluster central density have been suggested as candidates to be the second, or even third, parameter affecting the morphology of the HB (Fusi Pecci & Bellazzini, 1997; Catelan, 2008; Dotter et al., 2010; Gratton et al., 2010; McDonald & Zijlstra, 2015; D'Antona et al., 2002; Caloi & D'Antona, 2005, .etc.). Most of these parameters involve an effect on the mass of the stars which populate the cluster HB. Stars with smaller stellar mass are hotter in temperature and bluer in color. The HB stellar mass decreases as the cluster ages. At a certain age, He-rich star evolves faster and reach the Zero-Age Horizontal Branch (ZAHB) with lower mass. If the age and He content are the same, the mass of HB stars is fixed by the mass loss along the RGB. Although the RGB mass loss does not significantly affect the RGB evolutionary tracks, it determines the location of the stars on the HB, by tuning the stellar mass. Here we illustrate how helium content and the RGB mass loss affect the HB morphology in the case of 47Tuc.

The HB morphology with five different values of $\eta$ is displayed in Figure 3.15, Figure 3.16, and Figure 3.17 for the three sets of fitting parameters derived in Chap. 3.2.3. In each case we show isochrone with different metal/helium abundances ([Z=0.0056, Y=0.256], [Z=0.0055, Y=0.276], [Z=0.0056, Y=0.276], and [Z=0.0055 Y=0.296]). The isochrones with Z=0.0056 are calculated with [$\alpha$/Fe] ~0.4 and those with Z=0.0055 are calculated with [$\alpha$/Fe] ~0.2. The 47Tuc data (Sarajedini et al., 2007) are also plotted for comparison. In all the cases we show, the differences between the isochrone with [Z=0.0055, Y=0.276] (blue solid line) and the one with [Z=0.0056, Y=0.276] (orange dashed line) are negligible on the HB, even though they refer to a different $\alpha$-enhanced mixture. With the same RGB mass loss factor $\eta$, He-rich stars have their HB bluer, more extended (because of smaller stellar mass), and more luminous (because of larger He content in the envelope). For stars with larger mass loss efficiency $\eta$ during their RGB phase, their HB is bluer, fainter, and more extended, because of smaller stellar mass (hence smaller envelope mass, since the core mass does not vary significantly with the mass loss rate). Indeed, the effects of a higher He content and of a lower mass (no matter if it is the result of an older age or a larger RGB mass loss) on HB stars are difficult to distinguished by means of the color, but can be disentangled because the larger helium content makes the He-rich star slightly more luminous.

Table 3.5 lists the current mass, $M_{ZAHB}$, of the first HB star and the corresponding mass that has been lost $\Delta M^{RGB}$), in unit of $M_\odot$ . In the table we also show the HB mass range $\delta M_{HB}$ that produces the corresponding color extent of HB. All cases displayed in Figure 3.15 are itemized. With the same format, Table 3.6 lists the values for the models displayed in Figure 3.16, and Table 3.7 lists the values for the models displayed in Figure 3.17.

If one considers an uniform mass loss parameter $\eta$ for the two populations of 47Tuc ([Z=0.0056, Y=0.256], and [Z=0.0055, Y=0.276]), $\eta$ = 0.35 is the value that fits better the HB morphology in our best fit cases, as Figure 3.15, Figure 3.16, and Figure 3.17 illustrate. As shown in Table 3.5, Table 3.6 and Table 3.7, a RGB mass loss parameter of $\eta$ = 0.35 leads to a value of the mass lost in RGB that is between 0.172 $M_{\odot}$ –0.177 $M_{\odot}$ . This range is consistent with the results of Heyl et al. (2015) who studied the dynamics of white dwarf in 47Tuc, and concluded that the mass lost for the stars at the end of the RGB phase should be less than about 0.2 $M_{\odot}$ . Salaris, Cassisi & Pietrinferni (2016) also investigates the efficiency of the RGB mass loss in 47Tuc. By adopting age, distance, and reddening from the literature, they give a lower limit to the mass lost during the RGB of about 0.17 $M_{\odot}$ . The factor $\eta$ = 0.35 is lower than the value used by many authors. McDonald & Zijlstra (2015) investigate the RGB mass loss in GC and derives $\eta$ = 0.648 for 47Tuc. With the detailed chemical composition of the cluster, we have more precise fits and thus a more accurate $\eta$ parameter from the HB morphology modelling. A more efficient mass loss along the RGB of 47Tuc does not satisfy neither the above mass loss limit (0.2 $M_{\odot}$ ) nor it is able to provide good isochrone fits.

The LFs from the turn-off to the HB with a RGB mass loss parameter $\eta$ = 0.35 are displayed in Figure 3.18, Figure 3.19, and Figure 3.20 for the three sets of our best fitting parameters, respectively. For comparison, the LF of HST GC survey data (Sarajedini et al., 2007) is also plotted (black histogram filled with oblique lines) with the same binsize 0.05 mag. All model LFs are normalized to the total number of observed stars within a range of I magnitude (F814W) between 14 mag - 16 mag. The left panels of Figure 3.18, Figure 3.19, and Figure 3.20 show the LFs from the turn-off to the HB, for a 100% FG (red histogram), a 100% SG (blue histogram), and the percentage adopted in Chap. 3.2.2, 30% from FG and 70% from SG (orange histogram), respectively. With our best isochrone fitting parameters, age=12.00±0.2 Gyr, $(m-M)_0=13.22^{+0.02}_{-0.01}$, $E(V-I)=0.035^{-0.008}_{+0.005}$, $\eta$=0.35, and the population percentage obtained from literature (Carretta et al., 2009b; Milone et al., 2012), the model LF (orange histogram in each figure) show a very good agreement with the observed LF.

The three right panels in each figure are zooms that highlight the HB and RGBB regions. Comparing the number of observed stars that belong to the HB with that predicted by the models we see that the latter are about 2% $\sim$ 4% less than the former. The deficiency is mostly in the magnitude range between I=12.9 to I=13.1 mag. The discrepancy may be due to the possible existence

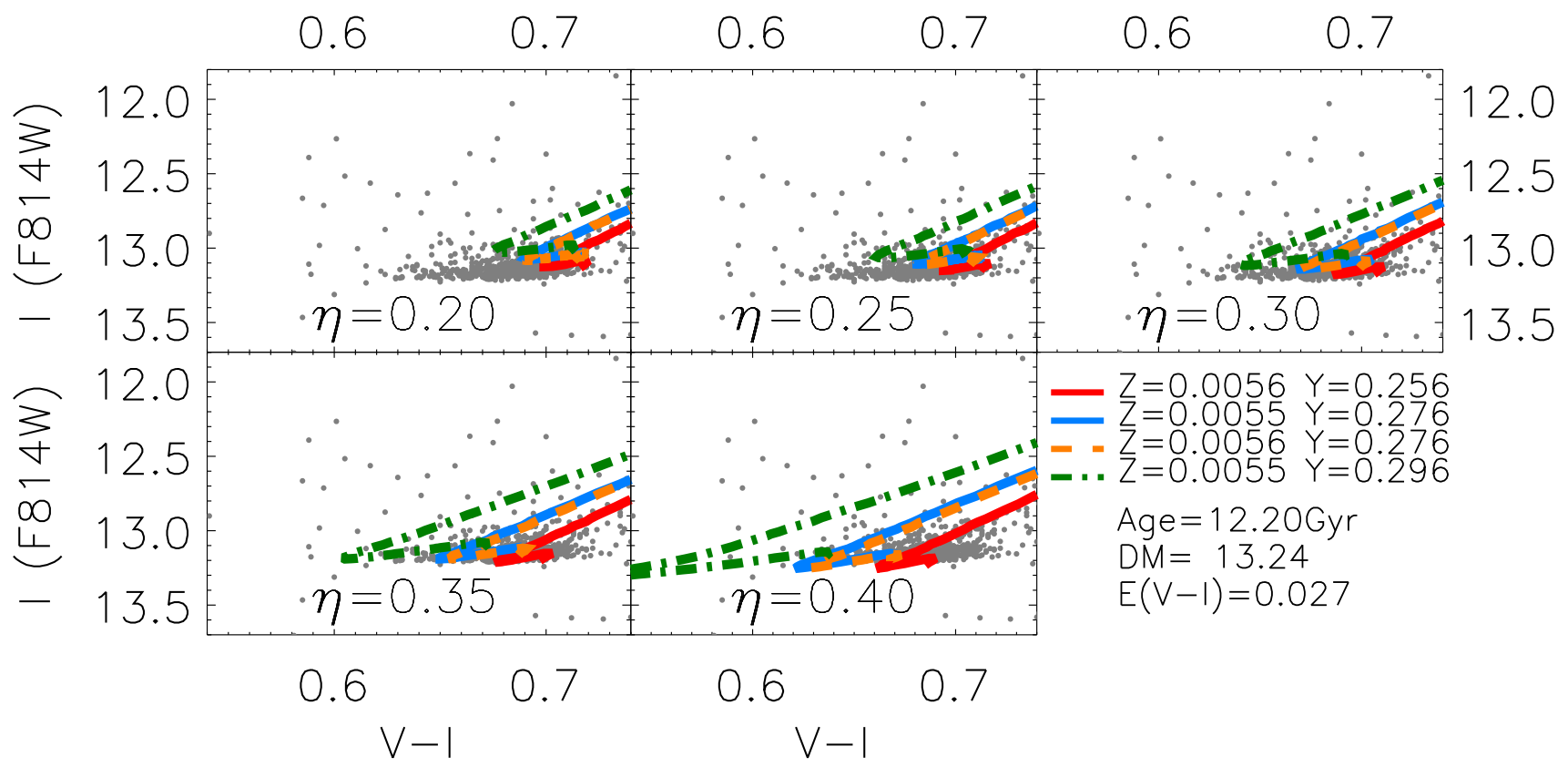

Figure 3.15: The Horizontal Branch morphology for different RGB mass loss parameters ($\eta$) and metal/helium abundances, with the same isochrone fitting parameters (age, $(\text{m-M})_0$, and E(V-I)) adopted in Figure 3.9. The red solid line, blue solid line, orange dashed line, and green dash-dot line represent isochrones of [Z=0.0056, Y=0.256], [Z=0.0055, Y=0.276], [Z=0.0056, Y=0.276], and [Z=0.0055, Y=0.296], respectively. The mass lost during the RGB in unit of $M_\odot$ for each $\eta$ and metal/helium abundance, is listed in Table 3.5

Table 3.5: The mass lost during the RGB in unit of $M_\odot$ for different $\eta$ and metal/helium abundance. The current mass of the first HB star is $M_{ZAHB}$, and $\Delta M^{RGB}$ represents the mass loss in unit of $M_\odot$. The HB mass range is itemized in the last column $\delta M_{HB}$. All values listed here are derived from isochrones with age=12.2 Gyr, $(m\text{-}M)_0$=13.24, and E(V-I)=0.027, as shown in Figure 3.15.

| Z | Y | $\eta$ | $M_{ZAHB}$ ( $M_\odot$ ) | $\Delta M_{RGB}$ ( $M_\odot$ ) | $\delta M_{HB}$ ( $M_\odot$ ) |
|---|---|---|---|---|---|
| 0.0056 | 0.256 | 0.20 | 0.791034 | 0.0953 | 0.0024 |
| | | 0.25 | 0.765362 | 0.1210 | 0.0034 |
| | | 0.30 | 0.738801 | 0.1475 | 0.0044 |
| | | 0.35 | 0.711251 | 0.1751 | 0.0053 |
| | | 0.40 | 0.682591 | 0.2037 | 0.0060 |
| 0.0055 | 0.276 | 0.20 | 0.753316 | 0.0960 | 0.0030 |
| | | 0.25 | 0.727353 | 0.1220 | 0.0040 |
| | | 0.30 | 0.700436 | 0.1489 | 0.0052 |
| | | 0.35 | 0.672459 | 0.1768 | 0.0061 |
| | | 0.40 | 0.643264 | 0.2061 | 0.0066 |
| 0.0056 | 0.276 | 0.20 | 0.758890 | 0.0955 | 0.0029 |
| | | 0.25 | 0.733089 | 0.1213 | 0.0040 |
| | | 0.30 | 0.706353 | 0.1480 | 0.0050 |
| | | 0.35 | 0.678573 | 0.1758 | 0.0060 |
| | | 0.40 | 0.649615 | 0.2048 | 0.0066 |
| 0.0055 | 0.296 | 0.20 | 0.722152 | 0.0961 | 0.0036 |
| | | 0.25 | 0.696078 | 0.1222 | 0.0048 |
| | | 0.30 | 0.668998 | 0.1492 | 0.0058 |
| | | 0.35 | 0.640787 | 0.1775 | 0.0065 |
| | | 0.40 | 0.611289 | 0.2070 | 0.0063 |

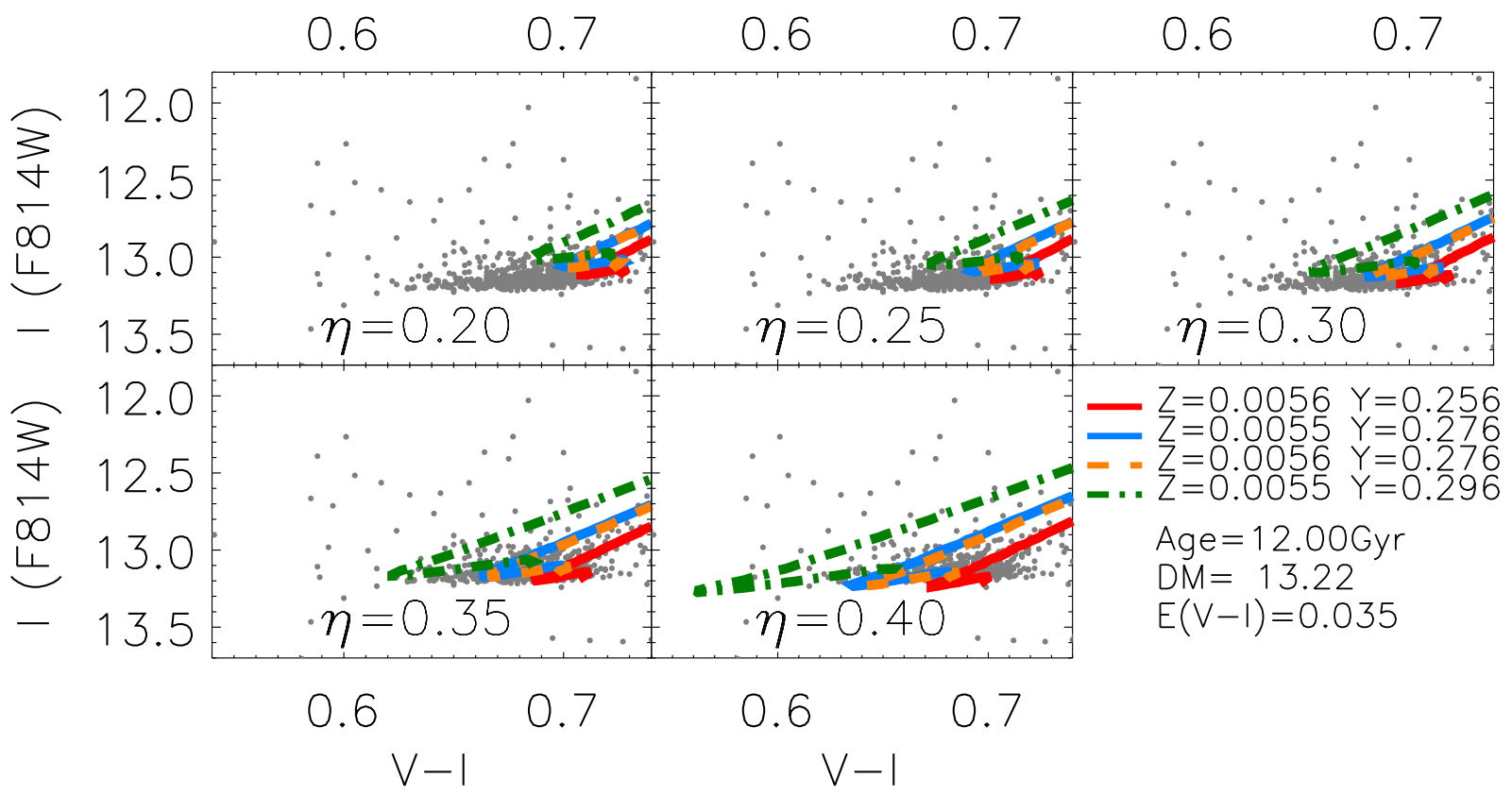

Figure 3.16: Horizontal branch morphology for different RGB mass loss parameters ($\eta$) and metal/helium abundances, with the same isochrone fitting parameters (age, $(m\text{-}M)_0$, and E(V-I)) as in Figure 3.10. The red solid line, blue solid line, orange dashed line, and green dash-dot line represent isochrones of [Z=0.0056, Y=0.256], [Z=0.0055, Y=0.276], [Z=0.0056, Y=0.276], and [Z=0.0055, Y=0.296], respectively. The mass lost during the RGB in unit of $M_\odot$ for each $\eta$ and metal/helium abundance is listed in Table 3.6

Table 3.6: The mass lost during the RGB in unit of $M_\odot$ for different $\eta$ and metal/helium abundance. The current mass of the first HB star is $M_{ZAHB}$, and $\Delta M^{RGB}$ represents its RGB mass loss in unit of $M_\odot$. The HB mass range is itemized in the last column $\delta M_{HB}$. All values listed here are derived from isochrones with age=12.0 Gyr, $(m-M)_0$=13.22, and E(V-I)=0.035, as shown on Figure 3.16.

| Z | Y | $\eta$ | $M_{ZAHB}$ ( $M_\odot$ ) | $\Delta M_{RGB}$ ( $M_\odot$ ) | $\delta M_{HB}$ ( $M_\odot$ ) |
|---|---|---|---|---|---|
| 0.0056 | 0.256 | 0.20 | 0.795832 | 0.0946 | 0.0023 |
| | | 0.25 | 0.770375 | 0.1201 | 0.0033 |
| | | 0.30 | 0.744052 | 0.1464 | 0.0044 |
| | | 0.35 | 0.716765 | 0.1737 | 0.0053 |
| | | 0.40 | 0.688402 | 0.2020 | 0.0059 |
| 0.0055 | 0.276 | 0.20 | 0.758027 | 0.0953 | 0.0029 |
| | | 0.25 | 0.732270 | 0.1211 | 0.0039 |
| | | 0.30 | 0.705582 | 0.1478 | 0.0051 |
| | | 0.35 | 0.677852 | 0.1755 | 0.0061 |
| | | 0.40 | 0.648949 | 0.2044 | 0.0067 |
| 0.0056 | 0.276 | 0.20 | 0.763649 | 0.0948 | 0.0028 |
| | | 0.25 | 0.738049 | 0.1204 | 0.0039 |
| | | 0.30 | 0.711533 | 0.1470 | 0.0050 |
| | | 0.35 | 0.683996 | 0.1745 | 0.0059 |
| | | 0.40 | 0.655309 | 0.2032 | 0.0066 |
| 0.0055 | 0.296 | 0.20 | 0.726732 | 0.0954 | 0.0035 |
| | | 0.25 | 0.700873 | 0.1213 | 0.0047 |
| | | 0.30 | 0.674033 | 0.1481 | 0.0058 |
| | | 0.35 | 0.646090 | 0.1760 | 0.0066 |
| | | 0.40 | 0.616896 | 0.2052 | 0.0067 |

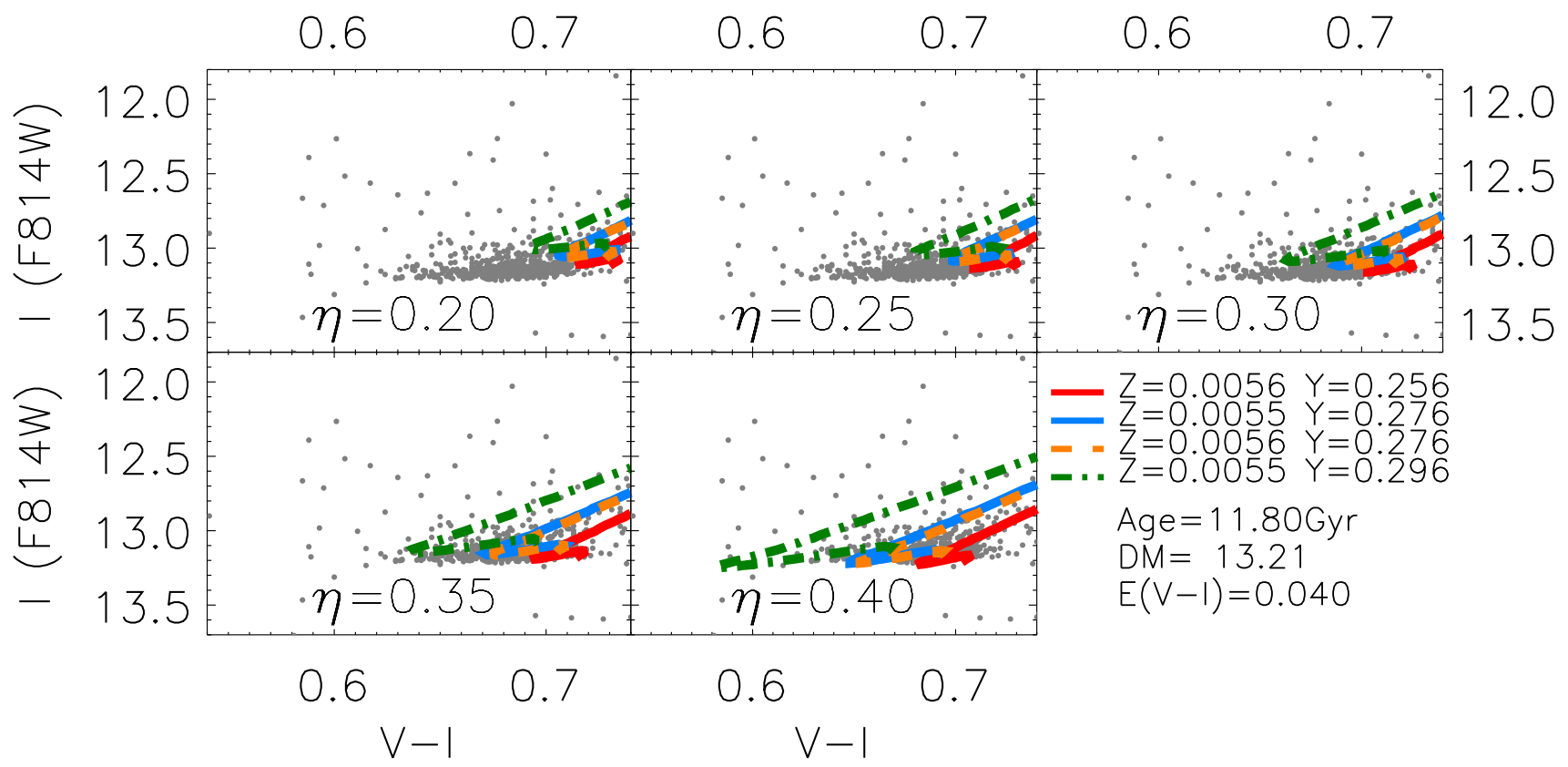

Figure 3.17: Horizontal branch morphology for different RGB mass loss factor ($\eta$) and metal/helium abundances, with the same isochrone fitting parameters (age, $(m\text{-}M)_0$, and E(V-I)) as in Figure 3.11. The red solid line, blue solid line, orange dashed line, and green dash-dot line represent isochrones of [Z=0.0056, Y=0.256], [Z=0.0055, Y=0.276], [Z=0.0056, Y=0.276], and [Z=0.0055, Y=0.296], respectively. The mass lost during the RGB in unit of $M_\odot$ for each $\eta$ and metal/helium abundance is listed in Table 3.7

Table 3.7: The mass lost during the RGB in unit of $M_\odot$ for different $\eta$ and metal/helium abundance. The current mass of the first HB star is $M_{ZAHB}$, and $\Delta M^{RGB}$ represents its RGB mass loss in unit of $M_\odot$. The HB mass range is itemized in the last column $\delta M_{HB}$. All values listed here are derived from isochrones with age=11.8 Gyr, $(m\text{-}M)_0$=13.21, and E(V-I)=0.040, as shown on Figure 3.17.

| Z | Y | $\eta$ | $M_{ZAHB}$ ( $M_\odot$ ) | $\Delta M_{RGB}$ ( $M_\odot$ ) | $\delta M_{HB}$ ( $M_\odot$ ) |
|---|---|---|---|---|---|
| 0.0056 | 0.256 | 0.20 | 0.800710 | 0.0939 | 0.0023 |
| | | 0.25 | 0.775472 | 0.1191 | 0.0033 |
| | | 0.30 | 0.749390 | 0.1452 | 0.0043 |
| | | 0.35 | 0.722373 | 0.1722 | 0.0053 |
| | | 0.40 | 0.694311 | 0.2003 | 0.0058 |
| 0.0055 | 0.276 | 0.20 | 0.762844 | 0.0947 | 0.0029 |
| | | 0.25 | 0.737287 | 0.1202 | 0.0039 |
| | | 0.30 | 0.710817 | 0.1467 | 0.0050 |
| | | 0.35 | 0.683329 | 0.1742 | 0.0061 |
| | | 0.40 | 0.654695 | 0.2028 | 0.0068 |
| 0.0056 | 0.276 | 0.20 | 0.768489 | 0.0942 | 0.0028 |
| | | 0.25 | 0.743093 | 0.1196 | 0.0039 |
| | | 0.30 | 0.716801 | 0.1459 | 0.0050 |
| | | 0.35 | 0.689510 | 0.1731 | 0.0059 |
| | | 0.40 | 0.661098 | 0.2016 | 0.0067 |
| 0.0055 | 0.296 | 0.20 | 0.731415 | 0.0947 | 0.0035 |
| | | 0.25 | 0.705769 | 0.1204 | 0.0047 |
| | | 0.30 | 0.679162 | 0.1470 | 0.0058 |
| | | 0.35 | 0.651479 | 0.1747 | 0.0067 |
| | | 0.40 | 0.622577 | 0.2036 | 0.0068 |

of another population with even higher He abundance and lower RGB mass loss efficiency (like the behavior of the isochrone with [Z=0.0055, Y=0.296] in the first three panels of Figure 3.15, Figure 3.16, and Figure 3.17). Indeed, Carretta et al. (2013) suggests that there are at least three different stellar populations in the giant branch of 47Tuc. Nevertheless we consider this result as a success of our new models because they may reproduce not only the observed location of the stars in the CM diagram over a magnitude range of about 13 mags, but also the stellar lifetimes in the post main sequence phases.

In addition, since the LF is directly proportional to the evolution time, the good agreement of LF between model and observation in Fig. 3.18, Fig. 3.19, and Fig. 3.20 indicates that the helium burning lifetime is correctly predicted in PARSEC . During the core He-burning phase in HB, hydrogen continues to burn in a shell at about the same rate as it did during the main sequence phase, the rate at which helium is burnt in the convective core determins the rate at which the star evolves (Chiosi, Bertelli & Bressan, 1992). Breathing convection in this phase increases the core helium burning lifetime which leads to a too much larger C-O core for the star, and our choice of core overshooting, $\Lambda_c = 0.5H_P$ (Bressan et al., 2012), reduces the extent of the convection which reproduce the data very well.

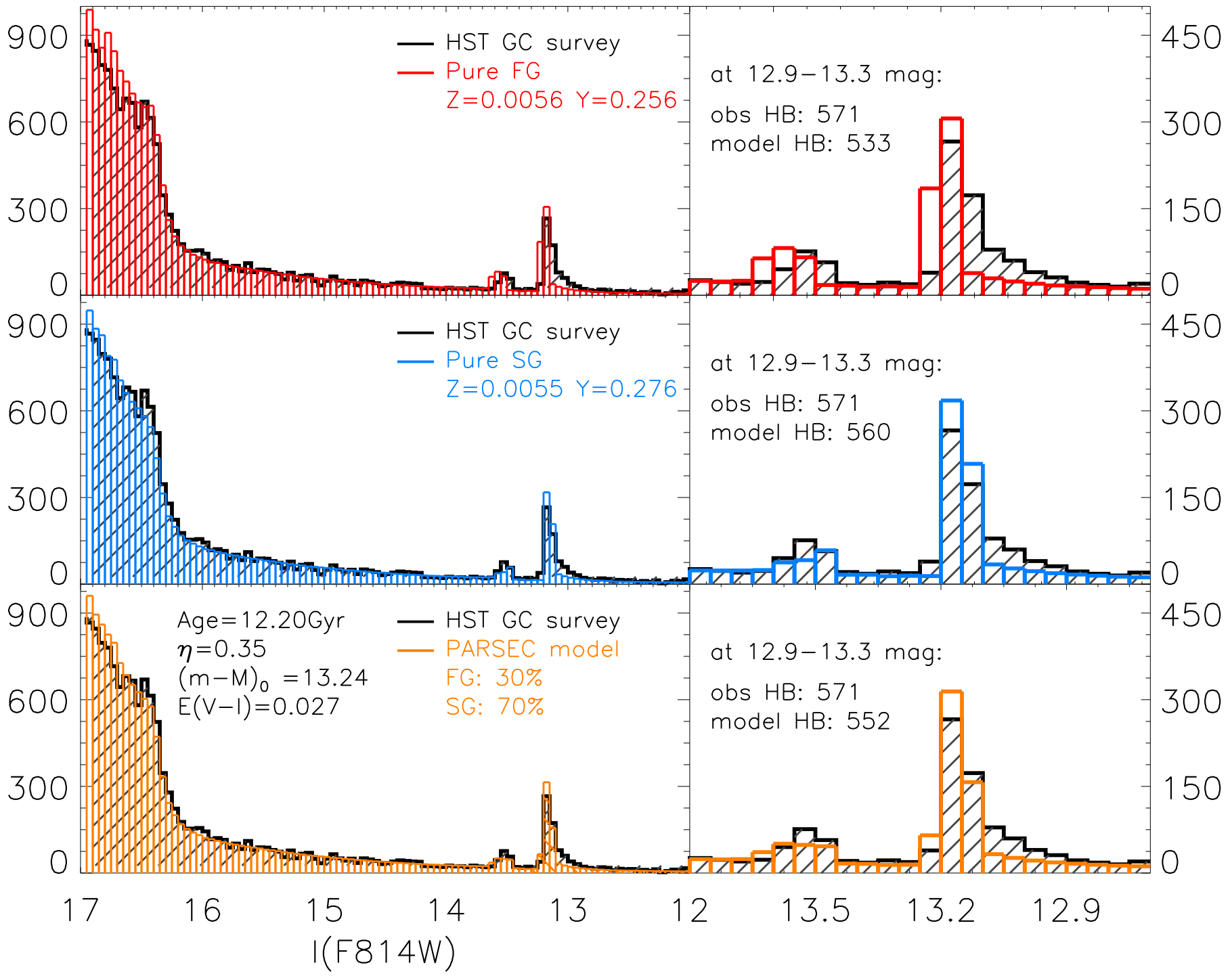

Figure 3.18: Comparison between the luminosity function of 47Tuc data (Sarajedini et al., 2007) and that derived from the new PARSEC models, from the turn-off to the HB. The Y axis represent the star counts in I magnitude (F814W). The black histogram filled with oblique lines is the data LF, whilst the red histogram in the upper panel, blue histogram in the middle panel, and orange histogram in the lower panel, represent 100% FG of 47Tuc [Z=0.0056, Y=0.256], 100% SG [Z=0.0055, Y=0.276], and their mix with 30% from the FG and 70% from the SG, respectively. The three panels on the right side show the LF of the RGBB and the HB region, for each population mixture. The total number of HB stars within 12.8 - 13.3 mag in the observations and in the models are also listed in the figure. All the model LFs are normalized to the total number of stars within 14 mag - 16 mag in I magnitude (F814W). The fitting parameters are: $\eta$=0.35, age=12.20 Gyr, $(m\text{-}M)_0$=13.24, and E(V-I)=0.027.

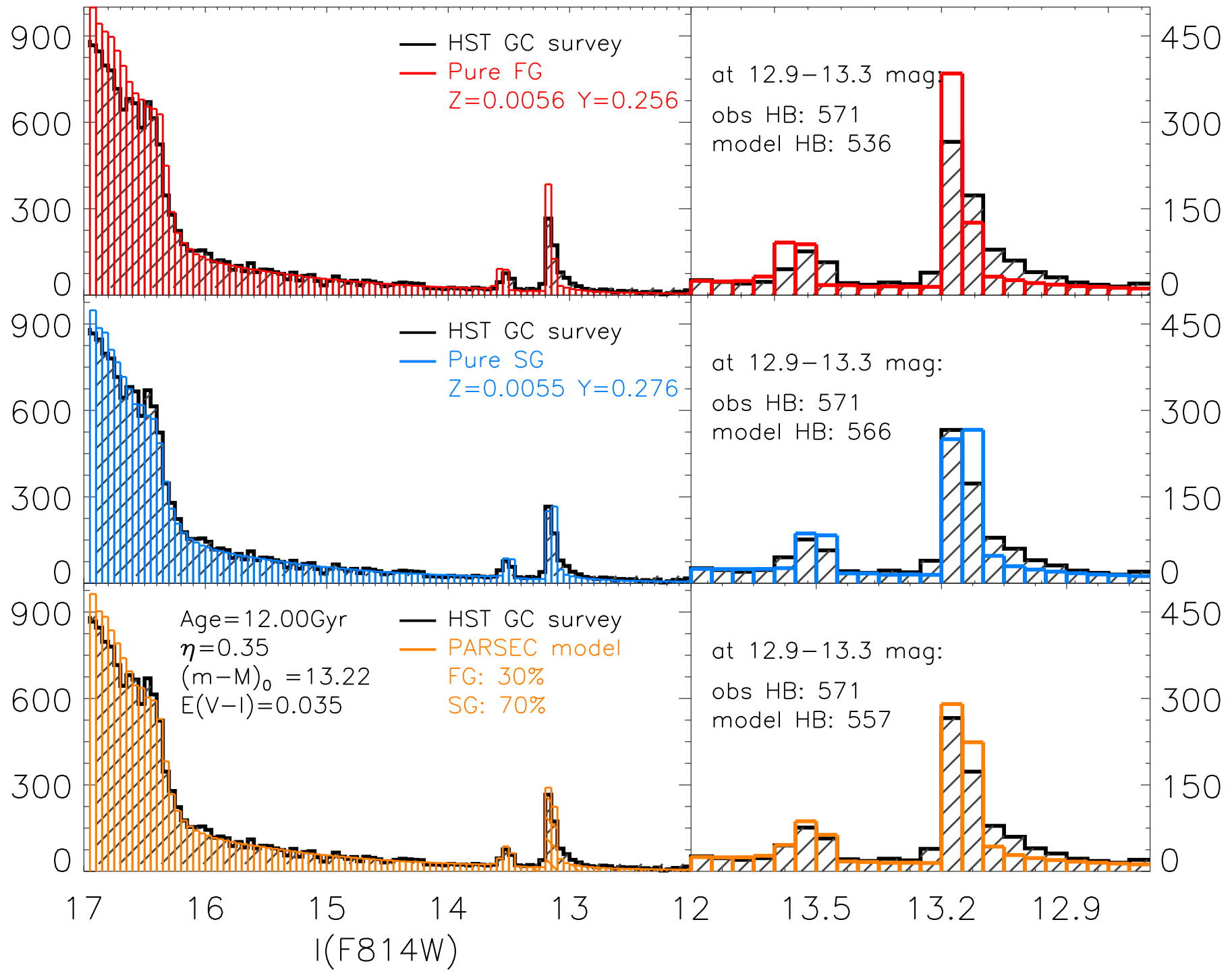

Figure 3.19: LF comparison from turn-off to HB as in Figure 3.18, but with fitting parameter age=12.00 Gyr, $(m\text{-}M)_0$=13.22, and E(V-I)=0.035.

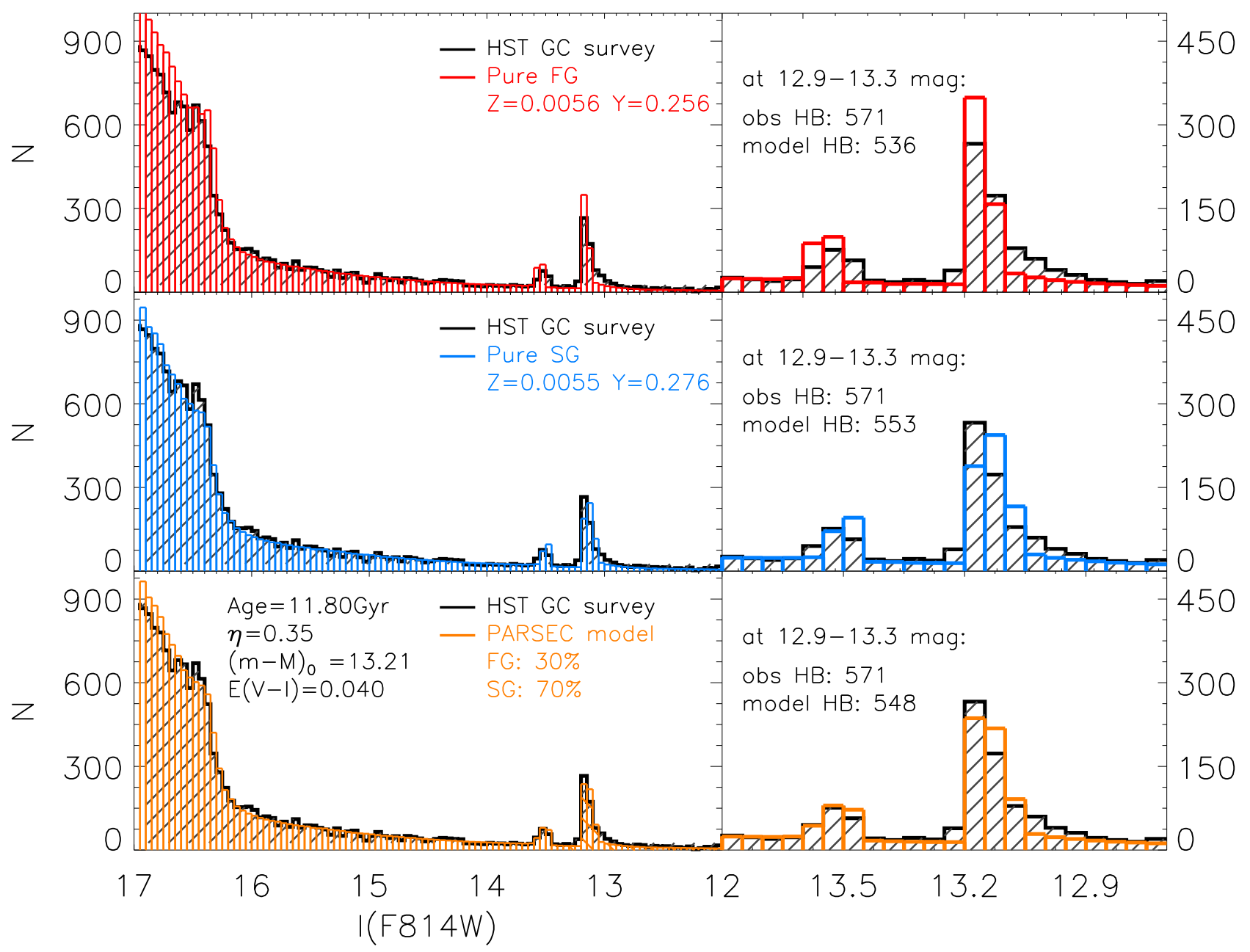

Figure 3.20: LF comparison from turn-off to HB as in Figure 3.18, but with fitting parameter age=11.80 Gyr, $(m\text{-}M)_0$=13.21, and E(V-I)=0.040.

# Chapter 4

# RGB bump comparison with other GC data and models

The new PARSEC $\alpha$ enhanced isochrones provide a very good fit of the color magnitude diagram of 47Tuc in all evolutionary stages from the lower main sequence to the HB. The location of the RGB bump shows that the efficiency of the envelope overshoot is quite significant, requiring EOV=0.5$H_P$. This can be considered a calibration of this phenomenon. We now use the calibrated EOV value to obtain *alpha*-enhanced isochrones of different metallicities. For this purpose we adopt the partition of heavy elements of the two stellar generations of 47Tuc ( [$\alpha$/Fe] ~ 0.4 and 0.2). The theoretical luminosities and effective temperatures along the isochrones are converted to magnitudes in different photometry systems. In Figure 4.1 we show two examples of the new isochrones transformed into the Gaia broad band photometrc system (Jordi et al., 2010).

An interesting application of this new set of isochrones is the comparison of the location of the RGB bump predicted by the models with the observed one in GCs with different metallicity.

## 4.1 Comparison with other models

We first compare the RGBB magnitude of our newly calibrated PARSEC models with other *alpha*-enhanced stellar tracks. Since the BaSTI (Pietrinferni et al., 2006, 2013) and DSEP (Dotter, 2007; Dotter et al., 2008) isochrones are publicly available online, we download the [$\alpha$/Fe] =0.4 isochrones at 13 Gyr from BaSTI Canonical Models database and DSEP web tool 2012 version. WE then compare the mean values of their absolute RGBB magnitude in the F606W (HST ACS/WFC) band, with our models. Figure 4.2 shows this comparison as a func-

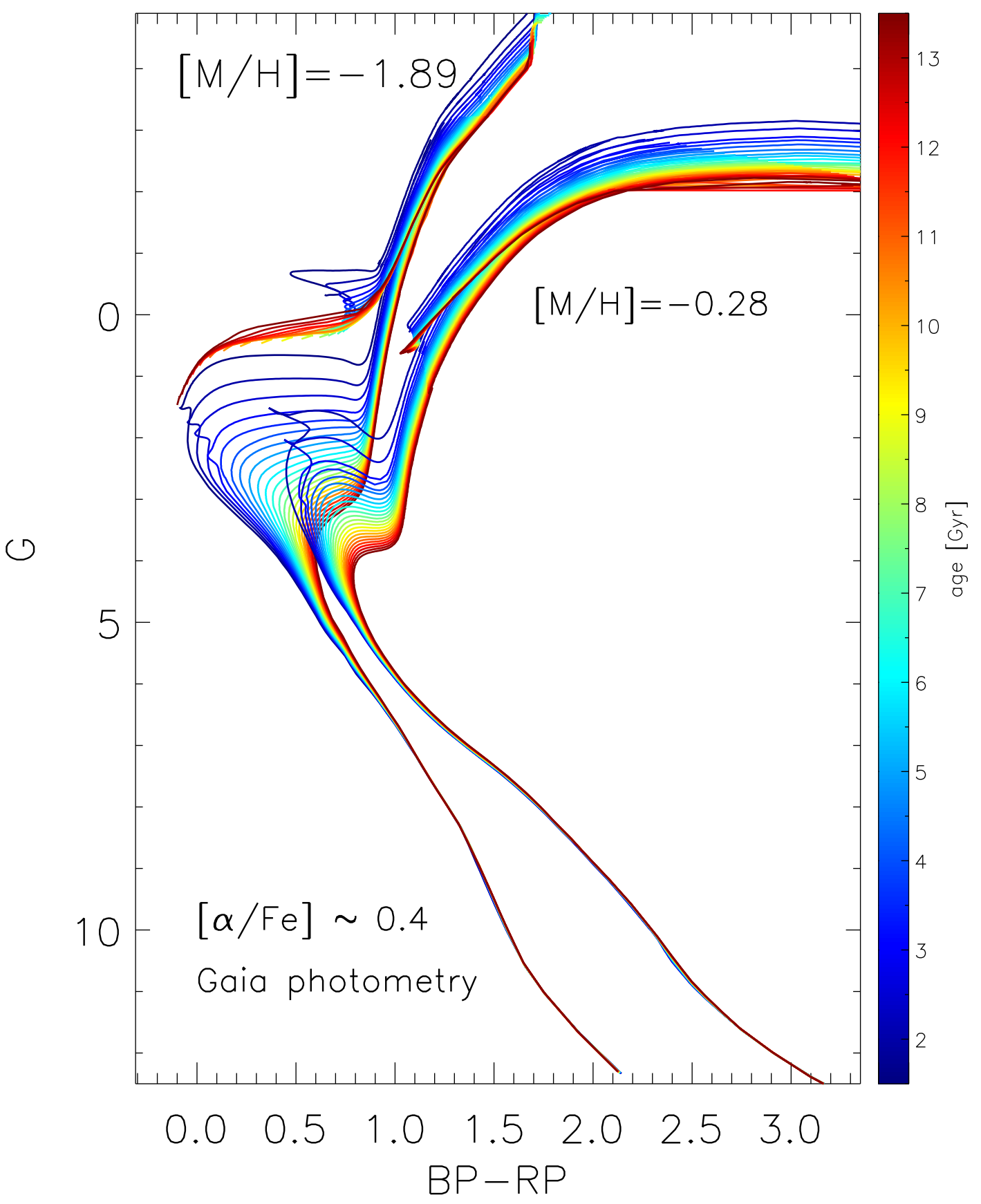

Figure 4.1: $\alpha$-enhanced isochrones ( [$\alpha$/Fe] ~ 0.4) with [M/H]=-1.89 (Z=0.0005) and [M/H]=-0.28 (Z=0.008). The age of the isochrones varies from 1 to 14 Gyr with a step of 0.5 Gyr. Different ages are displayed with different colors as illustrated in the color bar.

tion of the metallicity [M/H]. The model [M/H] is approximated by:

$$[M/H] = log\frac{Z/X}{Z_{\odot}/X_{\odot}} \tag{4.1}$$

As listed in Table 2.2 and Table 2.3, the solar metallicity in PARSEC is $Z_{\odot}$ = 0.01524 and $Z_{\odot}/X_{\odot}$ = 0.0207. Since DSEP models do not provide [M/H] directly but only [Fe/H] in their isochrones, we calculate [M/H] following Eq.4.1 with total metallicity Z, He content Y, and solar $Z_{\odot}/X_{\odot}$ taken from their models. Additionally, two PARSEC models with solar-scaled metal mixture ( [$\alpha$/Fe] =0), PARSEC v1.2S and PARSEC with EOV calibration from this work $\Lambda_e = 0.5H_p$, are also plotted. Compared with the new set of solar-scaled PARSEC model with $\Lambda_e = 0.5H_p$ (dark blue line with diamond), the $\alpha$-enhanced one (red line with triangle) is only slightly brighter as we have already discussed in Sec. 2. PARSEC v1.2S, with [$\alpha$/Fe] =0 and negligible EOV $\Lambda_e = 0.05H_p$, has a similar performance as BaSTI [$\alpha$/Fe] =0.4 model. We notice that the RGBB behavior of PARSEC v1.2S in this figure is different from Figure 3 of Joyce & Chaboyer (2015), which compares PARSEC v1.2S with other models. The reason for this disagreement is unclear to us.

Among the factors that may affect the brightness of the RGBB, as summarized in Sec. 3.2.2, we list the He content and the mixing efficiency. The Helium-to-metal enrichment law of the different models are different, as discussed in Sec. 2. PARSEC ($Y = 0.2485 + 1.78Z$) uses a slightly higher He abundance (~0.002) than the other two models (BaSTI: $Y = 0.245 + 1.4Z$, DSEP: $Y = 0.245 + 1.54Z$). Different model also adopts different mixing length parameters. The PARSEC mixing length parameter is $\alpha_{MLT} = 1.74$, BaSTI uses $\alpha_{MLT} = 1.913$, and DSEP adopts $\alpha_{MLT} = 1.938$. If all other parameters are the same, a lower He content and a larger mixing length should lead to a fainter RGBB. However in Figure 4.2 we see that BaSTI and DSEP RGBB are eventually brighter than PARSEC . Joyce & Chaboyer (2015) discusses that modifying the diffusion coefficients it is possible to make ≈0.2 mag difference on RGBB. This said, we remind that the larger effect on the RGBB magnitude is due to an efficient EOV. Our new $\alpha$-enhanced models are computed with the calibrated EOV parameter while, BaSTI and DSEP do not consider envelope overshooting. Compared to PARSEC v1.2S ($\Lambda_e = 0.05H_p$ for stars in the figure), the new model with [$\alpha$/Fe] =0 and $\Lambda_e = 0.5H_p$ shifts $M_{V,RGBB}$ down by about 0.35 mag. This brightness change is consistent with the work of Cassisi, Salaris & Bono (2002) who conclude that the difference should be of about 0.8 mag/$H_p$.

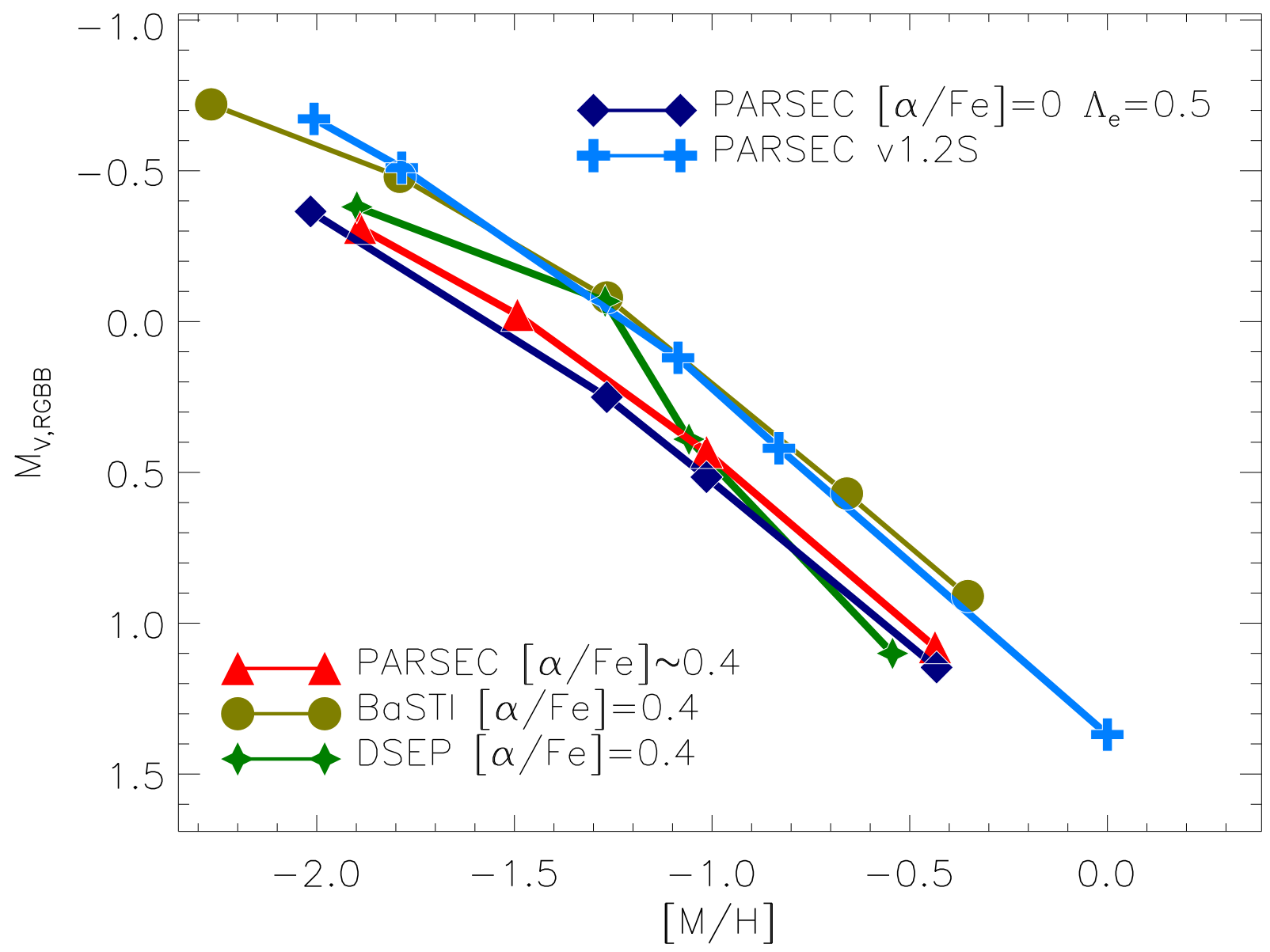

Figure 4.2: Comparison of the RGBB magnitude of different evolutionary tracks at 13Gyr. There are three different $\alpha$-enhanced models ( [$\alpha$/Fe] =0.4) in the figure: `PARSEC` (red line with triangle), BaSTI (yelow green line with dots), and DSEP model (green line with star). Other two sets of solar-scaled `PARSEC` models ( [$\alpha$/Fe] =0) are plotted for comparison: `PARSEC v1.2S` with negligible overshoot (light blue line with cross) and `PARSEC` with EOV calibration $\Lambda_e = 0.5H_p$ (dark blue line with diamond). The Y axis is the mean value of the absolute V magnitude (F606W in HST ACS/WFC) of the RGBB ($M_{V,RGBB}$), and X axis is metallicity [M/H].

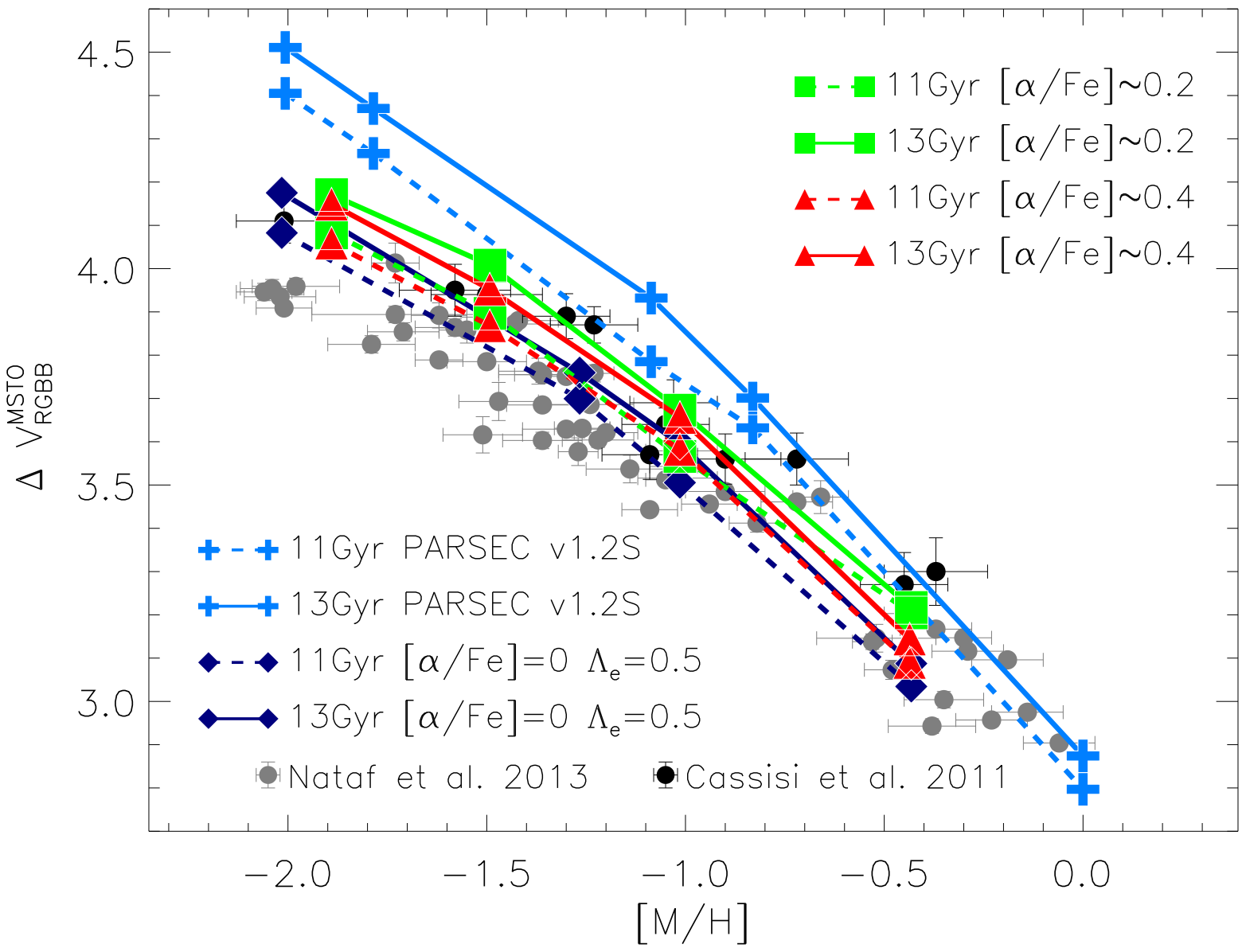

Figure 4.3: F606W magnitude difference between the MSTO and the RGBB ($\Delta V_{RGBB}^{MSTO}$) as a function of the metallicity [M/H]. Four different sets of theoretical $\Delta V_{RGBB}^{MSTO}$ value are plotted, at both 13Gyr (solid line) and 11Gyr (dashed line). Three of them are with new calibrated EOV $\Lambda_e = 0.5H_p$: [$\alpha$/Fe] ~ 0.4 (red lines with triangle), [$\alpha$/Fe] ~ 0.2 (green lines with square), and [$\alpha$/Fe] =0 (dark blue lines with diamond). Another one is from the standard `PARSEC v1.2S` (light blue lines with cross). The data are 55 clusters from Nataf et al. (2013, grey dots with error bar) and 12 clusters from Cassisi et al. (2011, black dots with error bar).

## 4.2 Comparison with other GC data

The RGBB of globular clusters, as already said in section 3.2.2, has been studied over 30 years since the 47Tuc RGBB was observed in 1985 (King, Da Costa & Demarque, 1985). However, there is a discrepancy between the observed brightness of RGBB and the model predictions: the model RGBB magnitude is about 0.2–0.4 mag brighter than the observed ones (Fusi Pecci et al., 1990; Cecco et al., 2010; Troisi et al., 2011). This discrepancy becomes more pronounced in metal-poor GCs (Cassisi et al., 2011).

In Figure 4.3 we compare our new $\alpha$-enhanced model with HST data from Nataf et al. (2013, 55 clusters) and Cassisi et al. (2011, 12 ,clusters). The models extend till [M/H]=-2. For comparison, two sets of models with solar-scaled metal partition, [$\alpha$/Fe] =0 ( `PARSEC v1.2S` and `PARSEC` with $\Lambda_e = 0.5H_p$) are also plotted. Here we use the magnitude difference between the RGBB and the main sequence turn-off (MSTO), $\Delta V^{MSTO}_{RGBB}$, as a reference for comparison between the theoretical magnitude of RGBB and the observed one. Unlike the absolute magnitude $M_{V,RGBB}$, $\Delta V^{MSTO}_{RGBB}$ is not affected by uncertainties in the distance modulus $(m\text{-}M)_0$ and extinction $A_V$ of the cluster. There are also works using the magnitude difference between HB and RGBB, $\Delta V^{RGBB}_{HB} = M_{V,RGBB} - M_{V,HB}$, as a way to avoid distance and extinction uncertainties (Fusi Pecci et al., 1990; Cassisi & Salaris, 1997; Cecco et al., 2010), but as we have elaborated in Sec. 3.2.4, the RGB mass loss together with different metal mixture and He content may affect the HB magnitude and thus make $\Delta V^{RGBB}_{HB}$ difficult to be interpreted. The only free parameter of the $\Delta V^{MSTO}_{RGBB}$ method is the age, if the composition of the cluster is fixed. In Figure 4.3 we compare the theoretical $\Delta V^{MSTO}_{RGBB}$ value at typical GC ages of 11Gyr and 13Gyr, with the observed value from Nataf et al. (2013) and (Cassisi et al., 2011). In Nataf et al. (2013) they fit polynomials to the upper main sequence of each GC at (F606W, F606W-F814W) plane and take the bluest point as the magnitude of MSTO. (Cassisi et al., 2011) derive the MSTO magnitude by fitting isochrone to the main sequence. To obtain the theoretical MSTO F606W magnitude in our model, we select the bluest point of the isochrone in the main sequence. The models of 13Gyr show larger difference between RGBB and MSTO $\Delta V^{MSTO}_{RGBB}$ than those at 11Gyr. Models with [$\alpha$/Fe] ~0.2 show a slightly greater $\Delta V^{MSTO}_{RGBB}$ value than the models computed with [$\alpha$/Fe] ~0.4. At the most metal-poor end the three sets of new models with $\Lambda_e = 0.5H_p$, [$\alpha$/Fe] =0, [$\alpha$/Fe] ~0.2 and [$\alpha$/Fe] ~0.4, are very similar since, when the total metallicity becomes very low, the effects of $\alpha$-enhancement are negligible. Compared to the previous `PARSEC` version `v1.2S`, the new models significantly improve the $\Delta V^{MSTO}_{RGBB}$ prediction. At the most metal-poor side, around [M/H]=-2.0, the new models are consistent with Cassisi et al. (2011) data (black dots), but are higher than the values derived by Nataf et al. (2013) (grey dots) by ~0.1 mag. We will discuss the possible reasons in the discussion chapter.

## 4.3 Discussion

The RGBB morphology of 47Tuc suggests that the envelope overshooting in the stellar models of low mass stars should amount to $\Lambda_e = 0.5H_p$. This calibration, together with the two metal partitions with an $\alpha$-enhancement typical of the 47Tuc populations ( [$\alpha$/Fe] ~0.4 and [$\alpha$/Fe] ~0.2), has been applied to compute models at other metallicities.

There long-lasting discrepancy between the observed and the predicted RGBB brightness (Fusi Pecci et al., 1990; Cecco et al., 2010; Troisi et al., 2011; Cassisi et al., 2011), seems to be solved by our new models that offer an overall better agreement with the data.

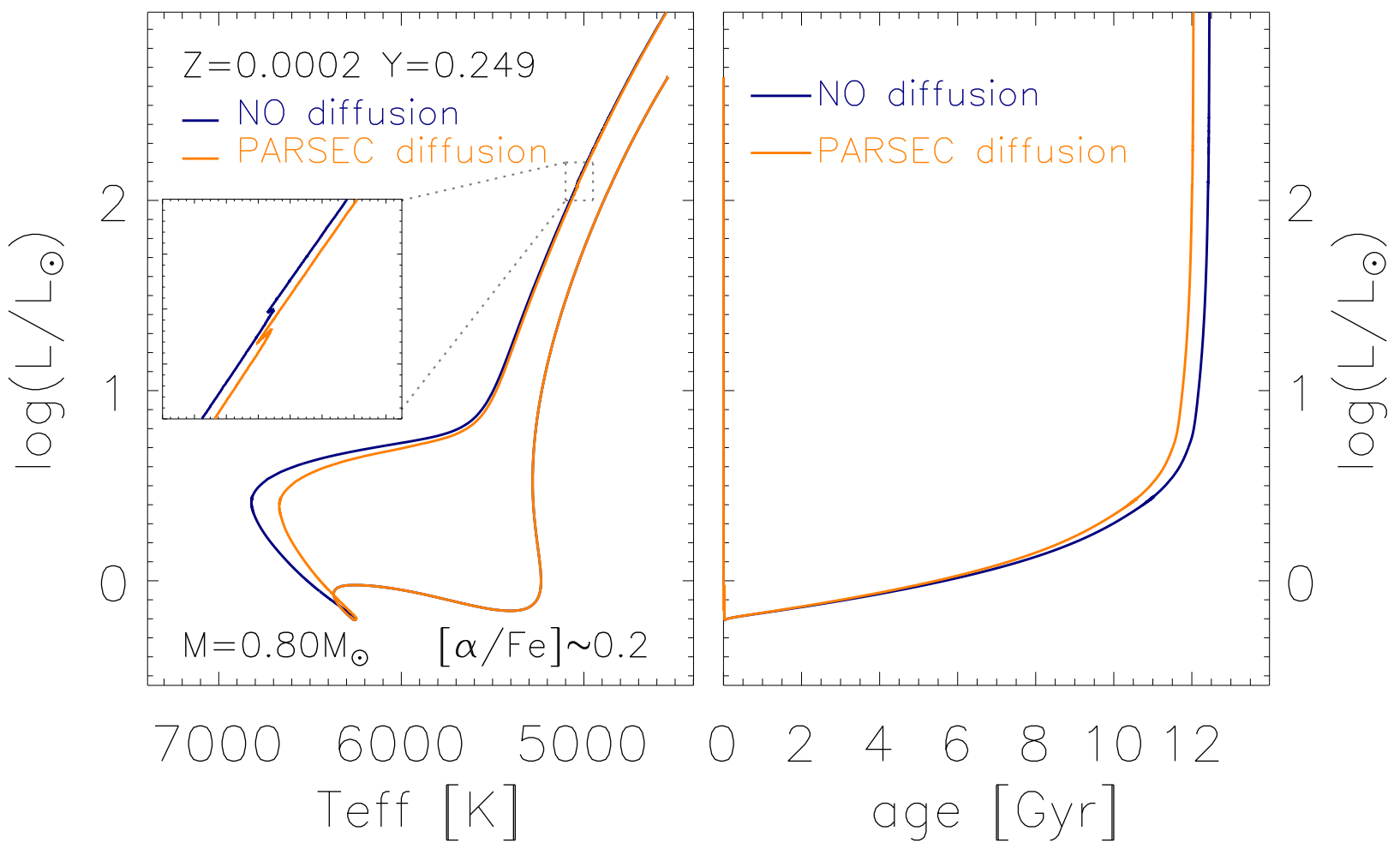

Figure 4.4: Comparison of [Z=0.0002 Y=0.249] tracks for a M=0.80 $M_{\odot}$ star with (orange line) and without (blue line) diffusion. The left panel shows the HRD of these two tracks, with the RGBB region zoomed in in the sub-figure. The right panel illustrates the luminosity evolution as the star ages.

However we note that in Figure 4.3 around [M/H]=-2.0 our model predicts $\Delta V_{RGBB}^{MSTO}$ about ~0.1 mag greater than the data points obtained by Nataf et al. (2013). If we consider a more He-rich model, the discrepancy will become even larger. There are works arguing that diffusion also affects the brightness of RGBB (eg. Michaud, Richer & Richard, 2010; Cassisi et al., 2011). Though the time-scale and surface gravity of RGB star are both very small, the atom diffusion still plays a role on the hydrogen burning shell where gravity is still large. Michaud, Richer & Richard (2010) conclude that atomic diffusion reduces about 0.02 dex the luminosity of RGBB. We also test the effects of diffusion on the RGBB morphology. Figure 4.4 displays the comparison between evolutionary tracks with and without diffusion. Both evolutionary tracks are with for [Z=0.0002 Y=0.249], [$\alpha$/Fe] ~0.2, and a stellar mass M=0.80 $M_{\odot}$ . The orange track is calculated with the standard `PARSEC` diffusion which includes pressure diffusion, temperature diffusion, and concentration diffusion as in Thoul, Bahcall & Loeb (1994), but is partially inhibited in the outermost region of the envelope (Bressan et al., 2013). While the blue line in this figure is a track without diffusion. As we can see from the right panel of Figure 4.4, diffusion shorten the main sequence life-time. As seen in the left panel, similar to that in Figure 1 of Michaud, Richer & Richard (2010), evolutionary tracks with diffusion show redder MSTO and slightly fainter RGBB. For isochrones obtaned from these two sets of evolutionary tracks, at 13 Gyr the RGBB with diffusion is 0.072 mag (F606W) fainter than the one without diffusion, and $\Delta V_{RGBB}^{MSTO}$ value is 0.008 mag (F606W) smaller. Thus inhibiting the diffusion along RGB phase will eventually makes the discrepancy more severe. Pietrinferni, Cassisi & Salaris (2010) conclude that the updated nuclear reaction rate for $^{14}\mathrm{N}\,(\mathrm{p}\,,\gamma)\,^{15}\mathrm{O}$ makes RGBB brighter by ~0.06 mag compared to the old rate. However we remind that we are already adopting the new rate (Imbriani et al., 2005) for this reaction (Table 2.1), To the best of our knowledge, there are two possible solutions to cover this 0.1 mag discrepancy: one is that EOV in metal-poor stars is even higher than our adopted value (e.g. $\Lambda_e = 0.7H_p$ suggested by Alongi et al., 1991). The other is that the opacity of metal-poor stars is larger than adopted in this work. This could be possible if the partition of heavy elements underestimates the abundance of some important elements.

# Chapter 5

# $\alpha$ enhanced metal mixture based on APOGEE ATLAS9

The Apache Point Observatory Galactic Evolution Experiment (APOGEE) is a large Galactic survey with high-resolution and high signal-to-noise infrared spectroscopy Majewski et al. (2015). An unified pipeline, the APOGEE Stellar Parameter and Chemical Abundances Pipeline (ASPCAP), is used to determine stellar chemical abundances and atmospheric parameters by comparing observed spectra to libraries of theoretical ones (Pérez et al., 2015). An APOGEE ATLAS9 atmosphere model (Mészáros et al., 2012) is the set of specific model atmospheres generated for APOGEE.

Since APOGEE covers the full range of the Galactic populations in the bulge, bar, disk, and halo, and offers a large dataset of spectra homogeneously analysed, it will be very useful for the community to have $\alpha$-enhanced stellar tracks and isochrones with their adopted $\alpha$ metal partition. For this purpose we calculate a set of [$\alpha$/Fe] =0.4 models based on APOGEE ATLAS9 atmosphere model.

## 5.1 Derive the opacity

In order to have an opacity grid with the metal mixture of interest, one should first generate a table with number density $N_i/N_Z$ and mass fraction $Z_i/Z_{tot}$ of each metal element. Following the abundance enhanced method used in APOGEE ATLAS9 atmosphere model (Meszaros, private communication), we first reproduce their solar elemental abundances and calculate the [M/H], [$\alpha$/M], [C/M] in every enhanced case from the models, then derive the corresponding number density.

The number density is obtained by the following calculation:

$$nabun = \begin{cases} \frac{ni}{ntot}; & for\ H,\ He \\ (log\frac{ni}{n_{tot}}) + [C/M]; & for\ carbon \\ (log\frac{ni}{n_{tot}}) + [\alpha/M]; & for\ \alpha\ elements \\ log\frac{ni}{n_{tot}}; & for\ other\ metal \end{cases} \tag{5.1}$$

where nabun is the value directly read from the atmosphere model file. [C/M] and [$\alpha$/M] are the labeled enhancement. From Equation 5.1 we get the "un-scaled" value of the element number ni. Then we derive the real number of each element by scaling ni with the scale factor scl:

$$\begin{cases} n_H = ni(H) \\ n_{He} = ni(He) \\ n_C = scl \times ni(C) \times 10^{[C/M]} \\ n_\alpha = scl \times \sum ni(\alpha) \times 10^{[\alpha/M]} \\ n_{other} = scl \times ni(other) \end{cases} \tag{5.2}$$

With this process, we can reproduce the abundance of the solar model they have adopted (Asplund et al 2005).

However, through the calculation, we have found two inaccuracies in the APOGEE ATLAS9 model:

i) The actual enhanced $\alpha$-elements in the model files are not as described in the paper (Mészáros et al., 2012) (8 alpha elements: O, Ne, Mg, Si, S, Ar, Ca, Ti), nor as specified in the website (6 elements: O, Ne, Mg, Si, S, Ca). Instead the $\alpha$-elements are the 6 elements: O, Mg, Si, S, Ca, Ti. ii) It looks that the labeled [M/H] is actually [Fe/H], [$\alpha$/M] is [$\alpha$/Fe] and [C/M] is [C/Fe].

Fig. 5.1 illustrates the comparison between the labeled [M/H], [$\alpha$/M], [C/M] and the real [Fe/H], [$\alpha$/Fe] , [C/Fe], calculated from the model.

We choose a metal partition with [$\alpha$/Fe] =0.4 and [C/M]=0 from APOGEE ATLAS9 and, after obtaining the corresponding opacities and EOS, we compute a set of preliminary evolutionary tracks. The metal partition has [$\alpha$/Fe] =0.4 compared with the adopted APOGEE ATLAS9 solar model but, if compared with the solar partition used in `PARSEC` , the value is [$\alpha$/Fe] =0.3785.

Table 5.1 list the number density $N_i/N_Z$ and mass fraction $Z_i/Z_{tot}$ for this APOGEE [$\alpha$/Fe] =0.4 partition of heavy elements. Fig. 5.2 displays the relative number density comparison among APOGEE [$\alpha$/Fe] =0.4 model, APOGEE solar model, and 47Tuc FG ( [$\alpha$/Fe] ~0.4).

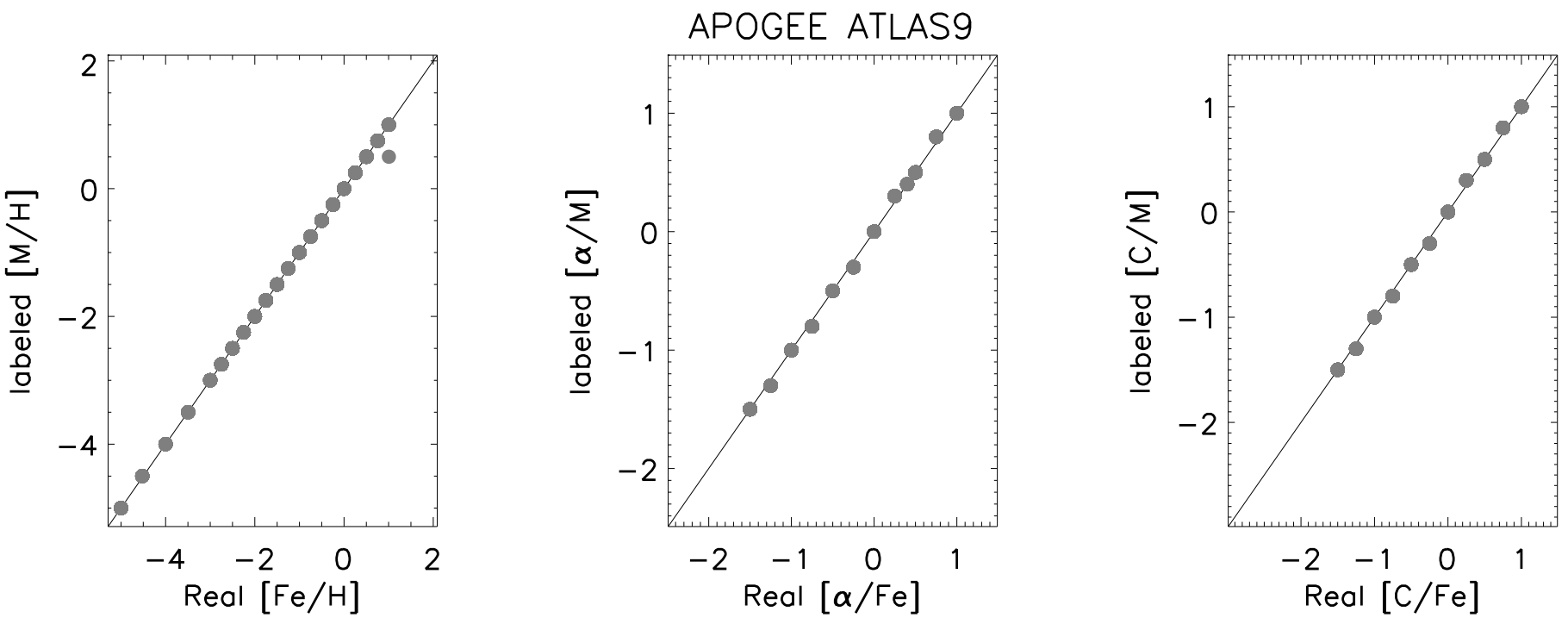

Figure 5.1: Comparison between the labeled [M/H] and real [Fe/H] (left panel) from the calculation, labeled [$\alpha$/M] and real [$\alpha$/Fe] (middle panel), labeled [C/M] and real [C/Fe] (right panel) of APOGEE ATLAS9 atmosphere model.

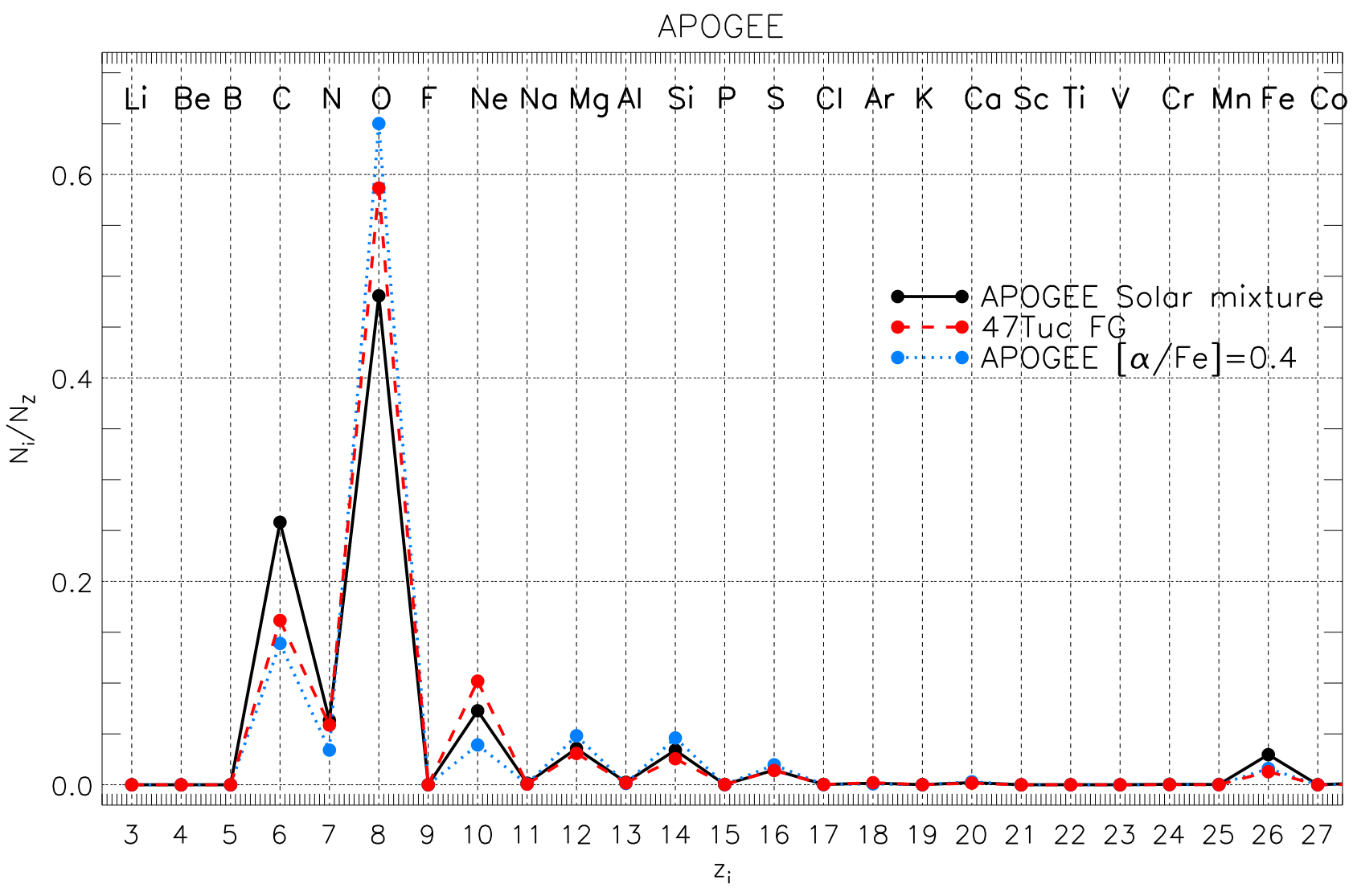

Figure 5.2: Relative metal mixture comparison among APOGEE solar model (black solid line), APOGEE [$\alpha$/Fe] =0.4 model (blue dotted line), and 47Tuc FG (red dashed line). The X axis is the atomic number of each element, the name of the element is displayed at the top of the figure. The Y axis shows the number density $N_i/N_Z$ (the number fraction of particular element over the total metal elements). Since the number density of elements with atomic number larger than Fe is very small, here we only illustrate the comparison from Li to Co. The number density details for all metal elements are listed in Table 5.1.

Table 5.1: [$\alpha$/Fe] =0.4 metal mixture from APOGEE atmosphere model.

| $Z_i$ | **Element** | $N_i/N_Z$ | $Z_i/Z_{tot}$ |
|---|---|---|---|
| 3 | Li | 1.125118E-07 | 1.789838E-07 |
| 4 | Be | 1.358245E-08 | 6.943298E-09 |
| 5 | B | 2.837775E-07 | 1.740407E-07 |
| 6 | C | 1.389562E-01 | 9.467165E-02 |
| 7 | N | 3.410971E-02 | 2.710027E-02 |
| 8 | O | 6.500967E-01 | 5.899821E-01 |
| 9 | F | 2.055313E-05 | 2.214894E-05 |
| 10 | Ne | 3.916318E-02 | 4.480950E-02 |
| 11 | Na | 8.374868E-04 | 1.092120E-03 |
| 12 | Mg | 4.819233E-02 | 6.645272E-02 |
| 13 | Al | 1.327022E-03 | 2.030966E-03 |
| 14 | Si | 4.602332E-02 | 7.331953E-02 |
| 15 | P | 1.296815E-04 | 2.278399E-04 |
| 16 | S | 1.962809E-02 | 3.569929E-02 |
| 17 | Cl | 1.790515E-04 | 3.601176E-04 |
| 18 | Ar | 8.571917E-04 | 1.942349E-03 |
| 19 | K | 6.805781E-05 | 1.509479E-04 |
| 20 | Ca | 2.903875E-03 | 6.601248E-03 |
| 21 | Sc | 8.374868E-07 | 2.135614E-06 |
| 22 | Ti | 1.129739E-04 | 3.068175E-04 |
| 23 | V | 5.662106E-06 | 1.636094E-05 |
| 24 | Cr | 2.471030E-04 | 7.288183E-04 |
| 25 | Mn | 1.389562E-04 | 4.330210E-04 |
| 26 | Fe | 1.595431E-02 | 5.054033E-02 |
| 27 | Co | 4.709534E-05 | 1.574330E-04 |
| 28 | Ni | 9.615635E-04 | 3.201937E-03 |
| 29 | Cu | 9.180746E-06 | 3.309272E-05 |
| 30 | Zn | 2.254125E-05 | 8.360387E-05 |
| 31 | Ga | 4.295146E-07 | 1.698541E-06 |
| 32 | Ge | 2.152673E-06 | 8.868589E-06 |
| 33 | As | 1.103768E-07 | 4.690756E-07 |
| 34 | Se | 1.210537E-06 | 5.423872E-06 |
| 35 | Br | 2.055313E-07 | 9.315754E-07 |
| 36 | Kr | 1.006417E-06 | 4.783928E-06 |
| 37 | Rb | 2.254125E-07 | 1.092798E-06 |

Continued on next page

**Table 5.1 – continued from previous page**

| $Z_i$ | **Element** | $N_i/N_Z$ | $Z_i/Z_{tot}$ |
|---|---|---|---|
| 38 | Sr | 4.709534E-07 | 2.340569E-06 |
| 39 | Y | 9.180746E-08 | 4.629843E-07 |
| 40 | Zr | 2.152673E-07 | 1.113894E-06 |
| 41 | Nb | 1.489286E-08 | 7.848404E-08 |
| 42 | Mo | 4.709534E-08 | 2.561576E-07 |
| 44 | Ru | 3.916318E-08 | 2.245141E-07 |
| 45 | Rh | 7.464109E-09 | 4.356868E-08 |
| 46 | Pd | 2.588082E-08 | 1.562438E-07 |
| 47 | Ag | 4.930352E-09 | 3.016689E-08 |
| 48 | Cd | 3.334095E-08 | 2.126129E-07 |
| 49 | In | 2.254125E-08 | 1.468070E-07 |
| 50 | Sn | 5.662106E-08 | 3.813382E-07 |
| 51 | Sb | 5.662106E-09 | 3.910541E-08 |
| 52 | Te | 8.767544E-08 | 6.347186E-07 |
| 53 | I | 1.832221E-08 | 1.318903E-07 |
| 54 | Xe | 9.837346E-08 | 7.326121E-07 |
| 55 | Cs | 6.652394E-09 | 5.015088E-08 |
| 56 | Ba | 8.374868E-08 | 6.523688E-07 |
| 57 | La | 7.636212E-09 | 6.016657E-08 |
| 58 | Ce | 2.837775E-08 | 2.255377E-07 |
| 59 | Pr | 2.152673E-09 | 1.720563E-08 |
| 60 | Nd | 1.595431E-08 | 1.305343E-07 |
| 62 | Sm | 5.662106E-09 | 4.829233E-08 |
| 63 | Eu | 1.874899E-09 | 1.616133E-08 |
| 64 | Gd | 7.292526E-09 | 6.504927E-08 |
| 65 | Tb | 1.078395E-09 | 9.721406E-09 |
| 66 | Dy | 7.814082E-09 | 7.202603E-08 |
| 67 | Ho | 1.832221E-09 | 1.714101E-08 |
| 68 | Er | 4.818124E-09 | 4.571214E-08 |
| 69 | Tm | 5.662106E-10 | 5.425671E-09 |
| 70 | Yb | 6.805781E-09 | 6.679964E-08 |
| 71 | Lu | 6.499470E-10 | 6.450469E-09 |
| 72 | Hf | 4.295146E-09 | 4.348619E-08 |
| 73 | Ta | 3.827172E-10 | 3.928162E-09 |
| 74 | W | 7.292526E-09 | 7.604653E-08 |
| 75 | Re | 9.615635E-10 | 1.015656E-08 |
| 76 | Os | 1.006417E-08 | 1.086005E-07 |
| 77 | Ir | 1.358245E-08 | 1.480899E-07 |

Continued on next page

**Table 5.1 – continued from previous page**

| $Z_i$ | **Element** | $N_i/N_Z$ | $Z_i/Z_{tot}$ |
|---|---|---|---|
| 78 | Pt | 2.471030E-08 | 2.734324E-07 |
| 79 | Au | 5.792659E-09 | 6.471848E-08 |
| 80 | Hg | 7.636212E-09 | 8.688804E-08 |
| 81 | Tl | 4.497570E-09 | 5.214136E-08 |
| 82 | Pb | 5.662106E-08 | 6.655256E-07 |
| 83 | Bi | 2.529170E-09 | 2.998067E-08 |
| 90 | Th | 6.499470E-10 | 8.554508E-09 |
| 92 | U | 1.709929E-10 | 2.308690E-09 |

## 5.2 Preliminary APOGEE [$\alpha$/Fe] =0.4 evolutionary tracks

Here we show the preliminary results of $\alpha$-enhanced tracks based on APOGEE ATLAS9.

Fig. 5.3 compares a model of M=0.85 $M_\odot$ of 47Tuc FG partition ( [$\alpha$/Fe] ~=0.4) with the coresponding APOGEE model ( [$\alpha$/Fe] =0.4) of the same metallicity. Since the APOGEE [$\alpha$/Fe] =0.4 is [$\alpha$/Fe] =0.3785 according to PARSEC solar value as already explained in Chap. 5.1, it shows a slightly cooler temperature compared with 47Tuc FG, in both the MSTO and the RGB phase (left panel of the figure). However, from the right panel we see that there is no notable difference in the luminosity evolution.

Fig. 5.4 and Fig. 5.5 show two sets of tracks with the APOGEE [$\alpha$/Fe] =0.4 partition, with [Z=0.0005, Y=0.249] and [Z=0.006, Y=0.259]. Stars with mass less than 3 $M_\odot$ are plotted. The full set of evolutionary tracks and isochrones with different [$\alpha$/Fe] metal partition will be available online after the full calculations are finshed.

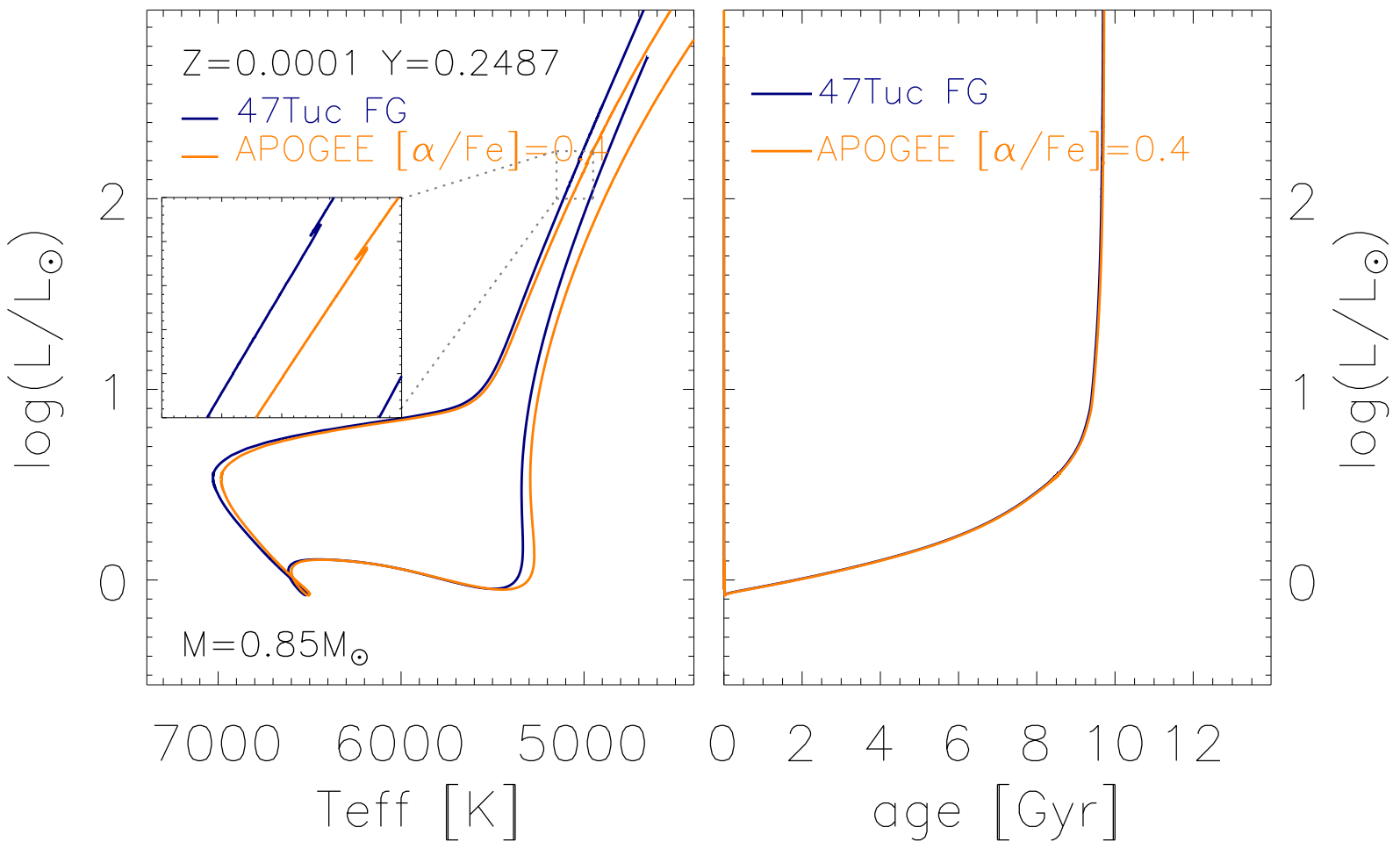

Figure 5.3: Evolutionary tracks with 47Tuc FG metal-mixture [$\alpha$/Fe] ~0.4 (blue line) and with APOGEE [$\alpha$/Fe] =0.4 mixture (orange line). Both of the tracks are for M=0.85 $M_\odot$ star and the total metallicity Z=0.0001. The left panel is HRD with sub figure zoom-in around the red giant branch region. The left panel shows how the luminosity of the star evolve with time.

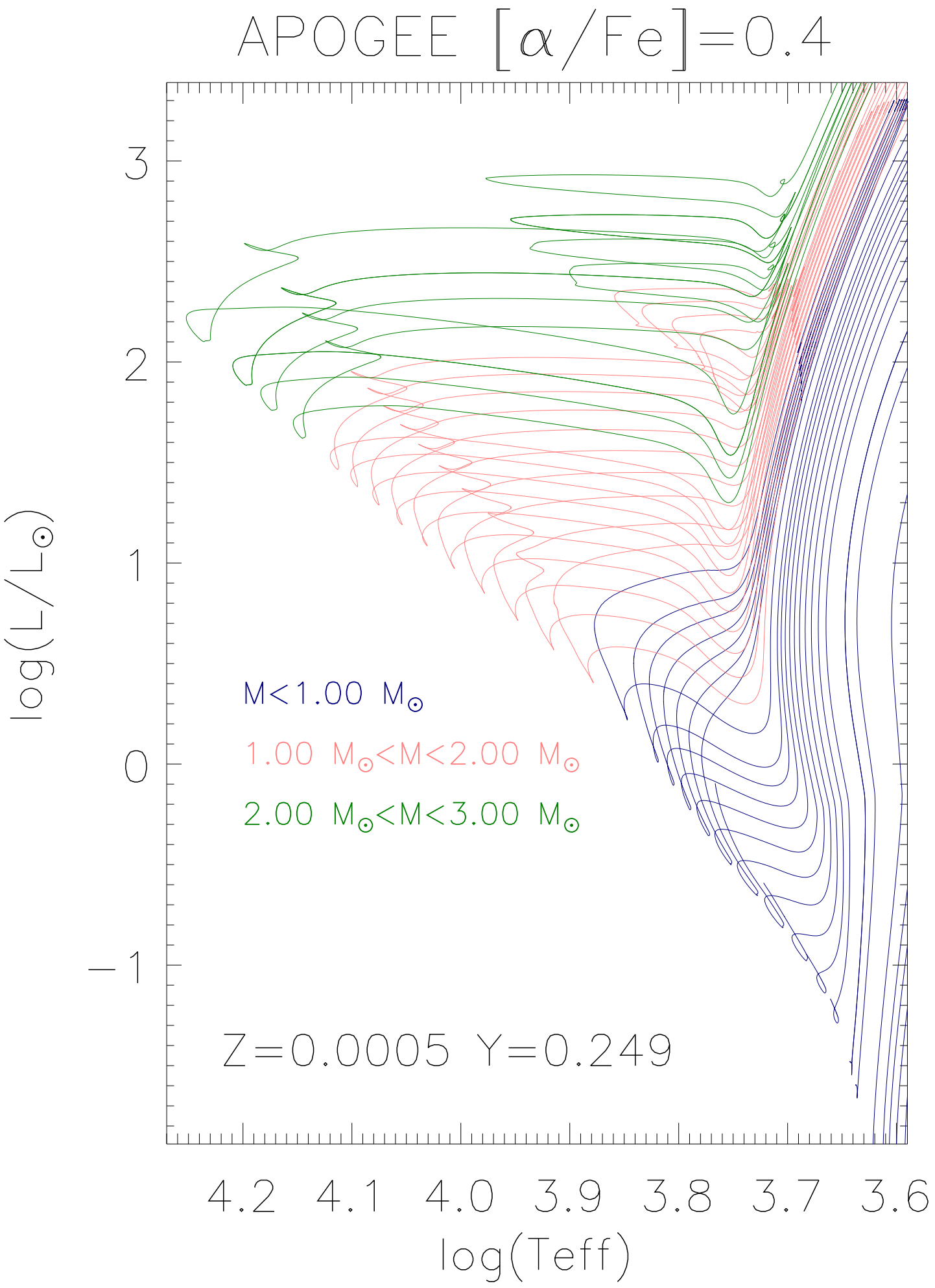

Figure 5.4: A HRD example of APOGEE [$\alpha$/Fe] =0.4 evolutionary tracks with [Z=0.0005, Y=0.249]

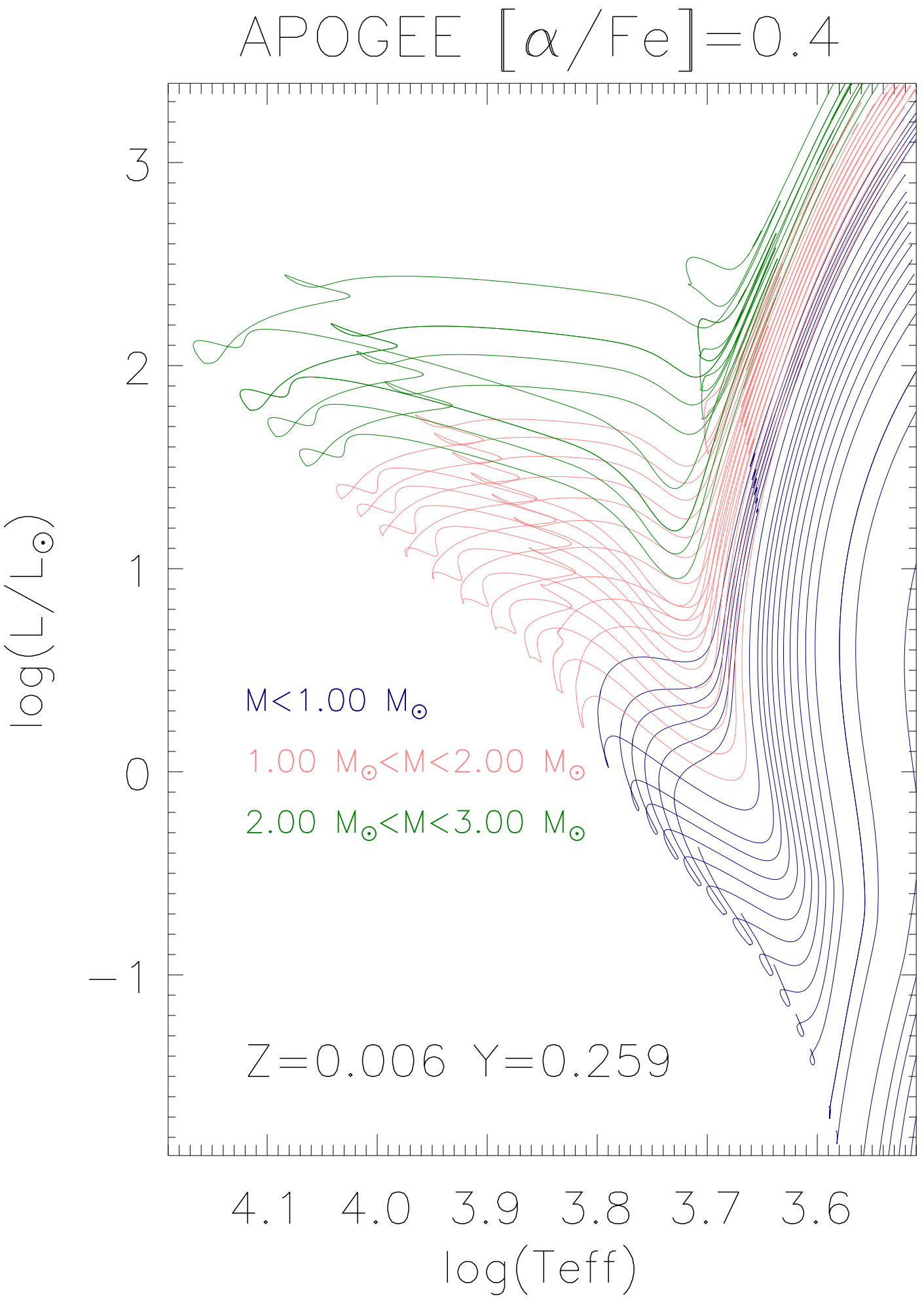

Figure 5.5: A HRD example of APOGEE [$\alpha$/Fe] =0.4 evolutionary tracks with [Z=0.006, Y=0.259]

# Chapter 6

# Test for various MLT

One of the most important quantity in current theory of stellar evolution is the mixing length parameter, $\alpha_{MLT}$, entering in the Mixing Length Theory (MLT). It sets the efficiency of convective energy transport in the external layers of the star where it is known that the real gradient deviates from the adiabatic one. The mixing length parameter is usually fixed by matching the current effective temperature and radius of the Sun. Then this single depth-independent value, obtained from a standard solar model, is applied to all other stellar models. In PARSEC this parameter is $\alpha_{MLT} = 1.74$ (See Chap. 2). However the mixing length theory is a local and time-independent theory, it assumes symmetry in the up- and down flows, and the convective energy flux is derived purely from local thermodynamical properties, ignoring any non-local properties of the flow. For all these reasons, non-local mixing length theories (e.g. Gough, 1977; Unno, Kondo & Xiong, 1985; Grossman, Narayan & Arnett, 1993; Deng, Xiong & Chan, 2006) have been studied and hydrodynamic calculations have been performed (e.g. Ludwig, Freytag & Steffen, 1999) to improve MLT.

Using 2D hydrodynamic model, Ludwig, Freytag & Steffen (1999) find that $\alpha_{MLT}$ varies significantly with the stellar parameters (from 1.3 to 1.8). The impact of a variable $\alpha_{MLT}$ on GC has been studied by (Freytag & Salaris, 1999). Recently Magic, Weiss & Asplund (2015) investigate MLT under full 3D radiative hydrodynamic (RHD) calculations, finding that $\alpha_{MLT}$ decreases towards higher Teff, lower surface gravity (log(g)) and higher metallicity. As already said the mixing length parameter changes the super-adiabatic structure of the star, influencing the position of the stars in the HR diagram. The larger $\alpha_{MLT}$ the more efficient is the convective energy transport and the hotter will be the star. The effect is larger at cooler temperatures because, in these conditions, a larger fraction of the total mass is in the super-adiabatic regime. Thus it is expected that the RGB location in the HRD is influenced by $\alpha_{MLT}$ more than the main sequence.

We have already showed in Chap. 3.2.3 that our isochrones with PHOENIX

bolometric correction were not perfectly matching the observations of 47Tuc in the RGB phase. Here we discuss how this discrepancy could depend from the MLT in our 1D model.

We first check the results of Magic, Weiss & Asplund (2015), i.e. we allow a variation of the MLT parameter following their suggested formalism. We apply the coefficients $a_i$ from Table B.1 of the appendix in Magic, Weiss & Asplund (2015) to our own solar MLT ($\alpha^{\odot}_{MLT}$ = 1.74), and introduce the dependence of the MLT from Teff and log(g) that they have described. With this new formalism for the MLT we compute new evolutionary tracks. The new tracks run significantly cooler than adopting the fixed MLT, with the effect being almost negligible near the base of the RGB but increasing as the star climbs the RGB. In principle the variable MLT taken as it is from Magic, Weiss & Asplund (2015) cannot solve the discrepancy of a too cool RGB, for 47Tuc. Actually the fit worsen significantly.

Another effect that may affect the effective temperature of the RGB is the so called density inversion problem (Harpaz, 1984). In 1D models the condition of hydrostatic equilibrium forces the appearance of a density inversion in the sub-photospheric region of the star where the opacity attains its maximum value. A large value of the opacity causes a large radiative gradient and also a large real gradient if convection is inefficient, causing a large radiative pressure gradient. In hydrostatic equilibrium, the total pressure gradient (radiative pressure plus gas pressure) must balance the local acceleration of gravity (which is negative) and, in these regions the gas pressure gradient become positive, i.e. the gas pressure must "decrease" with depth. Since temperature must always increase with depth in presence of an outgoing energy flux (negative temperature gradient), then it is the density that decreases. So in these layers density reaches a maximum then decreases and then increases again when the efficiency of convection becomes larger. This is shown in Fig. 6.1, adapted from Harpaz (1984) (their FIgure 2) where one may see that the density profile ($\rho$) is generally increasing toward the center of the star (smaller m/ $M_{\odot}$ ), while there is a density inversion around 0.87 m/ $M_{\odot}$ .

A way to suppress this density inversion is to use a mixing length (ML) that is proportional to the density scale height. In this way when the density has a maximum the density scale height diverges and the ML becomes very large, increasing the efficiency of convection. This shifts the real temperature gradient toward the adiabatic one and a density inversion never happens. In practice, to eliminate the density inversion one may force the real gradient to be limited (in modulus) by an upper limit that is easily derived from the condition that the derivative of the density with radius be negative. Girardi et al. 2000 used this criterion to eliminate the density inversion problem with the results being that all the RGB tracks were to hot with respect to the observed RGB locations. In PARSEC the density inversion is allowed and the predicted RGB location is fairly well reproduced.

Motivated by both these results, i.e. that with the variable MLT by Magic, Weiss & Asplund (2015) the RGB is too cool and that with the suppression of the density inversion it is too hot (Girardi et al. 2000), we analysed in more detail the results of Magic, Weiss & Asplund (2015) and we realized that in 3D models there should not be a “density inversion” Indeed, from the bottom left panel of Fig. 6.2, adapted from Fig. 11 in Magic, Weiss & Asplund (2015), we see that in their 3D model the density gradient is always $\nabla\rho > 0$, indicating that there is no density inversion. We notice that, the vertical velocity gradient $\nabla_{v,z}$ in the same figure reaches its largest value when $\nabla\rho$ has a minimum, indicating that convection is rendered more efficient by a significant increase of the turbulent velocities. This could partly explain how the density inversion is avoid. In order to give the correct explanation we need to use 3D models. This is one of the project for my near future.

For the moment we limited our investigation to the comparison of stellar evolutionary tracks computed with different hypothesis for the MLT and density inversion. We use the same composition [Z=0.0055, Y=0.276] and consider five different MLT cases: constant MLT 1.74: $\alpha_{MLT} = 1.74$; constant MLT1.77: $\alpha_{MLT} = 1.77$; variable MLT from Magic, Weiss & Asplund (2015): Var $\alpha_{MLT}$; constant MLT 1.74 with inhibited density inversion: $\nabla\rho > 0$; and variable MLT with inhibited density inversion: $\nabla\rho > 0$+Var $\alpha_{MLT}$. All these cases are compared in Fig. 6.3 and Fig. 6.4.

For the five MLT cases described above, Fig. 6.3 shows the Variation of $\alpha_{MLT}$ with surface temperature and with the evolutionary time while, Fig. 6.4 illustrates the corresponding location in the HRD. A constant larger MLT ($\alpha_{MLT} = 1.77$, red dotted-line) in Fig. 6.4 makes the star hotter, but the differences are relatively small as shown in the lower right panel. Along the giant branch, the MLT effect becomes more important as seen in the upper right panel, that displays the differences near the RGBB.

The track with variable MLT runs cooler and cooler, as already discussed, while the one with inhibited density inversion runs hotter and hotter. Both tracks are not able to fit the RGB of 47Tuc. On the other hand if one consider both variable MLT and inhibited density inversion, the track runs almost superimposed to the one with fixed MLT. This result, initially surprising, could be explained by the absence of density inversion in 3D models.

We thus conclude from this experiment that it is not possible to simply apply the formalism provided by Magic, Weiss & Asplund (2015) *without including also the 3D effects on the internal density structure.* This renders very difficult any application of 3D results to 1D models. On the other hand our results show that, once density inversion is inhibited as it happens in 3D models, applying the formalism provided by Magic, Weiss & Asplund (2015) actually is equivalent to use the standard MLT without density inversion inhibition.

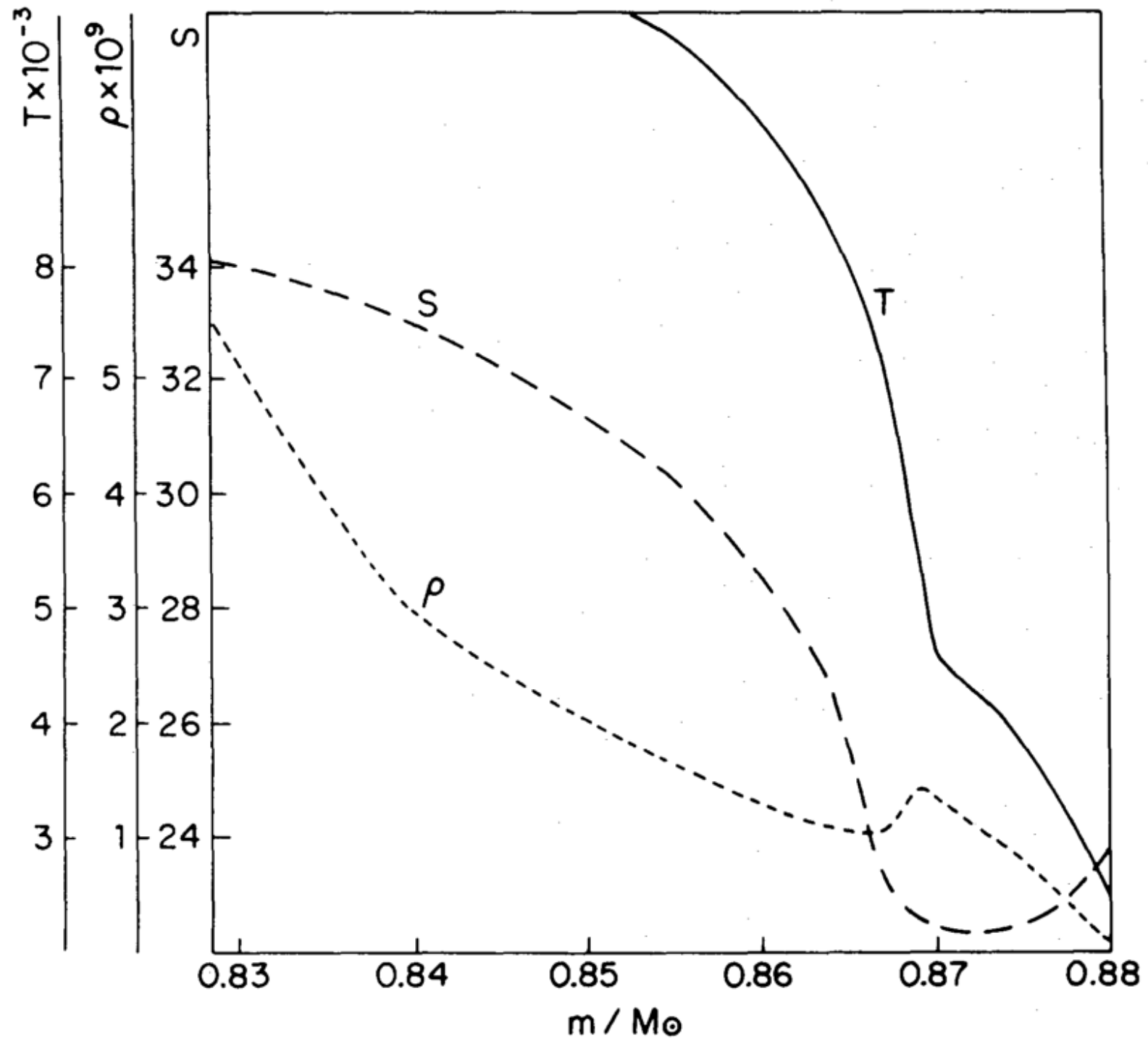

Figure 6.1: Figure 2 of Harpaz (1984) illustrating the density inversion problem. The original caption of this figure is: *The density, temperature and specific entropy in the outer part of the envelope versus mass.*

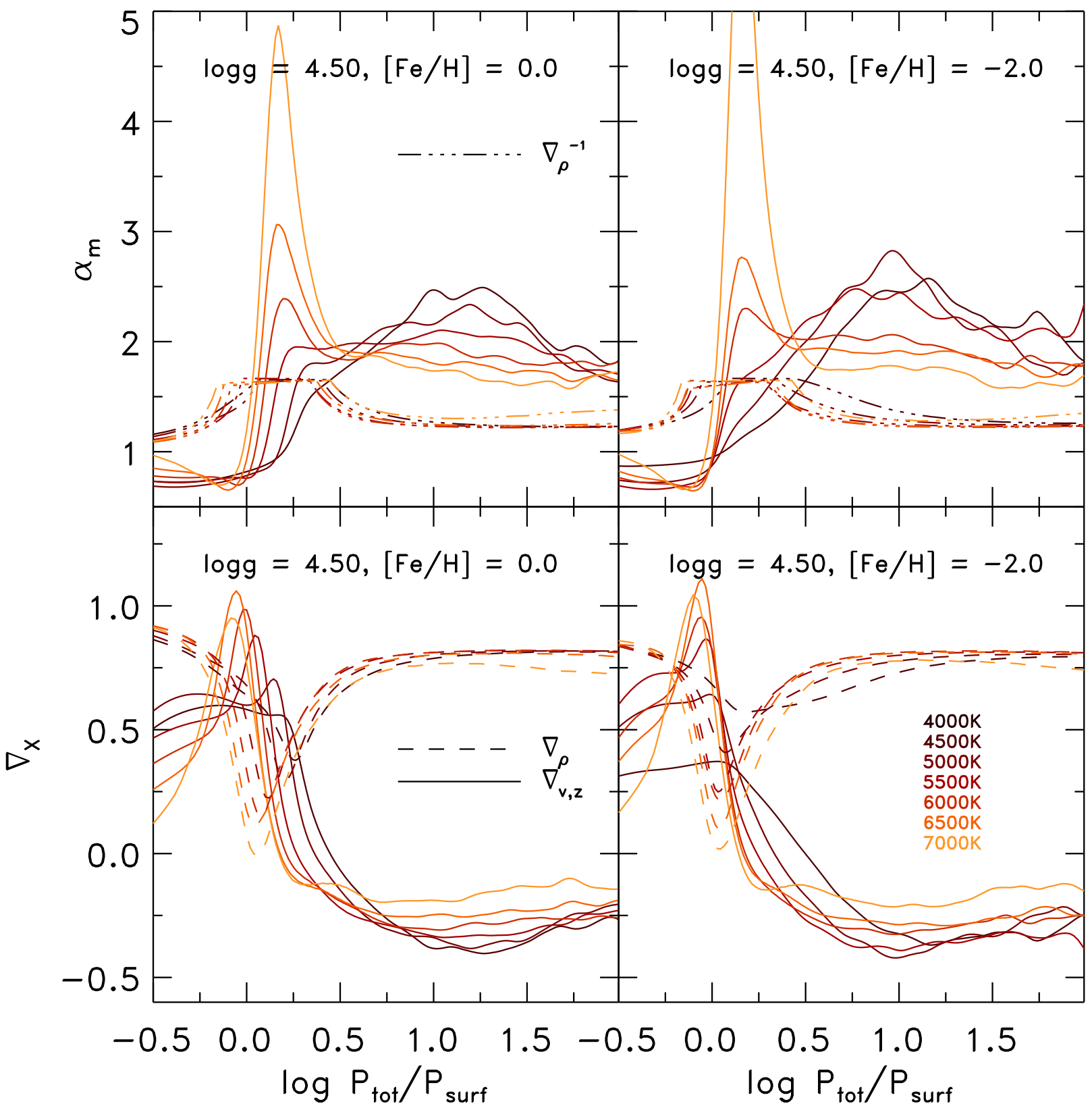

Figure 6.2: Figure 11 of Magic, Weiss & Asplund (2015). The lower bottom panel shows that the density gradient is always >0, indicating that there is no density inversion in their 3D models. The original caption of this figure is: *Bottom panel: the gradient for density, $\nabla\rho$, and vertical velocity, $\nabla_{v,z}$,(dashed and solid lines, respectively) for different stellar parameters.*

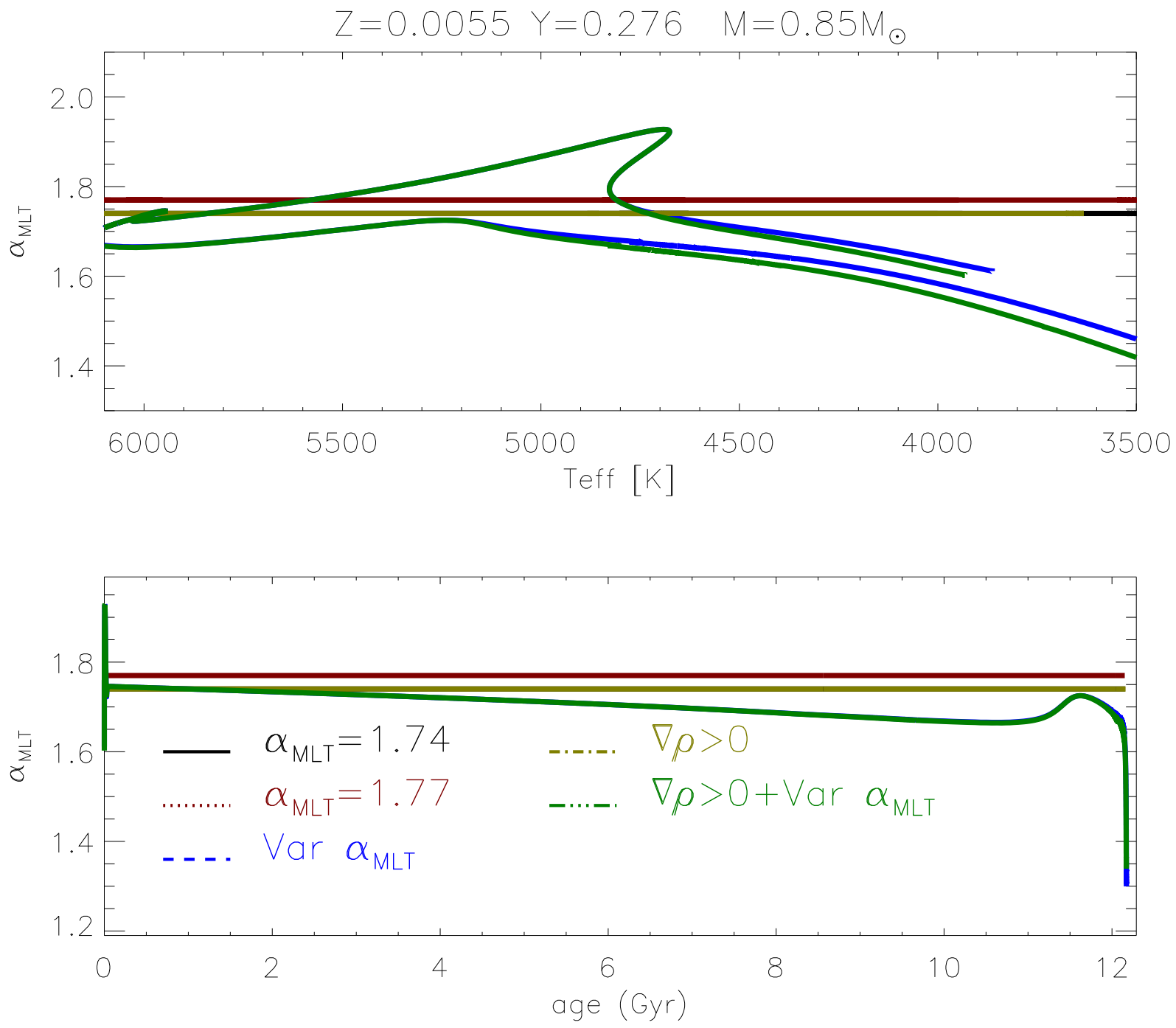

Figure 6.3: Variation of $\alpha_{MLT}$ versus Teff (upper panel) and age (bottom panel) for different cases.

While waiting for coming new more deep investigations on these aspects, these results allow us to be still confident on the old and apparently very robust mixing length theory.

> ’We will never know how to study by any means the chemical composition (of stars), or their mineralogical structure.’
>
> –AUGUSTE COMTE (1835)

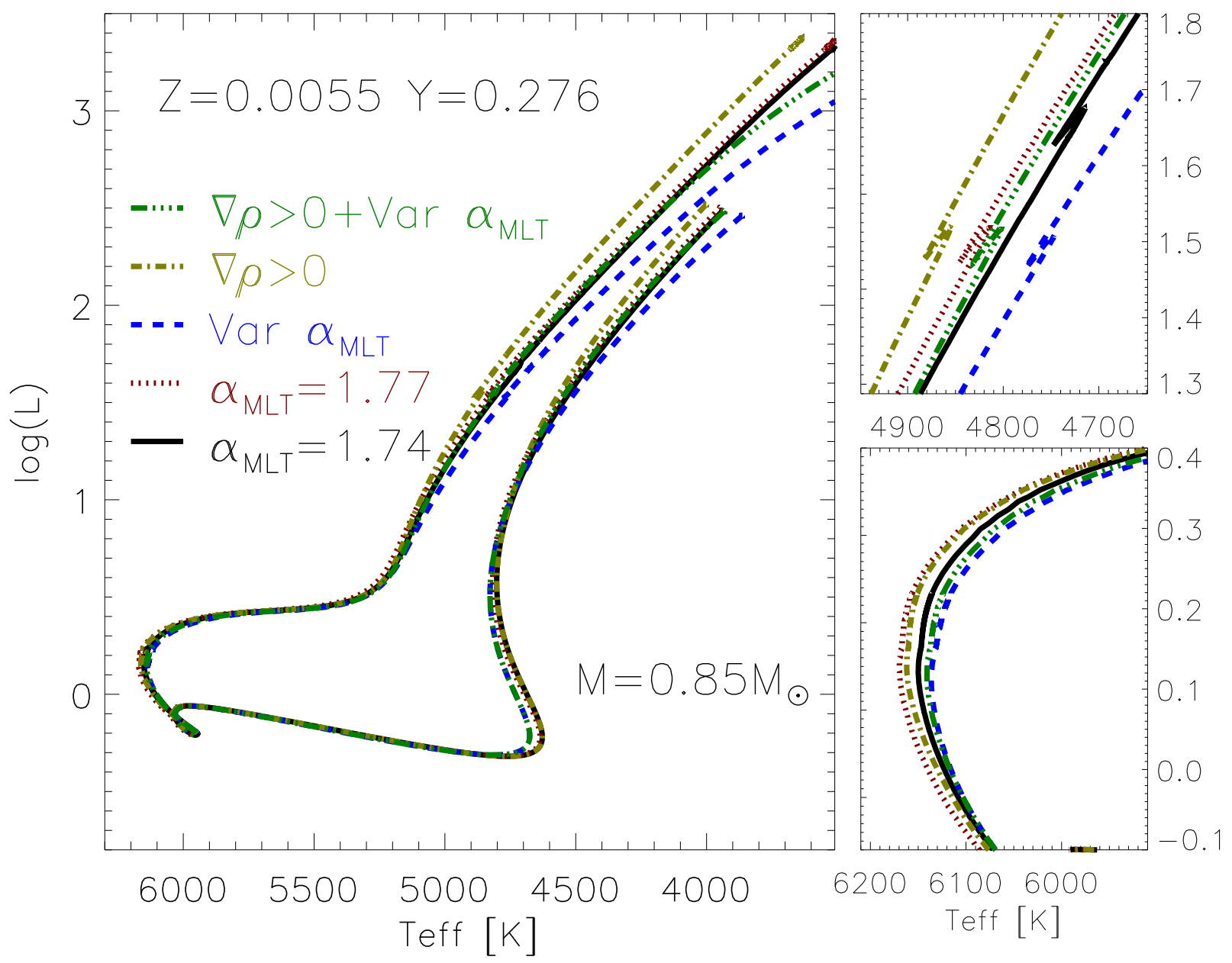

Figure 6.4: HRD comparison for a 0.85 $M_{\odot}$ star with different mixing length cases. The two subfigures at the right panel are the zoom-in around RGBB region and main sequence, respectively. Different MLT cases, take the upper right panel around RGBB for example, from left to right are tracks with density inversion inhibited ($\nabla\rho > 0$, dark yellow dash-dot line), with constant MLT $\alpha_{MLT} = 1.77$ (red dot line), with density inversion inhibited and various MLT ($\nabla\rho > 0$+Var $\alpha_{MLT}$, green dash-dot-dot line), with constant MLT $\alpha_{MLT} = 1.74$ (black solid line), and with various MLT (Var $\alpha_{MLT}$, blue dot line). All the tracks are calculated with [Z=0.0055, Y=0.276] and [$\alpha$/Fe] ~0.2.

# Part II:

# Lithium evolution

'We will never know how to study by any means the chemical composition (of stars), or their mineralogical structure.'

–Auguste Comte (1835)

# Chapter 7

# From Pre-Main Sequence to the Spite Plateau

## 7.1 Background

In this part we address the cosmological lithium problem on the ground of the Pre-Main Sequence (PMS) stellar physics, under the assumption that the standard BBN is correct. Hereafter when not otherwise specified for Li we refer to the isotope 7. The surface Li evolution during the PMS has been analyzed in stellar evolution models with solar metallicity by several authors (e.g. Dantona & Mazzitelli, 1984; Soderblom et al., 1993; D'Antona & Mazzitelli, 1994; Swenson et al., 1994; Ventura et al., 1998; Piau & Turck-Chièze, 2002; Tognelli, Degl'Innocenti & Prada Moroni, 2012), with the general conclusion that mixing efficiency plays a key role on Li depletion. Recent extensive studies in metal-rich young open clusters do show that the observed lithium abundance is modified during the PMS phase (Somers & Pinsonneault, 2014). In this respect, two points are worthy of consideration. On one side, it has been shown that overshoot at the base of the solar convective envelope is more efficient than hitherto believed (Christensen-Dalsgaard et al., 2011), and could be even more in the envelope of a PMS star. This would favour a more efficient Li depletion. On the other side, there is evidence that a residual mass accretion (also called late accretion) persists for several $10^7$ years and is observed up to the early main sequence phase (De Marchi et al., 2011). This may act to partially restore Li in metal-poor PMS stars. Molaro et al. (2012) first proposed that Li abundance both in the Population I and Population II stars could be modified due to the combined effects of efficient overshoot and late mass accretion during the PMS evolution.

Our stellar code `PARSEC`, as already introduced in Part I, is able to predict readily the evolution of stars with any chemical pattern of interest. In this part of the thesis we examine the working scenario proposed by Molaro et al. (2012) in a more systematic way.

The structure of this part is as follows. chapter 7.2 describes the theoretical pre-main sequence models, which include envelope overshooting, residual accretion and EUV photo-evaporation. chapter 7.3 describes the Li evolution in main sequence. chapter 7.5 presents the results of the models and the comparison with the observations. An ample discussion and the main conclusions are drawn in chapter 6.5.

## 7.2 Pre-Main Sequence Li evolution

Pre-main sequence is the direct continuation of the proto-stellar phase. Initially, as the young stellar object (YSO) evolves along its stellar birthline, its luminosity is mainly supported by an accretion process strong enough to maintain active deuterium fusion (Stahler, 1983). Once the accretion ceases, the proto-star, surrounded only by a residual disk, descends along its Hayashi line almost vertically in the Hertzsprung-Russell diagram (H-R diagram), undergoing a rapid gravitational contraction.

In this phase evolution of PMS stars critically depends on their mass, and lithium can be burned at different stages through the reaction $^7$Li(p,$\alpha$)$^4$He. Lithium is very fragile as the nuclear reaction rate $R_{nuclear}$ of $^7$Li(p,$\alpha$)$^4$He becomes efficient already at temperatures of a few million Kelvin. The effective Li burning temperature for PMS stars, i.e. that needed to consume Li in a timescale of $\sim\ 10^7\ yr$, is $\sim\ 4 \times 10^6\ K$. The $R_{nuclear}$ is adopted from JINA REACLIB database (Cyburt et al., 2010).

Very low mass stars with initial mass $m_0 < 0.06 M_\odot$ (e.g. most of the brown dwarfs) never reach this temperature. More massive PMS stars experience Li-burning, which initially affects the entire stellar structure as long as it is fully convective. Later, at the formation of the radiative core, the extent of Li burning can vary, depending on the mass of the star. Thereafter, the Li evolution is critically affected by the temperature at the base of the convective envelope, and the efficiency of overshoot at the base of the convective envelope begins to play a significant role.

We use the stellar evolution code `PARSEC` to calculate the PMS evolution of low mass stars with initial mass from 0.50 $M_\odot$ to 0.85 $M_\odot$, with metallicity $Z = 0.0001$ ($\sim\ [M/H] = -2.2$ dex), typical of POP II stars. The adopted helium abundance is $Y_p = 0.249$, based on the helium enrichment law (Bressan et al., 2012):

$$Y = Y_p + \frac{\triangle Y}{\triangle Z} Z = 0.2485 + 1.78Z \tag{7.1}$$

where $Y_p = 0.2485$ is the adopted primordial value and $\triangle Y / \triangle Z$ is the helium-to-metals enrichment ratio. Following the SBBN prediction, we set the initial lithium abundance A(Li)=2.72. A solar-model calibrated mixing length parameter $\alpha_{MLT} = 1.74$ is adopted.

### 7.2.1 Convective overshooting

As introduced in Chap. 3.2.2 of Part I, overshooting (OV) is the signature of the non-local convective mixing in the star that may occur at the borders of any convectively unstable region (Bressan et al., 2015). Efficient overshooting at the base of the convective envelope may significantly affect PMS surface Li depletion, because overshooting extends the mixed region into the hotter stellar interior.

Soon after the formation of the radiative core we assume that the overshooting at the base of the deep convective envelope is quite efficient, with $\Lambda_e \sim 1.5\ H_p$. This value is intermediate between the one suggested for the Sun (Christensen-Dalsgaard et al., 2011) and the larger values ($\Lambda_e = 2 - 4\ H_p$) suggested by Tang et al. (2014) for intermediate and massive stars. We will show in Chap. 7.5 that this value is not critical, since even assuming a maximum overshoot $\Lambda_e \sim 0.7\ H_p$, the model can well reproduce the data. The overshooting distance is computed using the natural logarithm of the pressure, downward from the Schwarzschild border ($P_{sch}$) at the bottom of the convective envelope, i.e. an underlying stable layer ($P_{lay}$) is mixed with the envelope if $\ln(P_{lay}) - \ln(P_{sch}) \leq \Lambda_e$. Then, as the star moves toward the zero age main sequence (ZAMS), we vary the overshooting efficiency proportionally to the mass size of the outer unstable convective region ($f_{cz}$), until the typical value estimated for the Sun $\Lambda_e = 0.3\ H_p$ (Christensen-Dalsgaard et al., 2011) is reached:

$$\Lambda_e = 0.3 + (1.5 - 0.3) * f_{cz} \tag{7.2}$$

This assumption produces an efficient photospheric Li depletion during the PMS phase, while preserving the solar constraints for the main sequence stars. Fig. 7.1 illustrates the temporal evolution of the size of the convective and overshoot regions for four values of the initial stellar mass, $m_0 = 0.5, 0.6, 0.7, 0.8\ M_\odot$ at metallicity Z=0.0001. The mass coordinate, on the Y axis, goes from 0, at the center of the star, to 1 at the surface. The stars are initially fully convective and the temperature at the base of the convective zone ($T_{bcz}$) is high enough to burn Li so efficiently to begin surface Li depletion. As the central region heats up, the base of the convective envelope moves outwards to the stellar surface, while $T_{bcz}$ decreases. When $T_{bcz}$ drops below the lithium effective burning threshold, the photospheric Li-depletion ceases, leading to an almost total or only partial Li depletion, depending on the stellar mass and on the overshooting efficiency.

Figure 7.2 illustrates the effect of efficient overshooting as the *solo* mechanism applied to the PMS phase for $m_0 = 0.60,\ 0.70,\ 0.80\ M_\odot$ stars. In standard model without overshoot (left panel), A(Li) almost remains the same as the primordial value, unless the convective zone in the star itself goes deep enough to burn Li, as for $m_0 = 0.60\ M_\odot$ star in the left panel. Conversely, assuming efficient envelope overshooting, Li abundance is found to be fully depleted over a few Myr in the same stellar models (right panel). It follows that if overshoot were the only process at work, a very low Li abundance should be measured at the surface of those stars.

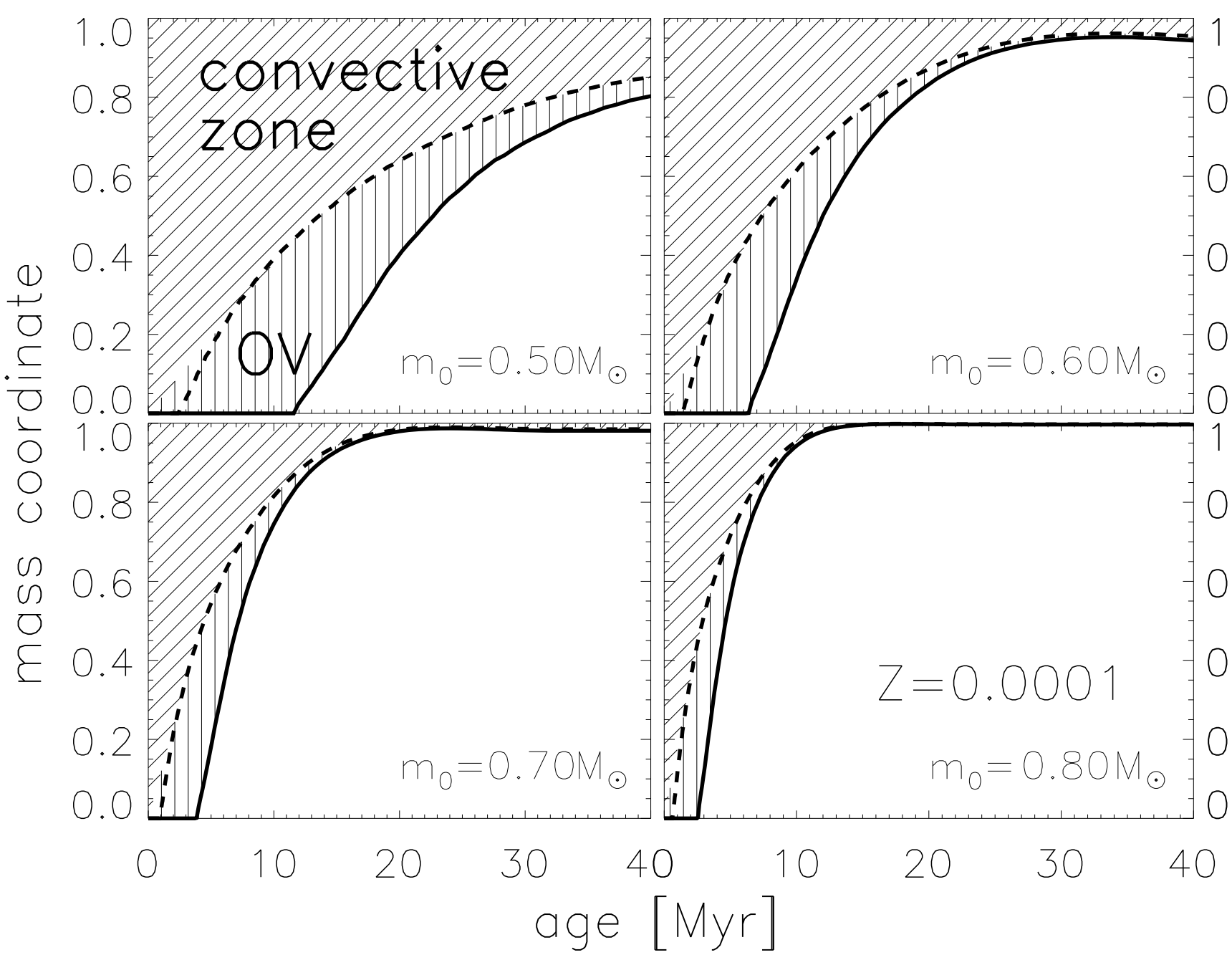

Figure 7.1: Kippenhahn diagrams for stars with initial metallicity Z=0.0001 and different initial masses. The hatched area corresponds the mass size of the convective zone, within it the area filled with vertical lines represents the contribution of the envelope overshooting. On the y-axis the mass coordinate of the convective zone starts from 0 (at the center of the star) which means the star is fully convective, while 1 means the base of the convective zone is at the surface of the star and there is no convection. As the star evolves, the base of the convective zone retreats towards the surface of the star, and the overshoot layer become shallower and shallower.

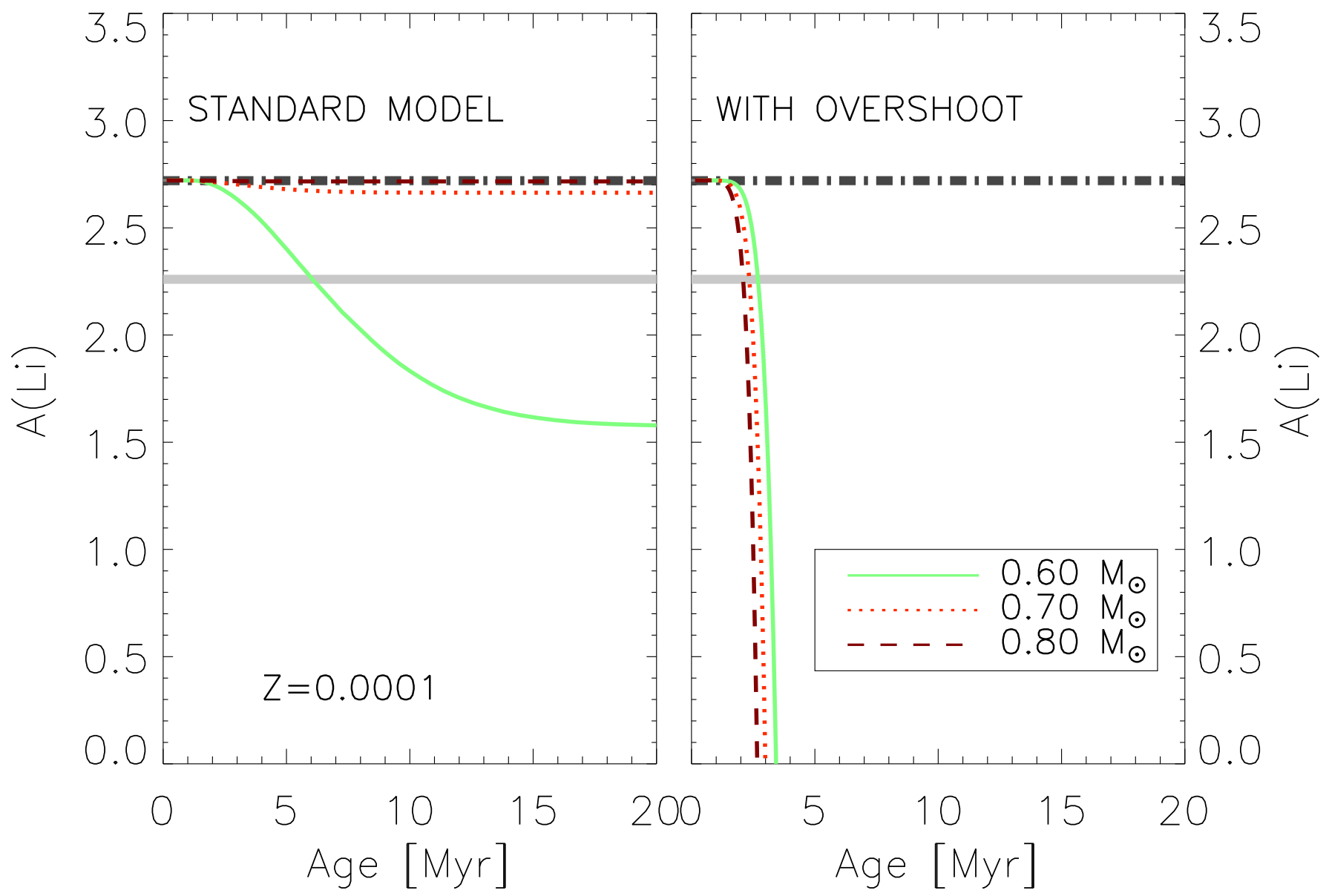

Figure 7.2: PMS Li evolution in the standard model (left panel) and model with envelope overshoot (right panel) for three stellar models under consideration. The two horizontal lines indicate primordial A(Li) (dark grey dot dashed line) and Spite plateau A(Li) (light grey solid line), respectively. Overshooting is the solo parameter tested, no other mechanism is applied.

### 7.2.2 Late mass accretion

Recent observations of late PMS stars – those that have already abandoned the stellar birth line and are joining the ZAMS – reveal that most of them show $H_\alpha$-line emission, which likely originates in a residual accretion process (De Marchi, Panagia & Romaniello, 2010; Spezzi et al., 2012). This suggests that the disk accretion may last much longer than previously believed: at least tens of Myr and even up to the early main sequence (De Marchi et al., 2011).

This late accretion (also known as residual accretion) is different from the one that maintains the star on the stellar birth line during the proto-stellar phase. The observed values indicate a median accretion rate of $\sim\ 10^{-8} M_\odot$ /yr for YSO T Tauri stars (TTS) with a steady decline as time proceeds (Espaillat et al., 2014). Most residual disks presented in Espaillat et al. (2014) and Gallardo, del Valle & Ruiz (2012) are associated with accretion rates at the level of $\sim\ 10^{-10} M_\odot$/yr - $10^{-8} M_\odot$/yr.

Since the accreting material keeps the initial Li abundance, its contribution cannot be neglected, especially if the photospheric value has been already depleted by another process. In this framework we modify our stellar evolution code PARSEC to account for the effect of such a residual accretion during the PMS phase.

We assume that, after the main accretion phase when the star leaves the stellar birth-line by consuming its internal deuterium, a residual accretion keeps going on. Since our PMS models are initially evolved at constant mass from a contracting configuration without nuclear burning, we assume that the residual accretion begins when deuterium burning ends, that is when the photospheric deuterium abundance drops to 1/10 of its initial value, and indicate this time as $\tau_{acc}$. We then apply the accretion to the models, starting at t=$\tau_{acc}$ and assume that it declines with time following a power law

$$\dot{M} = \dot{M}_0 \left( \frac{t}{\tau_{acc}} \right)^{-\eta} \qquad [M_\odot/\mathrm{yr}] \tag{7.3}$$

where $\dot{M}_0$ is the initial rate and $\eta$ is the parameter that specifies the rate of decline. In principle, the parameters $\dot{M}_0$, $\tau_{acc}$, and $\eta$ could be treated as adjustable parameters to be constrained with the observed rates, after properly considering a detailed PMS evolution that includes the main accretion phase. However, since in this work we perform an explorative analysis without a full description of the proto-star phase and the initial evolution with large accretion rates, we assume the residual accretion begins when deuterium burning ends as defined before.

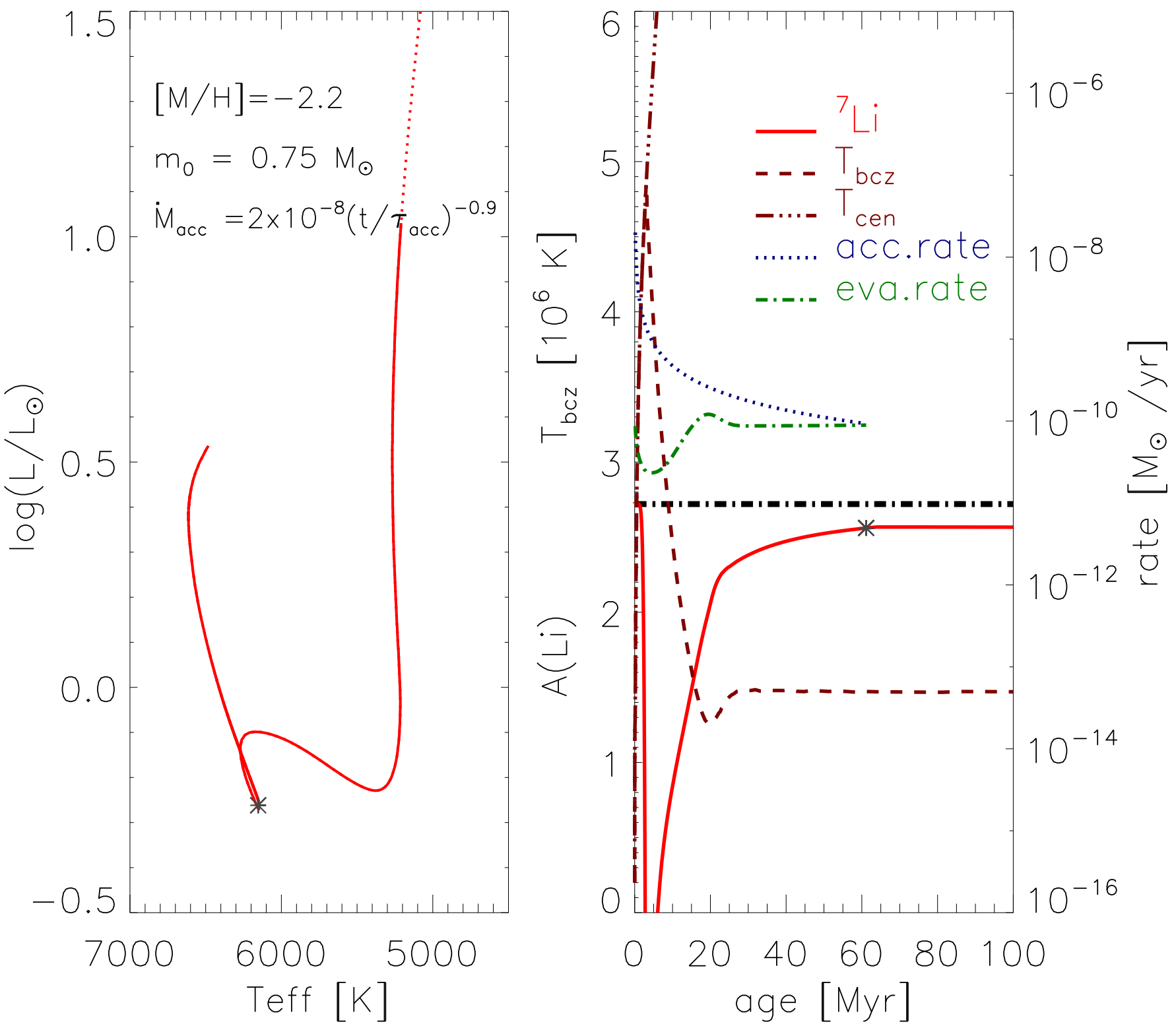

Figure 7.3: An example of the PMS lithium evolution for a star with $m_0 = 0.75 M_\odot$, [M/H]=-2.2. Left panel: Evolutionary track at constant mass in H-R diagram. The solid line starts at the end of the dotted-line-indicated deuterium burning phase. At this stage the accretion is reduced to $2 \times 10^{-8}$ $M_\odot$/yr. The asterisk marks the end of the residual accretion. Right panel: Li evolution starting from an initial abundance A(Li)=2.72 dex (horizontal black dot dashed line). The temperatures at the center ($T_{cen}$, dark red dot dot dashed line) and at the base of the convection zone ($T_{bcz}$, dark red dashed line) are also shown. The accretion (with rate drawn by the dark blue dotted line) is terminated by the EUV photo-evaporation (with rate drawn by the dark green dot dashed line) when the latter reaches the same value as the former.

Table 7.1 shows the age at the beginning of the residual accretion ($\tau_{acc}$) for different stellar masses. We note that the duration of the early large accretion, before $\tau_{acc}$, is around $1-2\ \times 10^5$ yr, hence negligible compared to the PMS lifetime (several $10^7$ yr). As to the initial accretion rate $\dot{M}_0$, we assume $\dot{M}_0 = 2{\times}10^{-8}\ M_\odot$/yr, which is a reasonable value close to the observed median accretion rate for young TTS (Hartmann et al., 1998). The exponent $\eta$ describes how fast the accretion rate declines with increasing age $t$. Here we set $\eta = 0.9$ to recover most of the observed rates from Espaillat et al. (2014); Gallardo, del Valle & Ruiz (2012); De Marchi et al. (2011).

We also assume that during this residual accretion phase the original material is accreted following Equation 7.3, irrespective of the geometry of the disk.

The right panel of Fig. 7.3 illustrates the effect of late accretion on the Li evolution, for a star with initial mass $m_0 = 0.75 M_\odot$. The accretion material falling onto the star contains lithium with the initial abundance. Even if accretion is very small after tens of million years and the primordial material is diluted, it restores the surface $^7$Li towards the undiluted initial value.

### 7.2.3 EUV photo-evaporation

Late accretion will last until the remaining gas reservoir is consumed or until some feedback mechanism from the star itself is able to clean the nearby disk. In this respect, Extreme UV (EUV) radiation (13.6 – 100 eV, 10 – 121 nm) photons could be particularly important in determining the end of the accretion process because they have energy high enough to heat the disk surface gas. The warm gas can escape from the gravitational potential of the central star and flow away in a wind (Dullemond et al., 2007). Although no actual evaporative flow observation is available yet, there is a large consensus that this radiation could significantly reduce the disk mass and shorten the lifetime of the disk. The existence of pre-transitional and transitional disks with a gap or a hole in the disk are taken as possible evidence of this process (Dullemond et al., 2007; Espaillat et al., 2014). The magnitude of the disk mass loss caused by EUV evaporation is given by the following relation (Dullemond et al., 2007):

$$\dot{M}_{EUV} \sim 4\times 10^{-10} \left(\frac{\Phi_{EUV}}{10^{41}s^{-1}}\right)^{0.5} \left(\frac{M_*}{M_\odot}\right)^{0.5} \quad [M_\odot \mathrm{yr}^{-1}] \tag{7.4}$$

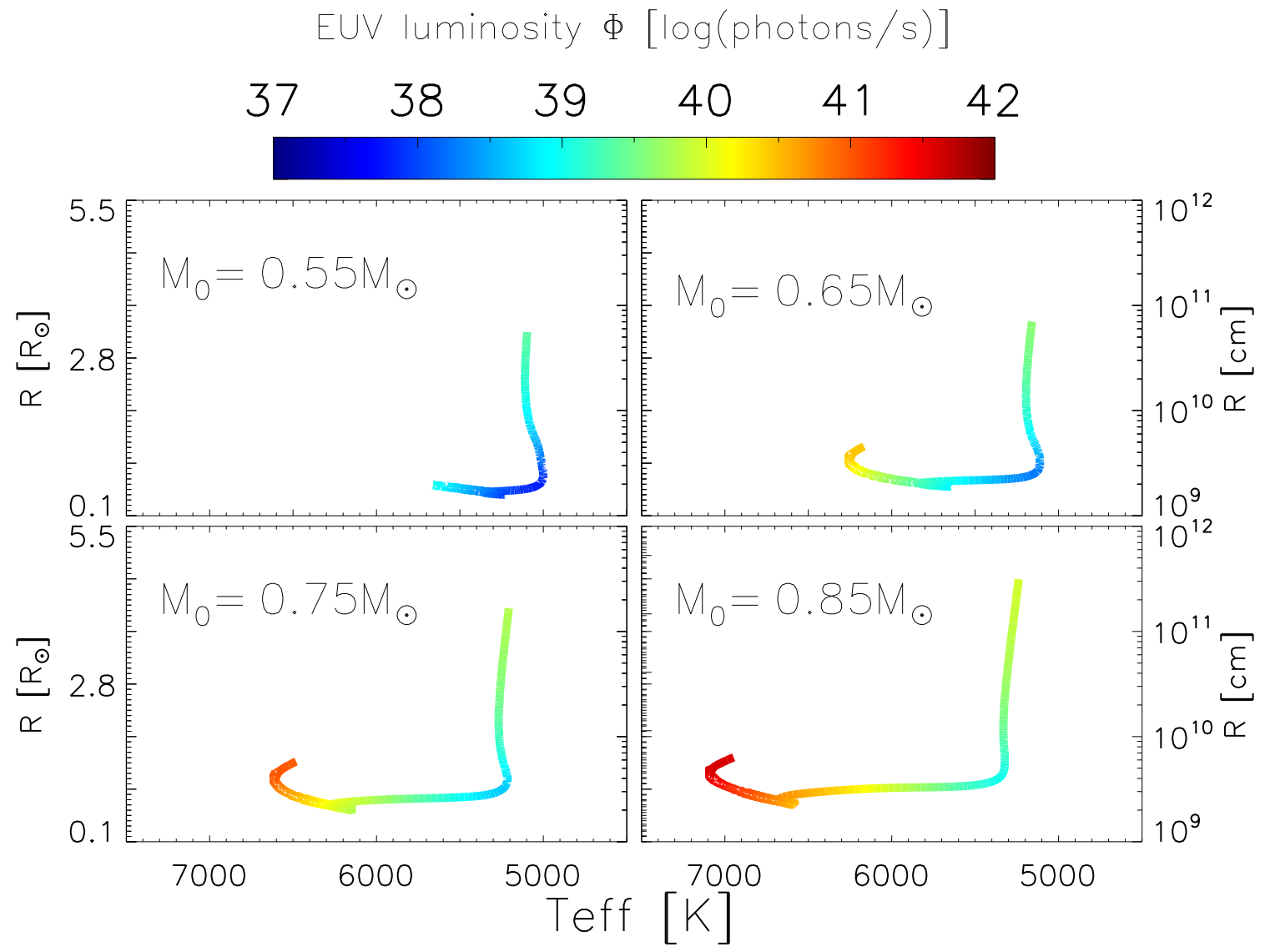

Figure 7.4: Evolution of the EUV photon luminosity from the central stars (color-coded according to logarithmic scale shown on the top bar) as a function of the effective temperature and radius, assuming black-body emission. Note the different units for the radius on the left and right Y-axes.

where $\Phi_{EUV}$ is the EUV photon luminosity [photons/s] produced by the central star. This effect can be easily included in PARSEC under the assumption that the stars emit as a black body at the given effective temperature $T_{\rm eff}$ In Fig. 7.4 we show the EUV luminosity evolution for stellar models with initial masses $m_0 = 0.55, 0.65, 0.75$, and $0.85\,M_\odot$. The photon luminosity $\Phi_{EUV}$ varies with the effective temperature and the radius of the central star, therefore being importantly affected by the stellar evolutionary phase. Most of the EUV luminosities range from $\Phi_{EUV} \sim 10^{38}$ photons/s to $\sim 10^{42}$ photons/s. The residual accretion does not change these values much because the residual accretion rate is never high enough to significantly increase the mass (and hence the evolution) of the star.

The EUV photo-evaporation rates obtained from Eq. (7.4) are shown in Fig. 7.5 (dot-dashed lines) for a grid of low-mass stars, and they are compared with the corresponding accretion rates (solid lines) obtained from Eq. (7.3). The EUV photo-evaporation rates increase with increasing initial mass. Furthermore, at any given initial mass they depend mainly on the evolution of the EUV luminosity because the total mass of the star is not significantly affected by the accretion rate. At relatively larger masses there is an minimum caused by the initial shrinking of the radius at almost constant effective temperature (Fig. 7.4). Thereafter, the star evolves at constant luminosity toward the ZAMS and, since the effective temperature increases, the fractional number of EUV photons increases rapidly, so does the evaporation rate. When the evaporation rate is larger than the accretion rate the accretion effectively stops. This point is marked with an asterisk in the figure and the terminal age is shown in table 7.1. Beyond this point the star evolves at constant mass. Figure 7.5 is meant to provide a schematic diagram of this mechanism for the first 100 Myr.

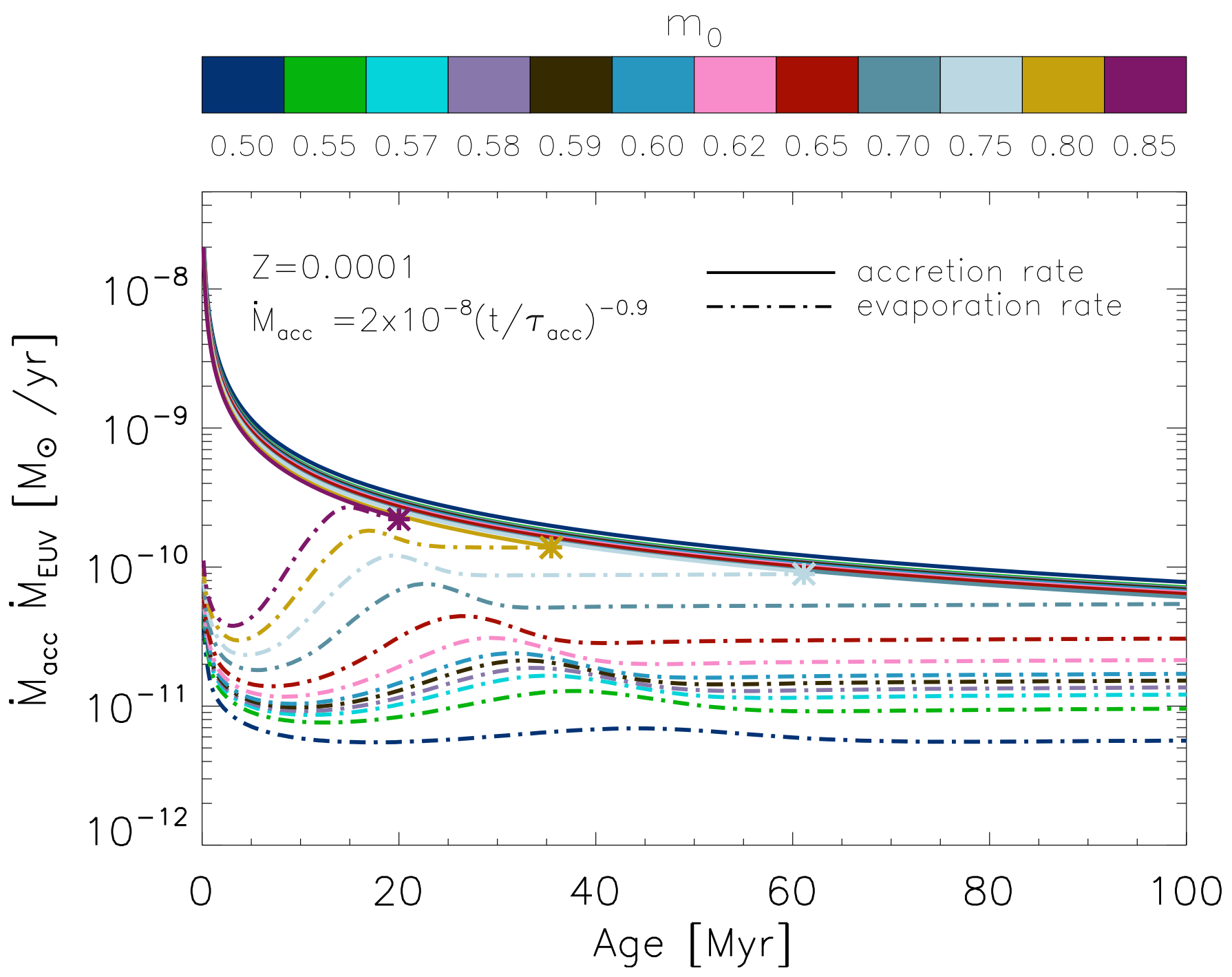

Figure 7.5: Evolution of the residual accretion rate during the first 100 Myr along the PMS evolution (solid lines) Different colors refer to different initial masses ($m_0$) as shown in the top color-bar. The residual accretion begins with the rate $2 \times 10^{-8}$ $M_{\odot}$/yr and drops following the Eq. (7.3). The EUV photo-evaporation (dash-dotted line) rates, calculated with Eq. (7.4) are also plotted. From bottom to top are stars with initial mass $0.50 M_{\odot}$ to $0.85$ $M_{\odot}$ as labeled in the color-bar. The asterisks mark the times when the accretion is terminated by the evaporation.

We would like to emphasize that the quenching of accretion is expected to be much more complex than our model description. For instance, we notice that only a fraction of the total EUV photons from the central star could reach the residual disk, if the latter maintains a small geometrical cross section. On the other hand, stellar activity in young star (e.g., accretion shock, magnetic field driven chromosphere activities, etc.), which is not included in our model, could also be a source of EUV photons and contribute to evaporation. This effect could balance the geometrical loss of EUV photons though the relative contribution between the two sources is not clear. The end of the late accretion phase could also depend on the mass of the residual disk before the EUV evaporation mechanism becomes effective. For metal-rich star-forming regions, disk masses between $0.01-0.2\ M_{\odot}$ during the burst accretion phase have been estimated by Hartmann et al. (1998), while Bodenheimer (2011) estimate disk masses from $0.5\ M_{\odot}$ down to $0.0001\ M_{\odot}$ in solar mass YSOs of the Taurus and Ophiuchus star-forming regions. The residual disk masses we are considering for the present work are even more uncertain than those of the main disk. Here we adopt the working hypothesis that the disk continues to provide material to the central star until EUV evaporation stops it.

### 7.2.4 A combined PMS model

A model that combines all the three effects previously discussed is shown in Fig. 7.3. The left panel shows the PMS evolutionary track of a star with mass $m_0 = 0.75 M_{\odot}$ in the HR diagram. The dotted line shows the PMS evolution at constant mass ups the PMS evolution at constant mass up to deuterium exhaustion. The latter point represents the end of the stellar birth-line, when the phase of large accretion terminates and the star begins the contraction at almost constant mass downward in the HR diagram. The solid line begins when the deuterium abundance reduces to 1/10 of its initial value. At this time we apply an accretion with a rate starting at $2\times10^{-8}\ M_{\odot}/yr$ (dark blue line in the right panel). This value is too low to prevent a rapid contraction and the central temperature $T_{cen}$ (brown dashed line) rises until surface Li depletion begins (red line).

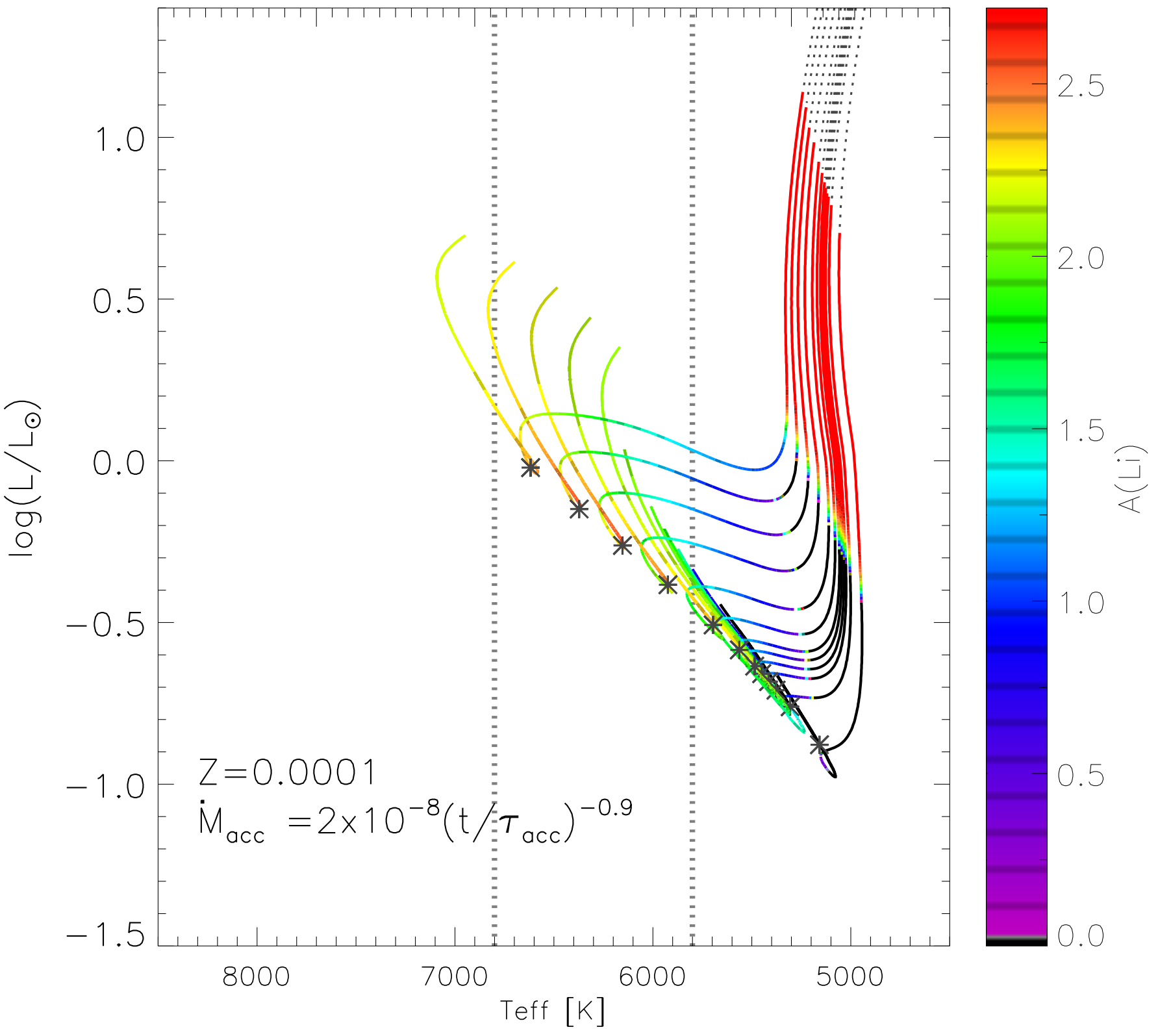

Figure 7.6: H-R diagram from PMS with overshoot, accretion, and photo-evaporation, to the end of the main sequence. The tracks are for stars with initial mass 0.85, 0.80, 0.75, 0.70, 0.65, 0.62, 0.60, 0.59, 0.58, 0.57, 0.55, and 0.50 $M_{\odot}$ from top to bottom. Black dashed lines at the upper-right side are Hayashi line with constant initial masses, which is not the real case for the late accretion. Solid lines are the stellar evolutionary tracks after deuterium burning with an initial lithium value A(Li)=2.72. Tracks are color-coded according to their surface lithium abundance, as indicated in the right vertical bar, starting from A(Li)=2.72 (in red) to A(Li) = 0 (violet). The black color corresponds to A(Li)< 0. The area between the two vertical dotted lines is the effective temperature range of the Spite plateau.

At this stage the star is fully convective and the surface Li depletion proceeds faster than the restoring effect of the accretion because of the high central temperature. After about five Myr, the core becomes radiative and the temperature at the base of the convective region $T_{bcz}$ (brown solid line) begins to decrease followed by a decline of the Li burning rate. Then the photospheric Li abundance begins to increase again, especially when $T_{bcz}$ falls below $4 \times 10^6$ K. This is because on one side Li nuclear burning quenches off and, on the other, the dilution of the infalling Li becomes weaker as the mass contained in the the convective envelope becomes smaller. The mass accreted in this phase is relatively little and only slightly affects the total mass of the star with negligible effects on its structure. In practice, the star evolves at constant mass.

The surface Li abundance steadily increases towards its primordial value, and will reach it if accretion does not terminate. This is due to the shrinking of the surface convective zone which becomes so small that even a residual accretion rate less than $10^{-10}$ $M_\odot$/yr is enough to engulf the thin surface convective layers. The interplay between the size of the convective regions and the evolution of the accretion rates are critical for reproducing a given pattern in the surface Li abundance as a function of the initial mass on the main sequence. At an age of about 60 Myr (see Table 7.1), the residual accretion is terminated by the EUV photo-evaporation.

The H-R diagram of the full sample of stars considered in this paper is shown Fig. 7.6. The area between the two vertical dotted lines illustrates the temperature range of the Spite plateau. The high end of this range is also the temperature threshold of the observable lithium. For stars with higher $T_{\rm eff}$ (the more massive ones), lithium is almost fully ionized in the stellar photosphere. Likewise in Fig. 7.3, the upper Hayashi lines before the completion of deuterium burning are drawn with dotted lines whilst the solid lines correspond to the stellar evolutionary tracks after deuterium burning. The surface abundance of Li is color-coded, starting from an initial abundance A(Li)=2.72. After the end of the accretion (marked with the asterisks), stars evolve with constant masses along main sequence and Li abundance no longer increases.

## 7.3 Main sequence Li evolution

During the main sequence phase Li is depleted by burning at the base of the convective zone because, even a low nuclear reaction rate at $T_{bcz} \sim 2 \times 10^6$ K, could cause significant depletion in a timescale of several Gys. However, for masses larger than $m_0 = 0.60 M_\odot$, Li burning is insignificant.

Another effect that could modify the photospheric Li abundance in low mass POP II stars is microscopic diffusion. This long-term stellar process cannot be observed directly in the star, and its efficiency needs to be calibrated. Microscopic diffusion shortens the main sequence lifetime and leads to a depletion of the surface elements. The solar model requires the inclusion of microscopic diffusion, otherwise even the age of the Sun could not be correctly produced. We calculated evolutionary models up to the end of the main sequence including pressure diffusion, temperature diffusion, and concentration diffusion (Thoul, Bahcall & Loeb, 1994). It has already been shown that microscopic diffusion without any correction for radiative levitation and surface convective turbulence cannot reproduce the observed photospheric abundances (Richard et al., 2002). In the case of Li, the surface convective zone of the stars at the plateau high temperature end (the relatively more massive stars) is so thin that, if microscopic diffusion is at work and not balanced by the radiative levitation, a too strong depletion is produced. For example, according to the model of Salaris & Weiss (2001) we can see that even at $T_{\rm eff} \sim 6200$ K the predicted Li depletion is too strong compared to the observed plateau.

Table 7.1: Relevant parameters for stars with different initial masses. $\tau_{acc}$ is the age at the beginning of the late accretion phase; $m_*$ is stellar mass in main sequence; $t_{end}$ is the age at the end of residual accretion; $\Delta n(Li)_{PMS}$ and $\Delta n(Li)_{MS}$ denote the decrement in the Li number density during pre-main sequence and main sequence, respectively. Note that the PMS Li depletion is stronger than the MS one.

| $m_0$ ($M_\odot$) | 0.57 | 0.58 | 0.59 | 0.60 | 0.62 | 0.65 | 0.70 | 0.75 | 0.80 |
|---|---|---|---|---|---|---|---|---|---|
| $\tau_{acc}$ ($10^5 yr$) | 1.876 | 1.873 | 1.834 | 1.787 | 1.731 | 1.689 | 1.595 | 1.560 | 1.433 |
| $m_*$ ($M_\odot$) | 0.616 | 0.625 | 0.633 | 0.641 | 0.658 | 0.685 | 0.729 | 0.775 | 0.821 |
| $t_{end}$ ($Myr$) | 585.55 | 518.96 | 460.83 | 399.76 | 309.84 | 211.46 | 110.98 | 61.14 | 35.46 |
| $\Delta n(Li)_{PMS}$ ($10^{-10}$) | 3.16 | 2.81 | 2.60 | 2.50 | 2.33 | 2.11 | 1.75 | 1.17 | 1.21 |
| $\Delta n(Li)_{MS}$ ($10^{-10}$) | 0.70 | 0.78 | 0.65 | 0.49 | 0.41 | 0.50 | 0.86 | 1.19 | 1.11 |

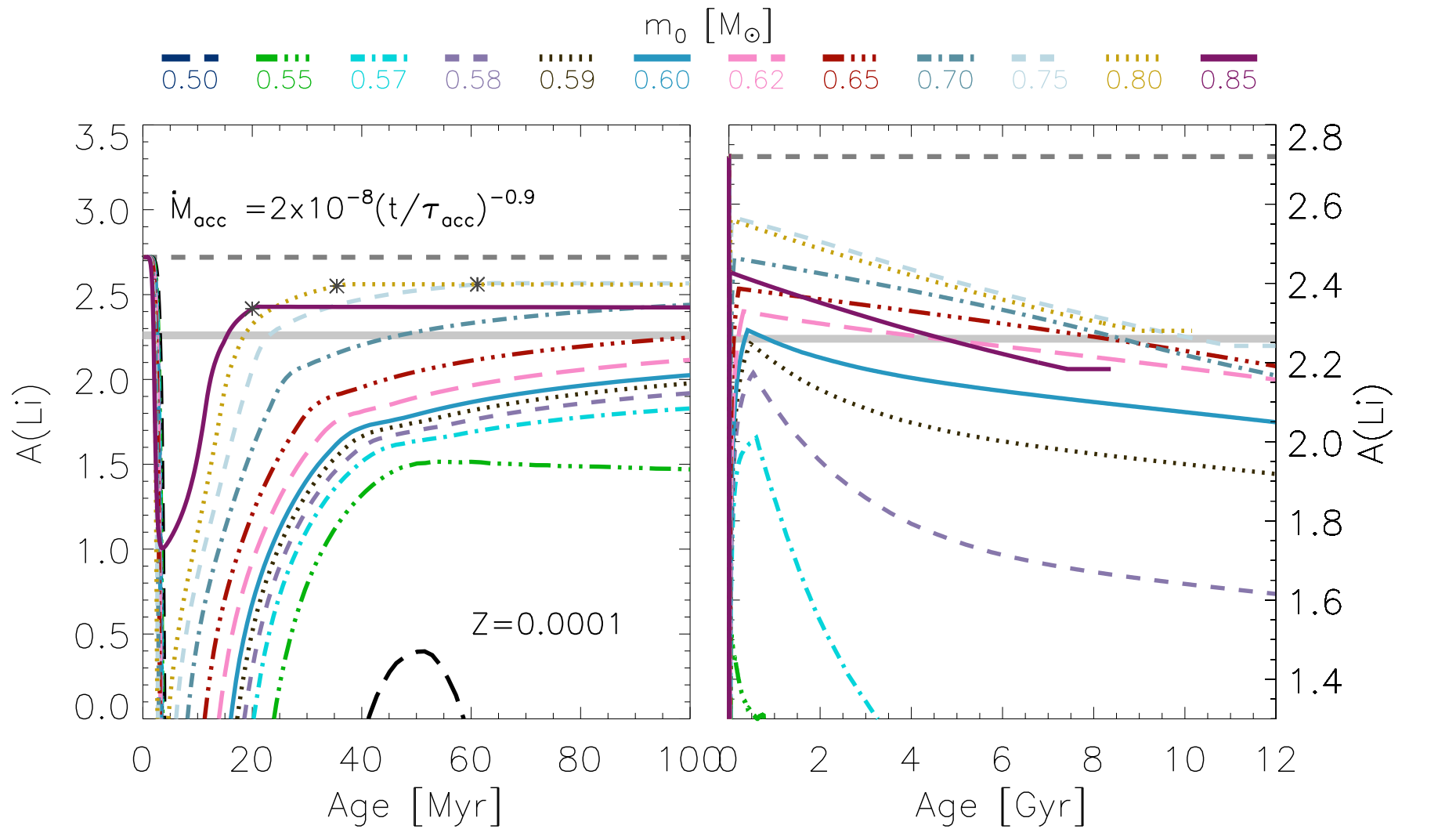

Figure 7.7: A(Li) as a function of the stellar age for stars with different initial masses. Different colors refer to different initial masses ($m_0$). The horizontal grey dashed line is the SBBN prediction (A(Li)=2.72), while the horizontal grey solid one indicates A(Li) for the Spite plateau. Left panel: Li abundance during the PMS phase during the first 100 Myr. The initial mass ($m_0$) of the models decreases from 0.85 $M_{\odot}$ to 0.50 $M_{\odot}$, from top to bottom in the rapid rising branch on the left of the diagram, as indicated in the legend. The asterisks mark the end of the accretion. Right panel: Li evolution up to an age of 12 Gyr, zooming in around the region around the Spite plateau. Color codes and the line styles are the same as in the left panel. Note that stars with initial masses greater than 0.80 $M_{\odot}$, have a main sequence life shorter than 10 Gyr, and they may have already evolved.

This does not happen with the standard PARSEC code because gravitational settling is always inhibited in the outermost region of the envelope, $\Delta M = 0.5\%$ of the stellar mass (Bressan et al., 2013). This choice of $\Delta M$ is made according to the suggestion by Chaboyer et al. (2001). They noticed that, while observations indicate that the relatively hot low mass stars at the turnoff of the globular cluster NGC 6397 show the same surface [Fe/H] abundance of the evolved RGB stars (in a later work Korn et al. (2007) present a difference Δ[Fe/H]~0.1 dex for the same cluster), models with gravitational settling that well reproduce the solar data predict that they should show a surface [Fe/H] abundance at least 0.28 dex lower. They thus conclude that gravitational settling should be inhibited in the outermost layers of such stars. In order to reconcile the predicted with the observed [Fe/H] abundance, they suggest the size of this layer should be $\Delta\ M \sim\ 0.5\% - 1\%\ M_{\odot}$ for a star of M$\leq 1\ M_{\odot}$. We checked that changing the standard PARSEC parameter in the range from $\Delta\ M = 0.1\%$ to $\Delta\ M = 1\%$ does not appreciably affect our results.

## 7.4 Results

We analyse the importance of overshoot and late accretion during the pre-main sequence phase of stellar evolution by computing a set of evolutionary tracks of initial stellar masses 0.85, 0.80, 0.75, 0.70, 0.65, 0.62, 0.60, 0.59, 0.58, 0.57, 0.55, and 0.50 $M_{\odot}$, assuming an efficient overshoot, as described in Sect. 7.2.1, and exploring different values for the parameters that describe the form of the residual accretion rate, $\dot{M}_0$ and $\eta$. We discuss below the results obtained assuming $\dot{M}_0 = 2 \times 10^{-8} M_{\odot}/\text{yr}$ and $\eta = 0.9$ in equation 7.3, which produces accretion rates compatible with the observed values. The models are calculated from the PMS till the end of the main sequence.

We will first focus on the evolution of Li abundance during the PMS phase. The surface Li abundance evolution of our selected models during the first 100 Myr of PMS phase is plotted in the left panel of Fig. 7.7. With our choice of the parameters, the initial Li evolution in these stars is regulated by overshooting. The convective overshooting is so efficient that the photospheric Li is significantly depleted after the first few Myr. The depletion caused by overshooting gets stronger at the lower masses. At later times the core becomes radiative and the convective envelope starts receding.

At subsequent stages Li evolution is governed by the competition between late accretion, dilution and nuclear burning at the bottom of the convective envelope. As soon as efficient Li burning ceases, accretion begins to restore the surface Li abundance. In the more massive stars the convective zone shrinks more rapidly and Li restoring is faster. Conversely, in lower mass stars where Li depletion is more efficient because of the deeper convective zones, the Li restoring time is longer.

Without EUV evaporation, more massive stars rapidly recover the initial surface Li abundance while in lower mass stars this process is slower because nuclear reactions will continue to burn the accreted Li and the convective dilution is larger. The effects of EUV evaporation strongly modify this picture and is regulated by the stellar mass. More massive stars evolve more rapidly toward the ZAMS and their EUV luminosity is able to stop accretion at earlier times, even before the initial abundance is completely restored (see Fig. 7.7). On the other hand, in lower mass stars where Li depletion is more pronounced, the EUV evaporation rates are lower and accretion, though decreasing, is still able to drive the surface Li abundance toward the initial value, until it is inhibited by evaporation. At even lower masses (e.g., $m_0 = 0.50M_\odot$), the recovery of the initial Li abundance is delayed or even inhibited by the deep convective zone which is still large when the model reaches the ZAMS. In this case Li restoring is prevented both by an efficient dilution, and possibly by the nuclear burning at the base of such deep convective envelopes.

In summary, our analysis indicates that Li evolution during the pre-main sequence phase may be regulated by the competition between the destroying effect of overshooting at early times, and the restoring effect of residual accretion, which is in turn modulated by EUV photo-evaporation quenching. This latter effect controls the end result, depending on the mass. At the higher stellar masses Li restoring is faster, but accretion quenching takes place earlier. At lower stellar masses the time-scale for Li restoring is longer, but the accretion lasts longer because the EUV photo-evaporation has a lower efficiency. The combination of these effects in stars with different initial masses tends to level out the Li abundance just after

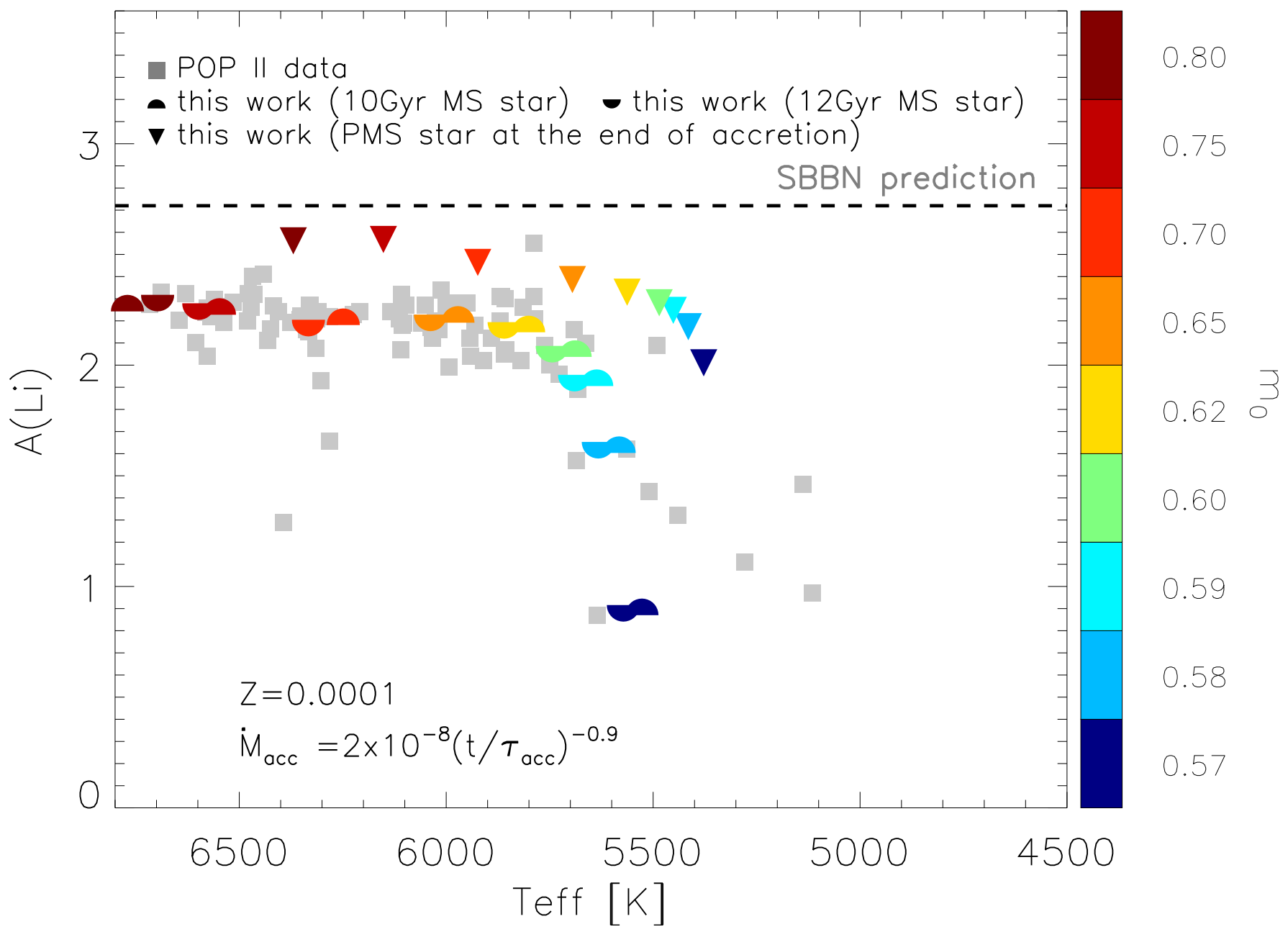

Figure 7.8: Our results in comparison with the lithium abundance measurements in POP II stars. The grey filled squares are POP II data from Molaro et al. (2012). Our predictions are shown for stars at the end of the late accretion phase (filled triangles), and on the main sequence at 10 Gyr (filled upper circle) and 12 Gyr (filled lower circle). Symbols are color-coded according to the initial stellar mass, from the left to the right are 0.80 $M_{\odot}$ to 0.57 $M_{\odot}$ as labeled. The black dashed line marks the primordial Li abundance according to the SBBN.

the PMS at a value that is already below the initial one.

Galactic halo POP II stars we observe today are about 10-12 Gyr old (Jofré & Weiss, 2011). In order to reproduce the observed Li abundance of these stars we have evolved our models up to the end of central hydrogen burning (till the turn-off phase). In this phase we also account for microscopic diffusion which is known to be a long term effect that can modify the photospheric element abundances. The evolution of the photospheric Li abundance during the main sequence is depicted in the right panel of Fig. 7.7. The predicted photospheric Li abundances are compared with those of the Galactic halo POP II stars in Fig. 7.8. We plot the abundances at the end of the accretion phase (filled triangles) and those at older ages, 10 Gyr (filled upper circles) and 12 Gyr (filled lower circles). Stars with initial mass $m_0 \leq 0.85 M_{\odot}$ are excluded because their effective temperatures on the MS are warmer than the observed ones (see Fig. 7.6), and also because their main sequence lifetimes are shorter than the relevant age range. For our standard choice of parameters, stars with initial mass from $m_0 = 0.62 M_\odot$ to $0.80 M_\odot$, nicely populate the Spite plateau (A(Li) $\approx$ 2.26).

Models with lower mass, $m_0 = 0.57 M_\odot$ to $0.60 M_\odot$, fall on the observed declining branch towards lower temperatures. We confirm that the first part of this branch is populated by low-mass main sequence stars. Indeed this is the signature of the strong Li depletion during the PMS phase (filled triangles), followed by further depletion during the main sequence evolution (filled upper and lower circles).

Both PMS Li depletion and MS diffusion contribute to make the total A(Li) decrease, with the former process playing the main role. Table 7.1 lists the Li number density decrease during PMS and MS, the former is much more than, or at least the same as, the latter, especially for the lower masses. The first part of Li declining branch is caused mainly by the PMS depletion.

The Spite plateau is populated by metal-poor main sequence stars with different metallicities. With the same overshooting and accretion parameters we can indeed reproduce the plateau, including the initial Li declining branch, over a wide range of metallicities, (from Z=0.00001 to Z=0.0005, that is from [M/H]=-3.2 to [M/H]=-1.5) as shown in Fig. 7.9.

Li abundance as a function of the stellar luminosity is also of great interest since it illustrates the evolution of the element. Lind et al. (2009) derive the lithium evolution from turn-off to giant branch in globular cluster NGC 6397, in Fig. 7.10 we compare their A(Li) vs. Log(L) result with our model. Model with Z=0.0001 (similar metallicity to NGC 6397) and mass range 0.57–0.75 $M_\odot$ are displayed between 12.6 Gyr and 13.8 Gyr. Up to the age of 13.8 Gyr, which is older than the Universe, stars with mass 0.65–0.73 $M_\odot$ are always with Li abundance of Spite Plateau at this metallicity. This indicates that if one observe a metal-poor star with this mass range, one could infer its Li abundance at Spite Plateau value. For M=0.75 $M_\odot$ star, its lithium abundance drops as it becomes brighter because the surface convective zone deepens before the first dredge-up as we have already introduced in Chap. 3.2.2. This trend is very similar to the NGC 6397 data at $log(L) \gtrsim 0.7$ show in the figure, thus we expect that stars in this luminosity are with the similar mass. In Fig. 7.11 we show another comparison on Li evolution for a more metallic GC M4 (NGC 6121). With the same pre-main sequence accretion rate as we applied to metallicity $0.00001 \leq Z \leq 0.0005$, Z=0.002 models show very good agreement with Li measurements (Mucciarelli et al., 2011) in M4 from the main sequence turn-off to RGB bump as a function of gravity log(g), even though at this metallicity the intrinsic Li is with an enriched value A(Li)=2.80 dex insterad of the BBN primordial one (A(Li)=2.72 dex). At the age of this cluster ($\sim$ 12 Gyr, Mucciarelli et al., 2011), stars with mass 0.80 $M_\odot$ contribute to the sub-giant branch and the turn-off.

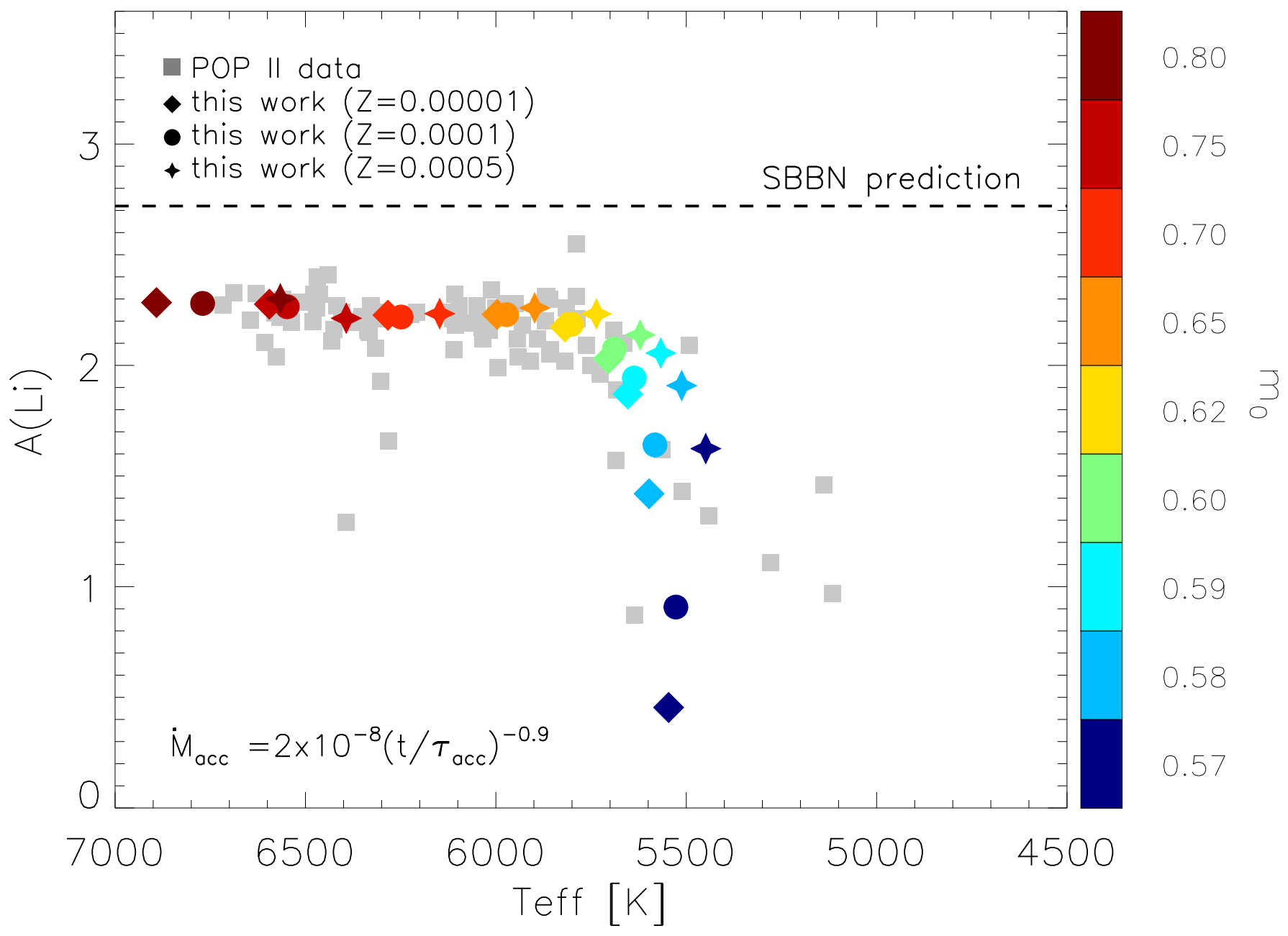

Figure 7.9: By applying the same parameter of envelope overshooting and accretion, we could reproduce the Spite plateau and the first Li decline branch for a wide range of metallicities (from Z=0.00001 to Z= 0.0005). The compared POP II data are the same as in figure 7.8. The model results are all main sequence stars at age 10 Gyr, with initial mass 0.80 $M_{\odot}$ to 0.57 $M_{\odot}$ from the left to the right for each metallicity as shown in the color-bar label.

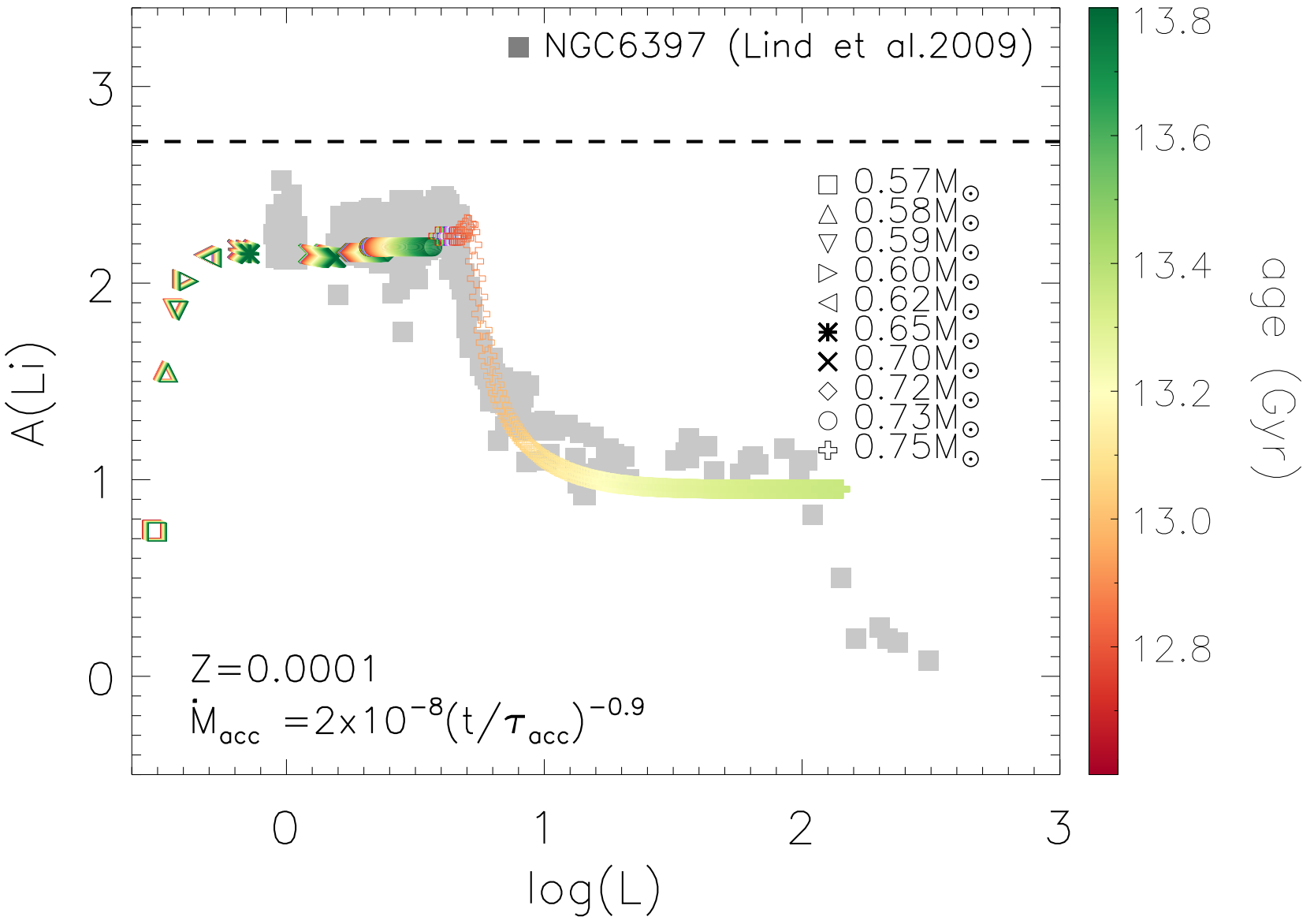

Figure 7.10: Lithium abundance as a function of luminosity log(L) for stars with mass 0.57–0.75 $M_\odot$ between 12.6 Gyr and 13.8 Gyr. The age of the stars is color-coded as shown in the color bar on the right side. Stars with different mass are mark with different symbols as shown in the legend. For comparison, data of NGC 6397 (Lind et al., 2009) is also plotted with grey filled squares.

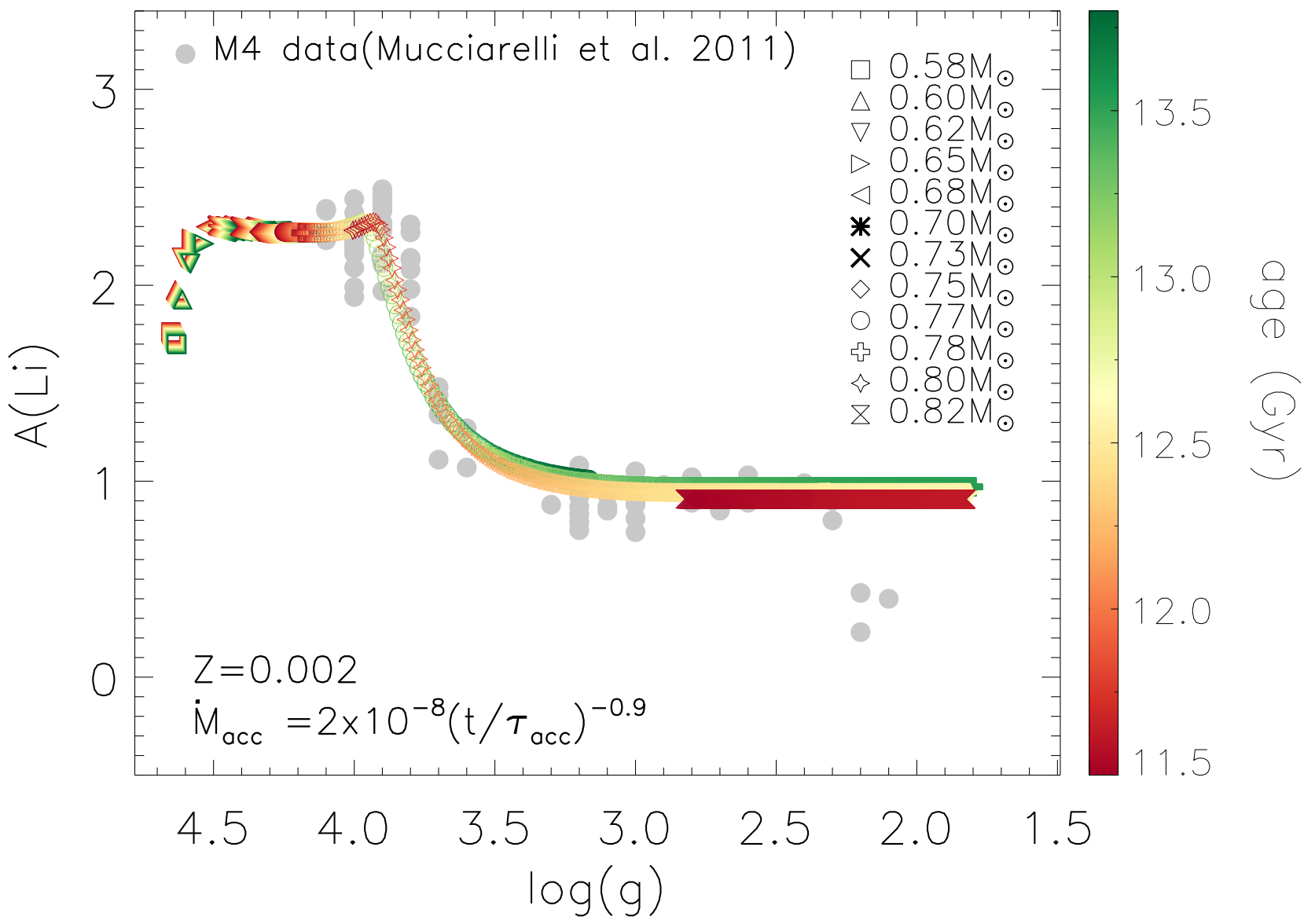

Figure 7.11: Lithium abundance as a function of surface gravity log(g) for stars with mass 0.58–0.82 $M_\odot$ between 11.5 Gyr and 13.8 Gyr. The age of the stars is color-coded as shown in the color bar on the right side. Stars with different mass are mark with different symbols as shown in the legend. For comparison, data of M4 (NGC 6121) (Mucciarelli et al., 2011) is also plotted with grey filled dots.

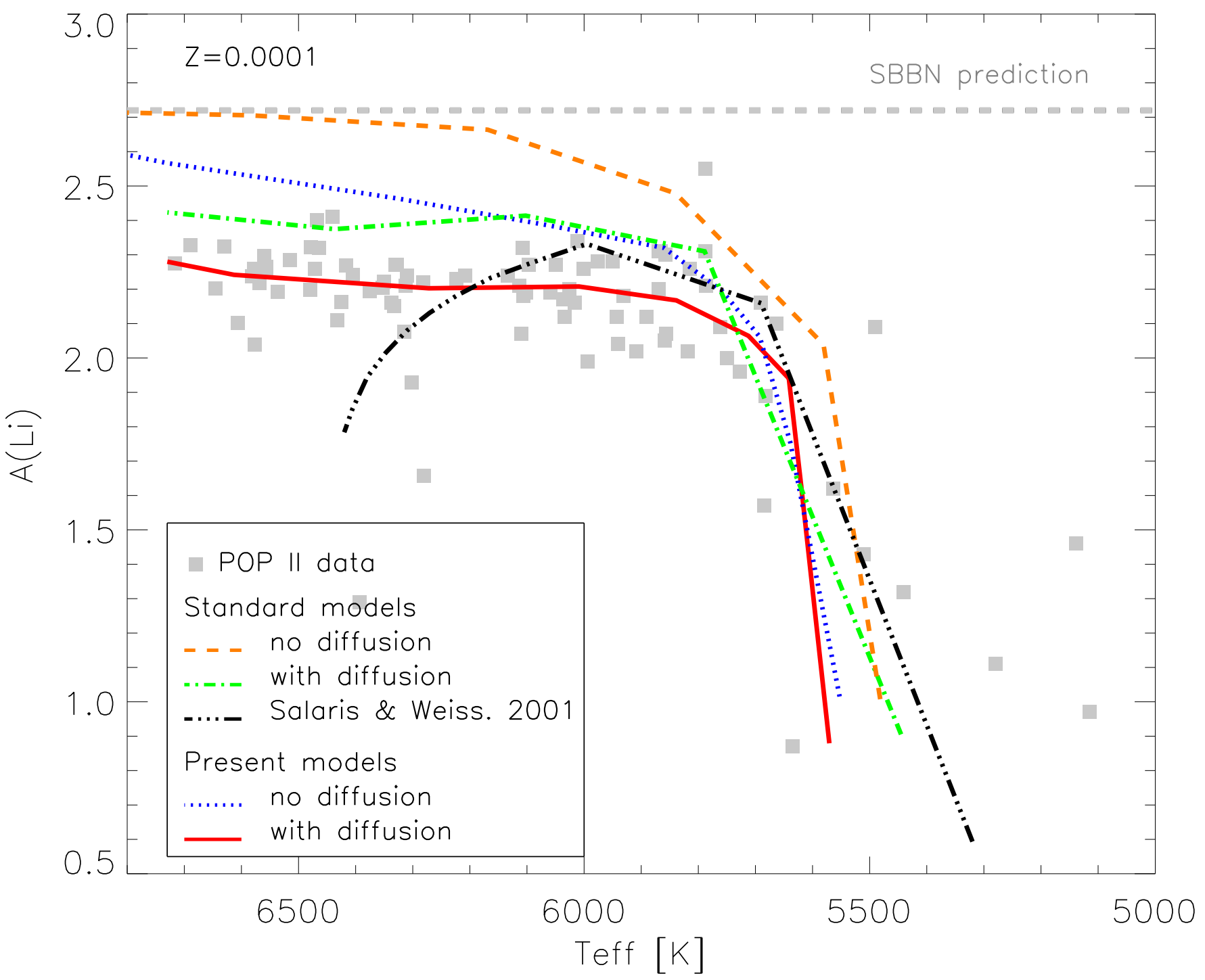

Figure 7.12: Comparison of the standard models and the present models for metallicity Z=0.0001 together with the POP II data and Salaris & Weiss (2001) pure diffusion model. Different sets of models are classified in the legend. The black dash dot dot line is extracted from Fig. 1 of Salaris & Weiss (2001) with [Fe/H]=-2.6 and age 12 Gyr.

## 7.5 Discussion and Conclusion

In the previous chapters we have shown how the effects of efficient envelope overshooting, residual accretion during the pre-main sequence phase, and microscopic diffusion during the main sequence may modify the photospheric Li abundance in low-metallicity stars. We have also considered the EUV photo-evaporation process which, by terminating the accretion phase, could introduce a sort of self-regulating process.

In order to give a more exhaustive picture of the importance of the various processes we computed more sets of models under different assumptions concerning the above physical mechanisms. We synthesize the results of these additional models in Fig. 7.12 where we compare the predicted Li abundances with the observed ones (see Fig. 7.8).

- **Standard models**

  We begin with two sets of models where none of the PMS effects have been considered. The first set (orange dashed line) is the standard model without microscopic diffusion while, in the second set of model (green dot-dashed line) microscopic diffusion is introduced as it is in the standard `PARSEC` model.

  The model without diffusion is clearly at variance with the observed data. On the other hand when diffusion is included the Li abundance reaches A(Li)=2.45 dex, producing a plateau somewhat higher than the observed one, by a factor of 1.6. The pure diffusion model of Salaris & Weiss (2001) is also plotted in this figure (extracted from figure 1 of Salaris & Weiss (2001) with [Fe/H]=-2.6 and age 12 Gyr). Contrary to them, our diffusive model does not produce the strong downturn at high effective temperatures. This is due to the inhibition of diffusion in the outer layers of the star which, as discussed in Sect. 7.3, is required to prevent a too strong sedimentation of helium and heavy elements at the stellar surface (Richard et al., 2002; Bressan et al., 2012).

  The plateau could eventually be reproduced by increasing the efficiency of diffusion strongly, we note however that, with the adopted diffusion parameters the solar model is very well reproduced (Bressan et al., 2012).

- **Present models**

  The present models, that include PMS effects and diffusion (red solid lines in Fig. 7.12), provide a good fit to the data. They reproduce both the observed plateau and the declining branch at low temperatures. As a test, we recomputed the same models but switching off microscopic diffusion during the main sequence phase (blue dotted line). These new models are identical in the PMS evolution and incorporate the Li nuclear burning during main sequence. The latter is mainly responsible for the behavior of Li depletion at low effective temperatures. They do not reach the plateau value, terminating somewhat at higher Li abundances, suggesting that some additional depletion is required.

In our model we do not consider stellar rotation. The additional effects of rotation on Li depletion have been discussed in many works, but mainly in the context of solar-type stars. During the PMS, rotation may affect the internal structure of the stars both because of the presence of the centrifugal force and the effects of rotational mixing at the base of the convective envelope. Eggenberger et al. (2012) conclude that by including the full treatment of rotation in PMS, 0.1 dex more Li depletion is obtained. During the evolution on the main sequence, rotational mixing tends to oppose to gravitational settling, and this effect should decrease the surface elements depletion with respect to non-rotating models. However in the case of Li, rotational mixing may be so efficient that Li is dragged to encounter temperatures which could efficiently destroy it by nuclear burning. This is the case of the slow rotator model of M=1.0 $M_{\odot}$ of Eggenberger, Maeder & Meynet (2010). Garcia Lopez, Rebolo & Martin (1994) find that because of the rotation-caused structural difference, stars with faster rotation in open clusters Pleiades and $\alpha$ Persei have higher Li abundance than the slower rotators. This results is confirmed by observations on the Pleiades (Jones et al., 1997; Gondoin, 2014), on M 34 (Gondoin, 2014), and NGC 2264 (Bouvier et al., 2016). On the contrary, Balachandran, Mallik & Lambert (2011) obtain lithium abundance from high resolution spectra and find that most massive rapidly rotating stars (at Teff¿6400 K) have a lower Li abundance than the slow rotators. In general it is found that rotation leads to a large spread in the main sequence Li abundance of solar-type stars at varying mass (Martin & Claret, 1996; Mendes, D'Antona & Mazzitelli, 1999), which could be difficult to reconcile with that observed in the Spite plateau of metal-poor stars.

It is important to stress that, as long as the stars are accreting at significant rates, deuterium burning keep their central temperature around $10^6$K. During this phase the stars evolve along the birth line increasing their masses (e.g. reaching 1 $M_{\odot}$ in 0.1 Myr for an accretion rate of $10^{-5}$ $M_{\odot}$ /yr as indicated in table 7.1), while preserving the initial Li abundance (Stahler, 1983). Once the accretion rates fall below several $10^{-8}$ $M_{\odot}$ /yr the stars abandon the stellar birth-line and evolve at almost constant mass. It is from this point that we consider the additional effects of envelope overshooting, which reduces the Li abundance, and of accretion, which tends to restore the original abundance. We have also shown that EUV evaporation of material falling into the stars can act as a natural self-regulating mechanism.

In the relatively more massive stars, where the original Li abundance could be more quickly restored, EUV evaporation is more efficient and quenches out accretion earlier than in lower mass stars, which thus have more time to accrete the pristine gas. These combined effects tend to level out the Li abundance in stars of different masses. At even lower masses, Li restoring is inhibited by nuclear burning in the deep convective envelopes. After the PMS, Li is slowly depleted by microscopic diffusion and nuclear burning which act during main sequence phase. For stars with $m_0 \geq 0.62 M_{\odot}$ diffusion drives the main Li depletion whilst, for $m_0 \leq 0.60 M_{\odot}$ stars, Li is significant burned at the base of the convective envelope.

In this scenario, the observed Spite plateau and its falling branch at low temperatures are well reproduced by our models, within a reasonable parametrization of the late accretion mechanism. Very efficient overshoot is only applied during the early PMS to favour strong Li depletion at the beginning of the evolution. We show in Fig. 7.13 that even for a maximum envelope overshoot $\Lambda_e = 0.7\ H_p$, which is the value calibrated from the RGB bump (Alongi et al., 1991), the plateau and its falling branch can be finally well reproduced for the same accretion rate. This indicates that a wide range of envelope overshoot could be applied to the PMS phase. Even for $\Lambda_e = 0.7\ H_p$ the PMS Li depletion is very strong; without accretion a very low Li abundance would be observed, e.g. for $m_0 = 0.70 M_\odot$ A(Li) drops to 1 dex before Li restoring. It would be important to directly observe whether the very early Li depletion phase exists or not, because it could confirm that efficient overshoot is at work. Indeed, at higher metallicities there is evidence of a significant drop of the Li abundance during the PMS phase (Somers & Pinsonneault, 2014). On the contrary, the lack of nearby metal-poor star forming regions prevents any observational test at low metallicity. This opportunity will be perhaps offered by the next generation telescopes (E-ELT, TMT, and GMT) which could observe the details of low mass metal-poor PMS stars in star forming regions like those of the Sagittarius dwarf irregular galaxy (metallicity at $Z \sim 0.0004$ (Momany et al., 2002)).

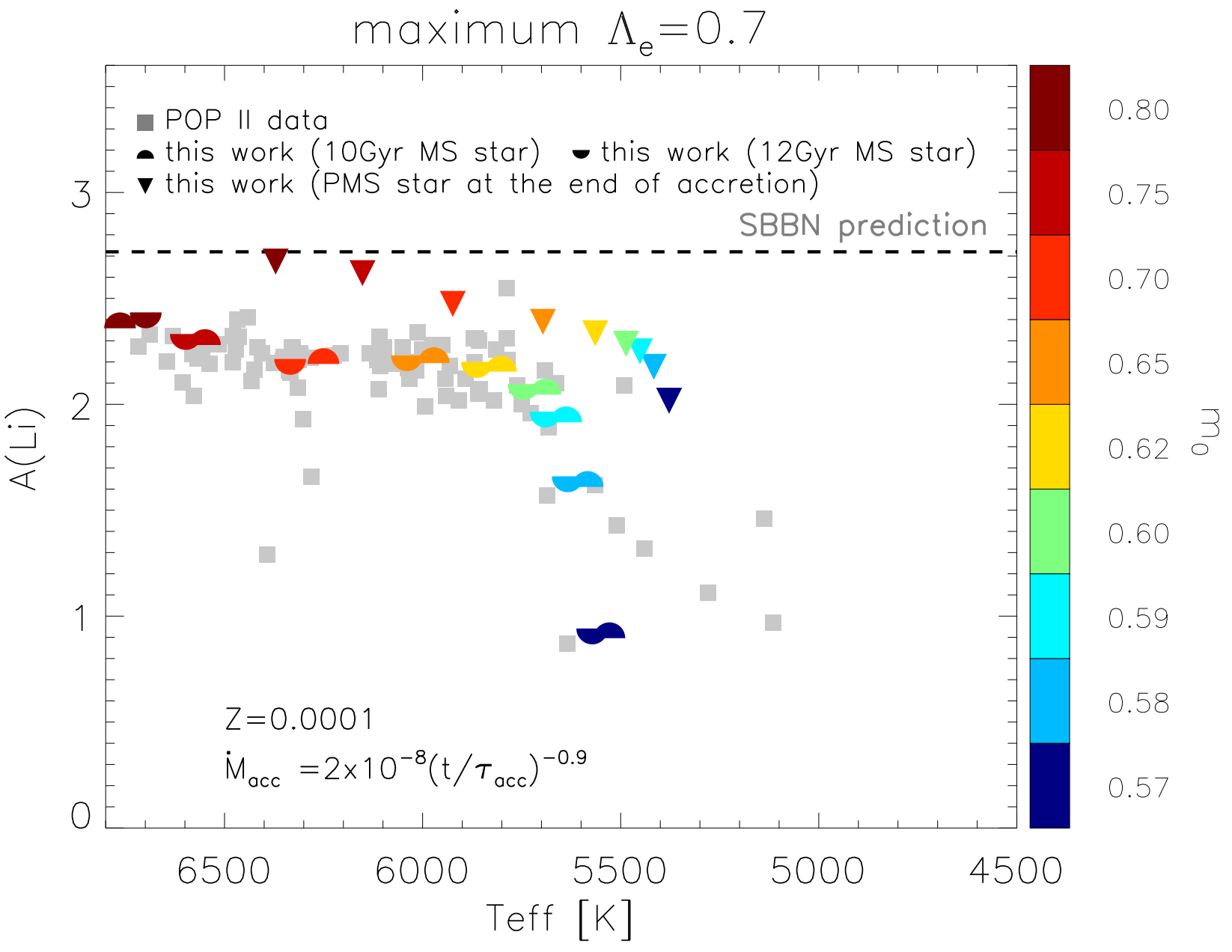

Figure 7.13: The same as in Fig. 7.8, but the models are calculated with maximum EOV $\Lambda_e$=0.7.

For stars with initial mass $m_0 = 0.62\ M_\odot - 0.80\ M_\odot$, ages in the range 10-12 Gyr, and a wide range of metallicities (Z=0.00001 to Z=0.0005), our final model is able to predict a present day abundance consistent with the Spite plateau, although starting from an initial value of A(Li)=2.72 as inferred from the baryon-to-photon ratio suggested by the CMB and by the deuterium measurements. Thus it offers a likely mechanism to solve the long standing *lithium problem*. It also reproduces the observed lithium drop at the low-temperatures. The proposed solution relies on stellar physics and evolution, assuming the validity of the current SBBN theory. Though this model is quite schematic and rough, it clearly suggests that the Spite plateau could be the net result of mechanisms more complex than those considered so far. In particular, a key role may be attributed to the interaction between the stars and their environments.

If the lithium abundance we presently observe in POP II stars has been restored in the stellar atmospheres by a residual tail of accretion after a phase of strong depletion, then other puzzling observations could be explained, namely:

- Li at very low metallicities.

  Sbordone et al. (2010) found in their observations that for the extreme metal-poor stars, Li abundance drops when [Fe/H]< −3. Caffau et al. (2011a) and Frebel et al. (2005) observed dwarfs with [Fe/H] at the lowest levels without any detectable Li. Hansen et al. (2014) detected Li in an un-evolved star with [Fe/H]=-4.8 at a level of A(Li) = 1.77 and anther recently discovered extremely metal poor dwarf SDSS J1742+2531 with [Fe/H] = -4.8 shows A(Li) < 1.8 (Bonifacio et al., 2015). The low Li abun-dance in these stars could be explained by a failed or weaker late accretion. For the most metal-poor stars, late accretion during the PMS phase might be inhibited and Li could not be restored. As we have already shown in Fig. 7.2 (right panel), if no accretion takes place after the initial Li is depleted by the convective overshoot, Li in these stars would correspond to a very low, or even undetectable, abundance. Thus the model presented here provides a possible key to account for this phenomenon, and we will discuss it in detail in a following paper.

- Spectroscopic binaries and outliers

  Our model might also provide an explanation to the Li discrepancy observed in metal-poor binary stars. The POP II spectroscopic binaries CS 22876-032 (González Hernández et al., 2008) and G 166-45 (Aoki, Ito & Tajitsu, 2012) are composed by dwarfs with effective temperatures characteristic of the Spite plateau, but with the primary stars showing slightly higher Li abundances. It could simply be an effect of the competition between the components in the late PMS accretion, with the primary star being the favorite one.

  The failure or increase of the accretion process could also provide an explanation for the few POP II $^{7}$Li depleted stars or the few POP II stars that exhibit a $^{7}$Li abundances at the SBBN prediction level, respectively. Among the latter there is BD +23 3912 with $A(Li)$=2.60 which stands out from the others in Bonifacio & Molaro (1997).

- Lithium behavior in POP I stars

  Li abundance at the solar system formation time, as obtained from meteorites, is $A$(Li)= 3.34 (Anders & Grevesse, 1989). The hot F stars of young open clusters never reach this meteoritic value (Ford et al., 2001) despite they should have an initial Li value even higher than the meteoritic one due to the Galactic Li increase with time. This Li drop cannot be explained by main sequence Li evolution because the stars are very young. A PMS Li modification could be the possible mechanism for the depletion. This will be discussed in detail in a following work dedicated to the metal-rich stars.

- Low lithium abundance in the planet host stars

  Israelian et al. (2004) and Gonzalez (2008) report that solar-type stars with massive planets are with lower lithium abundance than their siblings without detected massive planets. This result has also been confirmed in more recent works (e.g. Israelian et al., 2009; Gonzalez, 2014). To explain this phenomenon Bouvier (2008) and Eggenberger, Maeder & Meynet (2010) advance the hypothesis that giant exoplanets prefer stars with slow rotation rates on the ZAMS due to their longer disk lifetime during the PMS evolution. Then their ZAMS models with rotational mixing, which we already discussed in section 7.3, would indicate an enhanced lithium depletion, consistent with observations. Again it is worth noting that a strong PMS depletion at decreasing mass is predicted by solar metallicity models with rotation (Mendes, D'Antona & Mazzitelli, 1999), and it is not yet clear how this could be reconciled with the observations of Israelian et al. (2009) and Gonzalez (2014). In our model, this behavior could be explained if somehow the planet(s), located within the accreting disk, interfere with the late accretion process. The PMS accretion could end before being terminated by the EUV evaporation, inhibiting efficient lithium restoring. Thus these stars will show lower Li abundance than their analogues without detected planets. This effect could also explain the larger Li abundance scatter observed in metal-rich stars, which are more likely to host planets (Ida & Lin, 2004), with respect to what observed in metal poor stars.

# Chapter 8

# Pristine Li abundance for different metallicities

As already discussed in detail in the previous chapters, Li abundance in metal-poor main sequence stars shows a constant plateau and then it increases with metallicity as we approach solar metallicity (e.g. Lambert & Reddy, 2004; Chen et al., 2001). Meteorites, which carry information at the solar system formation time, show a lithium abundance $A$(Li)= 3.34 (Anders & Grevesse, 1989). This pattern reflects the $^7$Li enrichent history of the interstellar medium (ISM) (Romano et al., 2001), though its origin is not very well understood. Beside the primordial lithium produced by BBN, spallation of Galactic cosmic rays (GCRs), via the $\alpha$–$\alpha$ fusion process, could be responsible for $\sim$ 20–30% of the local Galactic abundance of $^7$Li (e.g. Reeves, 1970; Meneguzzi, Audouze & Reeves, 1971). The remaining unexplained fraction of the meteorites abundance has to originate from a stellar source, for instance, Molaro et al. (2016) report massive Be II ejecta from a Nova which results a significant $^7$Li enrichment. However the detailed mechanisms and contributions are still an open question.

Romano et al. (2001) consider five sources for $^7$Li enrichment in their Galactic chemical evolution model: the $\mu$-process in type II supernovae (SNe), AGB stars which undergo the hot bottom burning (HBB) process, low-mass red giants, novae, and GCRs. Their model well reproduces the lithium abundance in main sequence star observations. However, since they do not consider lithium evolution in PMS, the pristine value of BBN is set to A(Li)=2.2 dex as the same as the Spite plateau. Since Li is enriched from A(Li)=2.2 dex to the meteorites value instead of from A(Li)=2.72 dex (BBN value derived from CMB, Komatsu et al., 2011), they could overpredict the Li production.

In `PARSEC` the initial Li abundance has the BBN primordial A(Li)=2.72 dex for all POP II stars till [M/H]=-1. At higher metallicity it increases linearly in number density $n(Li)$ (not in A(Li)) up to the value of the meteorites A(Li)=3.34 dex (Anders & Grevesse, 1989), at solar metallicity. In Fig.8.1 we show our initial Li abundance as a function of metallicity [M/H]. Data of metal-poor (Charbonnel & Primas, 2005) and solar-metallicity dwarf stars (Chen et al., 2001) are plotted for comparison. Unlike the Galactic chemical evolution model of Romano et al. (2001) which is also displayed in the figure, our choice of the initial Li abundance ignores the detailed processes which cause the enrichment. There are three different lines of study that can help to calibrate this set of values. The most direct and difficult way is to observe Li in ISM at these metallicities. For instance, observations of interstellar $^7$Li in gas of Small Magellanic Cloud ([M/H] ~-0.5) is reported by Howk et al. (2012) with $A(Li)_{SMC} = 2.68 \pm 0.16$. One could also use the Galactic chemical evolution model and calculate the lithium contribution from every enrichment source as Romano et al. (2001) do. Finally, lithium abundance in PMS star also offers a solution to calibrate the initial value. The accretion rate and Li abundance for young stars around solar-metallicity have been recently observed in the Gaia-ESO survey (Lanzafame et al., 2015), and could be used to trace back their pristine value. As we have discussed in Chap. 7.5, this is one of our following work.

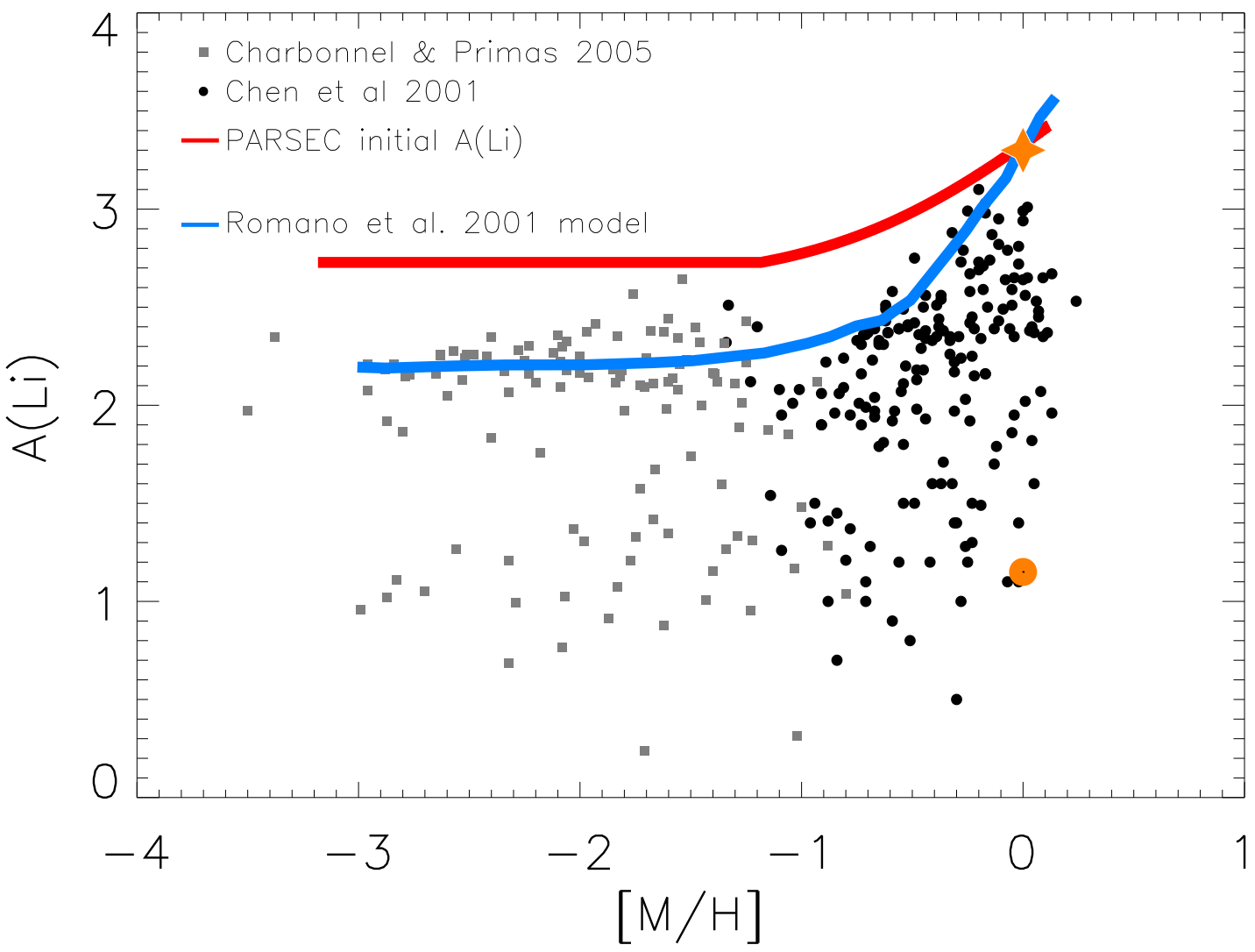

Figure 8.1: Initial Li abundance in PARSEC (red line) and Romano et al. (2001) (blue line) as a function of metallicity. Data for metal-poor (Charbonnel & Primas, 2005) and solar-metallicity dwarf stars (Chen et al., 2001) are also plotted with grey filled squares and black filled dots, respectively. Li abundance observed in meteorites is marked with filled orange star, and that of our Sun is with big orange filled dot.

# Part III:

# Summary and Outlook

In summary, during my Ph.D. study I have worked on two projects: building an $\alpha$ enhanced stellar evolutionary tracks and isochrones database for PARSEC (Fu et al., 2016), and proposing a new model to solve the cosmological Lithium problem (Fu et al., 2015). Both works are focused on the evolution of low mass stars.

Studies on globular clusters, Galactic bulge, halo, and thick disk call for stellar models with $\alpha$ enhancement because stars residing in them have $\alpha$-to-Iron number ratio larger than that of the Sun. This ratio, [$\alpha$/Fe] , not only affects the stellar features like the luminosity and effective temperature, but also echoes the formation history of the cluster/structure the stars are in. We extend the PARSEC models from solar-scaled composition to $\alpha$ enhanced mixtures, to offer a comprehensive database of evolutionary tracks to investigate these stars, to trace back their formation history, and to calibrate the IMF in different environments.

In order to ensure that the models are reliable one needs to calibrate them against a well-studied target. For solar-scaled models the target is the Sun. In this work we choose the globular cluster 47Tuc (NGC 104) to calibrate the $\alpha$-enhanced models. Its chemical composition, including the helium abundance is studied by many works. We collect detailed elemental abundances of this cluster and derive absolute metal mixtures for two populations: the first generation [Z=0.0056, Y=0.256] and the second generation [Z=0.0055, Y=0.276]. The relative number density for each metal element of these two generations is listed in Fig. 3.2 and 3.3. The $\alpha$-to-Iron ratio is [$\alpha$/Fe] =0.4071 ( [$\alpha$/Fe] ~0.4) for the first stellar generation and [$\alpha$/Fe] =0.1926 ( [$\alpha$/Fe] ~0.2) for the second stellar generation, respectively. We need to clarify that strictly speaking there are eight $\alpha$ elements: O, Ne, Mg, Si, S, Ca,Ar, and Ti. In our work in order to be consistent with the large Galactic survey APOGEE only six, out of eight, are considered in the total [$\alpha$/Fe] : O, Mg, Si, S, Ca, and Ti. However, this choice does not affect the final values of [$\alpha$/Fe] significantly. If all the eight $\alpha$ elements are considered, we obtain [$\alpha$/Fe] =0.4057 and [$\alpha$/Fe] =0.2277 for the two generations stars of 47Tuc, respectively.

We calculate evolutionary tracks and isochrones with these two $\alpha$-enhanced metal mixtures, and fit the color-magnitude diagram derived from HST/ACS data. The model envelope overshooting is then calibrated to the value $\Lambda_e$=0.5 $H_P$ in order to reproduce the RGB bump morphology in 47Tuc. After the calibration, the new $\alpha$-enhanced isochrones nicely fit the data from the low main sequence to the turn-off, giant branch, and the horizontal branch with age of 12.00±0.2 Gyr, distance modulus $(m-M)_0=13.22^{+0.02}_{-0.01}$, and reddening $E(V-I)=0.035^{-0.008}_{+0.005}$. By studying the morphology and luminosity function of the horizontal branch, we conclude that the mass loss in 47Tuc during the RGB phase is around 0.172 $M_\odot$ to 0.177 $M_\odot$ . We also find that the core helium burning lifetime is correctly predicted by the current version of PARSEC , which assumes a core overshooting of $\Lambda_C$=0.5 $H_P$ across the Schwarzschild boundary during central He burning.

The envelope overshooting calibration together with the $\alpha$-enhanced metal mixtures of 47Tuc are applied to other metallicities, till Z=0.0001 (Chap. 4). The RGB bump magnitude of the new $\alpha$-enhanced isochrones are compared with other stellar models and other globular cluster observations. We take $\Delta_{RGBB}^{MSTO}$, the magnitude difference between the main sequence turn-off and the RGB bump, as the reference quantity to be compared with the observations, in order to avoid uncertainties arising from the assu,ptions concerning distance and extinction. Our new models fit the data very well and significantly improve the prediction of RGB bump magnitude compared to previous models.

The second part of my Ph.D. work is on Lithium evolution in metal-poor stars.

Lithium is of my special interest among all elements. Fragile and scarce, sensitive and primitive, Lithium is one of the most complicated elements in stellar physics. Its abundance in stars, both on the main sequence phase and the giant branch phase, has plagued our current understanding of cosmology, stellar evolution, and metal sources of the interstellar medium. Among those problems, the cosmological Lithium problem is the most severe one.

$^7$Li, together with helium-4 ($^4$He), deuterium (D) and helium-3 ($^3$He), are the first elements created in our universe from the see of protons and neutrons. Following the Big Bang nucleosynthesis theory, their abundances in the early universe are based on a precise value of the cosmic ratio of baryons to photons. The measured "primordial" amounts of Helium and Deuterium match the predictions of the Standard Cosmological Model, but that of Lithium does not. $^7$Li observed in the old metal-poor stars shows an almost constant value which is called "Spite Plateau" but, instead of being near the value predicted by the Big Bang nucleosynthesis, it is three times lower. This discrepancy is referred to as the cosmological Lithium problem. For decades the community puzzled over where all the Lithium in stars had disappeared to. Several fields of study, including non-standard cosmology models, yet unknown aspects of particle physics, of nuclear physics, and of stellar physics, have been pursued to provide possible explanations to the Lithium deficit.

In Part II of the thesis we address this problem by introducing the interplay between the pre-main sequence star and the residual disk around it. In our model, at the early age of the star Lithium is first almost totally destroyed and later re-accumulated by the residual disk mass accretion. Specifically, $^7$Li could be significantly depleted by convective overshooting in the pre-main sequence phase, and then partially be restored in the stellar atmosphere by accretion of the residual disk. This accretion could be regulated by EUV photoevaporation. When stars evolve to the ages we observe, our model perfectly reproduce the observed Li abundance.

The Pre-main sequence is the stellar evolutionary phase right after the protostellar phase, during which the star descends along the Hayashi line in HR diagram with a rapid gravitational contraction. The surface convective zone of the star in this phase is very extended (at the beginning the star is fully convective) exposing the surface materials to hot temperatures inside the star, leading to a significant Lithium burning. The envelope overshooting we use in the pre-main sequence model varies with the total size of the convective zone and preserve the solar constraint when the star evolves to the main sequence. With this assumption, POP II stars deplete their Lithium from the Big Bang nucleosynthesis value (A(Li)=2.72 dex) in a few million years.

If this depletion were the only story during pre-main sequence, no Lithium could be observed in the metal-poor stars today. The Lithium we measure in POP II stars is restored by residual accretion from the disk around pre-main sequence stars. Though the main accretion ended before the pre-main sequence, the star continues to accrete gas with the pristine composition from the debris of the disk, at a tiny rate. This residual accretion is enough high to restore the surface Lithium toward the pristine value, till the Extremely UV photons from the star blows all disk material away. As the star evolves toward the main sequence, the surface convective zone becomes thinner and thinner, the restored Lithium can survive in the star since the temperature at the bottom of the convective zone is no longer high enough to burn it.

By considering the conventional nuclear burning and microscopic diffusion along the main sequence, we can reproduce the Spite plateau (A(Li)≈2.26) for stars with initial mass $m_0 = 0.62 - 0.80\ M_\odot$, and the Li declining branch for lower mass dwarfs, e.g, $m_0 = 0.57 - 0.60\ M_\odot$, for a wide range of metallicities (Z=0.00001 to Z=0.0005).

This environmental Li evolution model not only provides a solution to the cosmological Lithium problem, but also offers the possibility to interpret the decrease of Li abundance in extremely metal-poor stars, the Li disparities in spectroscopic binaries and the low Li abundance in planet hosting stars.

In the following work, I will complete the calculation and calibration of $\alpha$-enhanced models. Beside the $\alpha$-enhanced metal mixtures of 47Tuc, evolutionary tracks and isochrones based on $\alpha$-enhanced metal mixture derived from ATLAS9 APOGEE atmosphere model will also be provided. The full set of isochrones with chemical composition suitable for GCs and Galactic bulge/thick disk stars will be available online after the full calculations and calibration are performed.

I will also investigate how does the $\alpha$-enhancement affect Li abundance in stars. In a recently work Guiglion et al. (2016) show that Galactic thin disk (low [$\alpha$/Fe] ) and thick disk (high [$\alpha$/Fe] ) have distinct Li abundances. I will ad-dress this problem from a theoretical point of view in a following work, aimed at answering the following questions: i) Are different [$\alpha$/Fe] responsible for the dispersion of the Spite Plateau ? ii) Is Li a reliable reference for chemical tagging ? iii) Is Li a reliable reference for age tagging ?

I also plan to apply our model of pre-main sequence evolution to Population I stars, aiming to investigate Lithium abundance pattern in young open clusters and to calibrate the pristine Lithium abundance as discussed in Chap. 8. These models could be compared with the large set of pre-main sequence data that are being provided by The Gaia-ESO Survey, a large public spectroscopic survey systematically covering all the major components of the Milky Way. These data include homogeneous determination of Li abundances and accretion rates (Lanzafame et al., 2015) near very young star clusters. These new data will allow us to better understand the internal mixing processes and better describe Lithium evolution in young stellar objects at solar metallicity

In addition, I am going to build a fine grid of Lithium evolution models with PARSEC , that will provide Lithium abundances as a function of stellar age, mass, and metallicity. This grid will allow us to distinguish stars with anomalous Li abundance from the normal ones. Apart from its application to the bizarre Li-rich giants which can not be explained by standard stellar models (e.g. Brown et al., 1989; Ruchti et al., 2010; Reddy, Lambert & Prieto, 2006; Kirby et al., 2012, 2016), this grid will also be useful to search for planet-hosting stars since they show lower Li abundance than their analogues without planet(s) (e.g., Israelian et al., 2009; Gonzalez, 2014). Since most Li measurements at solar metallicity are obtained in field stars, the uncertainty of their distance makes the model difficult to be calibrated. Fortunately, the on-going *Gaia* mission will offer accurate distance information for these stars, so that stellar ages will be routinely derived with PARSEC isochrones. With the aforementioned new constraints from *Gaia*, the Li evolution model can be calibrated with both main sequence observations (Lambert & Reddy, 2004) and Galactic chemical evolution models (Romano et al., 2001).

# Acknowledgments

It is my great pleasure to have my Ph.D. study here in SISSA working with Prof. Alessandro Bressan. I am indebted to him for the patient and meticulous mentoring and for leading me into the details of stellar calculation. With his help I realized how wonderful it is to connect theory to observation. He has always encouraged me to have my own new ideas and to explore the unknown, and he is the one who helped me to find the way with his insight on science.

I am also grateful for the wisdom and guidance of my co-advisors. Dr. Paolo Molaro inspired me on the lithium research and enlightened me about the history of astronomy. I am impressed by his passion and broad knowledge. Dr. Léo Girardi helped me with the details on stellar calculation. Moreover he is the one who saved my life when I had my first night hiking on Teide volcano, if without him always staying behind me I would have got lost in the wilderness. Prof. Paola Marigo helped me through published journal articles and encouraged me a lot on research.

I am thankful for my dissertation referees, Dr. Angela Bragaglia and Dr. Francesca Primas, who offered me advices on the dissertation manuscript in great detail.

I would like to express my sincere gratitude to Dr. Francesca Primas, who offered me great help both on science and the career suggestion. Every discussion with her has benefited me a great deal.

I also thank the support of my family and friends. Thanks to my parents for always being supportive and accepting. Thanks to Jingjing, Lizhi, Yang, Marco, Isabella, Alessandro, Elio, Peppe, Ye, Xiaochuan, Xiao, Jing, Qingtao, and all my friends here who have helped me have such a joyful life. Thanks to the family of my landlady Maria for the warm hosting during the past four years, she is my teacher for Italian cuisine, Triestino culture, gardening, and is a life-long friend. Thank my pet friend Ares and Penny for always accompanying me. I am especially indebted to my boyfriend Zhiyu for his enormous support, thank him for sharing life and scientific interests with me.

Thank the fior di latte of gelateria Zampolli in town for always releasing me when I am in low spirits.

Thank this lovely city Trieste. I am really glad to have been a “Triestina” for the past four years.

# Publications, Awards, and Invited talks

### Refereed papers

1. Xiaoting Fu, Alessandro Bressan, Paolo Molaro, Paola Marigo,
   *"Lithium evolution in metal-poor stars: from pre-main sequence to the Spite plateau"*,
   Monthly Notices of the Royal Astronomical Society, 452, 3256-3265 (2015) Fu et al. (2015).

2. Xiaoting Fu, Alessandro Bressan, Paola Marigo, Léo Girardi, Josefina Montalbán, Yang Chen, Ambra Nanni.
   *"New `PARSEC` database of $\alpha$ enhanced stellar evolutionary tracks and isochrones for Gaia, I. Calibration with 47Tuc (NGC104) and the improvement on RGB bump"*,
   Monthly Notices of the Royal Astronomical Society, submitted. Fu et al. (2016).

### Conference proceedings

1. Xiaoting Fu, Alessandro Bressan, Paola Marigo, Léo Girardi, Josefina Montalbán, Yang Chen, Ambra Nanni, Antonio Lanza,
   *"New `PARSEC` database of $\alpha$ enhanced stellar evolutionary tracks and isochrones for Gaia"*,
   IAUS, 317, 300F, 2016.

2. Xiaoting Fu, Alessandro Bressan, Paolo Molaro, Paola Marigo,
   *"Lithium evolution from Pre-Main Sequence to the Spite plateau: an environmental solution to the cosmological lithium problem"*,
   IAUGA, 2254811F, 2015.

**Awards**

1. The best runner-up poster prize
   IAUS 298
   May 2013

2. Bronze poster prize
   ESO RUSPUTIN conference
   October 2014

**Invited talks**

1. *"New* `PARSEC` *database of alpha enhanced stellar evolutionary tracks and isochrones for Gaia"*,
   **IAU GA FM7**, Hawaii, USA, August, 2015

2. *"Lithium evolution from Pre-Main Sequence to the Spite plateau: an environmental solution to the cosmological lithium problem"*,
   **ESO**, Garching, Germany, November, 2015
   **KIAA**, Beijing, China, December, 2015

3. *"Lithium problems: on the main sequence and the red giant branch"*,
   **University of Science and Technology of China**, Hefei, China, December, 2015
   **Shanghai Astronomical Observatory**, Shanghai, China, December, 2014
   **Nanjing University**, Nanjing, China, December, 2014

**Selected press releases**

1. *"Top ten stories of 2015"*,
   Astronomy Now

2. *"Lost lithium destroyed by ancient stars"* ,
   Royal Astronomical society

3. *"Cosmological 'Lost' Lithium: an Environmental Solution"* ,
   SISSA